\documentstyle[11pt]{article} 

\begin{document}  

\begin{titlepage} 
\begin{flushright} 
June 20, 1996 
\end{flushright} 
 
\vskip 10pt
{\Large\bf 
\centerline {Relativistic Navigation: A Theoretical Foundation}}  
\vskip 20pt
\begin{center} 
Slava G. Turyshev\footnote{ On leave from Bogolyubov Institute for 
Theoretical Microphysics, Moscow State University,

\hskip 8pt  Moscow, 119899 Russia.
Electronic address: sgt@zeus.jpl.nasa.gov }  

\end{center}

\centerline{\sl Jet Propulsion Laboratory MS 301-230,}
\centerline{\sl California Institute of Technology}
\centerline{\sl 4800 Oak Grove Drive - Pasadena, 
CA 91109 - USA}

\begin{abstract}
{\small
We present a  theoretical foundation for    
relativistic astronomical measurements in  curved space-time.
In particular, we discuss a new iterative approach for describing the 
dynamics 
of  an isolated astronomical {\small N}-body system in  metric theories of
gravity. 
To do this, we generalize the  Fock-Chandrasekhar method of the weak-
field and 
slow-motion approximation ({\small WFSMA}) and develop a theory of
relativistic
reference frames  ({\small RF}s)  for a gravitationally bounded many-
extended-body 
problem. In any  proper {\small  RF}   constructed in the 
immediate vicinity of an arbitrary   body,   the {\small N}-body solutions
of the 
gravitational field equations  are formally presented as a sum of the
Riemann-flat 
inertial space-time, the gravitational field generated by the body itself,
the
unperturbed solutions for each body in the system  transformed  to the
coordinates 
of this proper {\small RF},  and  the gravitational 
interaction term.    
We develop the basic concept  of a general   {\small WFSMA} theory of the 
celestial {\small RF}s applicable to a wide class of  metric theories 
of gravity and  arbitrary model  of matter distribution.

We apply the proposed method to general relativity. 
Celestial bodies are described using a perfect fluid model; as such, 
they possess any number of internal mass and current multipole moments 
which explicitly characterize their internal structure.  The obtained  
relativistic corrections to the geodetic equations of motion arise 
because of a coupling of the bodies multiple moments to the surrounding 
gravitational field. The resulting relativistic transformations between 
the different {\small RF}s extend the Poincar\'e group to the motion of 
deformable self-gravitating bodies. 
Within the present accuracy of  astronomical  measurements we discuss
the 
properties of the 
Fermi-normal-like  proper {\small RF}   which is  defined in the 
immediate vicinity of the extended compact bodies.
We further generalize the proposed   approximation method and   include 
two Eddington parameters $(\gamma,\beta)$.  This generalized approach 
was used to derive the relativistic equations of satellite motion in 
the vicinity of the  extended bodies. 
Anticipating improvements in radio and
laser tracking technologies over the next few decades, we apply 
this method to spacecraft orbit determination. 
We emphasize the number of feasible  
relativistic gravity tests that may be performed within 
the context of the parameterized  {\small WFSMA}.
 Based on the planeto-centric 
equations of motion of a spacecraft around 
the planet, we suggested  a new null test of the Strong Equivalence 
Principle ({\small SEP}). 
The  experiment to measure the corresponding {\small SEP} violation
effect 
could be performed with the future {\it Mercury Orbiter} mission. 
We discuss other relativistic  effects including
the perihelion advance, redshift and geodetic precession of the orbiter's
orbital 
plane   
about Mercury, as well as the  possible future implementation of the
proposed 
formalism in software codes developed for solar-system orbit
determination.  
All the important calculations are completely documented  and 
the  references  contain  an extensive list of  cited literature.}

\end{abstract}

\end{titlepage}
 \setcounter{page}{2}
 {\Large \bf Contents}
\vskip 15pt
{\bf 0 \hskip 2mm Notations and Definitions\hfill 5}
 
{\bf 1 \hskip 2mm Introduction and Overview \hfill 5}
 
\hskip 6mm  1.1 \hskip 2mm The Motivation and the Structure of the
Report \dotfill \hskip 9mm 5

\hskip 6mm  1.2 \hskip 2mm The  Problem of  Relativistic 
Astronomical Measurements \dotfill \hskip 9mm 11    

\hskip 6mm  1.3 \hskip 2mm The Qualitative Description of 
the Astronomical Systems of Interest \dotfill \hskip 7mm 22  

\hskip 6mm  1.4 \hskip 2mm Different Methods of  
Constructing   the Proper RF \dotfill \hskip 7mm 24 

{\bf 2 \hskip 2mm Parametrized Post-Newtonian Metric Gravity \hfill 29}

\hskip 6mm  2.1 \hskip 2mm An Isolated One-Body Problem \dotfill \hskip
7mm 29  

\hskip 6mm  2.2 \hskip 2mm The Limitations of the 
Standard   PPN Formalism \dotfill \hskip 7mm 31

\hskip 14mm  2.2.1 \hskip 2mm The Simplified   Lagrangian Function 
of an Isolated  N-Body System \dotfill \hskip 7mm 33

\hskip 14mm  2.2.2 \hskip 2mm  The Simplified Barycentric 
Equations of Motion \dotfill \hskip 7mm 34

{\bf 3 \hskip 2mm WFSMA for an Isolated Astronomical    N-Body System
\hfill 36}

\hskip 6mm  3.1 \hskip 2mm The General Form of the N-Body 
Solution \dotfill \hskip 7mm 36

\hskip 6mm  3.2 \hskip 2mm The Post-Newtonian  KLQ  
Parameterization \dotfill \hskip 7mm 40

\hskip 14mm  3.2.1 \hskip 2mm  The Properties of the   
Coordinates Transformations in the   WFSMA \dotfill \hskip 7mm 41  

\hskip 14mm  3.2.2 \hskip 2mm  The Inverse
Transformations \dotfill \hskip 7mm 42  

\hskip 14mm  3.2.3 \hskip 2mm  The Coordinate Transformations Between
the Two Proper RFs\dotfill \hskip 7mm 44

\hskip 14mm  3.2.4 \hskip 2mm  The Notes on an Arbitrary Rotation of the
Spatial Axes \dotfill \hskip 7mm 45 

\hskip 6mm  3.3 \hskip 2mm The  Definition of the Proper 
RF \dotfill \hskip 7mm 47  

 {\bf 4 \hskip 2mm General   Relativity: 1. Solutions for the  Field 
Equations \hfill 51} 

\hskip 6mm  4.1 \hskip 2mm The Solution for the Interaction Term
\dotfill \hskip 7mm 51 

\hskip 6mm  4.2 \hskip 2mm The Solution of the Field Equations 
in the Proper  RF \dotfill \hskip 7mm 57

\hskip 6mm  4.3 \hskip 2mm Decomposition of the 
Fields in the Proper  RF \dotfill \hskip 7mm 59

{\bf 5 \hskip 2mm General  Relativity: 2. Transformations to the Proper RF 
\hfill 61} 

\hskip 6mm  5.1 \hskip 2mm Finding the Functions $K_A$ and
$Q^\alpha_A$
 \dotfill \hskip 7mm 62

\hskip 14mm  5.1.1 \hskip 2mm  Equations for the Functions $K_A$ and
$Q^\alpha_A$ \dotfill \hskip 7mm 62

\hskip 14mm  5.1.2 \hskip 2mm  The Solution for the Function $K_A$
\dotfill \hskip 7mm 63  

\hskip 14mm  5.1.3 \hskip 2mm  The Solution for the  Function
$Q^\alpha_A$ \dotfill \hskip 7mm 64

\hskip 6mm  5.2 \hskip 2mm Finding  the Function $L_A$ \dotfill \hskip
7mm 66

\hskip 14mm  5.2.1 \hskip 2mm  Equations for the Function $L_A$ \dotfill 
\hskip 7mm 67   

\hskip 14mm  5.2.2 \hskip 2mm  The Solution for the Function $L_A$
\dotfill \hskip 7mm 68 

\hskip 6mm  5.3 \hskip 2mm Equations of Motion for the Massive  Bodies
\dotfill \hskip 7mm 70 

\hskip 6mm  5.4 \hskip 2mm The Proper RF of the Small Self-Gravitating
Body \dotfill \hskip 7mm 72 

\hskip 6mm  5.5 \hskip 2mm The Fermi-Normal-Like Coordinates \dotfill 
\hskip 7mm 75  

{\bf 6 \hskip 2mm General Relativity: 3.  The Proper  RF  for the Extended 
Body \hfill 78}

\hskip 6mm  6.1 \hskip 2mm The Extended-Body Generalization \dotfill
\hskip 7mm 79 

\hskip 6mm  6.2 \hskip 2mm Conservation Laws in the Proper RF \dotfill 
\hskip 7mm 86

\hskip 6mm  6.3 \hskip 2mm The Solution for the Function $\delta
w^\alpha_{A_0}$ \dotfill \hskip 7mm 93

{\bf 7 \hskip 2mm Parameterized  proper   RF \hfill 96}

\hskip 6mm  7.1 \hskip 2mm Parameterized coordinate transformations
\dotfill \hskip 7mm 96 

\hskip 6mm  7.2 \hskip 2mm Equations of the Spacecraft Motion \dotfill 
\hskip 7mm 100

\hskip 6mm  7.3 \hskip 2mm Gravitational Experiment for Post 2000
Missions \dotfill \hskip 5mm 104

\hskip 14mm  7.3.1 \hskip 2mm  Mercury's Perihelion Advance \dotfill
\hskip 5mm 106 

\hskip 14mm  7.3.2 \hskip 2mm  The Redshift Experiment \dotfill \hskip
5mm 107

\hskip 14mm  7.3.3 \hskip 2mm  The SEP violation effect \dotfill \hskip
5mm 107 

\hskip 14mm  7.3.4 \hskip 2mm  The Precession Phenomena \dotfill \hskip
5mm 109

{\bf 8 \hskip 2mm Discussion: Relativistic Astronomical RFs.  \hfill 109}

\hskip 6mm  8.1 \hskip 2mm The Geocentric Proper RF \dotfill \hskip
7mm 111

\hskip 6mm  8.2 \hskip 2mm The Satellite Proper RF \dotfill \hskip 7mm
113

\hskip 6mm  8.3 \hskip 2mm The Topocentric Proper RF \dotfill \hskip
5mm 115

\hskip 6mm  8.4 \hskip 2mm Discussion \dotfill \hskip 5mm 117

{\bf Appendix A: Generalized Gravitational Potentials  \hfill 119} 

{\bf Appendix B:  Power Expansions of the General Geometric  
Quantities \hfill 120} 

\hskip 6mm B.1. Expansion for the Metric Tensor $g_{mn}$ \dotfill
\hskip 5mm 117 

\hskip 6mm  B.2. Expansion for the $ det \hskip 2pt \big[g_{mn}\big]$ 
and $g^{mn}$ \dotfill \hskip 5mm 121 

\hskip 6mm B.3. Expansion for the $\hat{g}^{mn}=\sqrt{-g} g^{mn}$ \dotfill 
\hskip 5mm 121  

\hskip 6mm  B.4. Expansion for the Gauge Conditions \dotfill 
\hskip 5mm 122  

\hskip 6mm  B.5. Expansion of the Christoffel Symbols 
\dotfill \hskip 5mm 123    

\hskip 6mm  B.6. Expansion  for  the Ricci Tensor $R_{mn}$ \dotfill \hskip
5mm 123

\hskip 6mm  B.7. Expansion for an Arbitrary Energy-Momentum 
Tensor  $T^{mn}$ \dotfill \hskip 5mm 124  

{\bf    Appendix C:  Transformation Laws of the Coordinate Base
Vectors\hfill 125}  

\hskip 6mm  C.1. Direct Transformation  of the Coordinate Base Vectors 
\dotfill \hskip 5mm 125

\hskip 6mm  C.2.  Transformation of the 
Background Metric $\gamma_{mn}$ \dotfill \hskip 5mm 126

\hskip 6mm  C.3. Inverse Transformation of the Coordinate 
Base Vectors \dotfill \hskip 5mm 128 

\hskip 6mm  C.4. Mutual Transformation Between the Two 
Quasi-Inertial RFs \dotfill \hskip 5mm 128 

{\bf    Appendix D:  Transformations of Some Physical 
Quantities and Solutions\hfill 129} 

\hskip 6mm  D.1. Transformation  of the Gauge 
Conditions \dotfill \hskip 5mm 129 

\hskip 6mm  D.2. Transformation  of
the    Ricci Tensor $R_{mn}$ \dotfill \hskip 5mm 130

\hskip 6mm  D.3. Transformation Law for an Arbitrary 
Energy-Momentum Tensor $T^{mn}$ \dotfill \hskip 5mm 131  

\hskip 6mm  D.4. Transformation of the Unperturbed Solutions
 $h^{(0)}_{mn}$ \dotfill \hskip 5mm 132    

\hskip 6mm  D.5. Transformation Rules for the Interaction 
Term $ h^{int}_{mn}$ \dotfill \hskip 5mm 134

\hskip 6mm  D.6. Transformation for 
the  Energy-Momentum Tensor of a Perfect Fluid \dotfill \hskip 5mm 134 

{\bf    Appendix E:  Transformations of the Gravitational Potentials 
\hfill 135}

{\bf    Appendix F:  Christoffel Symbols in the  Proper   {\small RF}$_A$
\hfill 140} 

\hskip 6mm    F.1. Christoffel Symbols with Respect
to the Background Metric $\gamma^A_{mn}$ \dotfill \hskip 5mm 140 

\hskip 6mm  F.2. Christoffel Symbols with Respect to
the  Riemann Metric $g^A_{mn}$ \dotfill \hskip 5mm 140  

{\bf    Appendix G:   The Component $g^A_{00}$ and the Riemann Tensor 
\hfill 142} 

\hskip 6mm  G.1. The Form of the Component $\gamma^A_{00}$  \dotfill
\hskip 5mm 142  

\hskip 6mm  G.2. Lemma \dotfill \hskip 5mm 143

\hskip 6mm  G.3. The Form of the `Inertial Friction' Term 
\dotfill \hskip 5mm 144 

\hskip 6mm  G.4. The Form of the Interaction Term 
\dotfill \hskip 5mm 145   

\hskip 6mm  G.5. The Form of the  Riemann Tensor in  the Proper 
{\small RF}$_A$ \dotfill \hskip 5mm  144   

{\bf Appendix H: Some Important Identities \hfill 147} 

{\bf Appendix I:  Astrophysical Parameters Used in the Report \hfill 150} 

{\bf   Acknowledgments \hfill 151}

{\bf    References \hfill 151} 
 
\vskip30cm
\vskip 11pt
\noindent {\Large \bf 0 \hskip 10pt Notations and Definitions}
\vskip 11pt\noindent 
In this paper the notations are the same as in 
(Landau \& Lifshitz, 1988). In particular, the small 
latin letters $n,m,k ...$ run
from 0 to 3 and greek letters $\alpha, \beta, \gamma,...$ 
run from 1 to 3; the italic capitals  {\small  \it A,B,C}  number the bodies
and run from 1 to {\small N}; the comma denotes a standard partial 
derivative and semicolon denotes a covariant derivative; repeated
indices imply an Einstein rule of summation;
round brackets surrounding indices denote  
symmetrization and square brackets denote 
anti-symmetrization. 
The geometrical units $c=$ {\small G} $=1$
are used throughout the paper where {\small G} is the universal 
gravitational constant; $c$ is the 
speed of light. We designate  $\epsilon_{\alpha\beta\delta}$ as 
the fully anti-symmetric Levi-Civita symbol $(\epsilon_{123}=1)$; 
the metric convention is accepted to be {\small $(+ - - -)$};
$\gamma_{mn} = diag${\small$ (1, -1, -1, -1)$} is the Minkowski 
metric in Cartesian coordinates of the  inertial {\small RF};
 $\gamma^A_{mn}(z^p_A)$ the Minkowski metric in the coordinates  
$(z^p_A)$ of the  
{\small RF}$_A$ which is constructed in the immediate vicinity of an
arbitrary 
  body {\small (A)};
$g_{mn}$ denotes the effective Riemann metric of the curved space-time;
and 
$g = det(g_{mn})$. 
To enable one to deal conveniently with sequences of many 
spatial indices, we shall use an abbreviated notation for `multi-indices' 
where an upper-case letter in curly brackets denotes a multi-index, 
while the corresponding lower-case  letter denotes its number
of indices, for example: {\small $\{P\}:=\mu_1\mu_2 \ldots \mu_p$, 
$S_{\{J\}}:= S_{\mu_1\mu_2\ldots \mu_j}$}. When needed, we also use 
{\small $\{L-1\}$}:  $\mu_1\mu_2\ldots \mu_{l-1}$, so that the tensor 
{\small $T_{a\{L-1\}}=T_{a\mu_1\mu_2\ldots \mu_{l-1}}$} has $l$
indices. We also  
denote $z^{\{L\}}= z^{\mu_1}z^{\mu_2}\ldots z^{\mu_l}$ and 
$\partial^{\{L\}}/\partial z^{\{L\}}=\partial^l/\partial z^{\mu_1}\partial
z^{\mu_2}\ldots
\partial z^{\mu_l}$. 
The explicit expression for the symmetric and trace-free {\small (STF)}
part of 
the 
tensor $T_{\{P\}}$ is given in 
(Thorne, 1980; Blanchet and Damour, 1986, 1989). For any
positive integer $l$ we shall denote   $l!=l(l-1)\cdot\ldots \cdot2\cdot1;
 l!!=l(l-2)\cdot\ldots\cdot( 2  \hskip 2pt {\rm or}  \hskip 2pt 1)$ as usual. 
A dot over any function means a differentiation with respect to time.

\section{Introduction and Overview}

\subsection{The Motivation and the Structure of the Report.}

The principal objective of spacecraft navigation 
is to determine the   present and future trajectory of a craft. 
This is usually done by measuring the 
spacecraft's coordinates  and then by correcting (fitting and adjusting)
the 
predicted spacecraft  trajectory using those measurements. 
There are three different types of measurements that are used in
spacecraft 
navigation:
radiometric (range and Doppler), very-long baseline interferometry
{\small (VLBI)}
and optical (Standish, 1995). As well as serving  navigation' needs, 
 high precision Doppler, and laser  and radio range  measurements of the
velocity 
of 
and the distance to  celestial bodies  and  spacecraft are presently  
the best ways to collect  
important information about relativistic gravity within the solar system.
Combined with   the technique of ground- and space-based  
 {\small  VLBI}, these  methods provide us with a unique
opportunity to explore the physical phenomena in  our 
universe with  very high precision. Most remarkable is the increase in   
accuracy of the
 modern {\small VLBI} observations   
especially in applications to problems of modern geodesy 
 (Soffel {\it et al.}, 1991; Herring, 1995).  Thus  the delay residuals are
presently of the order of 30-50 picoseconds (ps), 
which corresponds to an uncertainty in length of $\sim$ 1 cm. 
In the navigation of the interplanetary spacecraft, the short arcs of
spacecraft 
range and Doppler measurements, reduced with Earth orientation
information referred to the 
International Earth Rotation Service {\small (IERS)} celestial system, 
lead to a position determination in the extragalactic {\small RF}
with an accuracy of order $\sim 20$ milliarcseconds (mas).
At the same time,  the {\small VLBI} observations of the spacecraft with 
respect to an extragalactic radio-source enable one to measure directly
one 
component of the spacecraft position in this extragalactic {\small RF} to
an 
accuracy 
of about $\sim 5$ mas (Border {\it et al.}, 1982; Folkner {\it et al.}, 1994).
As a result, the  use of such a precise methods 
enables one to study the dynamics of   celestial
bodies and   spacecraft with an unprecedented   accuracy.

In addition  to these line-of-sight methods,
the computer revolution of the 1990's has revived   interest in the
classical 
approach 
for determining the gravity field based on the   spherical harmonics
representation. 
It is now believed that the use of the spherical harmonics to high degree
and 
order, where only high frequency 
noise is present in the raw Doppler residualis, is one of the 
best reduction approaches  becasue  it allows a fully three-dimensional
analysis.
Thus, the gravitational spherical harmonics of  the 
Earth gravity field is currently known up to the 70th degree 
and order   for the solutions based upon the spacecraft tracking data only,  
and to the 360th degree and order  with surface measurements  included
(Rapp {\it et al.}, 1991;  Nerem {\it et al.}, 1995). Let us mention  
that currently there is exists the possibility   
of determining the Venus gravity field to 120th degree and order  
(Konopliv {\it et al.}, 1995).   
It should be noted that the determination of the multipolar structure of
the Newtonian gravitational
field of the Earth and planets with such high resolution and
accuracy  enables one to take into account the relativistic corrections to
 the   gravitational  field of these bodies.
Then,  by using  modern  techniques of   data  reduction, one may generate  
 highly precise solutions which   have applications beyond that of  serving
the geodetic needs
(Hellings, 1986; Herring, 1996). 
For example, these very important results are  widely in  use  as  
the necessary foundation for studies of many modern scientific problems,
such as:

\begin{itemize}

\item[(i).] The problem of developing a more precise definition of the 
masses and the 
multipole structure of the Sun, Earth, other planets,
their satellites, and asteroids (Standish, 1992; Standish, 1994, Schubert
{\it et al.}, 1994;
Konopliv {\it et al.}, 1995). 
\item[(ii).] The establishment of better values for gravitational and other 
astronomical constants, as well as testing of the hypothesis of their
dependence
on time  which is predicted by a number of modern 
theories of gravity (Dirac, 1937;  Anderson {\it et al.}, 1986; Will, 1993). 
 \item[(iii).] The study of the dynamics and the evolution of the solar
system, 
aimed at a  better understanding of its metrological characteristics.
This will  help to solve some cosmogonical 
problems, such as:  determining whether or not there is a 
second asteroid  belt (the Kuiper belt)
behind the Saturnian orbit (Anderson {\it et al.}, 1986), 
giving better numerical estimates of the quantity  of  dark matter in the
solar system
(Braginsky, 1994; Anderson, {\it et al.}, 1995), 
and determining whether or not our Sun has a companion star.

\item[(iv).] Experimental tests of the modern gravitational theories 
in the  {\small WFSMA}  (Damour, 1983; Will, 1993; Lebach, {\it et al.},
1995;
Anderson {\it et al.}, 1996) including the establishment of upper 
limits on the  amplitude and energy density of  
gravitational radiation (Anderson {\it et al.}, 1986). The search for
gravitational 
waves, their detection and  studies of  mechanisms of  wave  
generation, as well as their propagation and interaction with matter. 
These studies  will  increase our knowledge  
of the early age of the universe, its cosmological evolution and 
the behavior of   stellar systems,  as well as further confirming the 
 hypothesis of the   existence of unseen matter in 
the universe (Anderson, {\it et al.}, 1995).  
\end{itemize}

 The modern approach to conducting these different scientific 
studies should be based upon the use of a well established common
relativistic framework
for both  collecting and  interpreting  astronomical observations.
Until recently, this task   had been done  
by taking into account only the post-Newtonian 
corrections to the solar static spherically-symmetric (Schwarzschild)
gravitational field.
The basic relativistic effects, such as   Mercury's perihelion advance, 
gravitational light deflection,  red-shift  and  
 time-delay  (the Shapiro effect) have been calculated with  post-
Newtonian accuracy 
by a number of authors and the corresponding  results are well-known
(Brumberg, 1972;
Misner {\it et al.}, 1973;  Will, 1993).  
It should be noted that during last 10 years  the   
precision  of theoretical predictions of satellite motion 
has increased considerably. This has happened  because some of the 
 leading static-field post-Newtonian  perturbations in the dynamics of 
the  planets, the Moon and   artificial satellites have been included in the 
equations of motion (eq.m.), and time and position
 transformation  (Moyer, 1971; Moyer 1981;  Dickey   {\it et al.}, 1989; 
Huang {\it et al.}, 1990;   Dickey {\it et al.}, 1994; 
Habib {\it et al.}, 1994;  Williams   {\it et al.}, 1996). 
However, due to enormous progress in the accuracy of astronomical 
observations at the present time, we must now   
 take into account the much smaller relativistic effects caused by the 
post-post-Newtonian corrections to the solar gravitational field  as well
as the 
post-Newtonian contributions from the lunar and  planets'  gravity.
Moreover, 
it is also well understood that the 
effects due to the  non-stationary behavior of the solar system
gravitational field as well as
its  deviation from   spherical symmetry should be also considered
(Kopejkin, 1988).
The successful  solution of these  problems requires a detailed 
critical review of  modern observational methods and the
development  of a consistent and physically well founded theory  of
relativistic celestial mechanics
and relativistic {\small RF}s. 
This theory should provide  one with reliable physical grounds for 
theoretical studies of the new 
relativistic gravitational phenomena as  well as meet  the  needs of
practical astronomy.

It has been long considered that such a theory already exists in the 
form of the parameterized post-Newtonian formalism {\small (PPN)}
(Nordevedt, 1968a,b;
Will, 1971;  Will \& Nordtvedt, 1972; Will, 1993). 
However, based on the  present understanding of the problem,
this point of view is not correct. Indeed,   the foundation of the {\small
PPN} formalism is
based upon the existence of an exclusive  set of inertial {\small RF}s. 
Usually, the origin of such a frame  either coincides with the solar system 
barycenter or it may be transformed to one by  the post-Galilean 
 coordinate transformations (Chandrasekhar \& Contopulos, 1967;
Kopejkin, 1988; Will, 1993).
The resultant  barycentric inertial {\small RF} is perfectly suited for 
analyzing both light ray propagation  in the  proximity  of the Sun
and  the motion of the planets around the Sun. However, it does not
address 
some  very practical  needs of modern astronomy, such as providing a
description of a satellite's
motion around the Earth (or other planet), studying properties of the
Earth's rotation, or
collecting and interpreting data  
from  satellite laser ranging {\small (SLR)},  lunar laser ranging {\small
(LLR)}, or 
ground-  or space-based  {\small VLBI}.
These difficulties are caused by the fact that the planet's center of mass, 
in general, does not move along the geodesic line.
The corresponding deviations are very small (Misner {\it et al.}, 1973;
Brumberg, 1972; Will, 1993) and a product of the coupling of the planet's
internal multipole moments   
to the external gravitational field. It is well-known that geodesic motion
in the 
 general theory of relativity, for example, can be is viewed as   free fall. 
Moreover,  in the immediate 
vicinity of the free-falling body one may introduce a {\it local} quasi-
inertial {\small RF}.
In this {\small RF} an external gravitational field should manifest
itself in the form of tidal forces only (Synge, 1960; Bertotti \& Grishchuk,
1990).
However, the {\small PPN} coordinate system, with its origin at the
center of mass of the planet,
does not satisfy  this last condition, and therefore it may not be treated
as a 
quasi-inertial {\small RF} (Kopejkin, 1988). 
However, from the practical point of view of  collecting and interpreting
experimental data, 
one needs to use a set of  {\small RF}s with well defined geometrical and
physical properties. 
Thus it has been shown that  a poor   choice of  coordinate
transformations for 
defining the proper {\small RF}  may lead to unnecessary 
complications in the  equations of motion. These equations may appear to 
contain  non-physical (or fictitious)  forces acting on the 
bodies in the system. Although, these forces are simply  a result of a
`bad' choice of   {\small RF},  their appearance in the equations of motion 
may make the  scientific interpretation of the collected results much
more difficult.  
For example, the term with an amplitude of about one meter in the 
relativistic theory of motion of the moon (Brumberg, 1958; Baierlein,
1967)
 has no real physical meaning when built on the basis of the proper
coordinates. 
The appearance of this term is an artifact of the choice of coordinates
and, therefore, the one-meter 
term is not observable (Soffel {\it et al.}, 1986; Kopejkin, 1988).
This example suggests that a clear understanding of the dynamic 
properties of 
a chosen coordinate {\small RF}  will help  make the separation between
physically
 measurable quantities and coordinate induced ones, and hence will
simplify the 
analysis of the data obtained.

>From this standpoint, the detailed construction of a
relativistic theory of  astronomical {\small RF}s is   greatly needed.
 It is especially important  because at present  almost 
all the astronomical  observations (such as optical, radio,
Doppler, laser etc.)  are   performed and/or  processesed   by  experimental
equipment 
placed on the Earth' surface.  Moreover, there is great demand 
for  a reliable relativistic navigation in outer
space for  near-future space missions such as the space-based
gravitational-wave astronomy.
Let us also note that there are near term plans for launching  
several drag-free satellites with GPS receivers onboard: 
Gravity Probe B {\small (GP-B)} (Bardas   {\it et al.}, 1989), {\it \small
LAGEOS III},
 a satellite test of the equivalence principle  {\small (STEP)}, and 
the {\it Mercury Orbiter} mission   which  has been
proposed by the European Space Agency as a cornerstone mission under 
the Horizon 2000 Plus program  (Anderson, Turyshev {\it et al.}, 1996).
There  exist   plans to include  the post-post-Newtonian contributions 
to the light propagation effects coming from the 
solar gravitational field and the post-Newtonian gravitational 
perturbations by the planets of the solar system (Klioner \& Kopejkin,
1992). 
In particular, one of the most promising 
projects is deploying in the Earth's orbit a precision optical
interferometer  {\small (POINTS)}.
This satellite will be designed to be able to measure the 
arcs between the pairs of stars separated on the sky by the right angle
with 
the  anticipated accuracy  on the order of a few  microarcseconds
($\mu$as) (Chandler \& Reasenberg, 1990).      
  These plans encourage the development  of an  orbit determination
algorithms which would 
enable one to process the data   with  the required relativistic   accuracy.
 This  alone   will require  substantial  work to be done in development of
a number
 of theoretical and practical questions, such as: 
\begin{itemize}
\item[(i).] The construction of a dynamic  inertial 
 {\small RF} and a more precise definition of the orbital 
elements of the Sun, Earth, moon,  planets and their satellites 
(Standish {\it et al.}, 1992, Chandler {\it et al.}, 1994; 
Dickey {\it et al.}, 1994;  Williams {\it et al.}, 1996; Standish, 1995). 
\item[(ii).] The construction of a kinematic 
inertial    {\small RF}, based on the observations of 
stars and quasars from   spaceborne astronomical observatories 
(Fukushima, 1991a; Standish {\it et al.}, 1992). 
\item[(iii).] The construction of a precise 
ephemeris for the motion of   bodies in the solar system
to  support   reliable navigation in the solar system  (Denisov {\it et al.},
1989;
Standish {\it et al.}, 1995; Standish, 1995). 
The construction of  precise radio-star catalogs for the spacecraft 
astroorientation and navigation  in outer space beyond the solar system.
\item[(iv).] The  comparison of   dynamic  and 
kinematic  inertial  {\small RF}s,  based on the observations of 
 spacecraft on the  background  of quasars, pulsars and  
radio-stars, as well as the verification of the zero-points of 
the coordinates in the inertial  {\small RF} (Jacobs {\it et al.}, 1993; 
Folkner {\it et al.}, 1994; Fukushima, 1995).
\end{itemize}

Therefore the motivation for this research is quite natural:
 In order to propose the necessary recommendations for corrections to
existing software codes, we will   re-examine the basic concepts of  
high-precision navigation in the solar system.  The principal goal of this
report is 
to provide one with a solid theoretical foundation for the relativistic 
astronomical measurements in the curved space-time. To reach this goal
we, 
by   using the 
methods of the {\small  WFSMA}, 
will   develop a new approach to the   relativistic  treatment of 
the satellite orbit determination problem. This approach will be based  
 upon a new theory of   coordinate  transformations  ({\it i.e.} the 
theory of the relativistic {\small RF}s)
and the measurement  models in the relativistic celestial mechanics.
The outline of the present report is  as follows:  

The next chapter in this Section contains a
brief historical introduction to the problem of motion of  
 {\small N}  weakly interacting self-gravitating extended bodies. 
To specify our theoretical studies we will present a qualitative
description of 
the astronomical {\small N}-body systems of interest. 
In order to provide a solid motivation for this research we  
will   analyze  the different methods used  to approach 
this problem and will present their advantages and the 
encountered difficulties. 

In the Section {\small 2} we discuss  the conventional {\small PPN}
barycentric 
approach which is based on the 
 solution to the gravitational one-body problem. Recognizing that  the
generalization 
of the obtained results  into a general case of motion 
of an arbitrary {\small N}-body system is not straightforward, we 
analyze the conditions  
necessary  to derive the  restricted
solution for the motion of the general {\small N}-body problem.
We also discuss   ways to obtain the complete 
multipolar solution to the problem in the general  case.

Section {\small 3} is devoted to a general description of the 
new method proposed   to overcome the above mentioned  problems. 
We discuss a new iterative approach to describing the  dynamics 
of  an isolated astronomical {\small N}-extended-body system in the 
metric theories of gravity.  
The {\small N}-body solution  of the gravitational
field equations    in the proper {\small RF}$_A$ originated
 in an arbitrary  body {\small (A)} is formally presented as a sum of the
four following terms: 
 (i)   $\gamma^A_{kl}$, which is the Riemann-flat inertial space-time:
$R^A_{mnst}(\gamma^A_{kl})=0$;
 (ii)    $h^{(0)A}_{mn}$, which is the gravitational field generated by the
body  {\small (A)} itself;
 (iii)  $h^{(0)B}_{mn}$ is the  perturbations caused by other bodies  in the
system ({\small $B\not=A$});
and, finally, (iv)  the gravitational interaction term  $h^{\it int}_{mn}$. 
 This method is presented in its most general  form and, 
hence, it is valid for a number of metric theories of gravity. 
We  discuss the general properties of the  post-Newtonian 
non-rotating coordinate  transformations and   present the
 straight, inverse and mutual coordinate transformations.
As a possible way of generalizing   the results obtained, 
we discuss the  use of the rotational coordinate transformations.
In addition, we discuss the necessary conditions   for
constructing a   proper {\small RF} with the well defined  dynamical
properties. Physically, these  conditions should 
provide one with a force  acting on the body in its proper {\small RF} such  
that the  body will be in the state of equilibrium. 
Mathematically, these conditions
required that the total dipole moment of the system of 
the fields produced  by matter, the field of inertia and the gravitational 
field taken jointly will vanish for all times. 

In Section {\small 4} we apply the proposed  formalism to the case  
 of   general relativity.  The celestial bodies are assumed to consist of
 a perfect fluid and possess any number of the   internal mass and 
current multiple moments which characterize the internal structure of
such
bodies. We present the physical and 
mathematical definitions of the proper {\small RF} in the {\small WFSMA}.
We find the explicit solution for the interaction term. This enables us to 
construct all the necessary expressions  for the metric tensor in both
the barycentric inertial and   arbitrarily parameterized 
proper quasi-inertial {\small RF}s.

In Section {\small 5} we  present the  general solution for the 
{\it global} and {\it local} problems, as well as showing the general
solution 
for the functions of the coordinate transformation in the case of  
bodies with a  weak external  gravitational field. 
In particular, within  the present accuracy of  radio  
measurements, we discuss the generalized
Fermi-normal-like  proper  {\small RF},  which is  defined in the 
immediate vicinity of such   extended  bodies.

 In Section {\small 6} we generalized the results obtained on the case 
of a system of {\small N} arbitrarily shaped and deformable extended
bodies. 
To do this, we study the existence of 
the conservation laws in the proper {\small RF}. It turns out that the 
existence  of these laws in {\small WFSMA} may be shown explicitly  in
the case 
of  well separated celestial bodies. This  allows us to evaluate the 
surface integrals on the boundaries of the domains occupied by the
celestial bodies
and present the explicit coordinate transformations between the 
different {\small RF}s in the {\small WFSMA} of the general relativity.
These results  are the extension of the post-Galilean transformations
obtained  by Chandrasekhar and Contopulos (1967) on the case of a system
of 
interacting celestial extended bodies. We discuss the properties of the 
corresponding quasi-group of motion and its application to the 
study of the dynamics of an arbitrary {\small N}-body gravitational
problem.

Section {\small 7} is devoted to the future relativity missions in the 
solar system. In order to provide the framework to study the relativistic
gravity 
for a number gravitational theories, our
previous derivations will be  generalized on the case of the tensor-scalar
theories. 
As a result, we   include in the analysis the two Eddington parameters
$(\gamma, \beta)$, which  
allows  us to develop a  parametric theory of the astronomical
{\small RF}. By analyzing the equations of motion  in the two-parameter
Fermi-normal-like 
{\small RF}, we  have obtained an interesting result: that 
although some terms in the planeto-centric eq.m. of the spacecraft around 
the planet  are zero for the case of general relativity, 
they may produce an observable effect  
in scalar-tensor theories. This  allows us to propose a new null 
test of the    {\small SEP}.
Also in  this Section we discuss the other 
relativistic gravitational experiments possible with the future 
{\it Mercury Orbiter}   mission which  has been
proposed by the European Space Agency as a cornerstone mission under 
the Horizon 2000 Plus program. 
The motivation for this research is to determine what  
scientific information may be obtained during this mission, how accurate
these 
measurements can be, and  what will be the significance of 
the  knowledge obtained. 
We  present there both quantitative  and
 qualitative analyses of   measurable effects  such as  Mercury's
perihelion advance,
the redshift experiment, and the precession phenomena of the Hermean
orbital plane.

In Section {\small 8} we present the hierarchy of the celestial 
 {\small RF}s, including the four frames that 
are widely in use for the practical needs of the modern relativistic 
astronomy.
Thus, in a compact and an explicit form we show the 
coordinate transformations between the barycentric and  the  geocentric
{\small RF}s,   between the geocentric and the satellite {\small RF}s, and 
between the geocentric and the  topocentric {\small RF}s.
This presentation contains the two Eddington parameters
$(\gamma,\beta)$,
which makes the obtained results valid for a wide class of metric
theories of gravity.
In  the  discussion   we   present a number of possible areas for the
immediate 
practical application of the   theory of astronomical {\small RF}s 
developed in this report.  We present our conclusions and 
recommendations for  future research on  relativistic gravity in the solar
system
and beyond.  

In order to   avoid  cumbersome calculations and to simplify the
presentation of 
the main results in the text, some expressions and 
intermediate relations  will be presented in Appendices. 
In Appendix A we present the generalized gravitational potentials.
 Appendix B is devoted to a   discussion of  the structure of the 
post-Newtonian power expansion of the general geometrical quantities 
such as the metric tensor $g_{mn}$, the Christoffel symbols and the
Riemann tensor $R_{mnkl}$  
in coordinates of an arbitrary {\small RF}  
with respect to small parameters. Appendix C  
contains the general theory of the relativistic coordinate 
transformations. We  discuss there the transformation of the base vectors 
for a different coordinate transition. In Appendix D we  present the
features of the
 transformations of a different  equations and  quantities, such as 
 the covariant gauge conditions, the Ricci tensor, 
the gravitational field solutions and the energy-momentum tensor.
The  transformation rules for the generalized gravitational potentials 
under the post-Newtonian coordinate transformation are presented 
in   Appendix E. The  Christoffel symbols in the 
proper {\small RF} are calculated in   Appendix F.
The calculation of  the form of the inertial part of the metric tensor 
in the proper {\small RF}, the form of the interaction term as well as 
the components of the Riemann tensor in this frame are 
presented in  Appendix G. 
In   Appendix H we present  some useful identities which 
are used in   Section {\small 6}   to 
study the existence of the conservation laws 
in an arbitrary proper {\small RF}. And, finally, in   Appendix I
we have presented the  astrophysical parameters used for estimations 
of the magnitudes of the gravitational effects in   Section {\small 7}.

\subsection{The  Problem of  Relativistic Astronomical Measurements.} 

Classical Newtonian mechanics
is based   upon the principles of   Euclidian geometry. 
The   physical experiments, within the accuracy  available at that time, 
had confirmed  the  two  basic postulates of this geometry: that time is 
absolute and homogeneous, and that space is also absolute and 
not only  homogeneous, but  also  isentropic. These properties of time and 
space were discovered   because, for the then known  
physical forces\footnote{There were only two known natural forces  at
this time: gravity and elasticity.
The first one was described by Newton's gravitational law and the second 
by Hooke's law.}, 
the corresponding eq.m. of the Newton's mechanics  
preserved their form under the Galilean group of motion. These properties 
may be  written for two different {\small RF}s moving relative to each
other 
with the constant  speed ${\vec v}$ as:
{}
$$t'=t+a, \qquad {\vec r}\hskip 2pt'={\vec r} -{\vec b}-{\vec v} t,
\eqno(1.1) $$
\noindent where the  parameters $a$ and ${\vec b}$ 
are the constant time-shift and the  displacement of the  origin of the
coordinate 
system respectively. This  form-invariancy   
suggested that, independent of the state of  motion of these 
{\small RF}s (they may be either at rest or uniformly moving  alone a
straight line
relative to each other), 
 all the mechanical phenomena will behave exactly the same way in any 
such  
{\small RF}. This principle has become known as the  principle of
relativity  
(Poincar\'e, 1904).  Note  that  the  transformations (1.1) are given  
in   Cartesian coordinates.
One may choose another  coordinate system {\small (CS)} in the same
{\small RF} 
without changing its state of motion (say ${\vec x}$)
by simply rotating of the  coordinate axes:  
$r^\alpha={\cal R}^\alpha_\beta x^\beta$, where  ${\cal R}^\alpha_\beta$ 
is a constant  orthogonal rotation matrix. 

Thus   Newton's mechanics had introduced into physics a notion of both an
absolute  
distance between two points in three-dimensional 
space, and of absolute time. In the other words,  he asserted that  time
and coordinates  
are directly measurable quantities.
Because of this, the theory of   gravitational measurements in celestial  
mechanics was based long upon the 
three laws of  Newton's mechanics and   coordinate transformations (1.1).
>From the practical point of view, there were two  astronomical {\small
RF}s of primary 
importance:
the barycentric frame {\small (BRF)}, which is related to the barycenter
of the solar system, and 
the geocentric frame {\small (GRF)}, whose origin coincides with the
Earth's center of mass. 
Because of the recent progress in the relativistic treatment of an isolated
{\small N}-body  system,
there is now clear and unambiguous  agreement on   an asymptotical
 {\small BRF} (which is valid  even through the post-Newtonian
level of {\small WFSMA}). By  assuming that the solar 
system  as  a whole is completely isolated, one
may put its barycentric {\small RF} to be non-accelerated (or to say `at
rest') and 
absolutely non-rotating. The latter condition implies: (i) the absence of
the centripetal
 and Coriolis forces (dynamical inertiality) and 
(ii) that the coordinate  directions to the remote light sources 
(such as quasars) must be  constant (kinematic  inertiality).
In addition, the absence of any external sources of gravity enables one to
consider only the 
proper (or `inertial') gravitational field of the solar system. 
As a result, such a {\small RF} was used for a long time as
the  basic tool to solve almost all the   problems in practical astronomy  
(even relativistic ones).

As far as the {\small GRF} is concerned, the situation turns out to be more
complicated. 
If one attempts to describe  
the {\it local} gravitational environment of some extended body 
from an {\small N}-body system, (for example, the Earth in the solar
system), 
first of all, based on the results of a study of the existence of 
the  energy-momentum  conservation laws, one generally defines the
barycentric inertial 
{\small RF}: $(t', {\vec r}\hskip 2pt') $. Then,  one  may introduce  a non-
rotational
accelerated  {\small GRF} $(t, {\vec r}) $ which is defined at the  center 
of mass of  the extended body under study by a
coordinate  transformation  similar to that of (1.1):
{}
$$t'=t, \qquad {\vec r}\hskip 2pt' = {\vec r} +{\vec r}_0(t), \eqno(1.2)$$
\noindent where ${\vec r}_0(t)$ is the Newtonian barycentric radius-
vector of the body.

To analyze  the gravitational environment of the body  under
consideration, one presents  
the effective potential in the body's vicinity in  the form: 
{}
$${\overline U}({\vec r})=U_0({\vec r})+U^{\rm tid}({\vec r}). \eqno(1.3a)$$ 
\noindent where $U_0$ is the   body's own gravitational potential. 
The influence  of the external bodies  in the chosen frame 
manifests itself in the form of   gravitational tidal forces only. The
corresponding
tidal  gravitational potential $U^{\rm tid}$  may 
be  given by:
{}
$$ U^{\rm tid}({\vec r} )=U^{\rm ext}({\vec r}_0+{\vec r}) 
-  U^{\rm ext}({\vec r}_0) - \Big({\vec r} \cdot{\vec \nabla} U^{\rm
ext}({\vec r}_0)\Big).\eqno(1.3b)$$
\noindent This  potential   is searched for as the solution of 
the usual  Poisson equation in the form:
{}
$$\bigtriangleup U^{\rm tid}=-4\pi \rho_0^{\rm ext} \eqno(1.4a)$$
\noindent with the boundary conditions 
{}
$${\vec \nabla} U^{\rm tid}({\vec r}_0)=0, \qquad U^{\rm tid}({\vec
r}_0)=0,
\eqno(1.4b)$$
\noindent where $\rho^{\rm ext}_0$ is the mass density of the external 
gravity in the vicinity of the body under study.
As a result, the theory of   astronomical observations become inseparable
from the 
problem of determining of the motion of   celestial bodies, because the  
 Newtonian eq.m. for the  body's center of mass is determined as follows:
$$\ddot {\vec r_0}= - {\vec \nabla} U^{\rm ext}({\vec r}_0). \eqno(1.5)$$ 

One may  also verify that, in the  proper {\small RF} for an 
extended body constructed this way, the body's own center of mass will be
at rest during the time 
of the  experiment.  Indeed, by integrating the {\it local} eq.m. of the 
Newtonian hydrodynamics (Fock, 1955): 
{}
$$ \rho_0{d{\vec v}\over dt}=-  \rho_0
{\vec \nabla} {\overline U}+ {\vec \nabla}  p, \eqno(1.6) $$
\noindent over the  body's compact volume, one   obtains the desired
result:
$\ddot{m}_0^\alpha=0$, where $m_0^\alpha$ is the body's first (dipole)
mass  moment.  
In the body's vicinity the external gravity produces 
negligibly small tidal  perturbations of the {\it local} motion, 
which are  presently well known (Standish {\it et al.}, 1992). 
This leads to so-called `quasi-inertial' properties of {\small GRF}.
The kinematic  advantage of these {\it local} coordinates $(t, {\vec r})$ is
that the {\small RF},
when obtained this way, moves with the considered body. Their  dynamic 
usefulness  comes from the fact that the coordinate transformations (1.2)
allow
one (to some extent)  to decouple the motion of the studied body from the
{\it global} dynamics
of the system as whole (Pars, 1965; Brumberg, 1972; Damour, Soffel \& Xu
(here and after (DSX)), 1991). 
These are the reasons why this proper  {\small RF} (or {\small GRF})
has become very useful  for  studying   {\it local} physics in a body's
vicinity. 

The situation changed drastically when, by generalizing  Faraday's  
thoughts on the electric and 
magnetic phenomena, Maxwell  discovered a set of equations describing
electro-magnetic field.
These equations successfully described   the two  then-`new' forces   
 corresponding to  electromagnetic and optical phenomena.  
However, it   turned out that the  famous Maxwell-Lorentz equations   of
electromagnetism were not
form-invariant under the Galilean transformations (1.1). This was an
indicator that either 
the laws of the Newtonian mechanics were incomplete or these
transformations were wrong.
>From the other side, recall that the  transformations (1.1) were  
a simple consequence of the laws of   Newton's mechanics.
It became clear that even if some other set of    equations  would be
substituted for these laws,  
the  transformations (1.1) may not provide a form-invariancy for this  
new set.
Thus, it became obvious that the principle of relativity  must  have a more
fundamental  character.
 In the Poincar\'e interpretation this principle was re-formulated so 
  that the physical laws should be the same for a two particular observers:
one being  at rest and the other one being in the state of steady-straight-
line motion 
so that there is no means  to   find out whether or not  the second observer
is moving.
The  significance of this principle was  that it stated that there are no
such  
things as  absolute space or time, and, moreover, it implied the
impossibility of an 
absolute motion in the general law of   nature.  

As we know now, the understanding of this theory sparked  a revolutionary
change in the 
course of   theoretical physics in the beginning of the 20th century.  
The  answer to this problem was given in a series of works by  Poincar\'e
(1904) and  Minkowski 
(1908)  (see also Lorentz {\it et al.}, 1923): 
that  space and time  must be  united together to 
form a  four-dimensional pseudo-Euclidian geometry.  
The coordinates of two points in this four-dimensional manifold   are  
denoted as $(ct, {\vec r}) \rightarrow   x^n \equiv  (x^0, x^\alpha)$, where
$n=0,1,2,3$ and 
$c$ is the speed of light.
The square of the geodesic  distance $ds^2$  between the  two infinitely 
close  points  of this space-time (interval) is given  by
the  four-dimensional  analog of the Pythagorean theorem:
$ds^2=\gamma_{mn}(x)dx^mdx^n $. 
The  function $\gamma_{mn}(x)$ is  the metric tensor which  has become
the main object to 
define the structure of studied space-time (Eisenhart, 1926). These
metric coefficients only 
(as referred to a particular coordinate system) together with  the
coordinate 
differentials will provide one  with  physically measurable  quantities. 
In  Cartesian coordinates of 
the Galilean (inertial) {\small RF},  for all the points of the 
pseudo-Euclidian space-time,  this metric function may be chosen in the
form of the 
Minkowski metric: $\gamma^{(0)}_{mn}=\hbox{\rm diag}(1, -1,-1,-1)$.
As a result of such a change, the coordinates  lost their absolute meaning 
and could not be used  for  direct physical observations. Even the
differentials 
do not have  physical sense, because they are not directly connected   with
either the 
distance between two points in the three-dimensional space or with the
temporal evolution of the 
physical processes.

By analyzing the Maxwell-Lorentz equations of the electromagnetic field 
and the  interval  
in the form  $ds^2=\gamma^{(0)}_{mn}dx^mdx^n,$  Poincar\'e was the first
to point  out that 
 the set of these field equations and the  quantity $ds^2$ are form-
invariant  under the 
Lorentz' transformations, which form the Poincar\'e group of motion:
{}
$$t'=\gamma\Big(t-{({\vec v}\cdot{\vec r})\over c^2}\Big), 
\qquad \gamma =\Big(1-{v^2\over c^2}\Big)^{-{1\over2}}, \eqno(1.7a)$$
{}
$${\vec r}\hskip 2pt '={\vec r}+ 
(\gamma-1){{\vec v}({\vec v}\cdot{\vec r})\over v^2}-\gamma {\vec v} t, 
\eqno(1.7b)$$
\noindent where ${\vec v}$ is the constant relative speed between the two
{\small RF}s.
Thus, the study of  electro-magnetic phenomena led to the  discovery of a
new
theory of the structure of  space-time.

The  form-invariancy of the metric tensor 
under the transformations (1.7) has suggested  a more general physical
property, namely:   for all the possible coordinate transformations
between the 
two arbitrary {\small RF}s,  which   preserve the 
form of the metric tensor $\gamma_{mn}$, the physical phenomena in both
obtained frames  will 
behave in exactly the same way. As a result, the   principle of 
relativity  becomes simply a consequence of the latter 
property. The next logical step was  to generalize the   equations of the
Newton's mechanics 
based on this new four-dimensional  relativistic treatment. 
The resultant set of   equations of motion has
become known as the  relativistic mechanics of   Poincar\'e (Sard, 1970).
This theory was formulated in a covariant form  which allows one to study
the 
physical processes in any physical {\small RF}. 
Note  that independent of Minkowski and Poincar\'e, Einstein  had also
formulated a new theory of  
space-time - the special theory of relativity (Lorentz {\it et al.}, 1923;  
 Landau \& Lifshitz, 1988). However, this theory  
was formulated based only on the Poincar\'e group of motion and was 
constrained   to the class of inertial {\small RF}s only. 

The discovery of the pseudo-Euclidian 
geometry had finally undermined any  
  absolute meaning of  finite time or finite distance and had substituted
instead a purely   relative one.
Now the  interval $ds^2$ - the square of the infinitesimal distance 
in four-dimensional manifold -  had become the only absolute quantity.
For example, based on the Lorentz transformations, 
the time in two different {\small RF}s was no longer the same, 
but rather  depended  on the relative speed between the frames:
{}
$$\Delta t' =\int_{t_0}^{t_1} dt 
\Big(1-{{\vec v}\hskip 1pt^2(t)\over c^2}\Big)^{1\over2}.\eqno(1.8a)  $$ 
\noindent Moreover, the length of an object    in two {\small RF}s 
was also no longer  invariant. Thus  a rod, which has a  length $dl_0$ 
in a  rest frame, will experience 
 the length contraction  in an inertially moving  frame
 in the direction ${\vec n}={\vec v}/v$, parallel to the speed of motion 
${\vec v}$:
{}
$$d{\vec l} ={\vec n}dl_0\Big(1-{{\vec v}\hskip 1pt^2\over
c^2}\Big)^{1\over2}.\eqno(1.8b)  $$ 
\noindent It should be stressed that the formulas (1.8) are simply the
consequence of the 
properties of the pseudo-Euclidian geometry. 
It should be noted that,  together with the properties of this geometry,
 the language of the `microscopic' (or field) description has appeared in  
theoretical physics  as the necessary tool for theoretical studies of  
physical 
processes. This `field' terminology  deals with the densities of  
physical quantities in a relativistic-coordinate-independent way, rather
than providing 
a coordinate-dependent 
(or {\small RF}-dependent) regular  `macroscopic' treatment and it
has become   a very powerful   substitution for the latter. 
As a result, for the special relativistic treatment of the gravitational 
observations,  contrary to Newtonian mechanics, one should always appeal
to the notion of 
the `proper' quasi-inertial {\small RF} of a body in order to correctly
define 
the body's  mass, its barycenter, and the intrinsic multipole moments.  

For a long time it was considered that the 
special theory of relativity and hence  the relativistic mechanics of
Poincar\'e, were 
theories which described  the physical processes solely in   different
inertial {\small RF}s 
(which may be linked to each other by the Lorentzian transformations
(1.7)).
>From the other side,   real astronomical phenomena   unavoidably involve 
 descriptions based on  non-inertial {\small RF}s,
which, by a misunderstanding (partially based on the Equivalence
Principle), was
considered as a prerogative  of the general theory of relativity only.
However, this is not true. Based on the discovery of the pseudo-Euclidean
space-time made by Poincar\'e and Minkowski,  one may use an infinite
class of  
admissible {\small RF}s both inertial and non-inertial in order to describe
the 
physical phenomena in the real world.
Indeed, the Riemann curvature tensor, which   defines the intrinsic
geometry
of  space-time,  is zero in any of these frames. 
However,  observing  any physical process  enables 
one to confidently distinguish the situations when an experiment is
performed 
in an inertial or in a non-inertial frame. 
This means  that the following generalized principle of relativity 
(Logunov, 1987) is valid:
Independent of the state of motion of the {\small RF}   chosen for the
experiment  (either inertial 
or non-inertial), one may define an infinite set of  other {\small RF}s 
for which  the physical phenomena will behave in exactly the  same way.
Moreover,  one may  not establish, by any means, in which  {\small RF} 
from this 
equivalent set the experiment is performed.   As a result, by defining 
the admissible coordinate transformations  which   leave the   
metric tensor in the chosen  {\small RF}  form-invariant, one  defines
the entire infinite set of   physically equivalent {\small RF}s.
Thus,   from Poincar\'e's  equations of   relativistic mechanics   
and the requirement of the form-invariancy of the metric tensor,
 one  may find another 
fundamental group of motion in the pseudo-Euclidian space-time, 
namely: the relativistic group of the uniformly-accelerated motion of a
monopole particle.
Indeed, for  a particle with   mass $m_0$ moving   under 
the influence of  a constant force ${\vec f}=(f, 0, 0)$ the law of motion 
is  given by: 
{}
$$t'=t, \qquad x'=x+{c^2\over a}
\Big[\Big(1+{a^2t^2\over c^2}\Big)^{1\over2}-1\Big],   \eqno(1.9) $$
\noindent where $a=f/m_0$ is the corresponding constant acceleration.  
The  interval  of the two-dimensional space-time in the co-moving
{\small RF} 
takes the form:
{}
$$ds^2={c^2dt^2\over 1+a^2t^2/ c^2} -{2at\hskip 4pt dtdx\over
\Big(1+a^2t^2/ c^2\Big)^{1\over2}}-
dx^2. \eqno(1.10) $$
\noindent From this  it is easy to show that the  corresponding 
two-parametric group of motion for the uniformly-accelerated {\small
RF}s
may be presented as follows:
{}
$$t'= \gamma\Big(t+t_0+{vx\over c^2}+{v\over a}\Big[\Big(1+{a^2t^2\over
c^2}\Big)^{1\over2}-1\Big]\Big),
\qquad \gamma =\Big(1-{v^2\over c^2}\Big)^{-{1\over2}},$$
$$x'=\gamma\Big(x+vt+{c^2\over a}\Big[\Big(1+{a^2t^2\over
c^2}\Big)^{1\over2}-1\Big]\Big)- $$
$$-{c^2\over a}\Bigg(\Bigg[1+ {a^2\over c^2-v^2}\Big(t+t_0+{vx\over
c^2}+{v\over a}
\Big[\Big(1+{a^2t^2\over c^2}\Big)^{1\over2}-
1\Big]\Big)^2\Bigg]^{1\over2}-1\Bigg)+x_0,
\eqno(1.11)$$
\noindent where $t_0$ and $x_0$ are the group constants. 

One can see that, in order to preserve the form-invariancy of the metric
tensor for the time translation
(given by a parameter $t_0$)  contrary to Poincar\'e group  (1.7), 
this nonlinear group of motion requires the transformation of  spatial
coordinates as well.
Thus, the non-inertiality of the {\small RF}  makes the 
physical analysis   more difficult than in the case of an inertial {\small
RF}s. 
The situation becomes even more complicated   
if one decides to describe the motion of an extended object.
This is because the bodies in this case, besides the `usual' Lorentzian 
relativistic contractions, will experience  other dynamic  
effects generated by the properties of the {\small RF} chosen for the 
analysis.
In practice, one is usually faced with the problem of extracting   the 
{\small RF}-induced  effects. 
As the  properties of the  pseudo-Euclidian space-time 
are well established, this problem may  be solved in a 
satisfactory manner  by   constructing a
quasi-inertial {\small RF} in the vicinity of the body under consideration.
The   vanishing of a Riemann curvature  leads to a maximum possible
number of the Killing vectors in this geometry  ({\small N}=10), which
enables us to   separate
 physically observable  and  coordinate-induced quantities in a
satisfactory manner.
Note  that the corresponding  theoretical methods of the classical
mechanics of Poincar\'e are 
presently  well tested in   different experimental situations and   they
are used extensively    in many areas of modern relativistic physics, such
as 
high-energy physics, theoretical astrophysics and solid state physics. 
Astronomers, however, previously had not fully accepted these methods
into real 
astronomical practice as there was little   observational 
data with the relativistic accuracy. 

This has  changed dramatically  during the last two decades, and now that
 the accuracy of  astronomical observations enables us to perform  studies
of the 
physical processes in the universe with much  higher precision, the  
problem of  relativistic gravitational measurements has become very
important.  
This has led to numerous    experiments  testing  different 
hypotheses which have lain the   foundations for a  number of 
recent theories of gravity (Will, 1993). 
Gravity, however,   remains the last yet unexplored frontier of the modern
theoretical physics
(Hawking \& Israel, 1987; Damour \& Sch$\ddot{\rm a}$fer, 1991; Damour
\& Taylor, 1992). 
This is mainly because of   weakness of the gravitational interaction in
the solar system  
presents great difficulties when planning and performing   gravitational
experiments. 
The other reason, is that  the discovery of the field equations of the
general 
theory of relativity has changed our physical conceptions once again.  
According to this theory, not only are space and time  united together by
forming a 4-dimensional Riemann manifold with the general metric tensor 
$g_{mn}$, 
but also it is   matter  which is   responsible for  generating  the
properties of this space-time.
In other words,   space-time tells matter how to move and  
matter tells the space-time how to curve (Misner {\it et al.}, 1973). 
There are many other gravitational theories currently under consideration,
but the   
metric theories of gravity have taken a special position 
among all the  possible theoretical models.  
The reason  is that, independent  of the many different 
principles at their foundations, the gravitational field  
in these theories affects  matter directly through the metric 
tensor of  Riemann space-time   $g_{mn}$, which is  
determined from the field equations of a particular theory of gravity. 
In contrast to Newtonian gravity, this tensor  
contains the properties of a particular gravitational theory as well as  
carrying the information about the gravitational field of the bodies
themselves. 
This property of the metric tensor enables one to analyze 
the motion of matter in one or another metric theory of
gravity  based only upon the underlying principles 
of modern theoretical physics.

The situation with  relativistic measurements has become even more
complicated. 
Because it is well known that  in the Riemann space-time  one can 
not have an explicit mathematical definition for the proper {\small RF}, 
it is permissible to introduce any coordinate system.
  As a rule, before solving these equations, four restrictions 
(coordinate or gauge conditions) must be imposed on the components of
the $g_{mn}$. 
These conditions  extract a particular subset from an infinite set of
space-time coordinates. 
Inside this subset the coordinates are linked by smooth  differentiable
transformations which do not 
change the coordinate conditions being chosen.
In general relativity, for example, there exists no absolute time or
Euclidean space. 
Besides, one may not, in the general case, introduce some `privileged' 
{\small RF}  in space-time. 
Contrary to the Newtonian theory of gravity, coordinates
in  curved space-time have no physical meaning and cannot be measured
directly by  astronomical 
observations.

Nevertheless, there are some special cases in which  one may speak about
privileged 
coordinates in general relativity. One  such case is space-time 
having a weak gravitational field and slowly moving  matter. The density
of the 
total non-linear Riemann metric tensor $g^{mn}$
of such  space-time may be linearized and presented as a sum of the 
density of the pseudo-Euclidian background metric $\gamma^{mn}$ plus 
the small perturbations caused by the physical gravitational field
$h^{mn}$:
$\sqrt{-g}g^{mn}=\sqrt{-\gamma}\gamma^{mn}+h^{mn}$. Then, in the 
Galilean inertial {\small RF}, such a space-time may be 
covered by coordinates which differ only slightly  from the absolute time
and
Cartesian space coordinates of Newtonian theory of gravity. We shall call
these space-time coordinates
quasi-cartesian. These quasi-cartesian  coordinates are the most
convenient coordinate system 
for developing a   relativistic theory of  
astronomical {\small RF}s inside the solar system. They are also used in 
the case of an
isolated astronomical system which consists of {\small N} well-
separated and extended bodies possessing
weak gravitational field and moving with slow orbital and rotational
velocities (such as our solar system).
 
The solution of the field equations of general relativity in {\small WFSMA}
for 
an isolated distribution of matter is presently well-known (Will, 1993). 
There have been a number attempts to describe the motion of  
different gravitationally-bounded astronomical systems. This  problem of
describing the motion of a system 
consisting of {\small N} massive monopole particles
was first considered by  Einstein {\it et al.} (1938), the rigid uniform
rotation of the bodies 
was included by Papapetrou (1948, 1951), Fock (1955),  etc. 
It was shown that the post-Newtonain equations of Einstein, Infeld and
Hoffmann ({\small EIH})
governing the motions of {\small N} mass points allow the same ten
classical integrals as the 
equations of Newtonian gravity, namely those expressing conservation of
energy, 
linear momentum, and the uniform motion of the center mass of the body. 
Moreover, Chandrasekhar \& Contopulos (1967) had shown   there   exists a
way to introduce 
the  notion of the `center of mass' of such a  system, which  enables one to
construct the 
barycentric inertial  {\small RF}$_0$.
Thus, by studying the  problem of the form-invariancy of 
the metric tensor and the corresponding post-Newtonian 
{\small EIH} eq.m., they had shown that both of these expressions  
are invariant under the following `post-Galilean' coordinate 
transformations  which establish a correspondence
between  frames with   uniform relative motion:
{}
$$t'=\Big(1+{v^2\over2c^2} + {3v^4\over8c^4})t- 
{ v_\mu y^\mu\over c^2}\Big(1 +{v^2\over2c^2}\Big) 
+{1\over 2c^4}\sum_Bm_Bn_\mu y^\mu +{\cal O}(c^{-6}),\eqno(1.12a)  $$
{}
$$x'^\alpha=  y^\alpha +\Big(1 +{v^2\over2c^2}\Big)v^\alpha t-
{v_\mu y^\mu\over 2c^2}v^\alpha  +{\sigma\over c^2}
\epsilon^\alpha_{\mu\beta}y^\mu v^\beta  +{\cal O}(c^{-4}), \eqno(1.12b)$$
  
\noindent where $v^\alpha$ is the constant velocity of the uniform
motion, 
$m_B$ is the post-Newtonian rest mass of the distribution of matter
under study 
and $\sigma$ is some arbitrary constant. One can see  that both of the
equations (1.12), 
contain  additional terms beyond those  obtained by expanding the Lorentz
transformation (1.7). The last term in the  eq. (1.12a) is the 
contribution which is unique to   general relativity and it is  this term
which gives  the 
transformation (1.12) its non-Lorenzian character. 
The other  additional term in   eq. (1.12b)  represents an arbitrary 
infinitesimal rotation which may be satisfactorily explained in terms of
the Poincar\'e group.
As a result, the obtained post-Galilean transformations  are
generalizations of the   
Lorentzian transformations  (1.7) in the gravitational  case. 
 
These  post-Galilean transformations (1.12) are of little use for
astronomical observations as they were  
obtained in order to demonstrate existence of the barycentric inertial
{\small RF}$_0$ and they are 
not suited for the construction of an astronomical {\small RF}  for even  
massive monopole bodies.
This is simply because  such   proper {\small RF}s  will  
generally  not be inertial, but rather quasi-inertial. 
Moreover, the expressions (1.12) do not  account  for the multipolar
structure of the 
extended bodies. However, we need some transformation that will work,
since,
  in order to present all the necessary 
expressions for the metric tensor and the equations of motion with the
same post-Newtonian accuracy,
 one must have  a physically grounded definition   of the 
transformation rules between the {\small RF}s. 
To find this transformation, one must expand the Newtonian contributions
in terms of the  intrinsic
mass and current multipole-moments  of the bodies (Damour, 1983, 1986).
The greater the required accuracy, the 
larger the number  of these 
terms which must be taken into account. It is known that  the fully
relativistic definition of these 
 moments may be  given in the proper quasi-inertial {\small RF} only. Such
a definition   replaces  
that which was given in the rest-frame of the one-body
problem\footnote{Note that, due to the breaking of the symmetry 
of the total Riemaninan space-time by realizing the $3+1$ split  (Thorne
{\it et al.}, 1988), 
these moments will not form  tensor quantities  with respect to general 
four-dimensional 
coordinate transformations in the {\small WFSMA}. 
Instead, these quantities will behave as  tensors under the sub-group of
this total
group of motion only, namely: the three-dimensional  rotation. 
This is similar to the situation in classical electrodynamics, where  
electric 
${\vec E}$ and magnetic ${\vec H}$ fields are not   true vectors, 
but  rather   components of the $4\times 4$ tensor 
of the electro-magnetic field $F_{mn}=({\vec E}\otimes {\vec H})$ (Landau
\& Lifshitz, 1988).}.
In presenting these transformations one should also 
take into account that, due to the non-linear character of the 
gravitational interaction, these moments are expected to interact  with  
external gravity, 
 changing the state of motion of the body itself. 
 Fock (1955) was the first to notice  that in order to find  the 
solution of the {\it global} problem (the motion of the {\small N}-body
system as a whole), 
the solution for the {\it local} gravitational problem (in the 
body's vicinity) is required. 
In addition, one must establish their correspondence by presenting 
the coordinate transformation by which the 
physical characteristics of motion and rotation are transformed 
from the coordinates of one {\small RF}  to another. 
Thus,   one must find   the solutions to the three following problems 
(Damour, 1987; DSX, 1991):

\begin{itemize}
\item[\small (1).]   \underline{  The {\it global} problem:} 
\begin{enumerate}
\item[(i).] We must  construct  the asymptotically  inertial {\small RF}.
\item[(ii).] We must  find  the barycentric inertial  {\small RF}$_0$ for
the system under study. 
This is primarily a problem of describing  the {\it global} 
translational motion of the bodies constituting the {\small N}-extended-
body  system
({\it i.e.} finding the geodesic structure of the space-time occupied by the 
whole system).  
\end{enumerate}
\item[\small (2).]    \underline{ The {\it local} problem:}  
\begin{enumerate}
 \item[(i).]  We must   establish   the properties of 
the gravitational environment in the proximity of each body in the system 
 ({\it i.e.} finding the geodesic structure of the {\it local} region of the
space-time
 in the body's gravitational domain). 
\item[(ii).] We must construct the  {\it local} effective rest frame  of
each body.
\item[(iii).]  We must study     the internal motion 
of matter inside the bodies as well as   establish    their explicit 
multipolar structure and  rotational motion.

\end{enumerate}
 \item[\small (3).]  \underline{The theory of the  {\small RF}s:} 
\begin{enumerate}
 \item[(i).]
We must find a way to describe the  mutual physical cross-interpre\-
tation 
 of the results obtained for the above two  problems ({\it i.e.}  
the fine mapping  of the  space-time).  
\end{enumerate}
\end{itemize}
 
\noindent 
Because the   solutions to  the first two problems will not be complete
without 
presenting the rules of the  coordinate transformations between 
the {\it global} and the {\it local} (or planeto-centric)   {\small RF}s
chosen 
for such an analysis, the theory of the astronomical 
{\small RF}s become inseparable from the 
problem of determining  the motion of the celestial bodies. 
>From the other side, if one attempts to describe the {\it global} dynamics
of the 
system of {\small N} arbitrarily shaped extended bodies, one 
will discover that, even in  {\small WFSMA},  
this solution will not be possible without
appropriate description of the gravitational environment 
in the  immediate vicinity of the bodies. 
 
Concerning the problem of  astronomical data reduction, first of all, one
must 
 find the connection between the coordinate quantities and the physically
observable ones.  Until quite recently, relativistic reduction of
astrometric observations 
was based on the use of the barycentric {\small RF} and  covariant
definitions of 
an observables (Zel'manov, 1956; Synge, 1960; Misner {\it et al.}, 1973;
Ivanitskaja, 1979 
 Soffel, 1989; Brumberg, 1991a,b; Nordtvedt, 1995). 
Thus, interval $ds^2$ in terms of observable coordinates $dX^{\overline
n}\equiv(cd\tau,dr^\alpha)$, 
is  taken to be diagonal and it  is usually presented in the  form pseudo-
Euclidian 
 Minkowski  space-time   in the Galilean {\small RF} as follows:
{}
$$ds^2=g_{mn}(x)dx^mdx^n= \gamma_{\overline{m}\hskip 1pt\overline{n}}
dX^{\overline m}dX^{\overline{n}}\equiv c^2d\tau^2-dr^2, \eqno(1.13) $$
\noindent where   the physical time $d\tau$ and the three-dimensional 
physically 
measurable distance $dr^2$ are given by: 
$$d\tau=\sqrt{g_{00}}dt+{g_{0\alpha}dx^\alpha\over  c\sqrt{g_{00}}},
\qquad
dr^2=\Big(-g_{\alpha\beta}+{g_{0\alpha} g_{0\beta}\over  g_{00}}\Big)
dx^\alpha dx^\beta.  \eqno(1.14)$$
\noindent In this method the directly measurable quantities by definition
are the 
  tetrad components (or base vectors) $\sigma^n$ of the null wave vector
of a 
photon projected onto a space-like
hypersurface being orthogonal to the four-velocity of an observer $u^n$: 
$\sigma^n=P^n_l k^l$,  where 
$P^m_n=\delta^m_n+u^m u_n$ is the projection operator which   
satisfies the following conditions: 
$P^\alpha_\alpha=2$ and $P^m_k P^k_n=P^m_n$.
By definition, the physically observable components of the vector
$\sigma^n$ in the 
locally orthonormalized tetrad basis of an observer,  has only the 
spatial components $\sigma^{\overline{\alpha}}$, while the temporal one
$\sigma^{\overline{0}}$ is 
equal to zero. Contraction of the componenets $\sigma^n$ with 
the basis vectors   $\lambda^{\overline p}_n$, 
that $\sigma^{\overline p}=\lambda^ {\overline p}_n\sigma^n$  is a
covariant quantity 
which is independent on the choise of the {\small RF}. 
This gives the procedure of relaying the 
coordinate quantities $dx^p$ to the observable ones $dX^{\overline
p}\equiv(cd\tau, dr^\alpha)$ as follows: 
$dX^{\overline p}=\lambda^{\overline p}_ndx^n$. From the equation (1.13), 
we can find the following relation:
{}
$$ds^2=\gamma_{{\overline p}\hskip 1pt{\overline n}}dX^{\overline
p}dX^{\overline n}=
\lambda^{\overline s}_m\lambda^{\overline p}_n
\gamma_{{\overline s}\hskip 1pt {\overline
p}}dx^mdx^n=g_{mn}dx^mdx^n,\eqno(1.15)$$ 
which provides one with the necessary equation for finding  
the components of the tetrad:
 {}
$$g_{mn}=\lambda^{\overline s}_m\lambda^{\overline
p}_n\gamma_{{\overline s}\hskip 1pt{\overline p}}.\eqno(1.16)$$ 
\noindent From this equation and with the help of  the relations $(1.14)$
one, 
in principle, may find all the necessary basis vectors  $\lambda^{\overline
p}_n$
( Logunov, 1987; Soffel, 1989). 
 Using this technique as well as the special methods of the Riemann
geometry, one may establish
the relationships between the  basis vectors and
transform the measurable components
$\sigma^{\overline{n}}=(0,\sigma^{\overline{\alpha}})$ to 
the coordinates of the barycentric {\small RF}.  However, the reduction
formula obtained this way,  
has been  proven to contain a non-observable  coordinate-induced 
contributions in the 
relativistic terms (Klioner \& Kopejkin, 1992). 
For example, the barycentric velocity of the astrometric spacecraft
orbiting the Earth is not 
directly observable and can not be derived with   necessary accuracy with
this barycentric 
method. To solve this and some other problems unavoidably 
arising in the solely barycentric approach, a consistent relativistic theory
of  
astronomical {\small RF}s is needed.
 
As we have mentioned earlier, a well defined  proper {\small RF} 
 must be  linked with  the  inertial {\small RF}$_0$    by relativistic 
coordinate transformations  which introduce  no spurious  terms into the
metric or the  equations 
of motion of the relativistic {\it local} problem. 
However, the precise definition of the quasi-inertial proper {\small RF} 
in a curved space-time 
(even  in the {\small WFSMA})  is not quite straightforward.
We  know  that   in  freely falling 
inertial frames,   the external gravitational field appears only in the form
of  tidal 
interactions.  Up to these tidal corrections,  
freely falling   bodies behave as if 
external gravity were absent (Synge, 1960; Bertotti \& Grishchuk 1990). 
The general theoretical consideration in this case is usually 
 based on the geodesic equation
{}
$$ {du^n\over ds}=-\Gamma^n_{kl}u^ku^l.\eqno(1.17)$$
\noindent This equation may be interpreted  
as if  on the  left side we have the four-acceleration of the particle,
while on the right side   is the force acting upon the particle. By careful
choice of the 
coordinates, one may make the Christoffel symbols $\Gamma^n_{kl}$
vanish in the 
immediate vicinity of the body's world-line, which will 
put this force equal to zero  (Fermi, 1922a,b; Landau \& Lifshitz, 1988).
This  allows one to use the  analogy of the inertial motion and, as a result,
the four-velocity
may be parameterized by the natural  parameter $s$ along the geodesic:
$u^n=a^n s+ b^n$, with $a^n$ and 
$b^n$ being the arbitrary constant parameters.  
The  analysis shows that in the vicinity of the world-line of the origin of
this 
well defined   {\small RF}$_A$,
the coordinate  transformation from the inertial 
{\small RF}$_0$  $(x^n)\equiv(x^0,x^\mu)$ to  the physically justified  
{\small RF}$_A$
$(y_A^n)\equiv(y^0_A,y^\mu_A)$   must have
the structure of a Taylor expansion with respect to the  powers of a
spatial coordinate 
$y^\mu_A$ (Manasse \& Misner, 1963; Manasse, 1963;
Misner {\it et al.}, 1973):
{}
$$ x^n=x^n_{A_0}(y^0_A)+{e_A}^n_\alpha(y^0_A) \cdot y^\alpha_A +
{1\over 2}{\Gamma_A}\hskip 1pt^n_{\alpha\beta}(y^0_A)
\cdot y^\alpha_A y^\beta_A+{\cal O}(y_A^3),\eqno(1.18)$$

\noindent where the function $x^n_{A_0}(y^0_A)$ represents the 
world-line's description of the origin of the coordinates 
$(y^p_A)$, the functions  ${e_A}^n_\alpha$ and  
${\Gamma_A}\hskip 1pt^n_{\alpha\beta}$ are coefficients of expansion.
This
relativistic transformation should replace  the post-Galilean 
transformations  (1.12), as well as 
the special-relativistic group of motion of the uniformly-accelerated
{\small RF}s (1.7),
allowing them both to be  generalized in the  case of a system of {\small
N} arbitrary 
extended self-gravitating bodies.  

It should be noted that the use of the approach depicted above was based
upon the geodesic 
equation (1.17), but, as we know,  extended bodies do not move along the
geodesic lines. Instead, the 
interaction of their  intrinsic mulipole moments with  external gravity 
causes 
 deviation of their motion from the geodesic. This means that  this
geodesic method is valid only 
for the case of  monopole structureless test particles. In order to 
provide the dynamic  definition for the proper {\small RF}, one should 
obtain the eq.m. of the extended bodies and require that the 
acceleration of the body  will vanish in it's proper {\small RF}. 
One way to do this is to   generalize the Fock-Chandrasekhar approach in 
derivation of the 
eq.m. for the extended bodies, which is based upon the 
equation of the conservation of the density of the 
energy-momentum tensor ${\hat T}^{mn}$ in the form: $\nabla_m{\hat
T}^{mn}=0$.
One may expect that the correct transformations will modify the
structure of the expressions 
(1.18)  in the  higher order terms of the spatial coordinates: $\sim
y^{\{K\}}_A,$ 
where $(k\ge 3)$. 

We should mention here that  in  the  scientific literature in addition   
to expression `reference frame' the  notion `coordinate  system' {\small
(CS)}
has recently come into a  wide  use 
   (Kopejkin, 1988; Brumberg \& Kopejkin, 1988a,b; DSX, 1991-1994).
This confusion in terminology partially came from  a 
misunderstanding of the basic principles of the 
theory of relativistic observables in the curved space-time developed by 
Zel'manov (1956). In accord to his   chronogeometric classification, one
should 
distinguish between these two physically different concepts. 
Thus, the {\small RF} is an arbitrary  set of   four coordinates 
chosen to define the position of the body under study. As we know, in order
to 
properly describe the motion of the {\small N}-body system, one should
have 
at least {\small N}$+1$ these {\small RF}s  (DSX, 1991). The {\small CS}
are the  
coordinates one may  choose to describe the physical processes in the 
vicinity of the body in its proper {\small RF}. A coordinate system is a 
particular code for labeling points in a {\small RF} by some numbers. 
However, once   the {\small RF} has been 
chosen, one may not   make the choice of the {\small CS} arbitrarily. In
order to 
introduce the {\small CS} one must   fulfill the chronogeometric
requirements,
which basically states or says that while introducing the {\small CS} 
one shouldn't change the state of motion 
of the {\small RF} already chosen for solution.  In the other words, the
choice of the 
{\small CS} should provide one with a new {\small RF} which should be
physically equivalent to the 
old one. In practice, one   usually may introduce  an infinite class of
{\small CS}s 
  without violating this   equivalency (Zel'manov, 1956; Logunov, 1987;
Denisov \& Turyshev, 1989).  
>From the other side, it is well-known that in the curved space-time there
are no 
inertial {\small RF}s even in the {\small WFSMA}, instead
one may introduce only quasi-inertial ones. Moreover, an non-optimal 
choice of the {\small CS}
may change the dynamic properties of the {\small RF} and may
significantly complicate   the eq.m. of the bodies, leading to the 
wrong conclusions (Kopejkin, 1988). This means that a clear physical
definition 
 for the {\small RF} is very important. Such a  definition should enable one   
 to study the form-invariancy of the corresponding metric tensor.
As a result, one may reconstruct the  group of motion, which   leaves this
metric 
tensor form-invariant, and  which will
provide one with the class of   admissible physically equivalent 
coordinate transformations in the {\small RF} of interest.
 We will keep this  relativistic  terminology, and  in our further 
discussion,   we will distinguish between the {\small CS} and the {\small
RF}.

As we noted before, the  properties of the proper {\small RF} should be
based primarily upon the  
structure of the metric tensor and the equations of motion of  the {\it
local} problem. 
For  practical reasons, in order to establish  the physical characteristics
of the  proper {\small RF}$_A$ 
constructed for a particular body {\small (A)} from the system, 
it is best to  use the  well-known properties of the freely falling  {\small
RF}s as a first approximation
when examining the interaction between the bodies. 
Thus the expected  properties, of a physically well defined  proper {\small
RF}$_A$
 may be expressed as follows:

\begin{itemize}

\item[(i).] The gravitational field solutions for both relativistic {\it
global} and {\it local} 
problems should be obtained  with the  same covariant gauge conditions.
At least up to the terms describing the motion of the mass monopoles, 
the metric tensor and the eq.m. of the {\it local} problem must not depend
on the 
`absolute' velocity  of the motion 
of the origin of  proper {\small RF}$_A$  relative to the inertial {\small
RF}$_0$. 
Both  the tensor and the eq.m. in this case may 
 admit the dependence  on the relative velocities of the bodies only (Fock,
1955; Kopejkin, 1988).
The body's own translational motion in its proper {\small RF} should 
vanish.

\item[(ii).]  This field in the {\it local} region must be made up of    
four physically different contributions, namely:
 the proper and  external gravitational fields, the field of inertia, and 
 the gravitational interaction term. 
The  proper gravitational field outside the body   should  be describable 
by the set of  mass and current intrinsic multipole moments 
 including the monopole, the dipole, etc. (Thorne \& Hartle, 1985; Kopejkin,
1988). 
The gravitational field of the external bodies  
must be presented  in the proper {\small RF}$_A$ solely in the form of 
 tidal terms generated by mass and current multipole moments 
of these bodies (Fermi, 1922a,b; Synge, 1960).
The field of inertia is due to the specific properties of the coordinate 
transformations chosen for the construction of this {\small RF}. The
interaction term 
describes the mutual coupling of the three above named terms.

\item[(iii).] Coordinate transformations between the different {\small
RF}s  
should be homogeneous functions omitting the infinite number of non-
singular 
partial derivatives. These functions should not violate the gauge
conditions chosen for the
problem  and must be  completely defined  by means of the {\it local}
gravitational 
field
at the origin of the coordinates of a particular quasi-inertial proper
{\small RF}. 
 \end{itemize}

It was considered for a long time that  
the physically adequate {\it local} {\small RF}  must physically resemble  
a frame  which falls freely 
in the background field created only by external bodies (Kopejkin, 1988). 
However, this is not true.  This effect is due to the  presence of the
gravitational interaction term,
which reflects the non-linear nature of  gravity. When describing the
motion of a monopole
particle, one may use this analogy  and describe the motion of the body as
if the external gravity 
were absent, but,   in the general case of the extended self-gravitating 
body, one must take into account the coupling of the body's 
intrinsic multipoles  to the external field. The existence of this coupling 
should be reflected
in the form of the transformation functions. 
As a result, one should not think that the `good' proper {\small RF} 
may be realized as a locally inertial {\small RF}  for a massless test body 
(Manasse \& Misner, 1963; Misner {\it et al.}, 1973; Ni, 1977;
Ni \& Zimmerman, 1978). Physically, we are looking for   a {\small RF} 
where one may effectively separate  the {\it local} physics from the
external gravitational environment. 
This is why we  would  like to apply such an elegant and simple Newtonian
tidal approach to the 
post-Newtonian physics of the {\small WFSMA}.
>From the mathematical standpoint,  we are looking for  a solution to the
{\it local} problem  for which the 
resultant space-time in the proper {\small RF}$_A$   
will be tangent to the total effective  space-time generated by all the
bodies in the system including the 
body {\small (A)}. It was shown that the solution with these properties 
could be found only at the 
immediate vicinity of the body and that the smaller  the 
Riemann curvature of the effective space-time, the further  out would be 
the boundary of validity of this solution (Brumberg \& Kopejkin, 1988).   
Note that the 
existence of a well defined  proper {\small RF}  has been more or less 
explicitly assumed by many authors (see, for example, 
Misner {\it et al.}, 1973;  Li \& Ni, 1978, 1979a,b; 
 Will, 1993; Nordtvedt, 1995).

\subsection {The Qualitative Description of the Astronomical Systems of
Interest. }
  
In order to provide a  quantitative 
description of the relativistic motion of an  astronomical  {\small N}-body
system, let us
first qualitatively define the small parameters involved in the description
of such a system.
It is known that, there are several major methods to study the dynamics
of such systems (Damour, 1983, 1986)
 depending on the relationships between the astrophysical parameters 
characterizing the orbital motion, rotation, gravitational field inside and
outside the bodies, 
their size, shape and internal structure, and the distance between the
bodies.
We shall investigate a structure of space-time for the case of a
gravitationally 
bounded and isolated distribution of matter.
We will restrict our attention to only   {\small N}-body systems such as
our solar system 
which have  slowly moving matter and weak gravitational 
field   both outside and inside the bodies.
 Let us assume that non-gravitational 
forces are absent, and that the bodies are well separated. 
Our assumptions are then   that the velocities of the orbital motion of the
bodies $v_B$ 
are non-relativistical ones, {\it i.e.} considerably smaller
 than the speed of light $c$: $v_B \ll c$, and that the any   two 
arbitrary bodies in the system are  at   distances
 $r_{BA_0}$ are    considerably greater  than their radii $L_A$ and 
$L_B$: $r_{BA_0} > L_A, L_B$. Note  that the motion  of the bodies 
 at    distances 
$r_{BA_0} \sim  r_{g_A}, r_{g_B}$, where $r_{g}$ is the gravitational 
radius of the body  has a highly  unpredictable 
character and will require very different mathematical techniques
(Shapiro \& Teukolsky, 1986a,b; Thorne, 1989).
Furthermore,   let us denote 
the following quantities for each body in the system: 
$m_B$ is the  mass of the body {\small (B)}, 
$r_{B_0}$ is the Newtonian barycentric radius-vector of this body; 
$L_B$ is its mean radius, $D_B$ is the minimal 
distance between the body under question and it's nearest companion 
in the system; ${\vec u}_B$ is the  internal 
(rotational ${\vec v}_{rot}$ and plus oscillatory ${\vec v}_{osc}$) velocity 
of the element of the body's matter in the   proper {\small RF}$_B$;
  $\omega_B$ is the frequency  of its rotation in  
 this {\small RF}$_B$; $I_B^{\{K\}}$ and $S_B^{\{K\}}$ are its   
internal mass and current moments of $k^{th}$ order respectively and,
finally,  
$M_0$ and $L_0$   denote the  mass and maximal diameter 
of the entire system. 

Then, making of use the definitions above, 
we will concentrate our attention on a solution of the problem 
of motion of such a gravitationally-bounded astronomical system 
of {\small N} extended bodies in the {\small WFSMA}. This approximation 
may be used successfully    if the system of interest   admits
the existence of the   following four groups 
of  small parameters induced by the {\it local}   and {\it global} 
 of the bodies in the system (denoted with the (l) and (g) subscripts
respectively): 

\begin{itemize}
\item[{\small (1)}.]  \underline{\it The shape and size induced 
parameters.} 
We  presume that for each body in the system  the following 
 parameters of a pure geometrical nature may be introduced:  

\begin{enumerate}
\item[(i).] $\delta_g \sim  sup[ \delta^B_g={L_B}/D_B]  \ll 1$, which 
describes the quasi-point structure 
of each body in the system; 
\item[(ii).] $\delta_l \sim sup[ \delta^B_l=I^{\{K\}}_B/m_BL_B^k] \ll 1$,
which 
  characterizes a dimensionless measure of the 
deviation of the distribution of the bodies' matter from a 
spherically-symmetric   distribution. 

\end{enumerate} 
 
\item[{\small (2)}.]   \underline{\it The special relativistic  parameters.} 
The orbital and rotational motion of the bodies in the system generates
the following
 dimensionless parameters: 
\begin{enumerate}
\item[(i).]  $\epsilon_g \sim  sup[\epsilon^B_g=v_B/c] \ll 1$,
characterizing the 
speed of the orbital motion of
the bodies; 
 
\item[(ii).] $\epsilon_l \sim sup[\epsilon^B_l=u_B/c] \sim
S^{\{1\}}_B/m_BL_Bc \sim \omega_B L_B/c \ll 1$, 
describing the slowness of the rotational motion of the bodies. 
\end{enumerate} 

\item[{\small (3)}.]   \underline{\it The general relativistic   parameters.} 
The gravitational field produced by the bodies in the system may be 
characterized as follows: 
 
\begin{enumerate}
\item[(i).] $\eta_g \sim sup[\eta^B_g=c^{-2}Gm_B/D_B] = r_{g_B}/D_B\ll
1$,  which  describes 
 the weakness of the gravitational field outside the bodies; 
\item[(ii).] $\eta_l  \sim sup[\eta^B_l=c^{-2}Gm_B/ L_B] = r_{g_B}/L_B\ll
1$,  which  describes 
the  weakness of the gravitational field inside the bodies. 
\end{enumerate}
 
\item[{\small (4)}.]  \underline{\it The background induced   parameters.} 
For an isolated system, the absence of  initial inhomogeneity 
of  space-time caused by  in-fallen radiation, 
external gravitational sources or    cosmological evolution  may
be characterized by the parameters: 
 
\begin{enumerate} 
\item[(i).]  $h \sim  ||  g^{<0>}_{mn} - \gamma_{mn}||/(M_0/L_0)\ll 1$,
which describes the smallness of the  maximal deviation of the
background 
metric  $g^{<0>}_{mn}$ from the Minkowskian metric $\gamma_{mn}$
everywhere in the system. 

\item[(ii).] $ \sigma\sim \dot{h}/\omega_B\ll 1$,
 which  describes the quasi-stationary behavior of the   background
metric. 
\end{enumerate}
\end{itemize} 

  We shall assume  that any  processes in the system 
may be considered  to be adiabatic $(\sim$ {\small 1} ${\rm yr})$ in
comparison to the 
characteristic time-scale of the cosmological evolution of this
background space-time
 $(\sim$ {\small $10^{10}$} ${\rm yr})$ (as  described by the Robertson-
Walker solution). 
Moreover, asymptotic regions of the isolated {\small N}-body system are
presumed to be in a state of  free fall. This means   that the influence of
the rest of 
matter will be  the  universe on the {\it local} dynamics is of order  
{\small $10^{-24}$},
while the relativistic gravitational perturbations in the system are
expected 
to be in the range of {\small $10^{-5}-10^{-21}$} (Will, 1993).
With these expected accuracies,   the influence of the rest of the matter
in the universe
on the {\it local} dynamics of the bodies in the system may safely be
neglected.  
Let us denote this  background  space-time as $\gamma_{mn}$. Although in
the general case 
this background metric may have   arbitrary properties, for the case of an
isolated 
system of  astronomical bodies and for the {\small WFSMA}, one may take
this metric in the form 
of   space-time with a constant curvature or introduce 
flat Minkowski  space-time in the vicinity of the 
system under consideration.   These assumptions are necessary   in order
to 
justify the existence  of a barycentric asymptotically  inertial {\small
RF}.
 
With these assumptions and consequences, the dependence on the 
background-induced parameters $h$ and $\sigma$ in the 
corresponding    eq.m. of the  extended  bodies   may be neglected. 
The  equations in this case  may be schematically  
presented  as follows (Damour, 1987):
{}
$${d^2x_B\over dt^2}={\cal F}^B\big[\delta_g,\delta_l;
\epsilon_g,\epsilon_l;
\eta_g, \eta_l\big].\eqno(1.19)$$ 
\noindent This expression may be formally expanded with respect to
powers of the remaining small parameters, which may be given by  
{}
$${d^2x_B\over dt^2}=\sum_{k,l,m,n,p,q\ge0} {\cal F}^B_{klmnpq}\cdot
\delta^k_g\delta^l_l\epsilon^m_g\epsilon^n_l \eta^p_g
\eta^q_l.\eqno(1.20)$$  
\noindent Depending on the relations between the  parameters in any
particular 
problem, there   exist  several basic approximation methods. Our approach 
 uses an assumption of a the weak gravitational field  
inside and outside the bodies as well as an assumption about the slowness
of the 
dynamic  processes in the system.
For this case  some of the parameters introduced above are linked by
equalities or 
inequalities. Thus, the first relation may be written as:
$\eta^B_g=\delta^B_g\eta^B_l,$
which automatically gives $\eta^B_g\leq\delta^B_g\ll1 $ or for the 
entire system $\eta_g\leq\delta_g\ll1  $.
Since we are  considering a gravitationally bounded {\small N}-body
system in the 
{\small WFSMA}, there 
should exist  relations linked by the virial theorem: $v^2_B/c^2 \sim
r_{g_B}/D_B$ and 
$v^2_{osc}/c^2 \sim r_{g_B}/L_B$ (Fock, 1955; Chandrasekhar, 1965), 
such that the parameters $\epsilon_g$ and $\eta_g$ 
are equivalent and connected by the following  relation:
 $\epsilon^2_g \sim \eta_g$. 
The parameters $\epsilon^B_l$ and $\eta^B_l$ are different and vary
from body to body in the system. 
One may also limit  the behavior of matter  forming
the bodies such that `arbitrary bodies'  must have slowly changing 
internal mulipole moments:
 $\dot{I}^{\{K\}}_B/ I^{\{K\}}_B \sim \epsilon^B_l \cdot k \omega_B$, 
$\dot{S}^{\{K\}}_B \sim \eta^B_l I^{\{K\}}_B$.
By assuming this, we exclude from this analysis such  systems 
where the bodies are rapidly changing their multipole structure 
with   time. Fortunately all the celestial bodies in our 
solar system satisfy  these conditions. 

Moreover, each body studied in this report will be supposed be  
 isolated, {\it i.e.} the  immediate vicinity of the body is 
devoid of matter and non-gravitational fields, and the distance $D_B$ 
(the scale of homogeneity of the space-time) is
 large compared with the body's size $L_B$. For such an isolated body, one
may
 split space-time up into three regions as measured in 
body's `instantaneous' proper {\small RF}$_B$ (Misner {\it et al.}, 1973; 
Thorne \& Hartle, 1985; Kopejkin, 1988): the 
{\it   local region}, which  contains a world-tube surrounding the
body and extending out to some radius $r_l > L_B$; the {\it buffer region}
 extending from radius $r_l$ to some large radius 
$r_0 < D_B$; and the {\it external region} located outside 
the distance $r_0$. In the {\it local} region the body's own gravitational 
field dominates; but in the external region gravitational fields of 
other bodies  become important. The buffer region is 
placed in the vicinity of the distance $r^* \sim D_B(m_B/M_0)^{1/3}$ 
from the body which is defined from the
 condition that the body's gravitational influence is approximately 
equal to the gravitational influence of the external masses. 
The buffer region plays the role of an asymptotically flat 
space-time region for the gravitational field of the 
body in question. In the other words,  the total three-dimensional 
volume $V_N$,  which is occupied 
by the {\small N}-body system under study, may be split  into 
 {\small N} non-intersecting domains defined around each body 
in the system plus the buffer domain $d_0$. 
The situation is similar to that in the problem of study the 
stellar  stability of the solar system (Gladman \& Duncan, 1990; Holman
\& Wisdom, 1993). 
Within each  domain $d_B$ where the gravitational influence of a
particular body 
{\small (B)} is dominant  over  external gravity, 
  the orbits of  massless test particles will be  stable and remain well 
inside this domain.
In the buffer domain, the trajectories of  particles are unstable.
As a result,   the set of small parameters defined above, in the 
case of the {\it local} problem,  should be 
supplemented by  another  parameter, namely the 
parameter of geodesic separation $\lambda_B=|y_B|/D_B<1$,
where  $L_B\leq y_B\leq r^*$ is the distance from the 
world-line of the body {\small (B)} to the 
current point of interest inside the domain $d_B$. 
This interpretation  enables us to evaluate the   surface integrals
 at the boundaries  of these interacting domains as well as to define 
the boundary of validity of the expansions with respect 
to the small parameter  $\lambda_B$.

\subsection{Different Methods of  Constructing   the Proper RF.}

The   metric approach in the theories of gravity permits one to choose 
any {\small RF}  to  describe  the gravitational environment around the 
body under question. As we know, a poor choice of the new coordinates 
may cause unreasonable complications in the  physical interpretations 
of the data obtained  (see the related  discussion in Kopejkin, 1988, Soffel
\&
Brumberg, 1991).  
Recently, several different attempts were made to remove these
complications  and 
consequently improve the present solution to the  {\small N}-body problem 
in  the {\small WFSMA} 
(see, for example, Ashby \& Bertotti, 1984, 1986;
Brumberg \& Kopejkin, 1988a,b, Kopejkin, 1988, Klioner, 1993; DSX, 1991-
94). 
Although these methods represent a significant improvement 
in our understanding of the general problem, not one of them gives 
a complete `receipe' to overcome the difficulties stated above. 

The methods differ  in their the physical and mathematical 
  treatment of the three problems, which constitute the general problem
of 
motion of a gravitationally-bounded astronomical {\small N}-body system 
(the {\it global} and the {\it local} problems and the theory of the {\small
RF}s). 
One   such method  was proposed   by Bertotti (1954) and has been further
developed 
in a number of publications by Ashby and Bertotti (1984, 1986), Bertotti
(1986),
 Ashby and Shahid-Saless (1990),
Shahid-Saless, Hellings and Ashby (1991), and Shahid-Saless (1992). 
An equivalent method was proposed and developed to the extent of
practical 
applications by  Fukushima  (1988, 1991a,b, 1995a,b).
In these works the `good' proper {\small RF} is constructed within the
first 
post-Newtonian approximation {\small (1PNA)} of general relativity for a
specific form of the {\small EIH} metric (Einstein {\it et al.}, 1938). 
The {\small EIH} metric was obtained   in the inertial 
{\small RF}$_0$ and describes the gravitational 
field only outside the bodies, which may be regarded as massive point
particles  
or spherically-symmetric and non-rotating extended bodies (Fock, 1955).

In the Bertotti-Fukushima method the construction of the {\it local}
{\small RF} is
based upon  finding the background external metric for the body under
consideration.
The external metric is obtained from the complete {\small EIH} metric by
dropping all 
of the divergent or undefined terms on the body's center of inertia world-
line.
Then, a {\it local} Fermi-normal-like frame (Fermi, 1922; Manasse \&
Misner, 1963; Misner {\it et. al.}, 1973)
is defined in the body's vicinity using the background
 metric with respect to which the body moves along geodesic.
After that, the coordinate transformation between the Fermi frame and 
background metric is obtained. The transformation is applied
to the complete {\small EIH} metric and, thus, the `good' proper {\small
RF} is obtained. The body's
gravitational field in this proper {\small RF} is spherically-symmetric
(Schwarzchild) and the gravitational field of  distant bodies appears only
through the curvature 
tensor of the background metric, {\it i.e.} through the tidal effects.

The Bertotti-Fukushima  method is conceptually simple. It confirms our 
expectations that the physically adequate proper {\small RF} exists
and gives an insight into structure of the transformations (1.18).
However, this method of construction of the Fermi normal coordinates
for massive bodies has some  drawbacks (Kopejkin, 1988), namely:
\begin{itemize}
\item[(i).] The background external metric was not derived by  solving  the
gravitational field equations.
\item[(ii).] There are physical and mathematical ambiguities in the way of 
constructing the external metric.
These ambiguities are caused by the terms   describing the   back-action
of the gravitational 
field of the body under consideration on the external gravity produced by
other bodies  (Thorne \& Hartle, 1985).
\item[(iii).] The method under review cannot be used for derivations of the
eq.m. of bodies,
{\it i.e.} their world-lines. A choice of the body's center of inertia world-
line as a geodesic is
justified only {\it a posteriori} and with the help of quite a different
technique (EIH, 1938;
Papapetrou, 1948, 1951; Brumberg, 1972; Damour, 1983; Thorne \& Hartle,
1985; Kopejkin, 1985, 1987).
\item[(iv).] The method has been elaborated only for the special case of
spherically-symmetric and non-rotating bodies. It is completely unclear
how one might construct 
the Fermi normal coordinates in real astronomical situations which are
considerably more
complicated. This method is inapplicable even to the   Earth  itself which
has oblateness 
and rotation which may not be ignored (Kopejkin, 1988).
\item[(v).] The proposed coordinate transformations between the {\small
RF}s 
are incomplete, which  significantly limits the applicability of the results
obtained
in  real astronomical practice.  
\end{itemize}

An important   method  of construction of 
the `good' proper {\small RF} was proposed  by
Thorne and Hartle (1985) (see also Fujimoto \& Grafarend (1986)) and
developed 
 to some extent by  Zhang (1985, 1986) and Suen (1986). 
The Thorne-Hartle method is conceptually  elegant and 
has produced  the largest corrections to the geodesic law of motion and
Fermi-Walker law of transport
(Misner {\it  et al.}, 1973).
The method consists of determining  the metric tensor from the Hilbert-
Einstein
equations under the condition  that one satisfies the properties of the 
well defined  proper {\small RF} 
which were mentioned above.
Thus, the metric in this method is derived entirely in the `good' proper
{\small RF}.
The solutions of the gravitational field equations are searched for in a
vacuum region of space-time under 
de-Donder (harmonic) gauge conditions in the body's neighborhood  where
the gravitational field is weak. The 
metric tensor is represented in the form of an expansion in powers of 
the small parameters $m_B/r, r/R$ etc., where $m_B$ is the body's mass,
$r$ is a distance from the body
and $R$ is an inhomogeneity scale (distance between the bodies).
The coefficients of the expansion are the internal and external   multipole
moments of the gravitational fields
created both by the body under consideration  and the  external gravity,
respectively.
In this method the  information about the properties of the chosen 
{\small RF} is 
completely contained in the set of these multipole moments.

Although the Thorne-Hartle  method  represents an important progression
in our understanding 
of the motion of un-isolated bodies and their interaction with the external
universe, and 
provides an important insight into the 
physical structure of a  multipole expansion  of the metric tensor in  
different {\small RF}s, 
it cannot  be used immediately in the ephemeris astronomy. The main
reasons for this are as follows:

\begin{itemize}
\item[(i).] The finding of the solutions of the Hilbert-Einstein field
equations and the matching of the 
asymptotic expansions were done  formally. Since the goal of the paper
was   to find the largest 
corrections to the laws of motion and precession only, the method does
not provide  a
complete multipole treatment of  extended bodies. 
As a result, the internal multipole moments are not presented as integrals
over the volumes of the 
sources and therefore have no clear physical meaning  (Kopejkin, 1988).
\item[(ii).] The authors have not presented the 
coordinate transformation between the {\small RF}s used for the analysis.
They have constructed only the `instantaneous' proper {\small RF}
which coincides with the body's center of inertia at a particular moment
of time. As time goes on the 
origin of the `instaneneous' proper {\small RF} propagates along a
geodesic, but, in the general case,
 the body's center of inertia world-line does not. The deviation 
from the geodesic is caused by the interaction of the body's own intrinsic
multipole moments
with the external gravity. This   leads to  a drifting of the  `instantaneous'
proper
{\small RF} from the   body's  center of inertia which is not acceptable for 
 astronomical practice (Soffel \& Brumberg, 1991; Williams {\it et al.},
1991). 
\end{itemize}

Another  method of constructing of the `good' proper {\small RF}
was proposed by D'Eath (1975a,b) (see also papers by Kates 
(1980a,b) and Damour (1983)). These 
papers are devoted to the derivation of the eq.m. 
of compact strongly gravitating astrophysical objects  such as  black
holes and 
neutron stars. The authors have applied an interesting 
mathematical method of matched asymptotic expansions, which was not  
developed to be used  in  practical astronomical applications for the more
common 
case of weakly gravitating bodies.
There have been many  works in which construction of the `good' proper
{\small RF}
has been accomplished   with the help of  infinitesimal transformations 
(Fukushima {\it et. al.},  1986; Hellings, 1986; Vincent, 1986).
Unfortunately, the methods used in these works may not be considered to
be satisfactory since they 
are based  upon heuristic principles rather than    exact theory (Kopejkin,
1988).

The critical breakthrough in  construction of a relativistic theory of  
{\small RF}s
   appropriate  for astronomical practice  was achieved by 
Brumberg and Kopejkin (for detailed description see: Kopejkin, (1985,
1987, 1988); 
Brumberg \& Kopejkin,  1988a,b; Voinov, 1990; Brumberg, (1991a,b, 1992); 
Klioner \& Kopejkin, 1992; Brumberg {\it et al.},  1993; Klioner,  1993; 
Klioner \& Voinov, 1993). 
The relativistic theory developed by  Brumberg and Kopejkin combined the
basic ideas of Fock (1955) -
the post-Newtonian approximation scheme; Thorne (1980) and Thorne \&
Hartle (1985)
- multipole formalism, and D'Eath (1975a,b),  Kates (1980a,b), and Kates
\& Madonna (1982) 
D'Eath \& Payne (1992) - matched asymptotic expansions.

The Brumberg-Kopejkin  method was the first  to develop the three sub-
problems of the 
gravitationally-bounded astronomical N-body system. 
The authors identify   the metric tensor of the relativistic {\it global}
problem with the solution of an isolated  distribution of matter in the 
inertial {\small RF} obtained in the {\small 1PNA}
 of general relativity (Fock, 1955; Brumberg, 1972; Will, 1993).
The solution   of the {\it local} problem is formally presented as 
an   isolated  one-body solution corrected by   electric-type and 
magnetic-type external multipole moments (Thorne, 1980). The  form
of these moments reflect  the properties of the proper {\small RF} chosen
for the 
analysis of the gravitational environment of the body under study. 
The structure of these moments as well as the post-Newtonian coordinate 
transformations between the inertial and the quasi-inertial  {\small RF}s 
are derived by  matching   both solutions in the body's neighborhood.  

This  method demonstrates a notable progression in the theory of  
astronomical  relativistic  {\small RF}s developed to
describe the   motion of a system of {\small N} extended  bodies in the
{\small WFSMA}. 
However, this  method  also has some drawbacks:

\begin{itemize}
\item[(i).] The authors have made {\it ad hoc} assumptions about the 
various multipole expansions of the metric tensor and coordinate
transformations  which are 
only partially   justified by some later consistency checks (DSX, 1991). 
 
\item[(ii).] The method to derive the solution to the Hilbert-Einstein 
gravitational field equations of the general theory of relativity 
 based on the Anderson-DeCanio approach (Anderson \& DeCanio, 1975; 
Anderson, J. L. {\it et al.}, 1982)   is not covariant.
In particular,  based only on  this method it is not possible
to derive the explicit solution to these field equations  
in an accelerated proper {\small RF} linked to the body's center of inertia.
As a result,  the introduced `external' multipole moments do  
not have a clear physical meaning.

\item[(iii).] The  obtained  relativistic coordinate transformation between
the 
different {\small RF}s is incomplete as it contains only contributions 
from the leading intrinsic multipoles of the body 
(the mass monopole and dipole and the current dipole).  
The contributions from the other intrinsic multipoles are hopelessly
mixed with the 
external moments in the structure metric tensor of the {\it local}
problem. 
Thus, the  transformation does not   take into
account the non-linear coupling of the body's own gravitational field  to  
external gravity even at the Newtonian level. As a result, the origin  of the
proper {\small RF} 
coincides with the  center of inertia of the body   at  a particular  moment
in time only, 
and, as time goes on, they  will drift  apart.
\item[(iv).]  The method under review does not provide 
us with the necessary microscopic description of the
relativistic phenomena in terms of  densities of the gravitational fields. 
Thus, the mass of the bodies, the momentum and the angular momentum
was never 
explicitly defined. The parameters, introduced to substitute these
quantities were never checked
whether or not  they correspond to the integral conservation laws in the
proper {\small RF}s
of the bodies.
In addition, the  mass  density of the gravitational field in the {\it local}
region 
at the Newtonian level is given solely by the  body's own mass density. But
the {\it local} 
gravitational field contains  tidal terms due to the  external bodies.
As a result, the theory does not admit a special-relativistic treatment of
the {\small N}-body problem
in the sense of the mechanics of Poincar\'e.  
 
\end{itemize}

Recently, a very powerful approach to this problem has been elaborated by
Damour, Soffel and Xu
(DSX, 1991-1994), Blanchet {\it et al.} (1995) and Damour \&
Vokrouhlick\'{y} (1995). 
It combines an elegant (`Maxwell-like') treatise of the space-time
metric in both the {\it global} and   {\it local} {\small RF}s with the
Blanchet-Damour 
multipole formalism (Blanchet \& Damour, 1986). This approach  allows
one to relate  
the multipole expansions of the gravitational field to the 
structure of the source of gravitation.
This method, though very promising and attractive, still requires 
extensive development to make it useful for practical  
astronomical  applications. 
Besides this, the method under review has some problems which 
should be worked out in more physically-grounded way. These include:   

\begin{itemize}
\item[(i).]  The  Blanchet-Damour `external' multipole moments were
defined in the 
rest frame of an idealized isolated distribution of matter, so they must 
be modified in order to take into account the non-inertiality of the proper
{\small RF}
as well as the interaction of the body's proper gravitational field with 
external gravity.
 \item[(ii).] The  proposed   relativistic coordinate transformation
between the 
different {\small RF}s is incomplete  because it does not take into account
the 
terms due to  interaction of the body's own gravitational field with 
external gravity.
Moreover, the suggested coordinate transformation 
completely neglects the   precession term and  does not include the terms 
due to interaction of the body's intrinsic multipoles   with the 
external gravity. 
This means that the proper {\small RF} constructed with these
transformations 
in the case of monopole structureless particles does not end  up with a 
{\small RF} 
defined on a geodesic line which is guaranteed by the Principle of
Equivalence. 
 It should be noted that 
the origin of the proper {\small RF}, in the  general case of   extended
bodies, 
coincides with the center of inertia of the 
{\it local} field  in the initial moment of time only, and it  drifts away  as
time progresses. 
 This leaves  the  quantities,
calculated with respect to   such a proper  {\small RF},   physically  ill-
defined
(Damour \& Vokrouhlick\'{y}, 1995).
 
\item[(iii).] The solutions of the Hilbert-Einstein equations in the
different {\small RF}s 
were obtained using non-covariant gauge conditions.  This does not provide 
one  with a clear understanding of what part of the solution of the {\it
local}
problem is due to the gravitational field, what is caused by the 
contribution of the inertial sector of the space-time, and how 
these two  interact with each other.

\item[(iv).] At this time, the method under review may not be extended 
for analysis
of the {\small WFSMA} of other metric theories of gravity. 
\end{itemize}

In light of this, the principle purpose of the present report is to develop a
classic field  approach to the problems of   astronomical measurements
in the {\small WFSMA} of a number of modern metric theories of gravity.  
This approach  will combine the well-established 
methods of the relativistic mechanics of Poincar\'e   with the 
Fock-Chandrasekhar  treatment of  the relativistic 
many-extended-body gravitational problem (Fock, 1955, 1957;
Chandrasekhar, 1965).  
One of the main goals of this research was to develop 
a foundation for extending  the applicability of the {\small PPN}
formalism, which has become a  very useful   framework for testing the 
metric theories of gravity.

 \section{Parametrized Post-Newtonian Metric Gravity.}

In this Section we will discuss the  status of the 
problem of constructing a  solution to the gravitational field equations 
for a gravitationally bound  astronomical {\small N}-body system. 
Within the accuracy of modern experimental 
techniques, the   {\small WFSMA}  
provides a useful starting point for testing the
 predictions of different metric
theories of gravity in the solar system. 
Following Fock (1955, 1957),  the perfect fluid 
is used most frequently   as the model of   matter distribution 
when describing   the   gravitational behavior of 
celestial bodies in this approximation.
The density of the corresponding energy-momentum  tensor
$\widehat{T}^{mn}$ 
 is as follows:
{}
$$\widehat{T}^{mn}=\sqrt{-g}\Big(\Big[\rho_{0}( 1 + \Pi) + 
p\Big]u^{m}  u^{n}  - p g^{mn} \Big), \eqno(2.1)$$ 

\noindent where $\rho_0$  is mass density of the ideal fluid in
 coordinates of the co-moving   {\small RF}, $u^k = {d z^k / d s}$ 
are the components of invariant four-velocity
of a fluid element, and $p(\rho)$ is the isentropic pressure 
connected with $\rho$ by an equation of state. The quantity $\rho\Pi$ is
the 
density of internal energy of an ideal fluid. The  definition of $\Pi$ is
 given by the equation based on the  first law of thermodynamics 
(Fock, 1955; Chandrasekhar, 1965;  Brumberg, 1972; Will, 1993):
{}
$$u^n\Big(\Pi_{;n}  + p\Big({1\over {\widehat \rho}}\Big)_{;n}\Big) =
0,\eqno(2.2)$$

\noindent where ${\widehat \rho}=\sqrt{-g}\rho_0u^0$ is the conserved
mass density. 
Given the energy-momentum tensor, one may proceed   to find  the
solutions of
the gravitational field equations for a particular relativistic theory of
gravity.  
The solution for an astronomical   {\small N}-body problem is the one of
most  practical interest. 
In the following  Sections we will discuss the  properties of the  solution
of 
an isolated one-body problem as well as the features  of construction of
the 
general solution for  the {\small N}-body problem in both  barycentric 
and planeto-centric {\small RF}s. 

\subsection{An Isolated One-Body Problem.}

The solution for the isolated one-body problem in  the {\small WFSMA} 
may be obtained from the linearized gravitational field equations
of a particular theory under study. 
As we mentioned above, a perturbative gravitational field $h_{(0)}^{mn}$   
in this case  is characterized by the deviation of the density of the 
general Riemmanian metric tensor $\sqrt{-g}g^{mn}$ from the background 
pseudo-Euclidian space-time $\gamma_{mn}$, which is considered to be a
zeroeth 
order\footnote{For  most  non-radiative problems  in   solar system 
dynamics, this tensor usually is  taken to be a  Minkowski  metric
(Damour, 1983, 1987; Will, 1993).}  approximation 
for the series of the  successive iterations:
$\sqrt{-g}g^{mn}-\sqrt{-\gamma}\gamma^{mn}=h_{(0)}^{mn}$, or
equivalently: 
{}
$$g_{mn}=\gamma_{mn}+h^{(0)}_{mn}. \eqno(2.3) $$ 

\noindent The search for the 
solution of the field equations is performed within a barycentric 
inertial {\small RF}$_0$ $(x^p)$ which is singled out by the 
Fock-Sommerfeld's boundary conditions   imposed on 
the $h^{(0)}_{mn}(z^p)$ and $\partial_k h^{(0)}_{mn}(z^p)$
(Fock, 1955; Damour, 1987; Will, 1993): 
{}
$$\lim_{{r \rightarrow \infty} } 
\Big[ h^{(0)}_{mn}(x^p); \hskip 2mm r \Big({\partial 
\over \partial x^0} h^{(0)}_{mn}(x^p)+ 
 {\partial \over \partial r} h^{(0)}_{mn}(x^p)\Big)\Big] 
\hskip 2mm \rightarrow \hskip 2mm 0, \eqno(2.4)$$
{}
$$ x^0 + r  = {\rm const}, \qquad r^2= -\gamma^{(0)}_{\mu\nu} x^\mu  x^\nu.
$$

In order to accumulate the features of many modern metric theories of 
gravity in one  theoretical scheme,  
to create a  versatile mechanism to plan   gravitational 
experiments, and  analyze the 
data obtained,  Nordtvedt and Will have proposed 
a parameterized post-Newtonian 
({\small PPN}) formalism (Nordtvedt, 1968a,b; 
Will,  1971; Will \& Nordtvedt, 1972).  
This formalism allows one to describe  the motion of 
  celestial bodies  for a wide class of metric 
theories of gravity within a
common framework.  The gravitational field in the  {\small PPN} 
formalism is
presumed to be generated by some isolated distribution of matter 
which is taken to be an ideal fluid (2.1). This field is represented by the
sum of  
gravitational potentials with   arbitrary coefficients:  the 
  {\small PPN}  parameters. The two-parameter  form of this  
tensor in four dimensions   may be written as follows:

$$h^{(0)}_{00}= -2U+2(\beta-\tau) U^2+2\Psi+ 2\tau(\Phi_2-\Phi_w)+
(1-2\nu) \chi,_{00}+{\cal O}(c^{-6}), \eqno(2.5a)$$
{}
$$h^{(0)}_{0\alpha}=(2\gamma+2-\nu-\tau)V_\alpha+ 
(\nu+\tau)W_\alpha+{\cal O}(c^{-5}),\eqno(2.5b)$$
{}
$$h^{(0)}_{\alpha\beta}=2\gamma_{\alpha\beta} (\gamma-\tau) U -
2\tau U_{\alpha\beta}+  {\cal O}(c^{-4}),\eqno(2.5c)$$

\noindent where $\gamma_{mn}$ is the Minkowski 
metric\footnote{Do not mix the post-Newtonian parameter $\gamma$ and  
the Minkowski metric tensor $\gamma_{mn}$. As necessary we will
distinguish the 
determinant $det ||\gamma_{mn}||$ with the special symbol.}. 
The generalized gravitational potentials are given
in  Appendix A.

Besides the  two Eddington  parameters 
$(\gamma,\beta)$, eq.(2.5) contains   two other parameters  $\nu$ and
$\tau$.  
The  parameter $\nu$ reflects the specific
choice of   gauge conditions. For the standard   
 {\small PPN}  gauge it is  given as  $\nu= {1\over 2}$, but for harmonic
gauge 
conditions one should choose $\nu=0$. The parameter $\tau$ 
describes  a possible pre-existing anisotropy of  space-time and  
corresponds to different  spatial coordinates  which may be  chosen for 
modelling the experimental situation. For example, the case $\tau=0$
corresponds to 
harmonic coordinates, while $\tau=1$ corresponds to the standard
(Schwarzschild) 
coordinates. A particular metric theory of gravity in this framework
with a  specific coordinate  
gauge $(\nu,\tau)$ may then be characterized by  means of 
 two above said  {\small PPN}  parameters $(\gamma, \beta)$, which are
uniquely prescribed 
for each particular theory under study.  
In the  standard  {\small PPN}  gauge 
(i.e. in the case when $\nu={1\over 2}, \tau=0$) these  parameters have
clear 
physical meaning.
The parameter $\gamma$ represents the measure of the curvature of the
space-time 
created by the unit rest mass; the parameter $\beta$ is the measure of
the 
non-linearity of the law of superposition of the gravitational fields in the
theory of gravity (or the measure of the metricity). 
Note that general relativity, when analyzed in  standard 
 {\small PPN}  gauge, gives: $\gamma=\beta=1$,
whereas, for  the Brans-Dicke theory, one has 
$\beta=1, \gamma= {1+\omega\over 2+\omega}$, where $\omega$ 
is an unspecified dimensionless parameter of the theory.

The properties of an isolated  one-body solution  are well-known. 
It has been shown (Lee {\it et al.}, 1974; Ni \& Zimmerman, 1978; Will,
1993)
that for an  isolated distribution of matter in  {\small WFSMA}   
 there exist  a set of inertial {\small RF}s  and ten integrals
of motion  corresponding to ten conservation laws. Therefore, 
it is possible to consistently define  the multipole moments
characterizing the
body under study. For  practical purposes one   chooses the inertial 
{\small RF} located in the center of mass of an 
isolated distribution of matter. By performing a power expansion 
of the potentials in terms of
spherical harmonics, one may obtain  the  
post-Newtonian  set of   `canonical' parameters (such as  unperturbed
irreducible 
mass $I^{\{L\}}_{A(0)}$ and current $S^{\{L\}}_{A(0)}$ 
multipole moments),
generated by the inertially moving extended 
self-gravitating body  {\small (A)} under consideration:
{}
$$I^{\{L\}}_{A(0)}=\Big[\int_A d^3z'_A {\hat t}^{00}_A(z'^p_A) 
z'^{\{L\}}_A\Big]^{\it STF}
\hskip -15pt, \qquad \hskip 10pt
 S^{\{L\}}_{A(0)} =\Big[\epsilon^{\mu_1}_{\hskip 4pt\beta\sigma}
\int_Ad^3z'_A 
z'^\beta_A{\hat t}^{0\sigma}_A(z'^p_A) 
z'^{\mu_2}_A...z'^{\mu_l}_A\Big]^{\it STF}\hskip -15pt, \eqno(2.6a)$$  
\noindent where ${\hat t}^{mn}_A$ are the components of the symmetric
density
of the energy-momentum tensor of matter and gravitational field taken
jointly.
As a result, the corresponding gravitational field $h_{(0)}^{mn}$ 
may be uniquely represented  in the external domain  as a functional of the
set of these moments. 
Schematically this may be expressed as:
{} 
$$h_{(0)A}^{mn}={\cal F}^{mn}\big[I^{\{L\}}_{A(0)},
S^{\{L\}}_{A(0)}\big],\eqno(2.6b)$$
\noindent  where the functional
dependence, in general, includes a non-local time dependence on the `past' 
history\footnote{Gravitational radiation problems are not within  
the scope of the present paper, and hence the  set of multipole moments
$(2.6a)$ are
used for both tensor and scalar-tensor theories.} of the 
moments (Blanchet {\it et al.}, 1995). However, by assuming   
that the  internal processes in the body are adiabatic, one may neglect
this non-local 
evolution. As a result,   an external observer may 
uniquely establish the gravitational field of this body through  
determination of these multipole moments, for example,  by studying the
geodesic 
motion of the test particles in orbit around this distribution of matter
(Misner {\it et al.}, 1973).

\subsection{The Limitations of the Standard PPN Formalism.}

It turns out that   the   generalization of the  results 
obtained for the one-body problem into a solution of the problem of motion
of 
an arbitrary {\small N}-body system is not   quite straightforward. 
Thus, the studies of the post-Newtonian motion of the extended 
bodies in the {\small PPN} formalism  begin  
by expanding the generalized gravitational potentials in the 
metric tensor and the corresponding eq.m of these bodies with respect 
the  deviation from   Newtonain motion. As a final result,  one needs 
to have the generalization of the expression $(2.6b)$ on the 
case of the  {\small N}-body problem. However, this 
generalization  is usually done by 
using  Galilean coordinate transformations  similar to those of (1.2) 
from the Newtonian mechanics (Fock, 1955; Will, 1993):
{}
$$x^0=y^0_B+{\cal O}(c^{-2}), \qquad  
x^\alpha=y^\alpha_{B_0}(y^0_B)+ y^\alpha_B+{\cal O}(c^{-2}), \eqno(2.7)$$
\noindent where $y^\alpha_{B_0}$ is the Newtonian  barycentric radius-
vector of the 
body {\small (B)} under study.
It was noted  that this accuracy is enough for the post-Newtonian terms
in 
these eq.m. (Brumberg, 1972),  but it is insufficient to account for 
the necessary special relativistic and the gravitational corrections. 
Thus, as we know, if the body is   spherically-symmetric in the proper 
{\small RF}, 
in the other frame it  will  experience both the Lorentzian  
contraction (linked to the relative  velocity between these frames)  
in the  direction of velocity between these {\small RF}s, and
gravitational  compression (or `Einsteinian' contraction which is linked
with the 
external gravity) (Kopejkin, 1987).  However,   the transformations (2.7) 
are ignored completely these Lorentzian and 
gravitational contractions, as well as the relativistic geodetic precession
and effects of the 
curvature of space-time. All these kinematic  and dynamic  effects  
appeare in 
the expressions for the metric tensor 
and eq.m. of the {\it local } problem, where they are shown as the terms
depending on  
both (i) the `absolute' velocity of the body's center of inertia with respect
to the 
barycentric inertial {\small RF}$_0$, and (ii) the absolute value and first
spatial
derivative of the external gravitational potential $U^{\rm ext}$. 
As a result, the relativistic eq.m. of the {\it local} problem differ
essentially 
from the Newtonian eq.m., which do not depend on the `absolute' velocity 
and contain only the second spatial derivative of $U^{\rm ext}$, {\it i.e.}
the tidal terms.  
The correct way to describe these 
phenomena is to use the appropriate coordinate transformations between
the 
different {\small RF}s in the {\small WFSMA}. These transformations
should generalize 
the expressions of the Poincar\'e group of motion (1.7) for  the problem of 
motion of the gravitationally bounded {\small N}-extended-body system.
However,  the 
standard {\small PPN} formalism was formulated once 
in the inertial {\small RF} and there is no way to  
 construct such a transformation for the  quasi-inertial 
proper frames of the bodies. This lack of transformation between 
the different {\small RF}s is  major 
limitation of this otherwise very useful  method.

Nevertheless, by the putting some additional restrictions on the shape and
internal structure of the 
bodies, one may generalize   the results presented above  in the case of an
 {\small N}-body system. The  assumption that the bodies posses only the
lowest
 multipole mass moments  considerably simplifies the problem. 
 It has been shown (Fock, 1955; Lee, Lightmann \& Ni, 1974; Ni \&
Zimmerman, 1978)
 that for an  isolated distribution of matter in  {\small WFSMA}   
it is possible to consistently define the lowest conserved 
multipole moments  such as the total rest mass of  the  
system $M_0$, it's center of mass $z^{\alpha}_0$, 
momentum $p^{\alpha}_0$ and the total angular momentum 
$S^{\alpha\beta}_0$ of the system. The definitions for the mass 
$M_0$ and coordinates of the center of mass of the
 body $z^{\alpha}_0$  in any inertial 
{\small RF} are given by the following formulae (for a more 
detailed analysis see Damour (1983) and  Will (1993)
 and references therein):
{}
$$M_0 = \int  d^3 x'  \hskip 1mm{\hat t}^{00}(x'^p),
\qquad z^{\alpha}_0(t) =  {1\over M_0 }  \int  d^3 x' 
\hskip 1mm  {\hat t}^{00}(x'^p)x'^{\alpha}, \eqno(2.8a)$$ 
\noindent where the energy density ${\hat t}^{00}(x'^p)$ of 
the matter and the gravitational field is given by:
$$  {\hat t}^{00}(x^p)= \widehat{\rho} \Big[ 1 +c^{-2}\Big(\Pi - {1\over2} 
U -   
{1\over2}v_\mu v^\mu\Big)+{\cal O}(c^{-4}) \Big], \eqno(2.8b)$$  
\noindent with $\widehat{\rho}$ being the conserved mass density.
In particular,  the center of  mass $z^{\alpha}_0$  moves in 
space with a constant velocity along a straight line:
 $z^\alpha_0(t) = p^\alpha_0 \cdot t + k^\alpha$,
where the constants $p^\alpha_0  = d z^{\alpha}_0 /d t$ 
and $k^\alpha_A$ are the body's momentum and center of 
inertia, respectively. Moreover, is was shown 
by Chandrasekhar \& Contopulos (1967) that, in the case of  point-like
massive 
particles, the  form of the metric tensor (2.5) and the corresponding
{\small EIH} 
eq.m.,  are invariant under the coordinate transformations (1.12). 
This form-invariancy   justifies the word `inertial' for harmonic {\small
RF}s 
constructed under the Fock-Sommerfeld boundary conditions (2.3). One
may choose from the set of 
inertial {\small RF} the barycentric inertial {\small RF}$_0$
for such a system. In this frame, the functions 
$z^\alpha_0$ must  equal zero for any moment of time. 
  This condition may be satisfied by applying the post-Galilean 
transformations (1.12) to the metric
(2.5),  where  the constant velocity  and displacement of the origin  
should be selected in a such a way that $p^\alpha_0$ and $k^\alpha$ 
equal zero (for details see Kopejkin, 1988; Will, 1993). 
The solar system barycentric {\small RF}$_0$, constructed using  
general relativity for the system of  point-like massive particles, 
is widely in use in   modern astronomical practice, for example, 
in the  construction of the planet ephemerides 
(Moyer, 1971; Lestrade and Chapront-Touze', 1982; Newhall {\it et al}.,
1983;
 Akim {\it et al}., 1986; Standish, 1995). Moreover, the coordinate time of 
the solar system barycentric (harmonic) {\small RF}$_0$  must be
considered as 
the {\small TDB} time scale, 
which is extensively used in modern astronomical practice (Fukushima,
1995a). 

\subsubsection{The Simplified   Lagrangian Function 
of an Isolated  $N$-Body System.}

In order  to extract the 
information about the gravitational field of an {\small N}-body system one
should
 study the motion of  light rays and   test bodies  in this gravitational
environment.
However, the standard methods of the {\small PPN} formalism (Will,
1993) do not 
enable  us to develop the correct theoretical model of the  
astrophysical measurements with the accuracy necessary to identify  the   
multipolar structure of the gravitational fields of the bodies. 
In particular, it was noted that taking into account the presence of any 
non-vanishing internal multipole moments of an extended body
significantly changes  its equations of motion due to the  
coupling of these intrinsic multipole moments of the body to the 
surrounding gravitational field. 
For example,  for a   neutral monopole test particle, the external 
gravitational field  completely defines the  feducial geodesic 
world-line  which this test body follows (Fock, 1955; Will, 1993). 
On the other hand, the equations of motion for   spinning  bodies   
contain an additional terms due to the coupling of the  body's spin  
to the external gravity through the Riemann curvature tensor
 (Papapertou, 1948, 1951; Barker \& O'Connel, 1975). 

An `absolute' limit of the {\small PPN} formalism  takes into account the 
lowest multipole moments of the bodies only,  such as the rest 
mass $m_A$ of the body {\small (A)}, its intrinsic spin moment 
$S^{\alpha\beta}_A$ and the quadrupole moment $I^{\alpha\beta}_A$. 
The general solution   with such  assumptions is  also known 
(see Damour (1986, 1987) and  references therein; and Turyshev (1990)).
In order to analyze the  motion of bodies in the solar system 
barycentric {\small RF}$_0$, one may obtain the restricted Lagrangian
function $L_N$ 
describing the motion of  {\small N}  self-gravitating bodies, which
 may be presented  as follows:
{} 
$$L_N=\sum_A^N{m_A\over 2}{v_A}_\mu{v_A}^\mu 
\Big(1-{1\over 4}{v_A}_\mu{v_A}^\mu\Big)- 
\sum_A^N\sum_{B\not=A}^N{m_A m_B\over r_{AB}}\Bigg({1\over2}
+(3+\gamma-4\beta)E_A-$$
{}
$$-(\gamma-\tau+{1\over 2}){v_A}_\mu {v_A}^\mu +
(\gamma-\tau+{3\over 4}){v_A}_\mu {v_B}^\mu -({1\over 4}+\tau)
{n_{AB}}_\lambda {n_{AB}}_\mu 
{v_A}^\lambda {v_B}^\mu+\tau ({n_{AB}}_\mu  {v_A}^\mu)^2+$$
{}
$$+{{n_{AB}}_\lambda\over r_{AB}} \Big[(\gamma+{1\over 2}){v_A}_\mu-
 (\gamma+1){v_B}_\mu\Big] S_A^{\mu\lambda} + {n_{AB}}_\lambda 
{n_{AB}}_\mu {I_A^{\lambda\mu}\over r_{AB}^2}\Bigg)+
(\beta-\tau-{1\over 2}) \sum_A^N m_A\Big(
\sum_{B\not=A}^N {m_B\over r_{AB}}\Big)^2- $$
{}
$$-\tau\sum_A^N\sum_{B\not=A}^N \sum_{C\not=A,B}^N
m_A m_Bm_C\Big[ {{n_{AB}}_\lambda\over 2r_{AB}^2}
(n_{BC}^\lambda+n_{CA}^\lambda)-{1\over r_{AB}r_{AC}}\Big]+
\sum_A^N m_A {\cal O}(c^{-6}), \eqno(2.9) $$ 

\noindent  where $m_A$ is the isolated rest mass of a body {\small
$(A)$},
the vector $r_A^\alpha$ is the barycentric radius-vector of this 
body, the vector $r_{AB}^\alpha=r_B^\alpha -r_A^\alpha$ is the 
vector directed from body {\small $(A)$} to body  {\small $(B)$}, and
the vector $n^\alpha_{AB}=r^\alpha_{AB}/r_{AB}$ is the usual notation 
for the unit vector along 
this direction. It should be noted that the expression (2.9) 
does not depend on the parameter $\nu$, which confirms that 
this parameter is the gauge parameter only. 
The tensor $I_A^{\mu\nu}$ is  the {\small STF} (Thorne, 1980) tensor of
the 
reduced quadrupole moment of 
body {\small $(A)$}  defined as:
{}
$$I_A^{\mu\nu}={1\over 2m_A}\int_A d^3z'_A\widehat{\rho}_A(z'^p_A) 
\Big(3z'^\mu_A z'^\nu_A-\gamma^{\mu\nu}z'_{A\beta} z'^\beta_A \Big). 
\eqno(2.10)$$

\noindent  The tensor $S_A^{\mu\nu}$ 
is the body's reduced intrinsic  {\small STF} spin moment  which is given
as:
{}
$$S_A^{\mu\nu}={1\over m_A}\int_A  d^3z'_A \widehat{\rho}_A(z'^p_A)
\big[{\underline v}_A^\mu z'^\nu_A-{\underline v}_A^\nu z'^\mu_A\big],
\eqno(2.11)$$

\noindent where ${\underline v}_A^\mu$ is the 
velocity of the intrinsic motion of matter in the body $(A)$. Finally, the 
quantity $E_A$ 
is the body's gravitational binding energy:
{}
$$E_A  = {1\over 2m_A}{\int\int}_A
 d^3z'_A d^3z''_A  {\widehat{\rho}_A(z'^p_A) 
\widehat{\rho}_A(z''^p_A)\over|z'^\nu_A-z''^\nu_A|}. \eqno(2.12)$$

Let us note that the Lagrangian function is obtained with 
condition that, in the proper {\small RF} of each body in the system, their
dipole mass moments
 vanish:
{}
$$I^{\{1\}}_A\equiv m^\alpha_A=\int  d^3z'_A {\hat t}^{00}_A(z'^p_A)
z'^\alpha_A=0, \eqno(2.13a)$$
\noindent where ${\hat t}^{00}_A$ defined by the following expression:
$$  {\hat t}^{00}(z^p_A)= \widehat{\rho}_A\Big[ 1 +c^{-2}\Big(\Pi -
{1\over2}  U_A -   
{1\over2}{\underline v}_\mu {\underline v}^\mu\Big)
+{\cal O}(c^{-4}) \Big]. \eqno(2.13b)$$  
\noindent The expression $(2.13a)$, together with the condition
${\dot m}^\alpha_A=0$, may be considered as an  indirect  post-Newtonian 
 definition of the proper {\small RF}$_A$ in the {\small PPN} formalism.

\subsubsection{The Simplified Barycentric Equations of Motion.}
In this part we will present the   barycentric equations of motion which  
follow  from 
the Lagrangian (2.9). The  assumption that bodies in the system possess
the 
lowest intrinsic multipole moments only  enables us to
obtain the corresponding simplified equations of motion. Thus, with the
help of the 
expressions  (2.9), for an arbitrary  body {\small (A)} these equations will
read as follows: 
{}
$$\ddot{r}^\alpha_A =\sum_{B\not=A} 
{{\cal M}_B \over r_{AB}^2}{\widehat{n}_{AB}}^\alpha +
\sum_{B\not=A} {m_B\over r_{AB}^2}\Big[{\cal A}^\alpha_{AB}+
{{{\cal B}^\alpha_{AB}}\over r_{AB}}+ {{\cal C}^\alpha_{AB} \over
r_{ab}^2}-$$
{}
$$-{n_{AB}^\alpha\over r_{AB}}\Big((2\beta+2\gamma-
2\tau+1)m_A+(2\beta+2\gamma-2\tau)m_B\Big)\Big]+$$
{}
$$+ \sum_{B\not=A}\sum_{C\not=A,B}m_Bm_C 
{\cal D}^\alpha_{ABC}+{\cal O}(c^{-6}), \eqno(2.14)$$

\noindent where, in order to account for the influence of the  
gravitational binding energy $E_B$, 
we have introduced the passive gravitational rest mass ${\cal M}_B$ 
(Nordtvedt, 1968b; Will, 1993)  as follows 
{}
$${\cal M}_B = m_B\Big(1+(3+\gamma-4\beta)E_B+{\cal O}(c^{-4})\Big).
\eqno(2.15)$$

\noindent The unit vector  $n_{AB}$ must also be   
corrected using the gravitational binding energy
and the  tensor of the quadrupole moment $I^{\alpha\beta}_A$ of the body
{\small (A)}
under question:
{}
$$\widehat{n}^{\alpha}_{AB} = n^{\alpha}_{AB}\Big(1+(3+\gamma-
4\beta)E_A+
5{n_{AB}}_\lambda {n_{AB}}_\mu {I^{\lambda\mu}_A\over r_{AB}^2}\Big)+
2{n_{AB}}_\beta {I^{\alpha\beta}_A\over r_{AB}^2}+{\cal O}(c^{-4}). 
\eqno(2.16)$$
 
\noindent The term ${\cal A}^\alpha_{AB}$ in the expression (2.14) 
is the orbital term,  which is given as follows:
{}
$${\cal A}^\alpha_{AB}= v^\alpha_{AB}
{n_{AB}}_\lambda\Big(v^\lambda_A-
(2\gamma-2\tau+1)  v^\lambda_{AB}\Big)+$$
{}
$$+ n^\alpha_{AB}\Big({v_A}_\lambda v^\lambda_A-
(\gamma+1+\tau) {v_{AB}}_\lambda v^\lambda_{AB}-
3\tau ({n_{AB}}_\lambda {v_{AB}}^\lambda)^2-
{3\over 2}({n_{AB}}_\lambda v^\lambda_B)^2\Big). \eqno(2.17)$$

\noindent The spin-orbital term $ {\cal B}^\alpha_{AB} $ has the form: 
{}
$$ {\cal B}^\alpha_{AB} =({3\over 2}+2\gamma){v_{AB}}_\lambda
(S^{\alpha\lambda}_A+S^{\alpha\lambda}_B)+{1\over2}{v_A}_\lambda
(S^{\alpha\lambda}_A-S^{\alpha\lambda}_B)+$$
{}
$$ +{3\over
2}(1+2\gamma){n_{AB}}_\lambda{v_{AB}}_\beta\Big[n_{AB}^\beta 
(S^{\alpha\lambda}_A+S^{\alpha\lambda}_B)-n^\alpha_{AB}  
(S^{\beta\lambda}_A+S^{\beta\lambda}_B)\Big]+$$
{}
$$+{3\over 2}{n_{AB}}_\lambda \Big[n_{AB}^\alpha 
({v_{A}}_\beta S^{\beta\lambda}_B-{v_B}_\beta S^{\beta\lambda}_A) + 
{n_{AB}}_\beta{v_{AB}}^\beta S^{\alpha\lambda}_B\Big]. \eqno(2.18)$$

\noindent  The term ${\cal C}^\alpha_{AB}$ is 
caused by the oblateness of the bodies in the system: 
{}
$$ {\cal C}^\alpha_{AB} =2{n_{AB}}_\beta I^{\alpha\beta}_B +
5n_{AB}^\alpha{n_{AB}}_\lambda {n_{AB}}_\mu I^{\lambda\mu}_B.
 \eqno(2.19) $$

\noindent And, finally, the contribution ${\cal D}^\alpha_{abc}$ 
to the equations of motion (2.14) of  body {\small (A)} (caused by the 
interaction of the other planets {\small (B}$\not=${\small A,
C}$\not=${\small A,B)} 
 with each other) is presented as:
{}
$${\cal D}^\alpha_{ABC}={n^\alpha_{AB}\over 
r_{AB}^2}\Big[(1-2\beta){1\over r_{BC}}- 
2(\beta+\gamma){1\over r_{ac}}\Big]+$$
{}
$$+\tau{{\cal P}^{\alpha\lambda}_{AB}
\over r^3_{AB}}({n_{bc}}_\lambda+{n_{ca}}_\lambda)  
+\tau{{n_{AB}}_\lambda\over r_{AB}^2}
{\Lambda^{\alpha\lambda}_{AC}\over r_{AC}}+
{1\over 2}(1+2\tau){{n_{BC}}_\lambda\over r_{BC}^2}
{\Lambda^{\alpha\lambda}_{ac}\over r_{ac}}+
2(1+\gamma){{n_{BC}}^\alpha\over r_{BC}^2r_{AB}}, \eqno(2.20)$$

\noindent where
$\Lambda_{AB}^{\mu\nu}=\eta^{\mu\nu}+n_{AB}^\mu n_{AB}^\nu$  
and ${\cal P}_{AB}^{\mu\nu}=\eta^{\mu\nu}+3n_{AB}^\mu n_{AB}^\nu$
is the projecting and the polarizing operators respectievly.

The metric tensor (2.5), the Lagrangian function (2.9) and the equations 
of motion (2.14)-(2.20)  define  the  
behavior  of the celestial  bodies in the post-Newtonian 
approximation in the {\small PPN} formalism. 
These equations may be simplified considerably  by
taking into account that 
the leading contribution to these equations is the 
solar gravitational field. With such an approximation, they are used to
produce the numerical codes in relativistic orbit determination 
formalisms for planets and satellites (Moyer, 1981; Huang {\it et al.},
1990;
Ries {\it et al.}, 1991; Standish {\it et al.}, 1992) as well as to
analyze   the gravitational experiments in the solar system
(Will, 1993; Pitjeva, 1993; Anderson {\it et al.}, 1996).
It should be noted here that  in the present numerical  algorithms  
for celestial mechanics problems (Moyer, 1971; Moyer, 1981; Brumberg,
1991;
Standish {\it et al.}, 1992;  Will, 1993) the bodies in the solar system 
are assumed to posses  the lowest post-Newtonian mass  moments only,
namely:
the rest masses and the quadrupole moments. 
The corresponding  barycentric inertial {\small RF}$_0$ defined in the
harmonic coordinates 
for the general relativity $(\gamma=\beta=1; \nu=\tau=0)$ has been 
adopted for the fundamental
planetary and lunar ephemerides (Newhall {\it et al}., 1983; 
Standish {\it et al.}, 1992).   

However, if one attempts to describe the {\it global} dynamics of the 
system of {\small N} arbitrarily shaped extended bodies, one 
will discover that even in  {\small WFSMA}  
this solution will not be possible without an
appropriate description of the gravitational environment 
in the  immediate vicinity of the bodies (Kopejkin, 1988; DSX, 1991). 
Thus, one needs to present the post-Newtonian definition for the proper
intrinsic multipole moments for the bodies, in order to describe   their
interaction with the 
surrounding gravitational field as well as to obtain the corresponding 
corrections to the laws of motion and precession of the extended bodies in
this system.
This could be done correctly  only by using  the theory of the 
 quasi-inertial proper {\small RF} with well defined dynamic and
kinematic properties.
In the next Section we will discuss a new perturbative method for finding
  the solution for  the relativistic {\small N}-extended body problem and 
will formulate the corresponding theory of  relativistic astronomical
{\small RF}s
in  curved space-time.

\section{WFSMA for an Isolated Astronomical  N-Body System.}

In this Section we will discuss the principles of  a new iterative method
for 
generating the solutions to an arbitrary   {\small N}-body gravitational 
problem in the {\small WFSMA}. This formalism will be  based upon the
construction of   proper 
 {\small RF}s  in the vicinities of  each body in the system. 
Such  frames are defined in the gravitational domain $d_B$, 
occupied by a particular body {\small (B)}.  
One may expect that,  in the immediate vicinity of this body, its proper 
gravitational field will dominate, while  the existence of 
the external gravity will   manifest   itself 
in the form of the tidal interaction only. Therefore, in  the case of 
the {\small WFSMA} in the closest proximity to the 
body under study, this proper {\small RF}  should resemble   the properties
of an 
inertial frame and the solution  for an isolated one-body problem
$h_{mn}^{(0)B}$   
should adequately represent the physical situation. 
However, if one decides to perform a physical  experiment at some
distance from the world-tube
of the body, one should consider the   existence of the external gravity as
well. This is true  because 
 external gravity  plays a more  significant role at 
large  distances from the body and this should be taken into account.
As we noticed earlier in Section {\small 1}, the physically adequate 
description  of this nature of gravity could be 
made in the well justified proper {\small RF} only.  
Let us mention that the dynamical properties of the inertical frames 
presently are well justified and correctly  modelled both physically  
 and  mathematically.
In particular, the properties of the barycentric inertial {\small RF}$_0$   
are based upon the properties of an {\small N}-body generalization  
of an unperturbed isolated one-body solution  of the gravitational field
equations 
in an inertial {\small RF} given by (2.5). These properties are well
established and 
widely in use in modern astronomical practice (Moyer, 1971; Moyer, 1981;  
Brumberg, 1991; Will, 1993). However, as  we discussed earlier, this
{\small N}-body 
generalization is   based on the  assumption that the 
bodies in the system possess the lowest intrinsic mass and current 
multipoles only. In order to account for the influence of a higher order
multipoles, the coordinate transformations to the proper {\small RF} are
necessary.
This proper {\small RF} should take into account both  Lorentzian and 
Einsteinian features of the motion of the extended bodies in the 
external gravitational field.
In the next chapter we will concentrate on   formulating  the basic
principles 
of a new method for constructing 
such   transformations for a wide class of metric theories of gravity. 

\subsection{The General Form of the $N$-Body Solution.}

In order to construct a general solution for the {\small N}-body problem 
in a metric theory of gravity, let us  make a few assumptions. First of all, 
let us assume that there  exists a background space-time $\gamma_{mn}$
with the   dynamic  and cosmological properties discusses in the Section
{\small 1}. 
Note that these properties do not forbid the existence of  in-coming and
out-going  gravitational
radiation. We will discuss  this  case further. 
We shall assume that the  solution of the gravitational field equations
$h_{mn}^{(0)}$ 
for an isolated unperturbed distribution of matter  is known and   
is given by the relations (2.5). 
We further  assume that for each body {\small (B)} in the 
system, one may establish a unique correspondence  
to each such   solution: {\small (B)} \hskip 2pt $\Leftrightarrow \hskip
2pt h_{mn}^{(0)B}$. 

With these assumptions, we may construct the total solution of the 
{\it global} problem $g_{mn}$ in an arbitrary  {\small RF} as a formal
tensorial 
sum of the background space-time metric $\gamma_{mn}$, the  
 unperturbed solutions $h_{mn}^{(0)B}$ plus the gravitational interaction
term
$h^{\it int}_{mn}$. 
Thus,  in the coordinates $x^p\equiv (x^0, x^\nu)$ 
of the barycentric inertial {\small RF}$_0$, 
one  may search for the desired total solution in the following form:
{}
$$g_{mn}(x^p) =\gamma_{mn}(x^p) + h_{mn}(x^p) = $$
$$= \gamma_{mn}(x^p) + 
\sum_{B=1}^N {{\partial y^k_B}\over{\partial x^m}}
{{\partial y^l_B}\over{\partial x^n}}
\hskip 0.5mm h^{(0)B}_{kl}(y^q_B (x^p)) + 
h^{\it int}_{mn}(x^p),  \eqno(3.1)$$

\noindent where  the coordinate transformation 
functions $y^q_B = y^q_B (x^p)$ are yet  to be determined.
The interaction term $h^{\it int}_{mn}$ will be discussed below.

In order to describe the matter distribution, let us assume that the 
corresponding Lagrangian function  $L^{tot}_M$ may be given as 
{}
$$L^{tot}_M = \sum_B^N L_M^{(0)B}+L_M^{int},$$
\noindent where $L_M^{int}$ is the Lagrangian describing  
interaction between the bodies.
Then, the total energy-momentum tensor of   matter  in the 
system may be presented as follows:
{}
$$T_{mn}(x^s) = 2{\delta L^{tot}_M\over\delta g^{mn}}=
\sum_B^N{{\partial y^k_B}
\over{\partial x^m}}{{\partial y^l_B}\over{\partial x^n}}
T^{B}_{kl}(y_{B} (x^s)) +2{\delta L^{int}_M\over\delta g^{mn}}.\eqno(3.2)$$ 

\noindent 
For the case of    compact and well separated bodies, we may take into
account that
the mutual gravitational interaction  between the bodies affects their 
distribution 
of matter  through the metric tensor only.
 Therefore  we can neglect the second term in the 
expression above\footnote{It is also true, if one recalls the 
 result that the interaction between the gravitational fields 
in the 1.5 post-Newtonian physics will appear in the $g_{00}$ component
of the 
metric tensor only and will have an ${\cal O}(c^{-4})$  order of
magnitude.}. Then
without any loss of   accuracy, we   obtain the  total  energy-momentum
tensor 
of the matter distribution in the  system in the following form:     
{}
$$T_{mn}(x^s) = \sum_{B=1}^N {{\partial y^k_B}
\over{\partial x^m}}{{\partial y^l_B}\over{\partial x^n}}
T^{B}_{kl}(y_{B} (x^s))(1+{\cal O}(c^{-4}))  = \sum_{B=1}^N {{\partial y^k_B}
\over{\partial x^m}}{{\partial y^l_B}\over{\partial x^n}}
T^B_{kl}(y_B(x^s)), \eqno(3.3)$$
 
\noindent where $T^B_{kl}$ is the  energy-momentum 
tensor\footnote{As a partial result of the representation (3.3) one can see
that 
  the Newtonian mass density $\rho_B$ 
of a particular body {\small (B)}   is defined in a sense of a three-
dimensional 
Dirac delta-function. Thus in the body's proper compact-support 
volume one will have: $\rho_B=m_B \delta(y^\nu_B)$, so that 
$$\int_a d^3y'_A \rho _B(y'^p_A) = m_B \delta_{AB},$$
\noindent where $\delta_{AB}$ is the three-dimensional Kronekker symbol 
($\delta^A_B=\delta_{AB}$;  $\delta_{AB}=1$ for $A=B$ and $=0,$ for
$A\not=B$).
Then in any {\small RF}$_A$  the total density ${\overline\rho}$ of whole 
{\small N} body system 
will be given by the expression ${\overline\rho}(y^p_A)=\sum^N_B \rho
_B(y^p_A)$.
This representation allows one to distinguish between the {\it local} and
integral
descriptions of the physical processes and, hence, provide correct  
 relativistic treatment of the problem of motion of  an astronomical 
{\small N}-body system.} 
 of a body {\small (B)} as seen by a co-moving observer. 

The  unperturbed solution   $h^{(0)B}_{mn}$ for the field equations  
in the {\small WFSMA} is  presented in a form of the double power series
with respect to 
two small scalar parameters: the  gravitational coupling constant {\small
G} and 
  the orders of $c^{-1}$. It is clear that a similar set of  small 
parameters may be used in order to construct an  iterative {\small N}-
body
solution at least at the post-Newtonian level in {\small WFSMA}. 
This means  that all the functions and fields involved in 
the perturbation scheme (such as the interaction 
term $h^{\it int}_{mn}$, the coordinate transformation 
functions $y^q_{B} = y^q_{B} (x^p)$, the energy-momentum tensor 
$T^{B}_{mn}$, etc.) are also power-expanded with 
 respect to these small scalar parameters. 
 At this point the actual form of the 
energy-momentum tensor $T_{mn}$ is not of great importance. We prefer
to keep this arbitrariness 
in our further calculations. The only restriction we will apply to the
possible 
form of this tensor is based on the physical expectations: we will 
 limit ourselves to such   tensors which have  the components   of the
following orders: 
$T_{00}\sim{\cal O}(1), T_{0\alpha}\sim {\cal O}(c^{-1}),
T_{\alpha\beta}\sim{\cal O}(c^{-2})$.
 
One may establish the properties of the solution  (3.1) with respect to  an 
arbitrary coordinate  transformation  simply by applying the basic rules 
of  tensorial coordinate transformations.
In particular,  in   the coordinates $y^p_A(x^q)\equiv   (y^0_A, y^\nu_A)$ of
an
arbitrary   proper {\small RF}$_A$   this tensor  
 will take the following form:
{}
$$g^A_{mn}(y^p_A) = {{\partial x^k}\over{\partial y^m_A}} 
{{\partial x^l}\over{\partial y^n_A}} \hskip 1mm 
g_{kl}(x^s(y^p_A)) = \gamma^A_{mn}(y^p_A) + h^A_{mn}(y^p_A) = $$
$$ = {{\partial x^k}\over{\partial y^m_A}}{{\partial x^l}
\over{\partial y^n_A}}\hskip1mm\Big( 
\gamma_{kl}(x^s(y^p_A)) + h^{\it int}_{kl}(x^s(y^p_A)) \Big) +$$
$$+ h^{(0)A}_{mn}(y^p_A) + \sum_{B\not=A} {{\partial y^k_B}
\over{\partial y^m_A}}{{\partial y^l_B}\over{\partial y^n_A}}
\hskip 0.5mm h^{(0)B}_{kl}(y^{s}_B (y^p_A)). \eqno(3.4)$$
\noindent The expression for  $T_{mn}(y^p_A)$ 
could be  obtained analogously from that given by equations (3.3). 
 To complete the formulation of the perturbative
 scheme we need to introduce the procedure  for  constructing
 the solutions for the various  unknown 
functions entering  expressions (3.1)-(3.4), including the four functions of
the coordinate 
transformations  $y^q_B = y^q_B(y^p_A)$ and the interaction term $h^{\it
int}_{mn}$.

We will construct the four functions of the coordinate transformations 
 by applying the relativistic theory of  
celestial {\small RF}s  in a curved space-time. 
 To do this we will use 
the most general form of the post-Newtonian non-rotating coordinate
 transformation between the barycentrical (inertial) 
coordinates $(x^p)$ and   the bodycentrical (quasi-inertial)  coordinates  
$(y^p_A)$:
{}
$$x^0 = y^0_A + c^{-2}K_A(y^0_A, y^\epsilon_A) + 
c^{-4} L_A(y^0_A, y^\epsilon_A) + O(c^{-6})y^0_A, \eqno(3.5a) $$
$$x^\alpha =  y^\alpha_A + y^\alpha_{A_0} (y^0_A) + 
c^{-2}Q^\alpha_A(y^0_A, y^\epsilon_A) + O(c^{-4})y^\alpha_A, 
\eqno(3.5b)$$ 

\noindent 
where $y^\alpha_{A_0} (y^0_A)$ is the Newtonian radius-vector 
of body {\small (A)}. 
The transformations (3.5)  should
complement  the post-Galilean coordinates transformations 
(1.12) in the case of the curved space-time generated by an arbitrary
{\small N}-body system. 
Note that the  transformations (3.5) are  presented as being
parametrized by the set of three unknown functions $K_A, L_A,
Q^\alpha_A$. 
This  is an example of that  which will be referred to as the {\small KLQ}
parameterization for the {\small WFSMA}. 
The  functions $K_A, L_A, Q^\alpha_A$ are expected to contain the 
information about 
the specific properties of the quasi-inertial {\small RF}$_A$    associated
with the body {\small (A)}.
The form of these functions will be determined  
by   the   iterative procedure for constructing the quasi-inertial proper 
{\small RF}$_A$. 

The way to construct the solution for the interaction term 
$h^{\it int}_{mn}$ is quite straightforward: It is sufficient to 
require that the metric tensor in the form of eqs.(3.1) or (3.4) 
will be the explicit solution   of the gravitational field equations in the 
corresponding {\small RF}. 
Note that the second term in the equation eq.(3.1) is linear 
with respect to the unperturbed solutions 
$h^{(0)B}_{kl}$ and the transformation functions between the different 
{\small RF}s  are determined  by the 
means of the external gravitational field in their origins.  
Only the interaction term   should contain the information about the 
dynamic  non-linearity of the gravitational interaction. The 
form of this term should  depend on the physical features of the  {\small
RF}s 
chosen for the analysis.
It should be noted that the search for the
solution in the barycentric {\small RF}$_0$ is  physically and
mathematically 
more appropriate then in the bodycentric one. 
Moreover,   to date   no analysis
 has been made  to propose a   covariant 
boundary condition  for the case of the non-inertial {\small RF} rather
then
 the `classical' Fock-Sommerfeld one. It is known that these conditions  
are applied to 
the entire gravitational field from the  system 
asymptotically at the infinitive distance from one and valid for the 
isolated distribution of matter. 
This means that making of use of the Fock-Sommerfeld
boundary conditions (Brumberg \& Kopejkin, 1988a; DSX,  
1991-1994) in a proper {\small RF}   is 
mathematically weakly founded, in order to find the general
 solution of the field equations  in this frame. Based on this 
conclusion we will perform the search for the $h^{\it int}_{mn}$ in the
coordinates of the barycentric 
inertial {\small RF}$_0$. 
 
By taking   into account that all the functions and fields 
in expressions  (3.1) - (3.5) are presented in the form of a 
power expansion with respect to the set of small parameters, 
one may   organize an iterative 
procedure in order to obtain the general solution for the problem.  
The two principle steps of   this procedure are the  
supplementary conditions necessary for the solution of the 
gravitational field equations, which may be expressed by both the 
covariant gauge conditions, and the   boundary conditions.

 In the proposed formalism these conditions are taken to be as follows:

  \underline{\it The covariant gauge conditions.} 
The  solutions  of the field equations are assumed to satisfy the 
covariant harmonical de Donder
gauge, which,  for an arbitrary {\small RF}$_B$,   may be written  
as follows: 
{}
$${\cal D}^B_{n}\Big(\sqrt{-g}_Bg^{mn}_B(y^p_B)\Big) = 0, \eqno(3.6)$$ 

\noindent 
where ${\cal D}^B_n$ is the covariant derivative with respect to the
metric 
$\gamma^B_{mn}(y^p_B)$  of the inertial Riemann-flat 
$\big(R^k_{nml}(\gamma^B_{mn}(y^p_B))=0\big)$ space-time in these 
coordinates\footnote{In Cartesian coordinates of the inertial Galilean 
{\small RF}$_0$  the flat metric $\gamma^B_{mn}$ can be chosen as 
$\gamma^{(0)}_{mn} = \hbox{diag}(1,-1,-1,-1)$, so that the Christoffel
 symbols $\Gamma^{k(0)}_{mn}  = 0$ all  vanish  and  
conditions (3.6)   take the more familiar form 
of the harmonic conditions 
{}
$$\partial^B_{n}\Big({\sqrt{-g}}_Bg^{mn}_B(y^p_B)\Big) = 0,$$
\noindent which  are equivalent to setting $\nu=\tau=0$ in the eqs.(2.5).}.
For most of the practically interesting problems in the {\small WFSMA} 
this  metric may be  represented in quasi-Cartesian coordinates 
as the sum of two tensors: the Minkowski metric 
$\gamma^{(0)}_{mn}$ and the field of  inertia $\phi_{mn}$:
{}
$$\gamma^B_{mn}(y^p_B) = {{\partial x^k}\over{\partial y^m_B}}
{{\partial x^l}\over{\partial y^n_B}}\gamma_{kl}(x^s(y^p_B))=
\gamma^{(0)}_{mn}+\phi^B_{mn}(y^p_B).\eqno(3.7)$$ 
\noindent Note that the term $\phi_{mn}$ appears to be 
parameterized by the coordinate transformation functions 
$K_A, L_A$ and $Q^\alpha_A$ defined in eqs.(3.5); 
thus we have $\phi_{mn}(y^p_A)=\phi_{mn}\big[K_A,L_A,Q^\alpha_A\big]$,
a formulation which will be referred to as the {\small KLQ} 
parameterization in the {\small WFSMA}. 

The advantage of  using of these gauge conditions  
is that they allow us to construct   the solutions to the field equations 
in a unique way without applying of the technique  of the, so-called, 
`external multipole moments' 
(Brumberg \& Kopejkin, 1988a; DSX, 1991).
The conditions of eqs.(3.6) do not fix  the harmonic 
{\small RF} in a unique way and,  in definition of coordinates of this
frame,  some arbitrariness
may still exist. Indeed, the coordinate transformation 
$ y'^p_B = y^p_B + \zeta^p_B(y^q_B)$ with the function $\zeta^p_B$ 
which  
satisfies to the  equation $g^{mn}(y^p_B) {\cal D}^B_m {\cal D}^B_n \hskip
1mm 
\zeta_B^p(y^q_B) = 0$ does not violate the chosen conditions  (3.6).
 In all the particular cases the remaining freedom of the harmonic 
 {\small RF} might be fixed by making the specific 
choice of the {\small CS} associated 
with the proper {\small RF}s  chosen for describing  the dynamics of the 
{\small N} bodies in the 
system\footnote{Or equivalently, by choosing some specific form of
$g_{mn}$
 (Thorne, 1980; Hellings, 1986; Fukushima, 1988) 
and the internal and `external' moments 
in vacuum power expansion of the metric tensor $g_{mn}$ in a set of 
multipoles (Kopejkin, 1988; DSX, 1991).}. 
 
\underline{\it The boundary conditions.} 
The search for the general solution for $h^{\it int}_{mn}(x^p)$ 
is performed in a barycentric inertial {\small RF}$_0$, 
which is singled out by the Fock-Sommerfeld's boundary conditions 
imposed on the $h_{mn}$ and $\partial_k h_{mn}$:
{}
$$\lim_{{r \rightarrow \infty} } 
\Big( h_{mn}(x^p); \hskip 2mm r\Big[{\partial \over \partial x^0}
 h_{mn}(x^p) + 
{\partial \over \partial r}  h_{mn}(x^p)\Big]\Big) 
\hskip 2mm \rightarrow \hskip 2mm 0, $$
$$ t + {r\over c} = {\it const},  \eqno(3.8a)$$

\noindent where $r^2 =  - \gamma^{(0)}_{\mu\nu} \hskip 1mm x^\mu
x^\nu$.
Note  that the conditions (3.8) 
must be satisfied along all past Minkowski light-cones.
Thus, these conditions   define the asymptotically Minkowskian space-
time
 in a weak sense, consistent with  the absence of any flux 
of gravitational radiation falling on the system from an external
 universe (Damour, 1983, 1986). 
Moreover, one assumes that  there exists such a quantity 
$h_{mn}^{max} = {\it const}$ (for the solar system this constant
is of order $\approx 10^{-5}$) for which the condition  
{}
$$ h_{mn}(x^p) \hskip 1mm < \hskip 1mm h_{mn}^{max},  \eqno(3.8b)$$

\noindent 
should be satisfied for each point $(\vec{x})$  inside 
the system : $|\hskip 1pt \vec{x}\hskip 1pt | \leq L_D$.   
Note  that any distribution of matter is considered 
isolated  if  conditions $(3.8)$ are
fulfilled in any inertial {\small RF} (Damour, 1983; Kopejkin, 1987, 1988).
 
By  making of use the conditions (3.8)  we
have an  opportunity to determine   the interaction term 
$h^{\it int}_{mn}(x^p)$ in a unique way while solving the  gravitational
field
equations of a metric theory of gravity.

\subsection{The Post-Newtonian  KLQ  Parameterization.}

It is well-known that for  practical   description of
 the translational and rotational motions of 
the {\small N}-body system one should introduce at least 
{\small $(N+1)$} different {\small RF}s   (Brumberg \& Kopejkin, 1988;
DSX, 1991).
It is desirable that one of these frames   be the inertial 
barycentric $(${\small RF}$_0)$  with coordinates denoted as
$(x^p)\equiv(x^0, x^\mu)$. 
The  origin of 
these coordinates is located at the center of the field of the entire
{\small N}-body system.
This particular {\small RF}  will be used to describe the {\it global}
dynamics of the 
whole system.  The other {\small N} frames  should be convenient 
for the description of  the {\it local} gravitational environment in the
immediate vicinity of 
the particular body {\small (B)} under consideration. The origins of
corresponding 
coordinate grids  $(y^p_B)\equiv(y^0_B, y^\mu_B)$, should be  associated
with the centers of 
the {\it local} fields of the interacting bodies of interest.

In this chapter  we will establish the general relationships 
describing the straight, inverse and mutual coordinate transformations
between the different quasi-inertial {\small RF}s.  We will show that in
the {\small WFSMA}, 
all these different types of coordinate transformations may be 
parametrized by the same set of functions $K_A,L_A$ and $Q^\alpha_A$. 
As a result, we will reconstruct in the general form of the post-
Newtonian  
non-linear group of motion of the background pseudo-Euclidean 
space-time for the {\small WFSMA}.  

\subsubsection{The Properties of the   Coordinates Transformations in the
WFSMA.}

As we mentioned above, in order to construct the  relativistic theory of
the 
{\small RF}s  in celestial mechanics, one should  not only solve
the  {\it global}  and {\it local} problems, but also one should establish 
the rules of the coordinate transformations between 
these solutions which belong to the different {\small RF}s . 
To do this, let us discuss   the expected physical and mathematical
properties 
of the  coordinate transformations given by the expressions (3.5) in the
form:
{}
$$x^0 = y^0_A + c^{-2}K_A(y^0_A, y^\epsilon_A) + 
c^{-4} L_A(y^0_A, y^\epsilon_A) + {\cal O}(c^{-6})y^0_A,  $$
$$x^\alpha =  y^\alpha_A + y^\alpha_{A_0} (y^0_A) + 
c^{-2}Q^\alpha_A(y^0_A,  y^\epsilon_A) + {\cal O}(c^{-4}) y^\alpha_A.$$

These coordinates are expected to cover   space-time in the 
immediate vicinity of the body under consideration. It is clear that such a 
mapping of the space-time
may be performed by both  the  barycentric and bodycentric coordinates. 
  This   suggests that   these coordinate  transformations should  be
reversible. 
The functions $K_A,L_A$ and $Q^\alpha_A$ should contain the information
about the specific 
physical properties of the {\small RF} chosen for the analysis.
It is generally believed that, in order to produce the transformations to
the  physically 
justified  proper {\small RF}, the following  properties of these functions
should be satisfied:

\begin{itemize}
\item[(i).] The functions $K_A, L_A, Q^\alpha_A$ should be completely
defined  by the means of the external gravitational 
field at the origin of the coordinate system of the
 proper {\small RF}$_A$   of body {\small (A)} for which the physically
adequate 
 proper {\small RF}  is constructed. These functions 
should not contain any terms caused by the pure gravitational 
field of the body  {\small (A)}  besides those  with  the coupling of the 
internal  multipole moments of the body {\small (A)} to the 
external  gravitation. 
 
\item[(ii).]   In order to  obtain   reversible transformations, the  
transformation functions should be homogeneous and infinitely
differentiable. Then,
based  on   assumptions about the properties of a well justified proper
{\small RF}
(given in the Section {\small 1}), 
the functions $K_A, L_A$ and $Q^\alpha_A$  
should admit an additional Taylor expansion in powers series of the
spatial coordinate $y^\mu_A$.
For convenience, these series may    originate  on the  world-line of the
center of the 
{\it local} field in the vicinity of the 
body {\small (A)}, so that these functions  could be expressed as follows:
{}
$$ f_A(y^0_A, y^\mu_A) = \sum^{\infty}_{l} {1\over l!}
{f_A}_{\{L\}}(y^0_A)\cdot y^{\{L\}}_A, \eqno(3.9)$$

\noindent where function $f_A(y^0_A, y^\epsilon_A)$ is any function from 
$K_A, L_A$ or $Q^\alpha_A$. As a result, the second 
derivatives taken from these functions will not depend on the order of the
derivative's
 application, namely:
{}
$$\big[{\partial\over\partial y^m_A}, 
 {\partial\over\partial y^n_A}\big] f_A(y^0_A, y^\epsilon_A) = 0, 
\eqno(3.10)$$

\noindent where the brackets are the usual notation  for the commutator:
$[a,b] = ab - ba. $  

\item[(iii).]  At the limit when the gravitation is absent $(${\small G}
$\rightarrow 0)$, 
the theory becomes Poincare-invariant and  
transformations $(3.5)$ should coincide with those of  Poincare' 
 (between two frames in uniform relative 
motion with a velocity ${\vec v}$ plus transition of 
origin and arbitrary rotation) which are given by the eqs.(1.7).  
\item[(iv).]  At the other limit, when the {\small N} $\rightarrow 1$, and
the problem 
may be described by the one-body gravitational field solution (2.5), the
transformations
should coincide with that of Chandrasekhar-Contopulos (1.12) for the
uniform motion between  the 
two {\small RF}s  in the isolated one body problem. 
 \item[(v).]  For the gravitational theories, whose foundations 
 are based upon the Equivalence Principle, the 
physical properties of constructed {\small RF}s  should be generic for all
the  
bodies in the system. Otherwise, the possible violation of this Principle
(which may be induced 
by the possible  dependence of the  gravitational coupling  on the
shape/size/composition
of the bodies), should be taken into account
while the proper {\small RF} is constructed.  
\end{itemize}

\subsubsection{The Inverse Transformations.}

The transformations  given by eq.(3.5) transform space and time 
coordinates from the barycentric space-time {\small RF}   $(x^p)$ to space
and 
time coordinates in the proper {\small RF}$_A$   $(y^p_A)$. However, in 
practice one needs to make the comparison between the proper time and
position in a 
different {\small RF}s   and hence it is necessary  to have the inverse
transformations to those of 
eq.(3.5) and the mutual transformations 
between the two proper quasi-inertial frames as well. The existence of 
the small parameters  and the assumptions  in  (3.9) and (3.10) make it
possible 
to generate these transformations in a general form as well as to
 construct the group of motion for the problems in the {\small WFSMA}.
 Thus, the general condition of the 
inreversibility of the transformations (3.5)   is given as usual :
{}
$$\hbox{det }||{\partial x^m\over \partial y^n_A}|| \not = 0. \eqno(3.11a)$$

\noindent The expressions $(B5)$ from  the  Appendix B  enable us to
present this condition in an arbitrary 
 {\small RF} obtained with the {\small WFSMA} as follows:
{}
$$\hbox{det }||{\partial  x^m\over \partial y^n_A}||  = 
1 +  {\partial\over\partial y^0_{A}} K_{A} (y^0_A,y^\nu_A) + $$
$$+ v^\lambda_{A_0}(y^0_{A})  v_{A_0}{_{{}_\lambda}}(y^0_{A})  + 
{\partial\over\partial y^\mu_{A}} Q^\alpha_{A} (y^0_A,y^\nu_A)  
+{\cal  O}(c^{-4})  \not = {\cal  O}(c^{-4}). \eqno(3.11b)$$

\noindent Note that  this  condition is satisfied for most of the 
problems in modern celestial mechanics. A similar analysis has been 
made by 
Brumberg \& Kopejkin (1989) for the dynamics of the planets in the solar
system.
It was shown that the determinant vanishes at the distance 
$r^* \sim c^2/|a_E|\sim 7.5 \cdot 10^{20} \hbox{ cm }$ from the center
of mass of the Earth. From this  it follows  that, in spite 
of an initial construction of geocentric 
{\small RF}  in the region lying inside the lunar orbit, it is possible to
smoothly 
(without intersecting) prolongate the spatial coordinates axes of the
geocentric {\small RF}  
for much larger distances beyond the orbit of Pluto.

We will search for the   post-Newtonian transformations which will be
inverse to those of 
 eq.(3.5)  in the following form:
{}
$$y^0_A = x^0 + c^{-2}\hat{K}_A(x^0, x^\epsilon)+ 
c^{-4}\hat{L}_A(x^0, x^\epsilon)+{\cal O}(c^{-6})x^0, \eqno(3.12a)$$
$$y^\alpha_A =  x^\alpha - y^\alpha_{A_0} (x^0) + 
c^{-2}\hat{Q}^\alpha_A(x^0, x^\epsilon)+ 
{\cal O}(c^{-4})x^\alpha.\eqno(3.12b)$$

\noindent where the functions $\hat{K}_A, \hat{L}_A$ and 
$\hat{Q}^\alpha_A$ are unknown at the moment. One can show that 
in the {\small WFSMA}, 
these  functions may be expressed  in terms of the functions 
$K_A, L_A, Q^{\alpha}_A$  written in a coordinates  $(x^p)$ of 
the barycentric {\small RF}$_0$. In order to find the expressions for the 
$\hat{K}_A, \hat{L}_A$ and $\hat{Q}^\alpha_A$ let us 
substitute the relations (3.5) into eqs.(3.12) and then expand 
the obtained relations with respect to the small parameters:
{\small G} $\sim  c^{-2}$. 
Thus for the spatial components we will obtain:
{} 
$$x^\alpha = x^\alpha - y^\alpha_{A_0} (x^0) + 
c^{-2}\hat{Q}^\alpha_A(x^0, x^\epsilon)+  
y^\alpha_{A_0}\Big(y^0_A(x^{0}, x^\alpha)\Big) + $$
$$ + c^{-2}Q^\alpha_A\Big(y^0_A(x^{0}, x^\alpha), y^\epsilon_A(x^{0}, 
x^\alpha)\Big) + O(c^{-4})y^\alpha_A. \eqno(3.13)$$

\noindent This equation   enables us to find the expression for the 
$\hat{Q}^\alpha_A(x^0, x^\epsilon)$  in terms of the functions
$ Q^\alpha_A$ and $\hat{K}_A$. By expressing the  arguments of 
the transformation functions 
$y^\alpha_{A_0} \Big(y^0_A(x^{0}, x^\alpha)\Big)$ and 
$Q^\alpha_A\Big(y^0_A(x^{0}, x^\alpha), y^\epsilon_A(x^{0}, 
x^\alpha)\Big)$ in terms of the coordinates $(x^p)$ and    
expanding   the obtained  relations in the power series of the small
parameter 
$c^{-1}$ we will get:  
{}
$$Q^\alpha_A\Big(y^0_A(x^{0}, x^\alpha), y^\epsilon_A(x^{0},
 x^\alpha)\Big) = 
Q^\alpha_A\Big(x^{0}, x^\alpha - y^\alpha_{A_0} (x^0)\Big) + 
{\cal O}(c^{-4}) x^\alpha, \eqno(3.14a)$$
$$y^\alpha_{A_0} \Big(y^0_A(x^{0}, x^\alpha)\Big) = 
y^\alpha_{A_0}\Big(x^0 + {c^{-2}\hat{K}_A(x^0, x^\epsilon)} + 
{\cal O}(c^{-4}) x^0\Big) = $$
{}
$$ = y^\alpha_{A_0}(x^0)  + v^\alpha_{A_0} (x^0) 
\cdot {c^{-2}\hat{K}_A(x^0, x^\epsilon)} 
 + {\cal O}(c^{-4}) x^\alpha, \eqno(3.14b)$$

\noindent where 
{}
$$ { d \over d y^0_{A}} y^\alpha_{A_0}\Big(y^0_A(x^{0}, 
x^\alpha)\Big) = v^\alpha_{A_0} (x^0) +
{\cal O}(c^{-4}) x^\alpha. $$ 

\noindent Then, by substituting  eqs.(3.14) into eqs.(3.13), 
we will obtain  the  expression for the 
function ${\hat{Q}^\alpha_A(x^0, x^\epsilon)}$:
{}
$${\hat{Q}^\alpha_A(x^0, x^\epsilon)} =
 -  Q^\alpha_A\Big(x^{0}, x^\alpha - y^\alpha_{A_0} (x^0)\Big)  
-v^\alpha_{A_0} (x^0) \cdot {\hat{K}_A(x^0, x^\epsilon)}  + 
{\cal O}(c^{-4}) x^\alpha. \eqno(3.15)$$

\noindent By repeating this procedure  for the temporal
 components of the transformations  $(3.12)$  we may obtain
the expressions for the functions ${\hat{K}_A(x^0, x^\epsilon)}$ 
and ${\hat{L}_A(x^0, x^\epsilon)}$ as well:
{}
$${\hat{K}_A(x^0, x^\epsilon)} = - K_A\Big(x^0, x^\epsilon -
 y^\epsilon_{A_0} (x^0)\Big) + {\cal O}(c^{-4}) x^0, \eqno(3.16)$$
{}
$${\hat{L}_A(x^0, x^\epsilon)} = - L_A\Big(x^0, x^\epsilon -
 y^\epsilon_{A_0} (x^0)\Big) -$$
{}
$$- \Big[\Big({\partial\over\partial x^0} + v^\nu_{A_0} 
(x^0){\partial\over\partial x^\nu}\Big) 
\cdot K_A\Big(x^0, x^\epsilon - y^\epsilon_{A_0} (x^0)\Big)\Big] 
\cdot {\hat{K}_A(x^0, x^\epsilon)} - $$
{}
$$ - {\partial\over\partial x^\nu}K_A
\Big(x^0, x^\epsilon - y^\epsilon_{A_0} (x^0)\Big) \cdot
{\hat{Q}^\alpha_A(x^0, x^\epsilon)} + {\cal O}(c^{-6}) x^0. \eqno(3.17)$$

\noindent Making of use the resulting expressions for the functions 
${\hat{K}_A}, {\hat{L}_A}$ and ${\hat{Q}^\alpha_A}$, which are
given by the relations (3.15)-(3.17), from  the equation (3.12) 
we finally   obtain   the
inverse transformations between  proper 
and barycentric kinematically non-rotating {\small RF}s  in the most
general form:
{}
$$y^0_A = x^0 - c^{-2}K_A\Big(x^0, x^\epsilon - y^\epsilon_{A_0} 
(x^0)\Big) + c^{-4}\Big[- L_A\Big(x^0, x^\epsilon - y^\epsilon_{A_0}
(x^0)\Big) + $$
{}
$$ +  {\partial\over\partial x^0}K_A\Big(x^0, 
x^\epsilon - y^\epsilon_{A_0} (x^0)\Big) \cdot 
K_A\Big(x^0, x^\epsilon - y^\epsilon_{A_0} (x^0)\Big) + $$
{}
$$ + {\partial\over\partial x^\nu}K_A\Big(x^0, x^\epsilon - 
y^\epsilon_{A_0} (x^0)\Big) \cdot
Q^\nu_A\Big(x^0, x^\epsilon - y^\epsilon_{A_0} (x^0)\Big)\Big] +
{\cal  O}(c^{-6}) x^0 \eqno(3.18a)$$ 
{}
$$y^\alpha_A =  x^\alpha - y^\alpha_{A_0} (x^0)+ $$
{}
$$+c^{-2}\Big[v^\alpha_{A_0} (x^0) \cdot K_A \Big(x^0, x^\epsilon - 
y^\epsilon_{A_0} (x^0)\Big)- Q^\alpha_A\Big(x^0, x^\epsilon - 
y^\epsilon_{A_0} (x^0)\Big) \Big] + {\cal  O}(c^{-4}) x^\alpha.
\eqno(3.18b)$$
 
\noindent Note  that the method used to derive the expressions (3.18)
corresponds to finding   
such a coordinate transformations $y^p_A=y^p_A(x^q)$ which  transform 
the space-time $\gamma^A_{mn}$ of the 
proper {\small RF}$_A$  to that of the barycentric inertial {\small
RF}$_0$ 
with the Minkowski metric $\gamma^{(0)}_{mn}$ in Cartesian coordinates. 
The latter may be presented   as follows:  
$ds^2=\gamma^A_{mn}(y^p_A) dy^m_Ady^n_A= c^2{dt}^2-d{\vec x}^2$. 

\subsubsection{The Coordinate Transformations Between the Two Proper
RFs.} 

The ability to make the power expansion with respect to the small 
parameters allows us to organize the 
iterative procedure for constructing the mutual coordinate
 transformation between the two different 
proper  {\small RF}s, namely {\small RF}$_A$   and  {\small RF}$_B$. 
The definition of the proper {\small RF} (3.5) was given based on 
the  clearly defined physical properties of the 
barycentric inertial {\small RF}$_0$ for the entire {\small N}-body
system. 
The transformation functions connecting the  two proper {\small RF}s  
are easy to find  by applying the same procedure  which was used for the
construction
 of the inverse transformation (3.18). Thus, by making of use the
expressions  (3.5) and  (3.18), we may  find the following relations for 
the mutual coordinate transformation:
{} 
$$y^0_B = y^0_A + c^{-2}K_{BA}(y^0_A, y^\epsilon_A) + 
c^{-4}L_{BA}(y^0_A, y^\epsilon_A) + {\cal  O}(c^{-6}) y^0_A, \eqno(3.19a)$$
$$y^\alpha_B =  y^\alpha_A + y^\alpha_{BA_0} (y^0_A) 
+ c^{-2}Q^\alpha_{BA}(y^0_A, y^\epsilon_A) + {\cal  O}(c^{-4})y^\alpha_A,
\eqno(3.19b)$$
\noindent where the 
functions $K_{BA}, L_{BA}, Q^\alpha_{BA}$ are given as follows:
{}
$$K_{BA}(y^0_A, y^\epsilon_A) = K_A(y^0_A, y^\epsilon_A) -
 K_B\Big(y^0_A,y^\epsilon_A + y^\epsilon_{BA_0} (y^0_A)\Big),
 \eqno(3.20a)$$
{}
$$ Q^\alpha_{BA}(y^0_A, y^\epsilon_A) = 
Q^\alpha_A(y^0_A, y^\epsilon_A) - 
Q^\alpha_B\Big(y^0_A, y^\epsilon_A + y^\epsilon_{BA_0} 
(y^0_A)\Big) -  v^\alpha_{B_0} (y^0_A)\cdot K_{BA}(y^0_A, y^\epsilon_A),
\eqno(3.20b)$$
{}
$$L_{BA}(y^0_A, y^\epsilon_A) = L_A(y^0_A, y^\epsilon_A) - 
L_B\Big(y^0_A,y^\epsilon_A + y^\epsilon_{BA_0} (y^0_A)\Big) - $$
{}
$$ - \Big[\Big({\partial\over\partial y^0_A} - v^\nu_{BA_0} 
(y^0_A){\partial\over\partial y^\nu_A}\Big) 
\cdot K_B\Big(y^0_A,y^\epsilon_A + y^\epsilon_{BA_0} 
(y^0_A)\Big)\Big] \cdot K_{BA}(y^0_A, y^\epsilon_A) -$$
{}
$$ - {\partial\over\partial y^\nu_A}K_B\Big(y^0_A,y^\epsilon_A + 
y^\epsilon_{BA_0} (y^0_A)\Big) \cdot
Q^\nu_{BA}(y^0_A, y^\epsilon_A). \eqno(3.20c)$$ 

The relations (3.5), (3.18)-(3.20)   represent  the necessary expressions
for developing the perturbation 
theory in the {\small WFSMA} for the problems
 of the dynamics of an astronomical gravitationally bounded system of  
{\small N} self-gravitating arbitrarily shaped extended bodies.  The 
transformations are presented 
in a functionally parameterized form by the  two scalar functions 
$K_A, L_A$ and one three-vector function $Q^\alpha_A$. 
 Assuming all the bodies in the system are described  by the 
same model of   matter, one may conclude
that the form of all these functions should be the same for any {\small
RF}.
This property of the  transformations 
 reflects the fact that a proper {\small RF}   may be  
 defined in a general way for each body  in the system. 
Moreover, one can see that the expressions  (3.19)-(3.20) represent the
group of motion which 
preserves the form-invariancy of the metric tensor $\gamma^A_{mn}$
of the background pseudo-Euclidian space-time for any proper {\small RF}. 
This means that   the 
{\small RF}s,    constructed this way, should be  equivalent and, hence, 
the physical  phenomena will behave exactly the same way in all of them.

\subsubsection{The Notes on an Arbitrary Rotation of the Spatial Axes.}

In this part  we will  show how one may generalize the results obtained on
the case of the 
transformations between dynamically rotational coordinate {\small RF}s.
The need for such a coordinate systems may appeared, for example,
 in the case when one will relate the {\small VLBI}, {\small LLR}
and the planetary ephemeris {\small RF}s as well as in the case of
relating 
the celestial and terrestrial frames (Folkner {\it et al}, 1994; Sovers \&
Jacobs, 1994).  
The most general form of post-Newtonian  transformations between the 
coordinates $(x^p)$ of 
the barycentric inertial {\small RF}$_0$  to those $(y^p_A)$ of the proper
{\small RF}$_A$  which are 
undergoing the rotational motion of the spatial axes with an arbitrary
time-dependent 
rotational matrix ${\cal R}_A^{\mu\nu}(y^0_A)$, may be  presented in the
following form:
{}
$$x^0 = y^0_A +  c^{-2}K_A\Big(y^0_A, {\cal R}_A^{^\epsilon}{_{{}_\nu}}
(y^0_A)\cdot y^\nu_A\Big) + 
c^{-4} L_A\Big(y^0_A, {\cal R}^\epsilon_{A\nu} (y^0_A)\cdot y^\nu_A\Big)
+ {\cal O}(c^{-6})y^0_A,\eqno(3.21a)  $$
$$x^\alpha =  y^\alpha_{A_0}(y^0_A) + {\cal R}^{\alpha}_{A\nu}
(y^0_A)\cdot y^\nu_A  + 
c^{-2}Q^\alpha_A\Big(y^0_A, {\cal R}^\epsilon_{A\nu} (y^0_A)\cdot
y^\nu_A\Big) + {\cal O}(c^{-4}) y^\alpha_A. 
\eqno(3.21b)$$  
\noindent The matrix ${\cal R}_A^{\mu\nu}(y^0_A)$  represents  both 
the rotation and the time-dependent   deformation of the spatial axes:
{}
$${\cal R}_A^{\mu\nu}(y^0_A)=\sigma_A^{\mu\nu}(y^0_A)
+\omega_A^{\mu\nu}(y^0_A), \eqno(3.22)$$

\noindent where the first term   is symmetric, 
$\sigma_A^{\mu\nu}=\sigma_A^{\nu\mu}$, and it represent  
the rescaling of the coordinates with respect to time. 
The second term is anti-symmetric, $\omega_A^{\mu\nu}=-
\omega_A^{\nu\mu}$,
and it describes the rotation of the spatial  axes of   
coordinate grid in the proper {\small RF}$_A$.  Besides this, the 
tensor $\omega_A^{\mu\nu}$ contains the information about the
precession
and nutation  of the 
spatial coordinates (Kopejkin, 1988; Fukushima, 1991; 
Folkner {\it et al}, 1994; Sovers \& Jacobs, 1994).   

In the case  when \hskip 2pt $\hbox{det } ||{\cal R}_A^{\mu\nu}||\not=0$,
one may   find the
inverse transformations to those given by expressions (3.21).  
To do this,  we may repeat  the same iterative procedure discussed above.
Making of use this method, one may easily obtain these inverse
transformations  in the following form: 
{}
$$y^0_A= x^0 -  K_{A}\Big(x^0, x^\epsilon- y^\epsilon_{A_0} (x^0)\Big) + 
\hat{L}''_{A}(x^0, x^\epsilon) +{\cal  O}(c^{-6})x^0,  \eqno(3.23a)$$
{}
$$y^\alpha_A = ({\cal R}^{-1}_A)^{\alpha}_\nu (x^0)\cdot\Big[ x^\nu -
y^\nu_{A_0} (x^0) +$$
{}
$$+ v^\nu_{A_0} (x^0)\cdot K_{A}\Big(x^0, x^\epsilon- y^\epsilon_{A_0}
(x^0)\Big)-
Q^\nu_A\Big(x^0, x^\epsilon - y^\epsilon_{A_0} (x^0)\Big) \Big]-$$
{}
$$- {d\over d x^0} ({\cal R}^{-1}_A)^\alpha_\nu(x^0) \cdot\left(x^\nu -
y^\nu_{A_0}
 (x^0)\right)K_{A}\Big(x^0, x^\epsilon- y^\epsilon_{A_0} (x^0)\Big) + {\cal 
O}(c^{-4})x^\alpha.
\eqno(3.23b)$$
\noindent where the function  $\hat{L}''_{A}$ is given as follows:
{}
$$\hat{L}''_{A}(x^0, x^\epsilon)  = - L_A\Big(x^0, x^\epsilon -
y^\epsilon_{A_0} (x^0)\Big) + $$
{}
$$ + {\cal R}^\nu_\lambda(x^0)\cdot {\partial\over\partial
x^\nu}K_A\Big(x^0, x^\epsilon - 
y^\epsilon_{A_0} (x^0)\Big) \cdot
Q^\lambda_A\Big(x^0, x^\epsilon - y^\epsilon_{A_0} (x^0)\Big)+$$
{}
$$+{1\over 2}\mu(x^0,x^\epsilon)\cdot K^2_{A}\Big(x^0, x^\epsilon-
y^\epsilon_{A_0}(x^0)\Big)+ 
{\cal  O}(c^{-6}) x^0 \eqno(3.23c)$$

\noindent with the differential operator $\mu(x^0,x^\epsilon)$ taking the
form:
{}
$$\mu(x^0,x^\epsilon)={\partial\over\partial x^0}+ 
\Big[v^\nu_{A_0}(x^0)\cdot\big(\delta^\mu_\nu-{\cal R}_A^{\mu\nu}\big)-
$$
{}
$$-{d\over d x^0} ({\cal R}^{-1}_A)_{\nu_\lambda}\cdot
{\cal R}_A^{\mu\nu}\cdot\Big(x^\lambda -
y^\lambda_{A_0}(x^0)\Big)\Big]\cdot
{\partial\over\partial x^\mu}+{\cal  O}(c^{-2}){\partial\over\partial x^0}.
\eqno(3.24a)$$

\noindent Note that   if we  neglect the rotation ({\it i.e.} will 
take the rotation matrix in the form of the Kronekker symbol, ${\cal
R}_{A\nu}^\mu (x^0)=\delta^\mu_\nu$ ), 
the differential operator eq.(3.24a) becomes: 
{}
$$\mu(x^0,x^\epsilon)= \big(1+ {\cal  O}(c^{-2})\big){\partial\over\partial
x^0} \eqno(3.24b)$$
and the coordinate transformations (3.23)  coincide with those 
of eq.(3.18) for the dynamically non-rotating case. 

For most  practical applications  in modern astronomy 
one may neglect the effects due to the time-dependent deformation of the
axes 
and assume that the body  is undergoing rigid three-dimensional rotation
with the rotational matrix 
 taken in the form: ${\cal
R}_A^{\mu\nu}(y^0_A)=\omega^{\mu\nu}_A(y^0_A)$.
In the proper {\small RF}$_A$   of an isolated rotating body, the  following 
equation describes the dynamic properties of the tensor ${\cal
R}_A^{\mu\nu}(y^0_A)$:
{}
$${d\over d y^0_A}{\cal R}_A^{\mu\nu}(y^0_A)= 
 \epsilon^{\mu\sigma\beta}{{\cal V}_A}_\sigma(y^0_A)\cdot{{\cal R}_{A
\beta}}^\nu(y^0_A).\eqno(3.25)$$
\noindent where   ${\cal
V}_A^\sigma=\epsilon^\sigma_{\rho\nu}\omega^{\rho\nu}_A$ is the 
vector of the angular velocity  of rotation of the body {\small (A)}.
Usually, for most of the 
problems in relativistic celestial mechanics,  
one assumes that the angular velocity of the rotation of the celestial
bodies is of the 
following order of magnitude:
${\cal V}_A^\sigma \sim {\cal O}(c^{-2}) v_{A_0}^\sigma$, 
where $v_{A_0}^\sigma$ is the barycentric velocity of 
the translational motion of the body {\small (A)} moving along its world-
line (DSX, 1991). 
Then, taking this condition into account, one may neglect the time
derivative terms
from the  transformation matrix in the relations (3.23) and make of use
the standard theory 
of a coordinate transformations with the rigid spatial rotation of the
proper 
{\small RF}$_A$. Otherwise, for a general case with an arbitrary rotation, 
 one should keep these terms in the post-Newtonian parts of the 
transformations (3.23).

Following the procedure depicted above, we may obtain the mutual
transformations 
between the coordinates of two rotating {\small RF}s. Furthermore, one
may extend the  
results obtained above to the case of the  non-uniform rotation of an
elastic body {\small (B)}
with the rotational matrix taken in a general form ${\cal
R}_B^{\mu\nu}(y^0_B, y^\epsilon_B)$. 
However, for the problems of  celestial mechanics in the {\small WFSMA}
this generality is not necessary. Moreover, in 1991 the  IAU have made the 
recommendation that, 
in order to avoid the appearance of the fictitious forces (Coriolis-like)
acting on a observer in the 
proper {\small RF}, all the coordinate transformations for the
astronomical applications should not 
introduce any rotation of the spatial axes at all (Fukushima, 1991;
Brumberg, 1991; 
Klioner, 1993). 
Because of this  reason  we will limit  ourselves   in our further
discussion 
solely to the case of the non-rotational coordinate transformations,
leaving the problem 
of rotation for  other publications.

\subsection{The  Definition of the Proper RF.} 

In this subsection we will finally present a  
way to find the  transformation functions necessary for constructing a  
proper {\small RF}
with the well defined physical properties. 
As one can see from the expressions (3.4), in the  {\small WFSMA}   
the main contribution to the geometrical properties of the proper {\small
RF}$_A$  
in the body's immediate vicinity   comes from its own gravitational field
$h^{(0)A}_{mn}$. 
Then, based on the Principle of Equivalence, the   
external gravitational influence should vanish at least to 
first order in the spatial coordinates (Synge, 1960; Manasse \& Misner,
1963).  
The  proper {\small RF}$_A$, constructed this way, should 
resemble the properties of a quasi-inertial (or Lorenzian) reference frame
and, as such, 
 will be well suited for discussing the physical experiments. 
Note  that the   tensors $h^{(0)B}_{mn}$ and $h^{\it int}_{mn}$ represent 
the real gravitational field which no coordinate transformation 
can eliminate everywhere in the system.
In the case of a massive monopole body, one can eliminate the influence of 
 external field on the  body's world-line only. However, for an 
arbitrarily shaped extended body, the coupling of the body's intrinsic 
multipole
moments to the surrounding gravitational field    changes the physical
picture significantly. 
This means that  the definition of the proper {\small RF} for  the extended
body must take  
into account this non-linear gravitational coupling. 

In order to suggest the   procedure for the choice of the 
coordinate transformations to the phyiscally adequate proper {\small
RF}$_A$, let us 
discuss the general structure of the solution $g^A_{mn}(y^p_A)$ given by 
expression (3.4). Thus, in the expressions for  
$g^A_{mn}$ one may easily separate the  four physically different terms. 
These terms are:
 
\begin{itemize}

\item[(i).]    The Riemann-flat contribution of the field of 
inertia $\gamma^A_{mn} $ given by expression (3.7). 
\item[(ii).]   The contribution   of the body's own gravitational field
$h^{(0)A}_{mn}$. 
 
\item[(iii).]  The term  due to the non-linear interaction of the proper 
gravitational field with an   
 external  field\footnote{This contribution is due to 
the  Newtonian potential  and the potential $\Phi_{2}$ in the 
expressions (2.5). These interaction terms   show  up as the   
coupling  of the body's intrinsic multipole moments with the external 
field.}.  
 
\item[(iv).] The term  describing the field of  the 
external sources of gravity. This term  comes from the 
transformed solutions $h^{(0)B}_{mn}$
and the interaction term $h^{\it int}_{mn}$. 

\end{itemize} 

\noindent The  first contribution depends on the external field   in the 
gravitational domain occupied by  the body {\small (A)} and 
appears to be `parametrized'  by the 
transformation  functions (3.5). Note that for any choice of these
functions, 
by the way it was constructed, the obtained  metric  
$g^A_{mn}$  satisfies  the gravitational field equations of the 
specific metric theory of  gravity under study.  
Furthermore, based on the   properties of the proper {\small RF}$_A$ 
discussed above, 
one may   expect that the functions $K_A, L_A, Q^\alpha_A$ should form  
a 
background  Riemann-flat inertial space-time $\gamma_{mn}^A$ 
in this {\small RF}  which will  be   tangent 
to the total gravitational field in the vicinity of 
the body   {\small $(A)$}'s world-line $\gamma_A$. Moreover, the  
difference of these fields should vanish to first order with respect to the 
spatial coordinates ({\it i.e.} the `external' dipole moment equals zero
(Thorne \& Hartle, 1985)). These conditions,   applied to  
moving test particles, are known as Fermi 
conditions (Fermi, 1922; Manasse \& Misner, 1963; Misner {\it et al.},
1973).  
 We have  extended the applicability of these conditions  
to the case of a system composed of {\small N} arbitrarily shaped 
extended celestial bodies.

In order to obtain the  functions $K_A, L_A$ and $Q^\alpha_A$ 
for the coordinate transformation eq.(3.5) we will 
introduce an iterative procedure which will be based on a 
 multipole power expansion with respect to the unperturbed spherical 
harmonics.  To demonstrate the use of these conditions,  
let us denote $H^A_{mn}(y^p_A)$   
as  the {\it local} gravitational field, {\it i.e.} the field 
which is formed from contributions (ii) and (iii) above. 
The    metric tensor in the {\it local} region 
in this case can be represented by the expression: 
$g^{(loc)}_{mn}(y^p_A)= \gamma_{mn}+ H^A_{mn}(y^p_A)$.
 Then the generalized Fermi conditions in the {\it local} 
region of body {\small (A)} (or in the immediate vicinity of 
its world-line $\gamma_A$) may be imposed on this {\it local} metric 
tensor   by the following equations:
{} 
$$ \lim_{\gamma\rightarrow \gamma_A} g_{mn}(y^p_A)     =  
\hskip 2mm g^{(loc)}_{mn}(y^p_A) \Big|_{\gamma_A},
 \eqno(3.26a)$$ 
$$ \lim_{\gamma\rightarrow \gamma_A}\Gamma^{k}_{mn}(y^p_A)    = 
\hskip 2mm \Gamma^{k(loc)}_{mn}(y^p_A) \Big|_{\gamma_A},
   \eqno(3.26b)$$
\noindent where $\gamma$ is the world-line of the point of interest and 
the quantities $\Gamma^{k(loc)}_{mn}(y^p_A)$  are the  Christoffel 
symbols calculated with respect to the {\it local} gravitational 
field $g^{(loc)}_{mn}(y^p_A)$.
Application of these conditions will  determine the  
 functions $K_A, L_A, Q^\alpha_A$ which are as yet unknown. 
Moreover, this procedure will enable us to derive the second-order 
ordinary differential equations for the functions 
$y^\alpha_{A_0} (y^0_A)$ 
and  $Q^\alpha_A (y^0_A, 0)$, or, in other words, to determine the 
equations of the perturbed motion of the center of the 
{\it local} field in the vicinity of  body {\small (A)}.

The relations (3.26)  summarize our expectations  
based on the Equivalence Principle 
about the {\it local} gravitational environment of the 
self-gravitating bodies.
By making  use of these equations, we will be able   to separate  
the {\it local} gravitational field from the external field   
in the immediate vicinity of the bodies.
However, these conditions  only allow  us to determine the transformation
functions for the free-falling massive monopoles  ({\it i.e.} only up to 
the second order with respect to the spatial coordinates). The
transformation 
 functions (3.5) in this case will depend only on the leading contributions
of the 
external gravitational potentials $U_B$ and $V^\alpha_B$ and  their first 
derivatives
 taken on the world-line of   body {\small (A)}. 
The results obtained will not   account for the  contribution of the 
multipolar interaction of the proper gravity with the external field in the
volume 
of the extended body. This accuracy is safficient  for taking into 
account  the terms describing  the interaction of the   
intrinsic quadrupole moments of the bodies  with the surrounding
gravitational field, but  
some more general  condition, in addition to eq.(3.26), must be applied
in order to account for the higher multipole structure of the bodies.

Thus, as we shall see later, the conditions  (3.26)   enable  
one  to obtain the complete solution for the Newtonian function $K_A$.
Functions $L_A$ and $Q^\alpha_A$ may be defined  up to the second order
with
 respect to the spatial point separation,
 namely  $L_A, Q^\alpha_A \sim O(|y^\alpha_A|^3) $, so the arbitrariness
 of  higher orders $(k \geq 3)$
in the spatial point separation will remain  in the transformation. 
In order to get the corrections to these functions up to 
$k^{th}$ order $ (k \geq 3)$ with respect to the powers of 
the spatial coordinate $y^\lambda_A$, 
one should use conditions which   contain the spatial 
derivatives of the metric tensor to   order  $(k-1)$. 
The mathematical methods of  modern theoretical physics  
generally consider   {\it local} geometrical quantities only and  
involve  second order differential equations. 
These equations alone  may not be very helpful  for 
constructing the remaining terms in functions $L_A, Q^\alpha_A$ 
up to the order $k \geq 2$ 
However, following Synge (1960), one may   apply  additional geometrical
constructions,  
such as properties of  the Riemann  tensor and  the Fermi-Walker 
transport law (Misner \& Manasse, 1963; Ni, 1977; Ni \& Zimmermann,
1978; 
Li \& Ni, 1978, 1979a,b). Another possibility is to   postulate   the
existence of 
so called   `external multipole moments'  (Thorne, 1980; 
 Blanchet \& Damour 1986; Brumberg \& Kopejkin, 1988a; DSX, 1991-
1994). 
However, those moments are defined through 
vacuum solutions of the Hilbert-Einstein field equations
of general  relativity in an inertial {\small RF}, while
 the influence of  external sources of gravity 
are ignored. The fact of defining the moments in this way is essentially 
equivalent to defining the structure of the proper {\small RF}
 for the body under question.   

The most natural approach to define the desirable 
  properties of the proper quasi-inertial {\small RF}s for the 
system of extended and deformable bodies  is to   
study the  motion of this system in an 
arbitrary {\small KLQ}-parametrized frame. There are two 
different ways to do that, namely: (i) to study the infinitesimal 
motion  of each element of the body, or (ii) to study the motion of a 
whole body with respect to an accelerated frame  attached, for example,
to the center of 
inertia of the {\it local} fields of matter, inertia, and   gravity. In our
method 
we will use the second way and  will study the dynamics of the 
body in its own {\small RF}. Our analysis will be directed toward finding  
the 
functions $K_A, L_A$ and $Q^\alpha_A$ with the condition  that the 
Riemann-flat inertial space-time $\gamma^A_{mn}(y^p_A)$ corresponding
to these functions 
will be tangent to the total Riemann metric $g_{mn}(y^p_A)$ of the entire
system
in the body's vicinity.
Physically, one expects that this inertial space-time will produce 
a `fictitious' (or inertial) force with the density ${\vec f}_{KLQ}$
acting on the body in its   proper {\small RF}. At the same time
the body  is under influence of the  
overall   real force due to the {\it local} fields of   matter and 
gravity with the density ${\vec f}_0$. Thus, the condition for finding the
transformation
functions $K_A, L_A$ and $Q^\alpha_A$ is conceptually simple: the
difference 
between these two densities ${\vec {\cal F}}={\vec f}_0-{\vec f}_{KLQ}$
should vanish 
after integration (or averaging) over the body's compact   volume:  
{}
$$ \delta {\vec  F} = \int_A d^3y'_A{\vec {\cal F}}=
\int_A d^3y'_A\Big( {\vec f}_0-{\vec f}_{KLQ}\Big)=0. \eqno(3.27)$$
 
\noindent Note that  the  notion of `the center of mass' in this case  
loses its practical value, and one  should substitute instead 
`the {\it local}  center of inertia'.  Thus, the  force ${\vec f}_{KLQ}$ should
provide the 
overall static equilibrium for the body under consideration in the
{\it local}  center of inertia, which is defined for all three  fields present
in the immediate vicinity of the body, namely:  matter, inertia and gravity.     
Let us mention here that in practice it is not possible to separate these 
two forces ${\vec f}_0$ and ${\vec f}_{KLQ}$ from each other. Fortunately, 
we will be able to obtain   the difference between them ${\vec {\cal F}}$. 
This will considerably simplify  the further analysis.

In order  to construct the necessary solution 
for the functions  $K_A, L_A$ and $Q^\alpha_A$  in a way, that will be 
valid for a wide class of    metric theories of gravity,
one must first analyze the conservation laws in an arbitrary 
{\small KLQ}-parameterized {\small RF}. This could be 
done based on the conservation law 
for the density of the total energy-momentum
tensor $\hat{T}^{mn}$ of the whole isolated  {\small N}-body system:
{}
$$\nabla^A_n\widehat{T}^{mn} (y^p_A) = 0, \eqno(3.28)$$
 \noindent where $\nabla^A_n$ is the covariant derivative with
 respect to total Rimannian metric $g^A_{mn} (y^p_A)$ in these
coordinates.
Then, by using a standard technique of  integration with  Killing vectors,
 one will have to integrate this equation 
over the  compact volume of the body {\small (A)} and 
 one can  obtain  the equations of motion of 
the extended body (Fock, 1957; Chandrasekhar, 1965; Will, 1993). 
Then the  necessary conditions, equivalent to those of (3.27), 
 may be formulated as the   requirement that the 
translational motion of the extended bodies vanish in their own   {\small
RF}s.
  This  corresponds  to the following conditions    applied to 
the  dipole mass moment  $m^\alpha_A\equiv I^{\{1\}}_A$:
{}
$${d^2m^\alpha_A\over d{y^0_A}^2}={dm^\alpha_A\over dy^0_A} 
=m^\alpha_A= 0,\eqno(3.29a)$$
\noindent where the quantity  $m^\alpha_A$ is calculated based on the
total
energy-momentum tensor matter, inertia and gravitational field taken
jointly 
(similar to the condition  eq.(2.12)).  
These conditions may also be presented in a different form. Indeed, if we 
require that the 
total momentum $P^\alpha_A$ of the {\it local} fields of matter, 
inertia and gravity in the vicinity of the 
extended body vanish, we will have the following   physically equivalent 
condition:
{}
$$ {dP^\alpha_A\over dy^0_A}=P^\alpha_A=0. \eqno(3.29b)$$
\noindent These conditions  finalize the  formulation of the basic
principles of 
construction of the relativistic theory of   celestial   {\small RF}s
in the {\small WFSMA}. 

 This    method is demonstrated to   be a useful 
 tool in  practical analytical and numerical calculations for a number of 
 a metric theories of gravity (Turyshev {\it et al.}, 1996). 
Thus the  properties in derivation of the unperturbed  solutions 
for a number of metric theories of gravity\footnote{The  solutions  
for an isolated distribution of matter {(the  \it global problem}) are well 
known and one may find their general properties   in Will (1993).} 
may be used in order to produce the general solution
 for the problem of motion of N-body system. 
In each particular case for a specific theory of gravity there  
exists the common strategy for constructing the  iterative procedure 
which may be expressed as follows:

\begin{itemize} 
\item[{\small I}.]  One should first choose the particular model  of the 
 matter distribution $T^{mn}$  and  
 define the small parameters  relevant to the particular 
problem under consideration. The next step is to perform the power
expansion with 
respect to these parameters for all the functions and 
fields entering the gravitational field equations of a particular metric
theory of
 gravity and by using the standard methods of the {\small WFSMA} 
(Fock, 1955; Will, 1993) to  find the unperturbed
 solution for an isolated distribution of matter $h^{(0)}_{mn}$. 

\item[{\small II}.] Then, by using the obtained unperturbed solutions and
the 
{\small WFSMA} theory  of the coordinate transformations (developed in 
Appendix B),
construct the general form of the solution for the total   metric tensor
from the 
{\it anzatz} eqs.(3.1)-(3.4). Then, by using the generalized 
de-Donder harmonical gauge and the Fock-Sommerfeld boundary conditions
(3.8),  
construct the interaction term $h^{\it int}_{mn}$  and to present the
solution in coordinates of 
inertial barycentric {\small RF}$_0$  and in an arbitrary, {\small KLQ}-
parameterized 
quasi-inertial one. 
 
\item[{\small III}.] In order to find the functions $K_A, L_A, Q^{\alpha}_A$
of 
the coordinate transformation to the 
coordinates of the proper {\small RF}$_A$   and fix the remaining
coordinate 
freedom, one should apply the procedure of constructing the proper
{\small RF}.  
First of all,   find the solution for these functions by implementing the 
conditions eqs.(3.26) in a {\it local region} of the body.
Then by generalizing the  obtained  result on the  case of an arbitrary
extended 
 body,   integrate the {\it local} conservation law (3.28) over of the
body's volume, in order to  obtain the general form of the  coordinate 
transformations from the conditions (3.29).

\item[{\small IV}.] In order to obtain the final multipolar solution for the
astronomical
{\small N}-body problem, one should  substitute the obtained
transformations into the 
generalized gravitational potentials. Then,  by making the expansion
of these quantities in the triple power series with respect to small
parameters (gravitational constant {\small G},
the inverse powers of the speed of light $c^{-1}$ and 
 parameter of the geodesic  separation $\lambda_A\sim
y^\nu_A/|y_{BA_0}|$), 
one will have  obtained the desired representation for the 
  metric tensor and the corresponding equations of motion.

\end{itemize}

In the following Sections we will discuss the application of the 
proposed perturbation formalism for the solution of the 
problem of motion of an arbitrary astronomical {\small N}-body system
in the general theory of  relativity.

\section{General   Relativity: 
1. Solutions for the  Field Equations.} 

 In this Section we will apply the iterative 
formalism discussed in the previous Section  for  constructing   the
 solutions for the problem of motion of the system of {\small N} extended
bodies 
in the theory of the  general relativity 
and a perfect fluid as a model for matter distribution. 
In this Section we will obtain the solution for the Hilbert-Einstein 
field equations and a perfect fluid model of matter distribution 
in its application for solving the problem of motion of {\small N} extended
self-gravitating 
bodies in the {\small WFSMA}. We will present these 
solutions in both barycentric inertial and proper quasi-inertial {\small
RF}s. 
To do this, we must obtain all the necessary 
transformation rules under the general coordinate transformations
discussed 
in the previous Section.  In order to simplify the discussion in this
Section, 
all these rules were obtained in a general form  and  are presented 
in the Appendices, which will be referred  to as necessary. 

The  gravitational  field equations  of the general theory of relativity
were discovered in 1915 and 
presented  by Einstein (1915a,b)    
 (for more details see Misner {\it et al.} (1973)) as follows:
{} 
$${\sqrt{ - g } \hskip 1mm  R_{mn}} = 
- {{8\pi G}\over c^4}\hskip 1mm \Big(\hat{T}_{mn} - 
{1\over2} g_{mn} \hat{T} \Big).   \eqno(4.1)$$     

\noindent  Let us mention that these equations were independently
obtained and studied also by   Hilbert (1915). 
At the present time there exists a  confidence that a 
relativistic theory  of the astronomical {\small RF}s
must be founded on the equations of the general theory of relativity (4.1). 
The mathematical elegance of the field equations  as well as 
the simplicity of the physical foundations of this theory made it
particular 
easy to perform and analyze the relativistic gravitational experiments. 
Thus, general relativity  has passed many serious tests both in 
the weak gravitational field of the solar system
(Will, 1993) and the   strong-gravitational-field test based on the 
data obtained from the continuous observations of the double 
pulsar {\small PSR} 1913+16 (Damour, 1987; Damour \& Taylor, 1992). 
It should be noted  that presently the analysis of high-precision 
measurements of the light deflection and the 
delay of propagation time of radio signals in the solar gravitational field, 
confirms   the {\small WFSMA} of the general theory of relativity with 
an accuracy of order 1.5 \% and 0.5 \% respectively. 
Concerning the practical applications, we must mention that most of the 
modern methods for the 
relativistic data reduction as well as   the solar system ephemerides  
are based upon the predictions of the 
 equations (4.1) with the perfect fluid model of matter (2.2). 
 This is why we begin the application of the 
  new  method for construction of the relativistic theory of the 
{\small RF}s in the {\small WFSMA}  
from the general theory of relativity.

\subsection{The Solution for the Interaction Term.}
Let us assume that the non-gravitational forces are absent, 
the bodies are well separated and the bodies matter may
be described  by the model of a perfect  fluid with the 
density of energy-momentum tensor $\hat{T}^{mn}$ 
given by the expressions (2.1)-(2.2).
 As we have previously discussed, all the field equations and the boundary
and initial conditions 
for this problem are much better defined mathematically in the
coordinates 
of the inertial {\small RF}$_0$, 
so it is quite natural to begin the discussion within this reference frame. 
In  Section {\small 2}, we  assumed that the general solution for the
gravitational
 field equations $g_{mn}$  in   coordinates  $(x^p)$
of the barycentric inertial {\small RF}$_0$  may be written as follows:
{}
$$g_{mn}(x^p) = \gamma_{mn}(x^p) + 
\sum_{B=1}^N {{\partial y^k_B}\over{\partial x^m}}
{{\partial y^l_B}\over{\partial x^n}}
\hskip 0.5mm h^{(0)B}_{kl}(y^q_{B} (x^p)) + 
h^{\it int}_{mn}(x^p). \eqno(4.2)$$

\noindent At this point we 
already have all the necessary `tools'  to construct the metric tensor
$g_{mn}(x^p)$. 
Let us recollect all the gained knowledge, which is  necessary to obtain
this tensor, namely:
\begin{itemize} 
\item[(i).] The unperturbed solution  for the  Hilbert-Einstein gravitational
field  
equations $h^{(0)B}_{kl}$ for an isolated distribution of matter with the
perfect fluid model   of matter distribution presented 
by the energy-momentum tensor $T^{mn}$ eq.(2.1), 
 in coordinates of inertial {\small RF}$_0$ has a simple form and 
in terms of the tensor $h^{(0)B}_{mn}$  it is given 
by the expressions  (2.5) with the conditions: 
$\gamma=\beta=1, \nu=\tau=0$. 
\item[(ii).] The general transformation rules of these solutions 
under the coordinate transformations (3.5)
with the transformation matrix as in eqs.$(C9)$ are 
established in the form of the relations $(D7)$. 
\item[(iii).] The  transformation properties of  the 
gravitational potentials, which were defined in   Appendix A,  are 
given by the expressions $(E9a), (E14a), (E15a)$ and $(E16a)$.
\end{itemize}
 By substituting all these expressions into the formula  (4.2)
we will obtain the  following expressions for the metric tensor $g_{mn}$  
in the coordinates $(x^p)$ of the barycentric inertial {\small RF}$_0$:
{}
$$g_{00}(x^0, x^\nu) = 1- 2\sum_{B} U_{B}(x^p) 
 + \sum_{B} \Bigg(2 U^2_B (x^p) + 2 \Psi_B (x^p) +
{\partial^2\over\partial {x^0}^2} \chi_B(x^p)+$$
{}
$$-
2 \int_B d^3x' \rho_B\Big(x^0, x'^\nu - 
y^\nu_{B_0}(x^0)\Big) {\partial \over\partial x'^\lambda} 
\Big[{Q^\lambda_{B} \Big(x^0, x^\nu-y^\nu_{B_0}
(x^0)\Big) - Q^\lambda_{B} \Big(x^0, x'^\nu-
y^\nu_{B_0}(x^0)\Big)\over{|x^\nu - x'^\nu|}} \Big]-$$
{}
$$ - 2 v_{B_0}{_{{}_\lambda}}(x^0)v^\lambda_{B_0}(x^0) 
\cdot  U_B(x^p) +
 v_{B_0}^\lambda(x^0)v^\beta_{B_0}(x^0) \cdot 
{\partial^2 \over \partial x^\lambda \partial x^\beta} 
\chi_B (x^p) +$$
{}
$$ + a^\lambda_{B_0}(x^0) \cdot {\partial \over \partial
 x^\lambda} \chi_B (x^p)  + 4{\partial\over{\partial x^0}}K_A\Big(x^0,
x^\nu_A - 
y^\nu_{B_0}(x^0)\Big) \cdot U_B(x^p) \Bigg)  + h^{{\it int}  <4>}_{00}(x^p)
 + {\cal O}(c^{-6}), \eqno(4.3a)$$ 
{}
$$g_{0\alpha}(x^0, x^\nu) = 4 \sum_{B} \gamma_{\alpha\lambda}
V^\lambda_B(x^p) + {\cal O}(c^{-5}), \eqno(4.3b)$$ 
{}
$$g_{\alpha\beta}(x^0, x^\nu) = \gamma_{\alpha\beta}
\Big( 1 + 2 \sum_{B}U_B(x^p) \Big)+{\cal O}(c^{-4}), 
\eqno(4.3c)$$ 

\noindent where interaction term $h^{{\it int} <4>} _{00}$ is the only term
which hasn't 
yet been specified. In order to find this  term, one should  use the  Hilbert-
Einstein 
field equations eq.(4.1)  written in 
the coordinates of inertial {\small RF}$_0$ and   expanded with 
respect to the small parameter $c^{-1}$.  

The necessary expansions for the Ricci tensor $R_{mn}$ eq.$(B9)$ and for
the modified 
energy-momentum tensor $S_{mn}$, which is  defined by 
eqs.$(B12$-$B13)$, are given correspondingly by the expressions $(D3)$
and $(D11)$ 
in this {\small CS}.
By making of use these expressions, one may obtain the linearized  
Hilbert-Einstein field equations for {\small N}-body system. 
Finally, by equating the expressions with the same 
orders of magnitude, with respect to powers of the small parameter $c^{-
1}$,
we will obtain the following  equations:
{} 
$$\gamma^{\nu\lambda}{\partial^2 \over \partial 
x^\nu\partial x^\lambda} g^{<2>}_{00}(x^p) = 
-  8\pi\sum_B \rho_B\big(y^q_B(x^p)\big)+{\cal O}(c^{-4}), \eqno(4.4a)$$
{} 
$$ \gamma^{\nu\lambda}{\partial^2 \over \partial 
x^\nu\partial x^\lambda}
g^{<2>}_{\alpha\beta}(x^p)  =  8\pi \gamma_{\alpha\beta}  
 \sum_B \rho_B\big(y^q_B(x^p)\big)+{\cal O}(c^{-4}), \eqno(4.4b)$$
{} 
$$\gamma^{\nu\lambda}{\partial^2 \over \partial 
x^\nu\partial x^\lambda}g^{<3>}_{0\alpha}(x^p)  = 
 - 16 \pi \hskip 1mm\gamma_{\alpha\mu} \sum_B 
\rho_B\big(y^q_B(x^p)\big) \cdot v^\mu(x^p)+{\cal O}(c^{-5}), 
\eqno(4.4c)$$
{} 
$$ \gamma^{\nu\lambda}{\partial^2 \over \partial 
x^\nu\partial x^\lambda} g^{<4>}_{00}(x^p) -
\gamma^{\lambda\mu}\gamma^{\nu\delta} g^{<2>}_{\lambda\nu}(x^p)
{\partial^2 \over \partial x^\mu \partial x^\delta} 
g^{<2>}_{00}(x^p)+ $$
{}
$$ + 
{\partial^2 \over \partial {x^0}^2} g^{<2>}_{00}(x^p) 
 - \gamma^{\lambda\nu}{\partial \over \partial 
x^\lambda} g^{<2>}_{00}(x^p)
{\partial \over \partial x^\nu} g^{<2>}_{00}(x^p) =$$
{}
$$ = -8 \pi \sum_B \rho_B\big(y^q_B(x^p)\big)\Big( \Pi\hskip 1mm
  - \hskip 1mm 2\sum_{B'}U_{B'} 
 - 2 v^\mu(x^p)v_\mu(x^p) + {3 p\over \rho}\Big)+{\cal O}(c^{-6})
\eqno(4.4d)$$

\noindent By substituting into these equations the expressions for 
the metric tensor $g_{mn}(x^p)$   given by 
the relations (4.3), one may see  that   first three 
equations from (4.4) are automatically satisfied for the components
 $g^{<2>}_{00}(x^p), g^{<2>}_{\alpha\beta}(x^p), 
g^{<3>}_{0\alpha}(x^p)$ of the metric tensor. However, the last equations
from this system
 eq.$(4.4b)$,   written for the component $g^{<4>}_{00}$, produce  
the necessary equation for  the determination of the 
interaction term $h^{{\it int} <4>} _{00}$
as follows:
{}
$$\gamma^{\mu\nu}{\partial^2 \over\partial x^\mu 
\partial x^\nu} \Bigg[ h^{{\it int} <4>} _{00}(x^p)-
 \sum_{B} \Bigg( 2{v_{B_0}}_\lambda(x^0)v^\lambda_{B_0}(x^0) 
\cdot  U_B(x^p) - $$
{}
$$-   v_{B_0}^\lambda(x^0)v^\beta_{B_0}(x^0) \cdot 
{\partial^2 \over\partial x^\lambda \partial x^\beta} 
\chi_B (x^p)  - 
 a^\lambda_{B_0}(x^0) \cdot {\partial \over 
\partial x^\lambda} \chi_B (x^p)+$$
{} 
$$+2\int_B d^3x' \rho_B\Big(x^0, x'^\nu - y^\nu_{B_0}(x^0)\Big)\times$$
{}
$$\times {\partial \over\partial x'^\lambda} 
\Big[{Q^\lambda_B\Big(x^0, x^\nu- y^\nu_{B_0}(x^0)\Big)
 - Q^\lambda_B \Big(x^0, x'^\nu-y^\nu_{B_0}(x^0)\Big)\over{|x^\nu -
x'^\nu|}}\Big]-$$
{} 
$$ - 4{\partial\over{\partial x^0}} K_A\big(x^0, x^\nu - 
y^\nu_{B_0}(x^0)\big) \cdot U_B(x^p)\Bigg)\Bigg] = $$
{}
$$= 4 \gamma^{\mu\nu} \sum_B {\partial\over \partial x^\mu} 
U_B (x^p) \sum_{B'} {\partial\over \partial x^\nu}
 U_{B'}(x^p).\eqno(4.5)$$ 

\noindent The general solution to this equation is easy to obtain and it 
may be written as follows: 
{}
$$h^{{\it int} <4>} _{00}(x^0,x^\nu) = \sum_B 
\Bigg[ 2v_{B_0}{_{{}_\lambda}}(x^0)v^\lambda_{B_0}(x^0) \cdot 
 U_B(x^p)-$$
{} 
$$-  v_{B_0}^\lambda(x^0)v^\beta_{B_0}(x^0) \cdot 
{\partial^2 \over\partial x^\lambda \partial x^\beta}
\chi_B (x^p) - $$
{} 
$$ -  a^\lambda_{B_0}(x^0) \cdot {\partial \over 
\partial x^\lambda} \chi_B (x^p)-
4{\partial\over{\partial x^0}} K_B\Big(x^0, x^\nu - 
y^\nu_{B_0}(x^0)\Big) \cdot U_B(x^p)+$$
{} 
 $$+2\int_Bd^3x'  \rho_B\Big(x^0, x'^\nu - y^\nu_{B_0}(x^0)\Big) 
{\partial \over\partial x'^\lambda} \Big[{Q^\lambda_B \Big(x^0, x^\nu-
y^\nu_{B_0}(x^0)
\Big) - Q^\lambda_B \Big(x^0, x'^\nu-y^\nu_{B_0}(x^0)\Big)\over{|x^\nu -
x'^\nu|}}\Big]+$$
{}
$$ +2\sum_{B'}\Bigg(U_B(x^p)  U_{B'}(x^p) - \int_B{d^3x'   \over{|x^\nu -
x'^\nu|}} 
\Big[\rho_B\Big(x^0, x'^\nu - y^\nu_{B_0}(x^0)\Big) U_{B'}(x^0, x'^\nu)  + $$
{}
$$+ 
\rho_{B'}\Big(x^0, x'^\nu - y^\nu_{B_0}(x^0)\Big)U_B(x^0, x'^\nu) \Big]
\Bigg)\Bigg] +  
W^{<4>}_{00}(x^0,x^\nu) + {\cal O}(c^{-6}), \eqno(4.6) $$ 

\noindent where   summations over both, {\small (B)} 
and {\small (B}$'${\small)}  are from 1 to {\small N}. 
The only requirement on the arbitrary function $W^{<4>}_{00}$ is that it 
should satisfy  the  ordinary Laplace equation: 
{}
$$\gamma^{\mu\nu}{\partial^2\over \partial x^\mu \partial x^\nu}
W^{<4>}_{00}(x^0,x^\nu) = {\cal O}(c^{-6}).\eqno(4.7)$$ 

\noindent The solution to the equation (4.7) 
has   terms with both possible asymptotic:  falling off at 
infinity $\sim {1/r^k}$, and   divergent $\sim  r^k$. 
The choice of the solution should be made in order to account for   
cosmological, galactic or gravitational wave contributions to the 
 behavior of the metric tensor $g_{mn}$ at  large distances from the
system.
If there is no incoming radiation falling on the system from outer space
and the background metric 
is accepted to be satisfied  for the  cosmological conditions of the
{\small PPN}  
gauge\footnote{The main requirement is that the cosmological evolution
of the 
background metric is described by the Robertson-Walker cosmological
solution
at   large distances from the system of the bodies under consideration
(Will, 1993)}, then the 
Fock-Sommerfeld boundary conditions eq.(3.8)  
enables us to choose the past-stationary and asymptotically-Minkowskian 
solution to the  field  equations  of general relativity (Damour, 1983).
However, for  further calculations we will retain
the function $W^{<4>}_{00}$ as unspecified. 

By substituting the obtained result for the interaction term 
$h^{{\it int} <4>}_{00}$   in the expression for the 
temporal component of the metric tensor eqs.$(4.3a)$, we 
could write the final solution for the Hilbert-Einstein field equations 
in  coordinates $(x^p)$ of the inertial barycentric {\small RF}$_0$ as
follows:  
{}
$$g_{00}(x^0, x^\nu) = 1 - 2\sum_{B} U_B(x^p)+
2 \Big(\sum_{B}U_B(x^p)\Big)^2 + $$
{}
$$+ \sum_{B} \Bigg(\hskip -2pt- 4 \Phi_{1B} (x^p)-
2\Phi_{3B}(x^p) - 6\Phi_{4B}(x^p) + {\partial^2\over\partial
 {x^0}^2} \chi_B(x^p)- $$
{}
$$-  4\int_B{d^3x'  \over{|x^\nu - x'^\nu|}}
\rho_B\Big(x^0, x'^\nu - y^\nu_{B_0}(x^0)\Big)
 \sum_{B'} U_{B'}(x^0, x'^\nu)\Bigg)+ W^{<4>}_{00}(x^0,x^\nu) 
+ {\cal O}(c^{-6}), \eqno(4.8a)$$
{}
$$g_{0\alpha}(x^0, x^\nu) = 4 \sum_{B} 
\gamma_{\alpha\lambda}V^\lambda_B(x^p) + {\cal O}(c^{-5}),
 \eqno(4.8b)$$
{}
$$g_{\alpha\beta}(x^0, x^\nu) = \gamma_{\alpha\beta}
 \Big( 1 + 2 \sum_{B}U_B(x^p) \Big) + {\cal O}(c^{-4}). \eqno(4.8c)$$ 

The obtained expressions (4.8) are  the usual form of the general solution
for the 
{\it global} problem in general relativity for the isolated distribution of
matter, which was 
first obtained by Fock (1957) (see also Fock, 1955; Damour, 1986;
Kopejkin, 1989; Will, 1993).
It is easy to see that the general solution for {\small N}-body problem  in
the barycentric 
inertial {\small RF}$_0$ eqs.(4.8) demonstrates the property of the linear
superposition of   unperturbed fields $h^{(0) B}_{mn}$ boosted by the
transformations (3.18)
in the components $g^{<2>}_{00}(x^p),g^{<2>}_{\alpha\beta}(x^p)$ and 
$g^{<3>}_{0\alpha}(x^p)$
 of the metric tensor.
The non-linear contribution due to the  motion of the bodies and their
gravitational  
interaction with each other appears beginning in the 
component $g^{<4>}_{00}(x^p)$ through the interaction term 
$h^{{\it int} <4>}_{00}$ which is given by the relation (4.6). One may note
that the 
interaction term contains three groups of terms with physically different 
origins,  namely: 
\begin{itemize}
\item[(i).] The first seven terms  are due to the boost  
of the isolated unperturbed solutions $h^{(0) B}_{mn}$
by the transformations (3.18). 
\item[(ii).] The eighth term is  due to the mutual gravitational interaction
between the bodies in 
the system.
\item[(iii).] The last term $W^{<4>}_{00}$ is caused by 
the possible inhomogeneity of the background  space-time.
\end{itemize}

It is clear that  the terms of the first group are frame-dependent (or
coordinate-dependent). 
Hence these terms are responsible for the coordinate dependence of the
quantity
$h^{{\it int} <4>}_{00}$ in general. This implies that this 
term depends on the properties of the proper coordinate system  chosen
for  description of the 
{\it internal} problem  in the vicinity of a body {\small (B)} in the system.
We can continue the analysis
of these terms in the barycentric inertial {\small RF}$_0$. However, for 
further calculations 
it will be more convenient to shift the discussion to  the  proper {\small
RF}$_A$.

The transformation properties of the interaction term are given by the
relations $(D9)$.
These relations suggest that in the first post-Newtonian 
approximation,  the form of the interaction term in the  coordinates
$(y^p_A)$ of the 
proper {\small RF}$_A$ could be obtained by taking into account the
transformation properties of the 
gravitational potentials only. Thus, by making of use the direct
transformations (3.5) with the 
transformation matrix (C1), one may write the interaction term 
$h^{{\it int} <4>}_{00}$ in the coordinates of the proper {\small RF}$_A$
as follows:
{}
$$h^{{\it int} <4>}_{00}(y^0_A, y^\nu_A) = h^{{\it int} <4>}_A(y^p_A)+ 
h^{{\it int} <4>}_{AB}(y^p_A)+$$
{}
$$+ h^{{\it int} <4>}_{B}(y^p_A)+h^{{\it int} <4>}_{BB'}(y^p_A)+
 W^{<4>}_{00}(y^p_A)+{\cal O}(c^{-6}), \eqno(4.9a)$$
\noindent where the following notations have been accepted:
{}
$$h^{{\it int} <4>}_{A}(y^p_A) = 2\int_Ad^3y'_A \rho_A (y_A^0, y_A'^\nu) 
{\partial \over\partial y'^\lambda_A}\Big[{ Q^\lambda_A (y_A^0, y_A^\nu)
-
 Q^\lambda_A (y_A^0, y_A'^\nu)\over{|y_A^\nu - y_A'^\nu|}}\Big]  -$$
{}
$$-2v_{A_0}{_{{}_\lambda}}(y_A^0)v^\lambda_{A_0}(y_A^0) \cdot 
 U_A(y_A^p)- v_{A_0}^\lambda(y_A^0)v^\beta_{A_0}(y_A^0) \cdot 
{\partial^2 \over\partial {y_A^\lambda} \partial {y_A^\beta}}
\chi_A (y_A^p) - $$
{} 
$$ -a^\lambda_{A_0}(y_A^0) \cdot {\partial \over 
\partial {y_A^\lambda}} \chi_A (y_A^p)-
4{\partial\over{\partial y_A^0}} K_A(y_A^0, y_A^\nu)\cdot U_A(y_A^p)
+{\cal O}(c^{-6}),  \eqno(4.9b)$$
{}
$$h^{{\it int} <4>} _{AB}(y^p_A)= 
4\sum_{B\not=A}\Bigg(U_A(y_A^p) U_{B}(y_A^p) - $$
{}
$$- \int_{A,B}{d^3y'_A \over{|y_A^\nu - y_A'^\nu|}}
\Big[\rho_A (y_A^0, y_A'^\nu) U_{B}(y_A^0, y_A'^\nu)  + 
\rho_B(y_A^0, y_A'^\nu)U_A(y_A^0, y_A'^\nu)\Big]\Bigg) + {\cal O}(c^{-6}),
\eqno(4.9c)$$
{}
$$h^{{\it int} <4>} _{B}(y^p_A)= \sum_{B\not=A}
\Bigg(2\int_Bd^3y'_A
\rho_B \Big(y_A^0, y_A'^\nu + y_{BA_0}^\nu(y_A^0)\Big)\times$$
{}
$$\times{\partial \over\partial y'^\lambda_A}
 \Big[{ Q^\lambda_B\Big(y_A^0, y_A^\nu+y_{BA_0}^\nu(y_A^0)\Big) - 
Q^\lambda_B \Big(y_A^0, y_A'^\nu+y_{BA_0}^\nu(y_A^0)\Big)
\over{|y^\nu_A - y'^\nu_A|}} \Big]+$$
{}
$$+2v_{B_0}{_{{}_\lambda}}(y_A^0)\Big(v^\lambda_{B_0}(y_A^0) -
2v^\beta_{A_0}(y_A^0)\Big)\cdot U_B(y_A^p)-$$
{}
$$-v_{B_0}^\lambda(y_A^0)v^\beta_{B_0}(y_A^0)\cdot 
{\partial^2 \over\partial {y_A^\lambda} \partial {y_A^\beta}}
\chi_B (y_A^p) -a^\lambda_{B_0}(y_A^0) \cdot {\partial \over 
\partial {y_A^\lambda}} \chi_B (y_A^p)-$$
{}
$$-4{\partial\over{\partial y_A^0}} K_B
\Big(y_A^0, y_A^\nu-y_{BA_0}^\nu(y_A^0)\Big)
\cdot U_B(y_A^p)\Bigg) +{\cal O}(c^{-6}), \eqno(4.9d)$$
{}
$$h^{{\it int} <4>} _{BB'}(y^p_A)= 
4\sum_{B\not=A}\sum_{B'\not=B}\Bigg(U_B(y_A^p) U_{B'}(y_A^p) - $$
{}
$$- \int_B{d^3y'_A \over{|y_A^\nu - y_A'^\nu|}}
\Big[\rho_B \Big(y_A^0, y_A'^\nu-y_{BA_0}^\nu(y_A^0)\Big) U_{B'}(y_A^0,
y_A'^\nu)  + $$
{}
$$+ \rho_{B'}\Big(y_A^0, y_A'^\nu-y_{B'A_0}^\nu(y_A^0)\Big)
U_B(y_A^0, y_A'^\nu)\Big]\Bigg) + {\cal O}(c^{-6}). \eqno(4.9e) $$ 
 
\noindent The physical meaning of these new functions is quite clear. 
The functions $h^{{\it int} <4>} _{A}$ and $h^{{\it int} <4>} _{B}$ are the 
post-Newtonian contributions  of the unperturbed solutions $h_{mn}^{(0)}$
for body {\small (A)} and 
all the rest of the bodies {\small (B}$\not=${\small A)} in 
the system, boosted by the transformations  (3.5), (3.20).
The function $h^{{\it int} <4>}_{AB}$ is the contribution  describing the 
gravitational interaction of the body {\small (A)} with the rest 
of the bodies in the system. And the last term, $h^{{\it int} <4>}_{BB'}$, is
the 
function, physically analogous to the previous one, but
 describing the gravitational field   generated by the gravitational 
interaction of the rest of the bodies in the system  {\small (B,
B}$'\not=${\small A)} 
 with each other  in the vicinity of the body {\small (A)}.

The  advantage    using   the   conditions (3.8) 
is that they provide an opportunity to determine   the interaction term 
$h^{\it int}_{mn}(x^p)$ in a unique way. It should be stressed that the 
 corresponding solution $g_{mn}(x^p)$ in the barycentric inertial {\small
RF}$_0$
 resembles  the form of the solution for an isolated one-body problem
(2.5).
The only change which should be made is to take into account the 
number of  bodies in the system: $\rho\rightarrow \sum_B\rho_B$, where
$\rho_B$
is the compact-support mass density of a body $(B)$ from the system. 
However, both the interaction  term $h^{\it int}_{mn}(y^p_a)$ and the total 
solution for the metric tensor $g_{mn}$ in the coordinates $(y^p_A)$ 
appear   to be `parameterized' by the arbitrary functions 
$K_A, L_A, Q^\alpha_A$. 
This result  reflects the covariancy of the gravitational field equations
as well as the well defined  transformation 
properties of the gauge conditions (3.6) used to derive the total solution.
This arbitrariness  suggests that one could choose any form of these
functions 
in order to describe the dynamics of the 
extended bodies in the system. However, as we noticed earlier, the 
unsuccessful choice of the proper {\small RF}$_A$  (or, equivalently, the
functions 
$K_A, L_A$ and $Q^\alpha_A$) may cause an unreasonable complication  in
the 
future physical interpretations of the results obtained.  

\subsection{The Solution of the Field Equations in the Proper  RF.}

Once  the interaction term $h^{{\it int} <4>}_{00}$ has been defined,  
one may easily obtain the form of the 
general solution to the Hilbert-Einstein
 field equations  $g_{mn}(y^p_A)$ in the coordinates  
of the proper {\small RF}$_A$.
This solution may be obtained directly from the tensor $g_{mn}(x^p)$ by
the 
regular tensorial transformation law as follows:
{}
$$g_{mn}(y^p_{A}) = {{\partial x^k}\over{\partial y^m_{A}}}
{{\partial x^l}\over{\partial y^n_{A}}} \hskip 1mm 
g_{kl}(x^s(y^p_A)) =  {{\partial x^k}\over{\partial y^m_{A}}}{{\partial x^l}
\over{\partial y^n_{A}}} \gamma_{kl}(x^s(y^p_A)) + $$
{}
$$+ h^{(0) A}_{mn}(y^p_{A}) + \sum_{B\not=A} {{\partial y^k_B}
\over{\partial y^m_{A}}}{{\partial y^l_B}\over{\partial y^n_{A}}}
\hskip 0.5mm h^{(0) B}_{kl} (y^{s}_{B} (y^p_{A})) +
{{\partial x^k}\over{\partial y^m_{A}}}{{\partial x^l}
\over{\partial y^n_{A}}}h^{\it int}_{kl}(x^s(y^p_{A})). \eqno(4.10)$$
 
  In order to obtain the final result for the metric tensor $g_{mn}$ in the 
coordinates of proper {\small RF}$_A$, we should establish and then make
use of 
the transformation properties of all the quantities  presented in the
expression (4.10).  
These quantities were obtained in Appendices, namely:
\begin{itemize}
\item[(i).] The transformation properties of the background Riemann-flat
metric $\gamma^A_{mn}$ 
in the coordinates $(y^p_A)$ are given by the relations $(C5)$. 
\item[(ii).] The transformations of the 
unperturbed solutions $h^{(0)B}_{mn}$  from the coordinates $(y_B^p)$ of
the proper {\small RF}$_B$  
to those of {\small RF}$_A$ are presented by the relations $(D8)$. 
\item[(iii).] The transformation properties 
of the interaction term $h^{\it int}_{kl}$ was established and 
discussed in the previous subection, where they were given by the 
relations (4.6) and (4.9).  
\item[(iv).] The transformation properties 
of all the potentials, which enter the above named formulae, are given by
the 
eqs.$(E9b), (E14b), (E15b), (E16b)$. 
\end{itemize}

By substituting all these 
quantities into the relations (4.10),  
we will obtain the components of the metric tensor $g_{mn}(y^p_A)$   
 in the  coordinates of the proper {\small RF}$_A$ as follows:  
{}
$$g_{00}(y^p_A) = 1 + 2 {\partial\over\partial
 y^0_{A}} K_{A} (y^0_A,y^\nu_A) + 
{v_{A_0}{_{{}_\beta}}(y^0_{A})} v_{A_0}^\beta(y^0_{A}) - 2\sum_{B}
 U_B(y^p_A)+ $$   
{} 
$$ + 2 {\partial\over\partial y^0_{A}} L_A (y^0_A,y^\nu_A) +
 \Big({ \partial\over\partial y^0_{A}}  K_A (y^0_A,y^\nu_A)\Big)^2   
+ 2  v_{A_0} {_{{}_\beta}}(y^0_A){\partial\over\partial y^0_A} 
Q^\beta_A(y^0_A,y^\nu_A)+$$ 
{} 
$$ + H^{<4>}_{00}(y^0_A, y^\nu_A) + {\cal O}(c^{-6}), \eqno(4.11a)$$
{}
$$g_{0\alpha}(y^p_{A}) = {\partial\over
\partial y^\alpha_{A}} L_{A} (y^0_A,y^\nu_A) -
 v_{A_0} {_{{}_\alpha}}(y^0_{A}){\partial\over\partial y^0_{A}}
 K_{A}(y^0_A,y^\nu_A) + $$ 
{}
$$ + {v_{A_0}}_\nu (y^0_{A}){\partial\over\partial 
y^\alpha_{A}}Q^\nu_{A} (y^0_A,y^\nu_A) + 
\gamma_{\alpha\nu}{\partial\over\partial y^0_{A}} 
Q^\nu_{A} (y^0_A,y^\nu_A)  + 4 \sum_{B}
\gamma_{\alpha\lambda}V^\lambda_B
(y^p_A) + {\cal O}(c^{-5}), \eqno(4.11b)$$ 
{}
$$g_{\alpha\beta}(y^p_{A}) = \gamma_{\alpha\beta} +
 {v_{A_0}{_{{}_\alpha}}(y^0_{A})} v_{A_0}{_{{}_\beta}}(y^0_{A}) + $$
{}
$$ +  \gamma_{\alpha\nu}{\partial\over\partial y^\beta_{A}}
 Q^\nu_{A} (y^0_A,y^\nu_A) +
 \gamma_{\beta\nu}{\partial\over\partial y^\alpha_{A}}
 Q^\nu_{A} (y^0_A,y^\nu_A) + 
2 \sum_B\gamma_{\alpha\beta}U_B(y^p_A) + {\cal O}(c^{-4}),
\eqno(4.11c)$$ 
\noindent where the post-Newtonian term 
$H^{<4>}_{00}$ in the component $g_{00}(y^p_A)$ of   equation 
(4.11a) denotes  the following expression: 
{}
$$ H^{<4>}_{00}(y^0_A, y^\nu_A)=h^{(0) <4>}_{00 A}(y^p_A)+
h^{{\it int} <4>} _{A}(y^p_A)+ h^{{\it int} <4>} _{AB}(y^p_A)+$$
{}
$$+\sum_{B\not=A} \Big[{{\partial y^k_B}
\over{\partial y^0_{A}}}{{\partial y^l_B}\over{\partial y^0_{A}}}
\hskip 0.5mm h^{(0) B}_{kl} (y^{s}_{B} (y^p_{A})) \Big]^{<4>}
+ h^{{\it int} <4>} _B(y^p_A)+h^{{\it int} <4>} _{BB'}(y^p_A)+
 W^{<4>}_{00}(y^p_A).\eqno(4.12a)$$
\noindent The latter expression may be presented in terms of the 
generalized gravitational potentials as follows:
$$H^{<4>}_{00}(y^0_A, y^\nu_A)=2\Big(\sum_{B} U_B(y^p_A)\Big)^2 +
\sum_{B}\Bigg(\hskip -2pt-4\Phi_{1B}(y^p_A)
- 2\Phi_{3B} (y^p_A) - 6\Phi_{4B} (y^p_A)- $$
{}
$$ - 4\int_B{d^3y'_A\over{|y^\nu_A - y'^\nu_A|}} \rho_B\Big(y^0_A,
y'^\nu_A + 
y^\nu_{BA_0}(y^0_A)\Big) \cdot \sum_{B'} U_{B'}(y^0_A, y'^\nu_A) +
 {\partial^2\over\partial {y^0_A}^2} \chi_B(y^p_A)+ $$
{}
$$ +2\int_Bd^3y'_A\rho_B\Big(y^0_A, y'^\nu_A + y^\nu_{BA_0}
(y^0_A)\Big){\partial \over\partial y'^\lambda_A}  \Big[{ Q^\lambda_{A}
(y^0_A,y^\nu_A) - 
Q^\lambda_A(y^0_A, y'^\nu_A)\over{|y^\nu_A - y'^\nu_A|}} \Big]-$$
{}
$$- 2v_{A_0}{_{{}_\lambda}}(y^0_A) v^\lambda_{A_0}(y^0_A) \cdot 
U_B(y^p_A) 
- v_{A_0}^\lambda(y^0_A)v^\beta_{A_0}(y^0_A) \cdot 
{\partial^2 \over \partial y^\lambda_A \partial y^\beta_A} 
\chi_B (y^p_A) - $$
{}
$$-a^\lambda_{A_0}(y^0_A) \cdot {\partial \over 
\partial y^\lambda_A} \chi_B (y^p_A)  
-4{\partial\over{\partial y^0_A}}K_A(y^0_A, y^\nu_A) 
\cdot  U_B(y^p_A)\Bigg) +W^{<4>}_{00}(y^p_A)+ {\cal  O}(c^{-6}).
\eqno(4.12b)$$

The first three terms in the expression  (4.12a) describe both the 
unperturbed gravitational field of the body {\small (A)}, 
boosted by the coordinate transformations 
(the terms $h^{(0) <4>}_{00 A}$ and $h^{{\it int} <4>} _A$)
and the gravitational field produced by the interaction of this field with 
one  produced by the rest of the bodies in the system (the term $h^{{\it
int} <4>} _{AB}$).
These are the terms  which  govern   the {\it local} gravitational
environment in
 the immediate vicinity of the 
body {\small (A)},  producing the major contribution to the equations of
motion  of the test particles 
orbiting this body. The next three terms in the expression $(4.12a)$ are
the terms  which 
are due to the boosted unperturbed gravitational fields produced by 
the rest of the bodies in the system, 
and the gravitational field caused by their  interaction with each other,
presented
in the coordinates of the proper {\small RF}$_A$. This external
gravitational field 
should appear in the equations of motion of the test particles around the
body {\small (A)}, 
written in the coordinates of the proper {\small RF}$_A$, in the form of a
tidal interaction only (Synge, 1960). 
Note  that the approach discussed here is the generalization of the
conception of the neutral 
test particle freely falling in the external gravitational field. It is known
that 
up to these tidal corrections,  a  freely falling  test particle will behave
as if 
external gravity is absent (Bertotti \& Grishchuk 1990).
In our case, the extended body {\small (A)} is not moving freely, instead,
as we will 
see later, it's internal multipole moments  are couples to the external
gravitational 
field through the terms $h^{{\it int} <4>} _{A}$ and $h^{{\it int} <4>}
_{AB}$. This 
coupling produces a force which is resulting in the deviation of the center
of mass  of this body 
from the support geodetic line  along which it would move
if it was   a neutral test particle (Denisov \& Turyshev, 1989).   
The presence of this term and it's significance for solving the {\it local}
problem  has been  pointed out by a number of authors (see, for instance,
Thorne \& Hartle (1985); Kopejkin (1987)), however, to our knowledge, 
the interaction term has never been previously presented in a closed 
relativistic form.

By straightforward calculation, one may  check that the 
obtained metric tensor $g_{mn}(y^p_A)$  satisfies 
 the Hilbert-Einstein field equations written in the coordinates of the
proper {\small RF}$_A$. 
To do this, let us note  that the covariant   de Donder gauge 
is singling out these coordinates  according to the conditions $(C2)$.
This  gives the expressions for the Ricci tensor $R_{mn}$ in the form of
eqs.$(C4)$. 
The modified energy-momentum tensor $S_{mn}$ in this coordinate
system is 
given by the expressions $(C12)$. 
By collecting all these expressions together,  one may obtain the 
linearized  Hilbert-Einstein field equations eq.(4.1) presented in 
the coordinates of the proper {\small RF}$_A$.
Finally, the substitution of the relations eqs.(4.11) in the obtained 
linarized equations will complete the proof of the correspondence  
between the  metric tensor  $g_{mn}(y^p_A)$ and the field equations.

Thus, the metric (4.11) is the {\small KLQ} parameterized solution of the 
Hilbert-Einstein gravitational field 
equations in the coordinates of the proper {\small RF}$_A$. The nature of
this result  is basically  
the post-Newtonian boost of the solution (4.8) (obtained in the inertial
{\small RF}$_0$) to the new 
non-inertial coordinate system defined in the vicinity of an arbitrary body
{\small (A)}. 
It is well known that the  Riemann   metric  tensor $g_{mn}(y^p_A)$
contains  
ten degrees of freedom and could not be transformed to the Minkowski
tensor  for the entire 
space-time by any choice of a coordinate transformation which has only 
four degrees of freedom. This transformation  could be done    at  
one point of the space-time only (Eisenhart, 1926) or along the geodesic 
line (Manasse \& Misner, 1963; Misner {\it et al.}, 1973; Landau \&
Lifshitz, 1988). 
Such a  {\small RF} is called a quasi-inertial or `locally Lorentzian frame'.
Our future discussion will be  based on the form of the metric tensor in
the proper 
{\small RF}$_A$ given by the relations (4.11).  
In the next Section we will implement the conditions for construction a
`good' 
quasi-Lorentzian proper {\small RF} as discussed in the Section {\small
II}, 
which will enable us to find the  unknown transformation functions $K_A,
L_A$ and $Q^\alpha_A$.

\subsection{Decomposition of the Fields in the Proper  RF.}
 
Concluding this Section we would like to emphasize that the solution 
to the Hilbert-Einstein field 
equations $g_{mn}$  in the vicinity of the body's {\small (A)} world-line in
the 
coordinates $(y^p_A)$ of it's proper {\small RF}$_A$
in the  first {\small WFSMA} may be 
decomposed  into the following three major groups:
{}
$$g_{mn}(y^p_A)=\gamma^A_{mn}(y^p_A)+  H^A_{mn}(y^p_A)+
H^B_{mn}(y^p_A), \eqno(4.13)$$

\noindent where  the notations for these groups and their meaning are 
as presented below:
\begin{itemize} 
\item[(i).] The first term, $\gamma^A_{mn}$,  is the {\it local} 
  inertial  (or Riemann-flat) field which is presented by the eqs.$(B4)$.
This term  is also  convenient to split into two parts 
as shown by the relation: 
{}
$$\gamma^A_{mn}(y^p_A)={{\partial x^k}\over{\partial y^m_{A}}}{{\partial
x^l}
\over{\partial y^n_{A}}}
\gamma_{kl}(x^s(y^p_A))=\gamma^{(0)}_{mn}(y^p_A)+
\gamma^{<PN>}_{mn}(y^p_A),\eqno(4.14)$$

\noindent where $\gamma^{(0)}_{mn}$ is the usual Minkowski 
metric in the coordinates of the proper {\small RF}$_A$. 
The second term here, $\gamma^{<PN>}_{mn}$ is the 
{\small KLQ}-parameterized post-Newtonian contribution 
to this {\it local} inertial field at the vicinity of the body's {\small (A)}
world-line.

\item[(ii).] The second term in eq.(4.13),  $H^A_{mn}$, is the 
{\it local} gravitational field, which is given as follows:
{}
$$H^A_{00}(y^p_A)=h^{(0) A}_{00}  + h^{{\it int} <4>} _{00 A}
+ h^{{\it int} <4>} _{00 AB} +{\cal O}(c^{-6}),$$
{}
$$H^A_{0\alpha}(y^p_A)=h^{(0) A}_{0\alpha}+{\cal O}(c^{-5}),\qquad 
H^A_{\alpha\beta}(y^p_A)=h^{(0) A}_{\alpha\beta} +{\cal O}(c^{-4}),
\eqno(4.15a)$$
\noindent  where  the  terms   $h^{(0) A}_{mn}$ are the components of the 
unperturbed proper gravitational 
field of the body {\small (A)}, the term $h^{{\it int} <4>} _{00 A}$ (given by
the eq.(4.9b)) 
is the contribution due to the  boost
of this unperturbed field to the accelerated coordinates of the proper
quasi-inertial {\small RF}$_A$,
and the last term, $h^{{\it int} <4>}_{AB}$ (which is presented by the
eq.(4.9c)), 
is caused by the interaction of the proper 
unperturbed gravitational field with the external gravitation. 
Thus, the component $H^{A <4>}_{00}$  has the following form:
{}
$$H^{A <4>}_{00}(y^p_A) = 2U^2_A(y^p_A) + 2 \Psi_{A}(y^p_A) 
 +  {\partial^2\over\partial {y^0_A}^2} 
\chi_A(y^p_A)+$$
{}
$$+2 \int_A d^3y'_A\rho_A (y_A^0, y_A'^\nu)
{\partial \over\partial y'^\lambda_A} 
\Big[{ Q^\lambda_A(y_A^0, y_A^\nu) - Q^\lambda_A(y_A^0, y_A'^\nu)
\over{|y_A^\nu - y_A'^\nu|}} \Big]  -$$
{}
$$-2v_{A_0}{_{{}_\lambda}}(y_A^0)v^\lambda_{A_0}(y_A^0) \cdot 
 U_A(y_A^p) - v_{A_0}^\lambda(y_A^0)v^\beta_{A_0}(y_A^0) \cdot 
{\partial^2 \over\partial {y_A^\lambda} \partial {y_A^\beta}}
\chi_A (y_A^p) - $$
{} 
$$ -a^\lambda_{A_0}(y_A^0) \cdot {\partial \over 
\partial {y_A^\lambda}} \chi_A (y_A^p)-
4{\partial\over{\partial y_A^0}} K_A(y_A^0, y_A^\nu)\cdot U_A(y_A^p)
+ $$
{}
$$+4\sum_{B\not=A}\Bigg(U_A(y_A^p) U_{B}(y_A^p) - 
\int_A{d^3y'_A \over{|y_A^\nu - y_A'^\nu|}}
\rho_A (y_A^0, y_A'^\nu) U_{B}(y_A^0, y_A'^\nu)\Bigg) + 
{\cal O}(c^{-6}), \eqno(4.15b)$$
\noindent where the subscript {\small  (A)}  for the integral sign means
that 
the integration is performed over the volume of that body for  
which mass density is integrated, namely:
$\int_Ad^3y'_A\rho_B=\delta_{AB}$.

\item[(iii).] The last term in the eq. (4.13), $H^B_{mn}$, 
is  the {\it external} gravitational field  presented as follows:
{}
$$H^B_{00}(y^p_A)=\sum_{B\not=A}  \Big[{{\partial y^k_B}
\over{\partial y^0_{A}}}{{\partial y^l_B}\over{\partial y^0_{A}}}
\hskip 0.5mm h^{(0) B}_{kl}(y^{s}_{B} (y^p_{A})) \Big]^{<4>}+$$
{}
$$+h^{{\it int} <4>} _{00 B}(y^p_A)+h^{{\it int} <4>} _{00 BB'}(y^p_A)+
 W^{<4>}_{00}(y^p_A)+ {\cal O}(c^{-6}),$$
{}
$$H^B_{\alpha0}(y^p_A)=\sum_{B\not=A}  {{\partial y^k_B}
\over{\partial y^\alpha_{A}}}{{\partial y^l_B}\over{\partial y^0_{A}}}
\hskip 0.5mm h^{(0) B}_{kl}(y^{s}_{B} (y^p_{A})) + {\cal O}(c^{-5}), $$
{}
$$H^B_{\alpha\beta}(y^p_A)=\sum_{B\not=A}  h^{(0)
B}_{\alpha\beta}(y^{s}_{B} (y^p_{A}))
 + {\cal O}(c^{-4}),  \eqno(4.16a)$$ 
\noindent where the first two terms in the   component $H^B_{00}$ 
are the result of the boost (see eq. $(4.9d)$ to the coordinates
 $(y^p_A)$ of the {\small RF}$_A$ of the unperturbed solutions  $h^{(0)
B}_{kl}$
for the bodies {\small (B)} in the system (besides {\small (A)}), 
the  third term, $h^{{\it int} <4>} _{00 BB'}$ 
(given by the eq.(4.9e)) is due to the mutual gravitational interactions 
of these external bodies with each other, and, finally, the last 
term is due to existing inhomogeneity of the 
background space-time in which the considered system is embedded. 
The component $H^{B <4>}_{00}(y^p_A)$
may be given as follows:
{}
$$H^{B <4>}_{00}(y^p_A) = \sum_{B\not=A}\Bigg(2 U_B(y^p_A)
\sum_{B'\not=A} U_{B'}(y^p_A)  
+ 2\Psi_B(y^p_A)  + {\partial^2\over\partial {y^0_A}^2} \chi_B(y^p_A)+$$
{}
$$+2 \int_Bd^3y'_A\rho_B \Big(y_A^0, y_A'^\nu+y_{BA_0}^\nu(y_A^0)\Big) 
{\partial \over\partial y'^\lambda_A} \Big[{Q^\lambda_{A}(y_A^0,
y_A^\nu)-
Q^\lambda_A(y_A^0, y_A'^\nu)\over{|y_A^\nu - y_A'^\nu|}} \Big]-$$
{}
$$-2v_{A_0}{_{{}_\lambda}}(y_A^0)v^\lambda_{A_0}(y_A^0)\cdot
U_B(y_A^p)
-v_{A_0}^\lambda(y_A^0)v^\beta_{A_0}(y_A^0)\cdot 
{\partial^2 \over\partial {y_A^\lambda} \partial {y_A^\beta}}
\chi_B (y_A^p) -$$
{}
$$- a^\lambda_{A_0}(y_A^0) \cdot {\partial \over 
\partial {y_A^\lambda}} \chi_B (y_A^p)  -4{\partial\over{\partial y_A^0}} 
K_A (y_A^0, y_A^\nu)
\cdot U_B(y_A^p)\Bigg) +  {\cal O}(c^{-6}), \eqno(4.16b)$$
\noindent  where the potential $\Phi_{2B}(y^p_A)$  entering the term
$\Psi_{B}(y^p_A)$
is defined as follows:
{}
$$\Phi_{2B}(y^p_A)= \int_B{d^3y'_A \over{|y_A^\nu - y_A'^\nu|}}
 \rho_B \Big(y_A^0, y_A'^\nu+y_{BA_0}^\nu(y_A^0)\Big)
 \sum_C U_C(y_A^0, y_A'^\nu) + {\cal O}(c^{-6}). \eqno(4.16c) $$ 

\end{itemize} 

The decomposition  presented by the eqs.(4.13)-(4.16), may be
successfully  continued to the 
next `post-post-Newtonian' order, however the obtained  accuracy is quite
sufficient for  
most   modern astronomical applications. The results obtained in this
Section  will 
 become a useful  tool in the next Section for  constructing a  proper
{\small RF}
with well defined physical properties. 

\section{General  Relativity: 2. Transformations to the Proper RF.} 

In this Section we will present   the construction of 
a `good' proper {\small RF} for an 
arbitrary body {\small (A)}.  This procedure should enable one to obtain 
the yet unknown transformation functions 
$K_A, L_A$ and $Q^\alpha_A$. It is clear that one may choose 
any form of these functions for the
description of the gravitational environment around the body under
question.  
The analysis  presented in the previous Section  shows by the results in
eqs.(4.12)  
that,  in order to solve the {\it local} problem,
 it is permissible to separate the contributions in the 
metric tensor $g_{mn}(y^p_A)$  into several terms.
First contribution is due to the inertial sector of the {\it local} space-
time, 
 the second  is produced by the body itself, 
the third term  is caused by the external sources of the 
gravitational field, and the last one is due to the 
interaction of the body's multipole moments with this 
external gravitational field. It is well known that  
if the body {\small (A)} is a neutral monopole test particle, this external 
gravitational field will define the parameters of the geodesic line  which
this 
test body will follow (Einstein {\it et al.}, 1938; Fock, 1957; Will, 1993).
The equations 
of motion for spinning    bodies are differ from the latter by additional
terms
due to coupling of the  body's spin  to the external gravitational field
(Papapetrou, 1948, 1951). 
It was noted that the presence of   non-vanishing internal multipole
moments of an extended bodies 
significantly changes their equations of motion and several attempts 
have made to account for these effects 
(see, for example, Ashby \& Bertotti, 1986; Shahid-Salees {\it et al.},
1991; 
Brumberg \& Kopejkin, 1988a; DSX, 1991-94). In this paper we will 
introduce a new approach based on the {\small KLQ} parameterization
discussed in the previous Section.
 
 The general idea for constructing the `good' {\small RF}$_A$ 
in terms of the functions $K_A, L_A, Q^\alpha_A$, 
is to choose these functions in such a way  
that the corresponding Riemann-flat inertial space-time
$\gamma_{mn}^A$  (which is    the 
background space-time for the proper {\small RF}$_A$) will be tangent 
to the total metric tensor $g_{mn}$ in the vicinity of the world-line of
the body {\small (A)}. 
These conditions when applied to 
inertially moving test particles are known as the Fermi conditions (Misner
{\it et al.}, 1973).  
 We would like to  extend the applicability of these conditions  
to the case of a system of  extended self-gravitating and arbitrarily
shaped celestial bodies. 
To do this, let us  recall that   relation  for the 
{\it local} gravitational field $g^{(loc)}_{mn}(y^p_A)$, 
which is based on the decomposition eqs.(4.13) 
may be given as follows:
{}
$$g^{(loc)}_{mn}(y^p_A)=\gamma^{(0)}_{mn}(y^p_A)+ H^A_{mn}(y^p_A).
\eqno(5.1)$$
\noindent Then the generalized Fermi conditions in the {\it local} region  
of body {\small (A)}
(or in the immediate vicinity of it's world-line $\gamma_A$)
may be introduced by the   equations (3.26) as follows:
{} 
$$ g_{mn}(y^p_A) \Big|_{\gamma_A}   =  
\hskip 2mm g^{(loc)}_{mn}(y^p_A) \Big|_{\gamma_A} + {\cal
O}(|y^\alpha_A|^2),
 \eqno(5.2a)$$ 
{}
$$ \Gamma^{k}_{mn}(y^p_A) \Big|_{\gamma_A}  = 
\hskip 2mm \Gamma^{k(loc)}_{mn}(y^p_A) \Big|_{\gamma_A} +  
{\cal O}(|y^\alpha_A|),   \eqno(5.2b)$$

\noindent where   
the quantities $\Gamma^{k(loc)}_{mn}(y^p_A)$  are the  Christoffel 
symbols calculated with respect to the {\it local} gravitational field
$g^{(loc)}_{mn}(y^p_A)$ 
given by the eq.(5.1).
These relations  summarize our expectations  based on the Equivalence
Principle 
about the {\it local} gravitational environment of self-gravitating 
and arbitrarily shaped extended bodies. These conditions enable us to
separate 
the {\it local} gravitational field from the {\it external} gravitation 
in the immediate vicinity of the body {\small (A)}. 
This separation is possible due to the remaining arbitrariness 
of the transformation functions $K_A, L_A$ and $Q^\alpha_A$. The
conditions eqs.(5.2) will  
give the differential equations for these functions, the solutions of which 
will correspond to the specific choice of the background 
inertial space-time in the proper {\small RF}$_A$.
To obtain these equations, one should  substitute the
relations for the metric tensor in the form (4.11)
in the expressions for the Christoffel symbols $(F2)$  and 
then make   use of the conditions (5.1). 

\subsection{Finding the Functions $K_A$ and $Q^\alpha_A.$} 
\subsubsection{Equations for the Functions $K_A$ and $Q^\alpha_A.$}

To obtain the equation for the function  $K_A$, one should substitute into
the conditions
(5.1) the relation for the component  $ \Gamma^{0}_{00}(y^p_{A})$ 
of the connection  coefficients given by eq.$(F2a)$. This will give the
following result:
{}
$$  \Big[{\partial \over \partial y^0_A}\Big({\partial \over \partial
 y^0_A}K_{A}(y^0_A, y^\nu_A) + 
{1\over2}v_{A_{0\epsilon}} {v^\epsilon_{A_0}} - \sum_{B\not= A}
 U_B(y^0_A, y^\nu_A)\Big)\Big]\Bigg|_{\gamma_A}
 = {\cal O}(c^{-5}). \hskip 15pt \eqno(5.3) $$

\noindent The components $\Gamma^{0}_{0\alpha}(y^p_{A})$ and 
$ \Gamma^{\alpha}_{00}(y^p_{A})$ which are given by the eqs.$(F2b),(F2d)$
correspondingly, will provide us with the following equation: 
{}
$$\Big[\hskip 1pt {a_{A_0}}_\alpha +\sum_{B\not= A} {\partial \over
\partial 
y^\alpha_A}U_B(y^0_A, y^\nu_A)\Big]\Bigg|_{\gamma_A}
 =  {\cal O}(c^{-4}). \hskip 48pt \eqno(5.4)$$

\noindent From the components  
$\Gamma^{\alpha}_{\beta\omega}(y^p_{A})$ of the connection coefficients
which are
given by eq.$(F2f)$, one may obtain the first equation for the function
$Q^\alpha_{A}$:
{}
$$ \Big[{\partial^2 \over \partial y^\beta_A \partial y^\omega_A}
Q^\alpha_{A}(y^0_A, y^\nu_A)  + 
\sum_{B\not= A} \Big(\delta ^\alpha_\beta {\partial \over 
\partial y^\omega_A} U_B (y^0_A, y^\nu_A) +$$
{}
$$+\delta^\alpha_\omega {\partial \over \partial y^\beta_A} U_B 
(y^0_A, y^\nu_A) - 
\gamma_{\beta\omega} \gamma^{\alpha\lambda}{\partial \over 
\partial y^\lambda_A} U_B (y^0_A, y^\nu_A)\Big)\Big]
\Bigg|_{\gamma_A} =  {\cal O}(c^{-4}).\eqno(5.5)$$

\noindent 
The components $\Gamma^{\alpha}_{0\beta}(y^p_{A})$  eq.$(F2e)$ will give
the second
and last equation for the second unknown transformation function:
{}
$$ \Big[{\partial^2 \over \partial y^0_A \partial y^\beta_A}Q^\alpha_{A}
(y^0_A, y^\nu_A) + \hskip 5pt v^\alpha_{A_0}{a_{A_0}}_\beta  + $$
{}
$$  + \sum_{B\not= A} \Big( 2   {\partial \over \partial y^\beta_A}
V^\alpha_B(y^0_A,y^\nu_A)-
2{\partial \over \partial {y^\mu_A}_\alpha} 
{V_B}_\beta (y^0_A, y^\nu_A)   
 + \delta ^\alpha_\beta {\partial \over \partial y^0_A} U_B (y^0_A,
 y^\nu_A)\Big]\Big) \Bigg|_{\gamma_A} =  {\cal O}(c^{-5}). \eqno(5.6)$$

\subsubsection{The Solution for the Function $K_A$.} 
In order to find the solutions to the differential 
equations above, let us first  denote the  limiting operation
 from the expressions (5.2) for any non-singular function $f(y^p_A)$ as
follows:
{}
$$  \Big<f\Big>_0\equiv \lim_{|{\vec y}_A| \rightarrow 0} f(y^0_A,
y^\nu_A)
= f(y^0_A, y^\nu_A)\Big|_{\gamma_A}. \eqno(5.7)$$    
\noindent  It is important to note  that   operation $(5.7)$   commutes
with
 the  time derivative, but not  with the   spatial derivative. 

Then, using this new notation, we may formally integrate the 
equation eq.(5.3) over time $y^0_A$ as follows: 
{}
$$\Big[{\partial \over \partial y^0_A}K_A(y^0_A, y^\nu_{A}) + 
{1\over2}v_{A_0}{_{{}_\mu}} {v^\mu_{A_0}} - \sum_{B\not= A} 
U_B(y^0_A, y^\nu_A)\Big] \Bigg|_{\gamma_A} = 
\zeta_{1A} (y^\nu_A), \eqno(5.8)$$
\noindent  where $\zeta_{1A}$ is an arbitrary function of the spatial
coordinates $y^\nu_A$.
To continue the solution, let us recall the  relation for the function $K_A$ 
given by the eq.$(C5b)$ in the following  form: 
{}
$$K_{A_{[0]}} (y^0_A,y^\nu_A) = P_{A} (y^0_A) -
{{v_{A_{[0]}}}_\mu(y^0_{A})} 
\cdot y^\mu_A +  {\cal O}(c^{-4}) y^0_A. \eqno(5.9)$$

\noindent where the subscript $({}_{[0]})$ denotes that the operation
$(5.7)$ 
was used to derive the result (5.9) for the functions $K_A$ and
$v_{A_{[0]}}$.
One may notice that the dependence on the spatial coordinate in this  
relation for $K_A$  disappears completely  
on a world-line of the body, so the function $ \zeta_{1A} (y^\nu_A)$ 
is a true constant, {\it i.e.}  
$\zeta_{1A} (y^\nu_A) =  \zeta_{1A} = const$. Then, from these two 
relations (5.8) and  (5.9), one may obtain the differential 
equation  for the function $P_A(y^0_A)$ as follows:
{}
$$ {\partial \over \partial y^0_A} P_A(y^0_A) =  {d\over dy^0_A
}P_A(y^0_A)=\sum_{B\not=A} 
\Big<U_B\Big>_0 -  {1\over2}{v_{A_{[0]}}}_\beta v^\beta_{A_{[0]}} + 
\zeta_{1A}. \eqno(5.10)$$

\noindent If we formally integrate this last equation over the time
 $y^0_A$ and for the function $K_A$ we will obtain 
the following final solution:
{}
$$K_{A_{[0]}}(y^0_A,y^\nu_A) = \int^{y^0_A} \hskip -10pt  dt'  
\Big[ \sum_{B\not=A} \Big<U_B\Big>_{0'}- 
{1\over2}v_{A_{[0]}}{_{{}_\nu}}v^\nu_{A_{[0]}} + \zeta^A_1\Big] - $$
{}
$$-{v_{A_{[0]}}}_\nu\cdot y^\nu_A +  {\cal O}(c^{-4})y^0_A. \eqno(5.11)$$
 
The equation eq.(5.4) provides us with the usual relation for the 
Newtonian acceleration   $a^\alpha_{A_{[0]}}$ 
of the center of inertia of a body {\small (A)} as follows:
{}
$$ a^\alpha_{A_{[0]}}(y^0_A) = - \hskip 2mm \gamma^{\alpha\nu}
\sum_{B\not= A} \Big<{\partial U_B\over \partial y^\nu_A}\Big>_0 +
  {\cal O}(c^{-4}).  \eqno(5.12)$$

Thus we have obtained the form of the first transformation 
function $K_A$ eq.(5.11), which describes the Newtonian corrections to
the proper time $y^0_A$.  
These corrections should be made in order to take into account the    
external gravitational field and the Lorentzian time contraction caused by
the
motion of the origin of the proper {\small RF}$_A$ with the velocity
$v^\nu_{A_{[0]}}$ 
relative to the inertial barycentric {\small RF}$_0$. This correction  
was first obtained by   D'Eath (1975a,b) by the method of matched 
asymptotic expansions while studying the motion of  black holes. 
In astronomical applications for the relativistic {\small VLBI}
measurements, 
this effect was independently obtained and studied by Hellings (1986).
The only new term in the expression eq.(5.11) is the constant $\zeta^A_1$,
which is the free parameter entering the  post-Poincar\'e group of motion. 
This parameter represents the possibility of the time shift 
in proper {\small RF}$_A$ and it is responsible for the energy
conservation 
in the immediate vicinity of the massive test particle moving along
the geodesic. The acceleration, eq.(5.12), is the contribution of the 
monopole into the equation of motion of the extended body. 
The contributions of the other multipoles to the results (5.11) and (5.12) 
will be obtained and discussed further.

\subsubsection{The Solution for the  Function $Q^\alpha_A$.} 

The solution for the function $Q^\alpha_A$   requires slightly more
sophisticated 
calculations. One may expect that the function $Q^\alpha_A$ 
behaves at least quadratically while approaching to the origin of 
the body's world-line, ({\it i.e.} $Q^\alpha_A \sim y^\mu y^\nu \cdot
f(y^0_A)$), where  
$f(y^0_A)$ is some time-dependent function. 
 Let us look at  the solution to the equation (5.5) for the 
function $Q^\alpha_A$  in a following form:
{}
$$Q^\alpha_A(y^0_A,y^\nu_A) = - \sum_{B\not=A} \Big[ c_1 \cdot
y^\alpha_A y^\mu_A\cdot\Big<{\partial U_B\over \partial
y^\mu_A}\Big>_0+
c_2 \gamma^{\alpha\sigma}\cdot {y_A}_\mu y^\mu_A \cdot\Big<{\partial
U_B\over \partial y^\sigma_A} \Big>_0  
\Big] + \Omega^\alpha_A(y^0_A,y^\nu_A), \eqno(5.13)$$

\noindent where $c_1$ and $c_2$ are the constants, 
unknown for the moment. The function $\Omega^\alpha_A(y^0_A,y^\nu_A)$ 
behaves linearly in the vicinity of the body's world-line: 
$\Omega^\alpha_A(y^0_A,y^\nu_A) \sim y^\mu \cdot f(y^0_A)$. 
By substituting the expression eq.(5.13) into equation (5.5), we
will find  that these constants are $c_1 = 1$, $c_2 = -1/2$ 
and that the function $\Omega^\alpha_A$ should satisfy  the equation: 
{}
$${\partial^2 \over\partial y^\beta_A \partial y^\lambda_A }
\Omega^\alpha_A(y^0_A,y^\nu_A) = 0. \eqno(5.14)$$

\noindent By making use of these  results, we may write the
 solution to the  equation eq.(5.5) as follows:
{}
$$Q^\alpha_A(y^0_A,y^\nu_A) = - \sum_{B\not=A} 
\Big[y^\alpha_A y^\beta_A\cdot\Big<{\partial U_B\over \partial
y^\beta_A}\Big>_0- 
{1\over2}\gamma^{\alpha\sigma} {y_A}_\beta  y^\beta_A 
\cdot\Big<{\partial U_B\over \partial y^\sigma_A}\Big>_0 
\Big] + \Omega^\alpha_A(y^0_A,y^\nu_A) \eqno(5.15)$$

\noindent Further calculations require  somewhat more sophisticated
approach.
After   some algebra   the equation eq.(5.6)
 might be rewritten as follows:
{}
$$\Big[{\partial \over \partial y^0_A}\Big( \gamma_{\alpha\nu}
{\partial\over\partial y^\beta_{A}} Q^\nu_{A} (y^0,y^\nu)  +
 \gamma_{\beta\nu}{\partial\over\partial y^\alpha_{A}} 
Q^\nu_{A} (y^0,y^\nu)  + $$
{}
$$  + {v_{A_0}}_\alpha {v_{A_0}}_\beta  +
2 \gamma_{\alpha\beta}\sum_{B\not=A}U_B(y^0_A, y^\nu_A) \Big)\Big]
\Bigg|_{\gamma_A} =     {\cal O}(c^{-5}).\eqno(5.16)$$ 

\noindent By integrating this equation (5.16)  over the time $y^0_A$, 
we   obtain: 
{}
$$ g^{A <2>}_{\alpha\beta}(y^0_A, y^\nu_{A}) \Bigg|_{\gamma_A} = 
\Big[ {v_{A_0}}_\alpha {v_{A_0}}_\beta   
\hskip 5pt + 2 \gamma_{\alpha\beta}\sum_{B\not=A}U_B(y^0_A,
y^\nu_A)+ $$
{}
$$ + \gamma_{\alpha\nu}{\partial\over\partial y^\beta_{A}} Q^\nu_{A}
 (y^0,y^\nu) + 
 \gamma_{\beta\nu}{\partial\over\partial y^\alpha_{A}} Q^\nu_{A}
 (y^0,y^\nu) \Big]
\Bigg|_{\gamma_A} = \sigma^A_{\alpha\beta} = 
{\rm const}.  \eqno(5.17)$$

\noindent Then the function $\Omega^\alpha_A$ from the eq.(4.2.9) 
may be represented in the following form:
{}
$$\Omega^\alpha_A(y^0_A,y^\nu_A) = - \sum_{B\not=A}   
y^\alpha_A\cdot \Big<U_B\Big>_0  -
{1\over2} v^\alpha_{A_{[0]}} {v_{A_{[0]}}}_\mu \cdot y^\mu_A  + 
F^\alpha_A(y^0_A,y^\nu_A), \eqno(5.18)$$

\noindent where $F^\alpha_A$ is some unknown function. 
By substituting the expression eq.(5.18) into the equation 
eq.(5.17) we will define the function $Q^\alpha_A$ as follows:
{}
$$Q^\alpha_A(y^0_A,y^\nu_A) = - \sum_{B\not=A} \Big[ y^\alpha_A 
y^\beta_A\cdot\Big<{\partial U_B\over \partial y^\beta_A}\Big>_0 - 
{1\over2} \gamma^{\alpha\sigma}{y_A}_\beta  y^\beta_A \cdot
\Big<{\partial U_B\over \partial y^\sigma_A}\Big>_0 
 + y^\alpha_A\cdot \Big<U_B\Big>_0\Big] -$$
{}
$$ - {1\over2} v^\alpha_{A_{[0]}} v_{A_{[0]}}{_{{}_\beta}} \cdot y^\beta_A 
+ 
 F^\alpha_A(y^0_A,y^\nu_A),  \eqno(5.19)$$ 

\noindent with the condition on the  function 
$F^\alpha_A(y^0_A,y^\nu_A)$:
{}
$$\gamma_{\nu\beta}{\partial \over \partial y^\alpha_A} 
F^\nu_A(y^0_A,y^\nu_A) +
\gamma_{\nu\alpha}{\partial \over \partial y^\beta_A} 
F^\nu_A(y^0_A,y^\nu_A) = \sigma^A_{\alpha\beta}. \eqno(5.20)$$

\noindent From the expressions eqs.(5.6),(5.18),(5.20)  one may write 
the equation for the function $F^\nu_A$ in the vicinity of
the body   {\small (A)}'s  world-line as:
{}
$${\partial^2 \over \partial y^0_A \partial y^\beta_A}
 F^\alpha_A(y^0_A,y^\nu_A) = 
{1\over2}\Big[a^\alpha_{A_{[0]}} v_{A_{[0]}} {_{{}_\beta}} -
v^\alpha_{A_{[0]}} 
a_{A_{[0]}} {_{{}_\beta}}\Big] +2 \sum_{B\not=A}\Big[\Big<\partial^\alpha
V_{B \beta}\Big>_0 - 
\Big<\partial _\beta V^{\alpha}_B\Big>_0\Big]. \eqno(5.21)$$

\noindent This  last equation, eq.(5.21), may be solved together
 with eq.(5.20) as follows:
{}
$$F^\alpha_A(y^0_A,y^\nu_A) =  {y_A}_\beta\int^{y^0_A} \hskip -10pt dt' 
\Big[{1\over2} a^{[\alpha}_{A_{[0]}} v_{A_{[0]}}^{\beta]}  + 
2 \sum_{B\not=A} \Big<\partial^{[\alpha} V^{\beta]}_B\Big>_0\Big] + $$
{}
$$ + {f^\alpha_A}_\beta \cdot y^\beta_A + w^\alpha_{A_{[0]}} (y^0_A), 
\eqno(5.22)$$

\noindent where the 
constants $\sigma^A_{\alpha\beta}$ and $f^A_{\alpha\beta}$
 are connected as: 
$f^A_{\alpha\beta} + f^A_{\beta\alpha} = \sigma^A_{\alpha\beta}$.
The time-dependent function $w^\alpha_{A_{[0]}} (y^0_A)$ is unknown at
the 
moment. 

Finally, by collecting the obtained relations from the eqs.(5.19),(5.22), 
we will obtain the final solution for the second transformation function,
$Q^\alpha_A$,
is as follows:  
{} 
$$Q^\alpha_{A_{[0]}}(y^0_A,y^\nu_A) = - \sum_{B\not=A} \Big[ y^\alpha_A 
y^\beta_A\cdot\Big<{\partial U_B\over \partial y^\beta_A}\Big>_0- 
{1\over2} \gamma^{\alpha\sigma}{y_A}_\beta  y^\beta_A \cdot
\Big< {\partial U_B\over \partial y^\sigma_A}\Big>_0  
 + y^\alpha_A\cdot \Big<U_B\Big>_0\Big]+$$
{}
$$  +   {y_A}_\beta\int^{y^0_A} \hskip -10pt dt' \Big[ {1\over2}
a^{[\alpha}_{A_{[0]}}
 v_{A_{[0]}}^{\beta]} + 
2 \sum_{B\not=A} \Big<\partial^{[\alpha} V^{\beta]}_B\Big>_0 \Big]   - $$
{}
$$ - {1\over2} v^\alpha_{A_{[0]}} {v_{A_{[0]}}}_\beta  \cdot y^{  \beta}_A 
+ {f^\alpha_A}_\beta \cdot  y^\beta_A + w^\alpha_{A_{[0]}} (y^0_A)
+ {\cal O} (c^{-4}) y^\alpha_A +{\cal O} (|y^\alpha_A|^3). \eqno(5.23)$$

Thus we have obtained the second transformation function, $Q^\alpha_A$,
which is the first function to describe the post-Newtonian coordinate
transformation to the proper {\small RF}  of a moving massive monopole
body. 
The only function which is still unknown in the expression (5.23) is the 
function $w^\alpha_{A_{[0]}}$, which defines the post-Newtonian
correction 
to the radius-vector $y^\alpha_{A_{[0]}}$. This time-dependent function
will be obtained later.  Besides the usual Lorentzian  
terms of the length contraction (caused by the velocity of motion of the
coordinate origin), 
the expression above contains terms caused purely by  gravity. 
The first two terms are due to the acceleration
of the proper {\small RF}$_A$  caused by the external gravitational field.
The third term is the length contraction caused by the external
gravitational 
field. The fourth term with the integral sign is the generalization of the 
expression for geodesic and Thomas precession of the coordinate axis (see
Thomas, 1927).
The similar expression  was obtained by D'Eath (1975a,b). In  astronomical
practice,
this result  was introduced by the Brumberg \& Kopejkin (1988) (see also:
 Ries {\it et al.}, 1991;   DSX, 1991). The obtained relation is different 
from the previous results in that it contains a generalized 
representation of the term containing the 
precessions. In particular, the obtained relation  is defined explicitly 
and does not contain an arbitrary multiplier $q$ as in the Brumberg-
Kopejkin method.
This suggests, that the precession term should always be present in the
expressions 
for the coordinate transformations and neglecting this term will
correspond  
to the {\small RF}, which is deviating from the 
geodesic world-line even for the massless test particles and  will 
lead to   the {\small SEP} violation.  
 In addition to this, the expression (5.23) has an arbitrary group parameter 
${f^\alpha_A}_\beta$. 
This parameter represents the angular momentum conservation law   at
the 
immediate vicinity of the world-line of the body {\small (A)} in its 
proper {\small RF}$_A$.
Besides this, we have studied separately the post-Newtonian part of the 
radius-vector of the body {\small (A)} $w^\alpha_{A_{[0]}}$, which was
never done before.  
Similarly to the case of the function $K_A$ (5.11) and the Newtonian eq.m.
(5.12), 
  the contributions of the other multipoles to the result (5.23) 
will be obtained and discussed in the next Section.

\subsection{Finding  the Function $L_A$.}
In this chapter we will consider the problem of finding the 
function $L_A$ which is the last unknown   function for the
transformations  (3.5).
This function corresponds to the post-Newtonian correction to the 
transformation of   barycentric time to   time in the proper {\small RF}.
As we shall see,   this function will depend on the model of matter
distribution 
taken to describe the internal structure of the bodies in the system.
In contrast to the functions $K_A$ and $Q^\alpha_A$, the analog of the
function $L_A$
has never previously been obtained, which makes the results presented    
in this chapter particularly interesting.   

\subsubsection{Equations for the Function $L_A$.}    

The relations  $(F2)$ and  conditions eqs.(5.1) enable us to 
obtain  the equations for the function $L_A$.
Thus, from the components  $\Gamma^{0}_{\alpha\beta}(y^p_{A})$ which 
are given by eq.$(F2c)$, we will have the first equation for this 
function as follows:
{}
$$ \Big[{\partial^2 \over \partial y^\alpha_A \partial y^\beta_A}
\Big(\hskip 1pt L_{A}(y^0_A, y^\nu_A) \hskip 3pt + \hskip 3pt
 {v_{A_0}}_\lambda Q^\lambda_{A}(y^0_A, y^\nu_A)\hskip 2pt\Big)
  +  $$
{}
$$ + \sum_{B\not= A} \Big( 2  \gamma _{\beta\lambda} 
{\partial \over \partial y^\alpha_A} V^\lambda_B (y^0_A, y^\nu_A) + 
2\gamma _{\alpha\lambda}{\partial \over \partial y^\beta_A} 
V^\lambda_B (y^0_A, y^\nu_A)  - $$
{}
$$-  \gamma _{\alpha\beta} {\partial \over \partial y^0_A} U_B 
(y^0_A,  y^\nu_A)\Big)\Big]\Bigg|_{\gamma_A} =  
   {\cal O}(c^{-6}). \eqno(5.24)$$

\noindent The second necessary equation may be obtained from 
the expression for the components $\Gamma^{0}_{0\alpha}
(y^p_{A})$ eq.$(F2b)$ by simply making of use the 
solution for the function $K_A$ given by the eq.(5.11) and
the result for the acceleration of the center of mass eq.(5.12).
This equation has the following form: 
{}
$$\Big[{\partial \over \partial y^\alpha_A} \Bigg( {\partial \over
 \partial y^0_A} L_{A}(y^0_A, y^\nu_A) +  
{1\over2}\Big({\partial \over \partial y^0_A}K_{A}(y^0_A, 
y^\nu_A)\Big)^2 + $$
{}
$$+{v_{A_0}}_\epsilon{\partial \over \partial y^0_A}Q^\epsilon_{A}
(y^0_A, y^\nu_A) + {1\over2}\sum_{B\not=A}H^{B <4>}_{00}(y^0_A,
y^\nu_A)+$$
{}
$$ +  2 \zeta^A_1 \cdot  U_A(y^0_A, y^\nu_A)\Bigg)
\Big]\Bigg|_{\gamma_A}=
   {\cal O}(c^{-6}), \eqno(5.25)$$

\noindent where the function $H^{B <4>}_{00}$ is given by the eq.$(4.16b)$. 
>From the relations for the components $\Gamma^{\alpha}_{00}(y^p_{A})$
eq.$(F2d)$ and with the help of the 
expression eqs.(5.11), (5.12), (5.17) and eq.(5.24)  one
may  obtain:
{}
$$ \Big[{\partial \over \partial y^0_A}\Bigg(\gamma^{\alpha\lambda}
{\partial \over \partial y^\lambda_A}\Big(
L_{A}(y^0_A, y^\nu_A) +  {v_{A_0}}_\nu Q^\nu_{A}(y^0_A, y^\nu_A)\Big) -
$$
{}
$$-v^\alpha_{A_0}{\partial \over \partial y^0_A} K_{A}(y^0_A, y^\nu_A) 
 + {\partial \over \partial y^0_A}Q^\alpha_{A}(y^0_A, y^\nu_A) + $$
{}
$$+ 4 \sum_{B\not= A}V^\alpha_B(y^0_A, y^\nu_A)
\Bigg)\Big]\Bigg|_{\gamma_A} 
 = \Big(\sigma^{\alpha\mu}_A - 2 \gamma^{\alpha\mu} \zeta^A_1\Big)
 \cdot 
\Big<{\partial U_A\over \partial y^\lambda_A}\Big>_0 + {\cal O}(c^{-6}).
 \eqno(5.26)$$

\noindent This last equation   may be formally integrated over time as
follows:
{}
$$ \Big[\gamma^{\alpha\lambda}{\partial \over \partial y^\lambda_A}
\Big(L_{A}(y^0_A, y^\nu_A) +  {v_{A_0}}_\nu Q^\nu_A(y^0_A,
y^\nu_A)\Big)-
v^\alpha_{A_0}{\partial \over \partial y^0_A} K_A(y^0_A, y^\nu_A)  + $$
{}
$$+{\partial \over \partial y^0_A}Q^\alpha_{A}(y^0_A, y^\nu_A) + 
 4 \sum_{B\not= A}V^\alpha_B(y^0_A, y^\nu_A)
\Big]\Bigg|_{\gamma_A}=$$
{}
$$= \Big(\sigma^{\alpha\mu}_A - 2 \gamma^{\alpha\mu} \zeta^A_1\Big)
 \cdot\int^{y^0_A} \hskip -7pt d t' \cdot \Big<{\partial U_A\over \partial 
y^\lambda_A}\Big>_{0'} + \sigma^\alpha_A +
{\cal O}(c^{-6}), \eqno(5.27)$$

\noindent 
where we have separated the integrating  constant $\sigma^\alpha_A$.
Using the relations for the components $\Gamma^{0}_{00}(y^p_{A})$
eq.$(F2a)$  
and the solutions  (5.11),(5.12) and  (5.27), one may 
obtain  the last equation for the function $L_A$ as given below:
{}
$$\Big[{\partial \over \partial y^0_A} \Bigg({\partial \over \partial
y^0_A}
 L_{A}(y^0_A, y^\nu_A)  + 
{1\over2}\Big({\partial \over \partial y^0_A}K_{A}(y^0_A, y^\nu_A) 
\Big)^2 + $$
{}
$$+{v_{A_0}}_\epsilon {\partial \over \partial y^0_A}Q^\epsilon_{A}
(y^0_A, y^\nu_A) +{1\over2} \sum_{B\not=A}H^{B <4>}_{00}(y^0_A,
y^\nu_A) 
\Bigg)\Big]\Bigg|_{\gamma_A} = $$
{}
$$=\Big(\sigma^{\mu\lambda}_A - 2 \gamma^{\mu\lambda}
\zeta^A_1\Big) 
 \Big<{\partial U_A\over \partial y^\mu_A}\Big>_0 \int^{y^0_A} 
\hskip -7pt d t' \cdot  
\Big<{\partial U_A\over \partial y^\lambda_A}\Big>_{0'}\hskip 1mm + $$
{}
$$  + \sigma ^\mu_A\cdot \Big<{\partial U_A\over \partial
y^\mu_A}\Big>_0 
 - 2 \zeta^A_1 \cdot {\partial  \over \partial y^0_A}\Big<U_A\Big>_0 + 
   {\cal O}(c^{-7}). \eqno(5.28)$$

Thus, we have obtained four equations necessary to  determine  the 
last unknown transformation function $L_A$, namely:  
eqs.(5.24)-(5.26)  and (5.28). 

\subsubsection{The Solution for the Function $L_A$.}  
The determination of the functions $K_A$ and $Q^\alpha_A$ helps
us find the solution for the function $L_A$ as well. In order 
to do this, let us look for the function $L_A$ in the  following form:
{}
$$L_A(y^0_A,y^\nu_A) =  \sum_{B\not=A} \Big[ k_1 \cdot {y_A}_\beta
 y^\beta_A \cdot \Big<{\partial U_B\over \partial y^0_A} \Big>_0 +  
k_2  \cdot y^\lambda_A  y^\beta_A \cdot \Big<{\partial_\lambda} 
{V_B}_\beta\Big>_0 + $$
{}
$$ + k_3 \cdot  {v_{A_{[0]}}}_\beta \Big(y^\beta_A  y^\lambda_A \cdot 
\Big< {\partial U_B\over \partial y^\lambda_A}\Big>_0-
{1\over2} \gamma^{\beta\sigma}y_{A \lambda} y^\lambda_A \cdot
\Big<{\partial U_B\over \partial y^\sigma_A}\Big>_0
 \Big) \Big] + {B^A_1} (y^0_A,y^\nu_A).\eqno(5.29)$$

\noindent Then from the equation eq.(5.24) one may easily obtain 
the  unknown constants $k_1, k_2,  k_3$ and the 
condition on the function ${B^A_1}$ as follows:
{}
$$k_1 = {1\over2}, \hskip 15pt k_2 = - 2, \hskip 15pt k_3 = 1;$$
{}
$$ {\partial^2 \over\partial y^\beta_A \partial y^\lambda_A }
{B^A_1}(y^0_A,y^\nu_A) = 0. \eqno(5.30)$$

The unknown function ${B^A_1}$ may  be determined from the equation 
eq.(5.27), by making use of the expressions eqs.(5.11),(5.23) and (5.24).  
Thus,  the intermediate solution for the function $L_A$ may be presented
as follows:
{}
$$L_A(y^0_A,y^\nu_A) =  \sum_{B\not=A} \Big[ {1\over2} {y_A}_\beta  
 y^\beta_A \cdot \Big<{\partial U_B\over \partial y^0_A}\Big>_0 - 
2 y^\lambda_A  y^\beta_A \cdot \Big<{\partial_\lambda}
{V_B}_\beta\Big>_0 + $$
{}
$$+ {v_{A_{[0]}}}_\beta \Big(y^\beta_A  y^\lambda_A \cdot 
\Big< {\partial U_B\over \partial y^\lambda_A}\Big>_0-
{1\over2} \gamma^{\beta\sigma}y_{A \lambda} y^\lambda_A \cdot
\Big<{\partial U_B\over \partial y^\sigma_A}\Big>_0
 \Big) \Big] + $$
{}
$$+ {v_{A_{[0]}}}_\beta  y_{A_\lambda} \int^{y^0_A} \hskip -8pt dt' 
\Big[ {1\over2} a^{[\lambda}_{A_{[0]}} v_{A_{[0]}}^{\beta]} + 
2 \sum_{B\not=A} \Big<\partial^{[\lambda} V^{\beta]}_B\Big>_{0'} \Big]  + 
$$
{}
$$ + y_{A_\beta} \Big[ 2 v^\beta_{A_{[0]}}\sum_{B\not=A}
\Big<U_B\Big>_0  -
 4\sum_{B\not=A} \Big<V^{ \beta}_B\Big>_0 
 - \dot {w}^\beta_{A_{[0]}}(y^0_A) + v^\beta_{A_{[0]}}\cdot{{\zeta_1}_A} +
$$
{}
$$ + \Big(\sigma^{\beta\lambda}_A - 2 \gamma^{\beta\lambda} 
\zeta^A_1\Big) 
\int^{y^0_A} \hskip -7pt dt'\cdot \Big<{\partial U_A\over \partial
 y^\lambda_A}\Big>_{0'} +  \sigma^\beta_{A} -
{v_{A_{[0]}}}_\lambda\cdot f^{\beta\lambda}_{A}\Big] + {B^A_2}(y^0_A),
 \eqno(5.31)$$

\noindent where $\sigma_A$ is a constant, and the unknown time-
dependent 
function ${B^A_2}$ may be obtained from the equation eq.(5.28). 
In order to do this, let us first integrate the equation  eq.(5.28) 
over the time $y^0_A$:
{}
$$ \Big[ {\partial \over \partial y^0_A} L_{A}(y^0_A, y^\nu_A)  + 
{1\over2}\Big({\partial \over \partial y^0_A}K_{A}(y^0_A, y^\nu_A)
 \Big)^2 + $$
{}
$$+{v_{A_0}}_\epsilon {\partial \over \partial y^0_A}Q^\epsilon_{A}
(y^0_A, y^\nu_A)  + 
{1\over2} H^{B <4>}_{00}(y^0_A, y^\nu_A) \Big]\Bigg|_{\gamma_A} = $$
{}
$$ = \Big(\sigma^{\mu\lambda}_A - 2 \gamma^{\mu\lambda}
\zeta^A_1\Big) 
\int^{y^0_A} \hskip -7pt d t' \cdot \Big<{\partial U_A\over \partial
y^\mu_A}
\Big>_{0'}\int^{t'} \hskip -7pt d t''\cdot  
\Big<{\partial U_A\over \partial y^\lambda_A}\Big>_{0''} \hskip 1mm + $$
{}
$$  + \sigma ^\mu_A\cdot \int^{y^0_A}\hskip -7pt d t'\Big<{\partial
U_A\over
 \partial y^\mu_A}\Big>_{0'}  -
 2 \zeta^A_1 \cdot  \Big<U_A\Big>_0 + \zeta^A_2  +    {\cal O}(c^{-7}).
\eqno(5.32)$$

\noindent Then, the function ${{B^A_2}}$ may be determined  from the
equation (5.32)
 with the help of eqs.(5.11), (5.23), (5.31) in the following form: 
{}
$$ {B_2}_A(y^0_A) = \int^{y^0_A} \hskip -10pt dt' \Big[ - 
{1\over2} \sum_{B\not=A}\Big<H^{B <4> }_{00}\Big>_{0'} - 
{1\over2}\Big(\sum_{B\not=A} \Big<U_B\Big>_{0'} -
{1\over2}{v_{A_{[0]}}}_\beta v^\beta_{A_{[0]}}  + \zeta^A_1\Big)^2 + $$
{}
$$ + \Big(\sigma^{\mu\lambda}_A - 2 \gamma^{\mu\lambda}
\zeta^A_1\Big) 
\int^{t'} \hskip -7pt d t'' \Big<{\partial U_A\over \partial
y^\mu_A}\Big>_{0''}
\int^{t''} \hskip -7pt d t'''
\Big<{\partial U_A\over \partial y^\lambda_A}\Big>_{0'''} \hskip 1mm + $$
{}
$$  + \sigma ^\mu_A\cdot \int^{t'} \hskip -7pt d t''\Big<{\partial U_A\over
 \partial y^\mu_A}\Big>_{0''}  -
 2 \zeta^A_1 \cdot  \Big<U_A\Big>_{0'}  + {\zeta^A_2} -
v_{A_{[0]}}{_{{}_\mu}}
\dot{w}^\mu_{A_{[0]}}(t')\Big].
\eqno(5.33)$$

\noindent Finally, by collecting  results obtained 
eqs.(5.31) and (5.33) together, we will get the following expression for
the transformation 
function $L_A$ in the  coordinates $(y^p_A)$ of quasi-inertial  {\small
RF}$_A$:
{}
$$L_A(y^0_A,y^\nu_A) =  \sum_{B\not=A} \Big[ {1\over2} {y_A}_\beta 
 y^\beta_A \cdot \Big<{\partial U_B\over \partial y^0_A}\Big>_0 - 
2 y^\lambda_A  y^\beta_A \cdot \Big<{\partial_\lambda}
{V_B}_\beta\Big>_0 + $$
{}
$$+ {v_{A_{[0]}}}_\beta \Big(y^\beta_A  y^\lambda_A \cdot 
\Big< {\partial U_B\over \partial y^\lambda_A}\Big>_0-
{1\over2} \gamma^{\beta\sigma}y_{A \lambda} y^\lambda_A \cdot
\Big<{\partial U_B\over \partial y^\sigma_A}\Big>_0
 \Big) \Big] + $$
{}
$$+ {v_{A_{[0]}}}_\beta  y_{A_\lambda} \int^{y^0_A} \hskip -8pt dt' 
\Big[ {1\over2} a^{[\lambda}_{A_{[0]}} v_{A_{[0]}}^{\beta]} + 
2 \sum_{B\not=A} \Big<\partial^{[\lambda} V^{\beta]}_B\Big>_{0'} \Big]  + 
$$
{}
$$ + y_{A_\beta} \Big[ 2 v^\beta_{A_{[0]}}\sum_{B\not=A}
\Big<U_B\Big>_0  -
 4\sum_{B\not=A} \Big<V^{ \beta}_B\Big>_0 
 - \dot {w}^\beta_{A_{[0]}} + v^\beta_{A_{[0]}}\cdot{{\zeta_1}_A} + $$
 {}
$$ + \sigma^\beta_{A} - {v_{A_{[0]}}}_\lambda  \cdot f^{\beta\lambda}_{A}
+ 
\Big(\sigma^{\beta\lambda}_A - 2 \gamma^{\beta\lambda} \zeta^A_1\Big) 
\int^{y^0_A} \hskip -7pt dt'\cdot \Big<{\partial U_A\over \partial
y^\lambda_A}
\Big>_{0'} \Big] + $$
{}
$$ + \int^{y^0_A} \hskip -10pt dt' \Big[ -{1\over2} \sum_{B\not=A}
\Big<H^{B <4>}_{00}\Big>_{0'}   - {1\over2}\Big(\sum_{B\not=A} 
\Big<U_B\Big>_{0'} - 
 {1\over2 }{v_{A_{[0]}}}_\beta v^\beta_{A_{[0]}}  + \zeta^A_1\Big)^2 - $$
{}
$$- v_{A_{[0]}}{_{{}_\mu}}\dot{w}^\mu_{A_{[0]}}(t') + 
\Big(\sigma^{\mu\lambda}_A - 2 \gamma^{\mu\lambda} \zeta^A_1\Big) 
\int^{t'} \hskip -7pt d t'' \Big<{\partial U_A\over \partial
y^\mu_A}\Big>_{0''}
\int^{t''} \hskip -7pt dt'''
\Big<{\partial U_A\over \partial y^\lambda_A}\Big>_{0'''}+$$
{}
$$  + {\zeta^A_2} - 2 \zeta^A_1 \cdot  \Big<U_A\Big>_{0'}   + 
\sigma ^\mu_A\cdot \int^{t'} \hskip -7pt d t''\Big<{\partial U_A\over 
\partial y^\mu_A}\Big>_{0''}\Big] + {\cal O}(c^{-6})y^0_A+ {\cal
O}(|y^\alpha_A|^3).\eqno(5.34)$$

Thus we have obtained the last  function  $L_A$ for the post-Newtonian
transformation
in the {\small WFSMA}.
Notice that this function is the only one which  depends on the 
model of matter chosen for description of the bodies in the system
through the 
term $H^{B <4>}_{00}$.
This function contains two new  parameters of the group of motion,
namely: 
parameter ${\zeta^A_2}$, which is the extension of the 
Newtonian parameter ${\zeta^A_1}$ on the post-Newtonian order, and the
parameter 
$\sigma^\alpha_A $, which  represents the time-dependent Poincar\'e
rotation.
The  function $L_A$ demonstrates  the non-linear character of the 
obtained group of motion. This non-linearity is due to the interaction 
of the proper gravitational field of the body {\small (A)} with the external
gravitation. Thus, the  Newtonian potential $U_A$ and its gradients 
influence the dynamic of proper {\small RF}$_A$ in the case when some  
of the parameters from the ten parametric group 
$(\zeta^A_1, \zeta^A_2; \sigma^\alpha_A; f^{\alpha\beta}_A)$ 
are not zero.  

It is worth noting that some parts of the
expression (5.34) were obtained by  D'Eath (1975a,b) whose method has
been used 
in the Brumberg-Kopejkin formalism (Brumberg \& Kopejkin, 1988a,b). 
However, this is the first time  the function $L_A$ 
has been obtained in the form of the expression above.
This function describes the post-Newtonian corrections to the proper time
and, 
besides the usual Lorentzian contributions, it contains the purely 
gravitational terms caused by the external gravitational field.
The only unknown function in this expression is the function
$w^\alpha_{A_{[0]}}$,
which will be discussed in the following subsection. Let us mention that
the knowledge of the 
function $L_A$ will required for analyzing   the  results of the proposed
post-Newtonian 
redshift experiment planned for the   {\it Solar Probe} mission (Anderson,
1989). 
This effect to the necessary accuracy was studied by Krisher (1993), who
had
formulated the frequency shift of the spacecraft clock to order $c^{-4}$.  
However, his formulation is appeared to be very simplified and does not
include the 
dynamical effects due to proper accelerated motion of the spacecraft
 in the close proximity to the Sun, which is   the crucial phase of the 
experiment. We believe that the correct 
derivation of the corresponding effect should be based upon the
relativistic theory of the 
astronomical {\small RF}s, so that the function $L_A$ (5.34) will provide
one with 
all the required corrections including both kinematical and dynamical
effects.
Moreover, in the Section {\small 7} we will
obtain the parameterized form of this function which will enable one to   
include in the analysis an alternative   tensor-scalar theories of gravity. 

\subsection{Equations of Motion for the Massive  Bodies.}

By finding the form of the  function $L_A$, we  
determined almost all of the functions for the 
coordinate transformation between {\small RF}s.
However, one quantity  still remains unspecified: the function 
$w^\alpha_{A_{[0]}}$ in expressions (5.23),(5.34). 
This function might be obtained from the last unused equation, namely the
equation eq.(5.25). 
By substituting the relations obtained for the functions $K_A, Q^\alpha_A$
given by 
eqs.(5.11), (5.23) into the equation eq.(5.25), 
and making of use the expression for the 
function $L_A$ given by eqs.(5.34), one obtains 
 the following ordinary differential equation for 
the last unknown function $w^\alpha_{A_{[0]}}$: 
{}
$$\ddot{w}^\alpha_{A_{[0]}}(y^0_A) = \sum_{B\not=A} 
\Big({1\over2}\gamma^{\alpha\sigma}\Big< {\partial H^{B <4>}_{00}\over
\partial y^\sigma_A}\Big>_0
+ v^\alpha_{A_{[0]}}\Big<{\partial U_B\over \partial y^0_A}\Big>_0- 
4   \Big<{\partial  V^{ \alpha}_B\over \partial y^0_A} \Big>_0\Big) - $$
{}
$$-{1\over2}v^\alpha_{A_{[0]}}{v_{A_{[0]}}}_\beta  a^\beta_{A_{[0]}} + 
a^\alpha_{A_{[0]}}\sum_{B\not=A} \Big<U_B \Big>_0  + $$
{}
$$ + {a_{A_{[0]}}}_\lambda  \int^{y^0_A} \hskip -10pt dt' 
\Big[ {1\over2} a^{[\alpha}_{A_{[0]}} v^{\lambda]}_{A_{[0]}} 
  +2 \sum_{B\not=A} \Big<\partial^{[\alpha} V^{\lambda]}_B\Big>_{0'}
\Big]-  
  {a_{A_{[0]}}}_\lambda  \cdot f^{\alpha\lambda}_{A} +
\sigma ^{\alpha\mu}_A\cdot \Big<{\partial U_A\over \partial
y^\mu_A}\Big>_0
 + {\cal O}(c^{-6}). \eqno(5.35)$$

\noindent 
We may check that this equation  is the post-Newtonian part of the 
acceleration $A^\alpha_{A_{[0]}}$ of the center of the field of the 
body {\small (A)} with respect to the barycenter written in it's proper
coordinate system.
If we perform  the coordinate 
transformation from the coordinates $(y^p_A)$ of the proper {\small
RF}$_A$ 
to those   $(x^p)$ of the inertial barycentric {\small RF}$_0$ of
 all the functions and potentials  entering in the equation 
eq.(5.35),  we  obtain the well known geodesic equation for the test body
written in the coordinates $(x^p)$ of the barycentric inertial {\small
RF}$_0$.
To do this, let us first combine the two parts of the   
acceleration $A^\alpha_{A_{[0]}}$ as follows:
{}
$$A^\alpha_{A_{[0]}}(y^0_A)=a^\alpha_{A_{[0]}}(y^0_A) + 
\ddot{w}^\alpha_{A_{[0]}}(y^0_A) + O(c^{-6}),\eqno(5.36)$$

\noindent where the terms in this equation are given by the 
relation (5.12) and (5.35).  
Then, by using the transformation rules  from the
 Appendix E, we may obtain the following result for the
acceleration $A^\alpha_{A_{[0]}}$ transformed into the coordinates of the
inertial 
barycentric {\small RF}$_0$:
{}
$$A^\alpha_{A_{[0]}}(x^p) = \sum_{B\not=A}\Big[ - 
{\partial^\alpha} U_B (x^p)\cdot 
\Big( 1 - {v_{A_{[0]}}}_\beta  v^\beta_{A_{[0]}} -
4 \sum_{B'\not=A} U_{B'} (x^p)\Big) -  $$
{}
$$ - 3 v^\alpha_{A_{[0]}}{\partial_0}U_{B} (x^p) - 
4 v^\alpha_{A_{[0]}}v^\lambda_{A_{[0]}} {\partial_\lambda} U_{B'} (x^p) 
 -  4  {\partial_0} V^\alpha_{B} (x^p)  + $$
{}
$$+4 {v_{A_{[0]}}}_\lambda  \Big( 
{\partial^\alpha}V^\lambda_{B}(x^p) -
 {\partial^\lambda} V^\alpha_{B}(x^p)\Big) - 2 {\partial^\alpha} 
\Phi_{1B}(x^p) - 
2 {\partial^\alpha} \Phi_{2B}(x^p) -$$
{}
$$ - {\partial^\alpha} \Phi_{3B}(x^p) - 3{\partial^\alpha} 
\Phi_{4B}(x^p) +
{1\over2}\gamma^{\alpha\nu}{\partial^3 \over \partial x^\nu 
\partial {x^0}^2 }  \chi_B (x^p) \Big]\Bigg|_{\gamma_A} +   {\cal O}(c^{-6}),
\eqno(5.37)$$

\noindent where the  quantities in the right hand side of this expression
are 
taken at the  world-line of the test body {\small (A)}. 
The equation (5.37)  is the usual form of the geodesic equation 
(Will, 1993; Brumberg, 1991) in the  coordinates
of an inertial {\small RF}$_0$. This result proves the previous conclusion
 that the relation  (5.35)  is also the geodesic equation, simply 
written in the coordinates of the proper quasi-inertial {\small RF}$_A$.

\subsection{The Proper RF of the Small Self-Gravitating Body.} 

In this part we will discuss  the transformation functions for the 
massive rotating test body with the small proper dimensions obtained 
in the previous parts of this Section. 
In order to do this, let us note  that the generalized Fermi conditions  
eqs.(5.2) involve  the first derivatives from the metric tensor, which gave
us the 
differential equations of the second order on the transformation 
functions $K_A, L_A, Q^\alpha_A$. The expected form of the 
post-Newtonian expansions of the metric tensor in the proper {\small
RF}$_A$, which 
resulted in the condition $(B3a)$, enabled us (with the help of 
the conditions (5.2)) to obtain the complete solution for the function
$K_A$. 
However, the functions $L$ and $Q^\alpha$ were only defined up to 
the second order with respect to the spatial point separation,
 namely: $L_A, Q^\alpha_A \sim {\cal O}(|y^\alpha_A|^3)$.  
This means that the arbitrariness due to the highest orders 
of the spatial point separation caused by the multipoles of higher 
than quadrupole $(k \geq 3)$ orders should be included in the expressions
for 
these functions. 
Taking these notes into account, we should include in the final 
expressions for these functions   the higher order terms with    
respect the spatial points separation. 
Then, the  solutions for these functions, presented by the 
relations (5.23) and (5.34) respectively,  should be 
extended  as follows:
{}
$${\widehat Q}^\alpha_{A_{[0]}}  (y^0_A, y^\nu_A) =
Q^\alpha_{A_{[0]}}(y^0_A, y^\nu_A) + 
\sum^{k}_{l \ge 3} {Q^\alpha_A}_{\{L\}} (y^0_A) \cdot 
y^{\{L\}}_A +  {\cal O}(|y^\nu_A|^{k+1}),\eqno(5.38a)$$
{}
$${\widehat  L}_{A_{[0]}}(y^0_A, y^\nu_A) = L_{A_{[0]}}(y^0_A, y^\nu_A) + 
\sum^{k}_{l\ge 3} {L_A}_{\{L\}}(y^0_A) \cdot y^{\{L\}}_A+
 {\cal O}(|y^\nu_A|^{k+1}). \eqno(5.38b)$$

As a result, the   post-Newtonian  
dynamically non-rotating coordinate transformations from the
coordinates 
of  barycentrical  inertial  {\small RF}$_0$  to those  
 of the proper quasi-inertial {\small RF}$_A$    will take 
the following form:
{}
$$x^0 = y^0_A + c^{-2}K_{A_{[0]}}(y^0_A, y^\epsilon_A) + 
c^{-4} {\widehat L}_{A_{[0]}}(y^0_A, y^\epsilon_A) + {\cal O}(c^{-6})y^0_A,
\eqno(5.40a)  $$
$$x^\alpha =  y^\alpha_A + y^\alpha_{A_{[0]}} (y^0_A) + 
c^{-2} {\widehat Q}^\alpha_{A_{[0]}}(y^0_A, y^\epsilon_A) + {\cal O}(c^{-
4})y^\alpha_A.
\eqno(5.40b)$$
With the transformation functions $K_A, Q^\alpha_A$ and $L_A$ given as
follows:
{}
$$K_{A_{[0]}}(y^0_A,y^\nu_A)  =\int^{y^0_A} \hskip -10pt  dt'  
\Big[ \sum_{B\not=A} \Big<U_B\Big>_{0'}-  
{1\over2}{v_{A_{[0]}}}_\nu v^\nu_{A_{[0]}} + \zeta^A_1\Big]  
-{v_{A_{[0]}}}_\nu\cdot y^\nu_A +  {\cal O}(c^{-4})y^0_A, \eqno(5.41a)$$
{}
$${\widehat Q}^\alpha_{A_{[0]}}(y^0_A,y^\nu_A) = - \sum_{B\not=A} \Big[
y^\alpha_A 
 y^\beta_A \cdot \Big<{\partial U_B\over \partial y^\beta_A}\Big>_0- 
{1\over2} \gamma^{\alpha\sigma}{y_A}_\beta  y^\beta_A 
\cdot\Big<{\partial U_B\over\partial y^\sigma_A} \Big>_0 
 + y^\alpha_A\cdot \Big<U_B\Big>_0\Big]+$$
{}
$$+{y_A}_\beta \int^{y^0_A} \hskip -10pt dt' \Big[ {1\over2}  
a^{[\alpha}_{A_{[0]}}
 v_{A_{[0]}}^{\beta]}   + 
2 \sum_{B\not=A}  \Big< \partial^{[\alpha} V^{\beta]}_B \Big>_{0'}\Big]  
- {1\over2} v^\alpha_{A_{[0]}} {v_{A_{[0]}}}_\beta \cdot y^{\beta}_A+$$ 
{}
$$+ {f_A^\alpha}_\beta \cdot  y^\beta_A  + w^\alpha_{A_{[0]}}(y^0_A) +
\sum^k_{l \ge3}  {Q^\alpha_A}_{\{L\}} (y^0_A) \cdot 
y^{\{L\}}_A + {\cal O}(|y^\nu_A|^{k+1})+ {\cal O} (c^{-4}) y^\alpha_A,  
\eqno(5.41b)$$ 
{}
$${\widehat L}_{A_{[0]}}(y^0_A,y^\nu_A) =  \sum_{B\not=A} \Big(
{1\over2} {y_A}_\beta 
y^\beta_A \cdot \Big<{\partial U_B\over \partial y^0_A} \Big>_0 - 
2 y^\lambda_A  y^\beta_A \cdot \Big<{\partial_\lambda}
V_{B\beta}\Big>_0 +$$
{}
$$+ {v_{A_{[0]}}}_\beta \Big[y^\beta_A  y^\lambda_A \cdot 
\Big<{\partial U_B\over \partial y^\lambda_A} \Big>_0- 
{1\over2}\gamma^{\beta\sigma} y_{A_\lambda}y^\lambda_A 
\cdot\Big<{\partial U_B\over \partial y^\sigma_A} \Big>_0\Big] \Big) +$$
{}
$$+{v_{A_{[0]}}}_\beta y_{A_\lambda} \int^{y^0_A} \hskip -10pt dt'
 \Big[ {1\over2} a^{[\lambda}_{A_{[0]}} v_{A_{[0]}}^{\beta]} + 
2 \sum_{B\not=A} \Big<\partial^{[\lambda}  V^{\beta]}_B\Big>_{0'} \Big]  + 
$$
{}
$$ + y_{A_\beta} \Big[ 2
v^\beta_{A_{[0]}}\sum_{B\not=A}\Big<U_B\Big>_0
 - 4\sum_{B\not=A} \Big<V^{\beta}_B\Big>_0
 - \dot {w}^\beta_{A_{[0]}}(y^0_A) + v^\beta_{A_{[0]}}\cdot\zeta^A_1 + $$
{}
$$ + \sigma^\beta_{A} - {v_{A_{[0]}}}_\lambda  \cdot f^{\beta\lambda}_{A}
+ 
\Big(\sigma^{\beta\lambda}_A - 2 \gamma^{\beta\lambda} \zeta^A_1\Big) 
\int^{y^0_A} \hskip -7pt dt'\cdot \Big<{\partial U_A\over \partial
y^\lambda_A}
\Big>_{0'} \Big] + $$
{}
$$ +\int^{y^0_A} \hskip -10pt dt' \Big[ -
\sum_{B\not=A}\Big<W_B\Big>_{0'}  - 
{1\over2}\Big(\sum_{B\not=A} \Big<U_B\Big>_{0'} -
{1\over2}{v_{A_{[0]}}}_\beta 
v^\beta_{A_{[0]}}+\zeta^A_1\Big)^2-$$
{}
$$ - v_{A_{[0]}}{_{{}_\mu}}\dot{w}^\mu_{A_{[0]}}(t') + 
\Big(\sigma^{\mu\lambda}_A - 2 \gamma^{\mu\lambda} \zeta^A_1\Big) 
\int^{t'} \hskip -7pt d t'' \Big<{\partial U_A\over \partial
y^\mu_A}\Big>_{0''}
\int^{t''} \hskip -7pt dt'''
\Big<{\partial U_A\over \partial y^\lambda_A}\Big>_{0'''} \hskip 1mm + $$
{}
$$+ {\zeta^A_2}   - 2 \zeta^A_1 \cdot  \Big<U_A\Big>_{0'}  
 + \sigma ^\mu_A\cdot \int^{t'} \hskip -7pt d t''\Big<{\partial U_A\over 
\partial y^\mu_A}\Big>_{0''}\Big] + $$
{}
$$+ \sum^{k}_{l \ge3}  {L_A}_{\{L\}}(y^0_A) \cdot y^{\{L\}}_A+
   {\cal O}(|y^\nu_A|^{k+1}) +{\cal O}(c^{-6}).\eqno(5.41c)$$ 

\noindent with the equations for both time-dependent functions 
$y^\alpha_{A_{[0]}}(y^0_A)$ and $w^\alpha_{A_{[0]}}(y^0_A)$  
defined by the equations (5.12) and (5.35) respectively.

At this point we are ready to present the general form 
of the metric tensor in the proper {\small RF}$_A$ defined with 
the generalized Fermi conditions. Thus, by substituting 
the solutions obtained for the functions $K_A, L_A$ and $Q^\alpha_A$ 
 into general form of the metric tensor
$g_{mn}(y^p_A)$  in a  proper {\small RF}$_A$ given by the relations
eqs.(4.11), 
we will obtain this tensor in the following form: 
{}
$$g^A_{00}(y^p_A) = \hskip 5pt 1  
 -2 \Bigg(\sum_BU_B(y^0_A, y^\nu_A)- \sum_{B\not=A}\Big[
y^\mu_A \Big<{\partial U_B\over \partial y^\mu_A}\Big>_0   +
\Big<U_B\Big>_0 \Big]+
\zeta_{1A}\Bigg) + $$
{}
$$ + 2 \Bigg(\sum_B W_B(y^0_A, y^\nu_A) - \sum_{B\not=A} 
\Big[y^\mu_A \Big<{\partial W_B\over \partial y^\mu_A}\Big>_0   +
\Big<W_B\Big>_0 \Big]+ 
\zeta_{2A}\Bigg) + $$
{}
$$ + y^\delta_A y^\beta_A \cdot \Big[ \gamma_{\delta\beta} 
\hskip 1mm {a_{A_{[0]}}}_\lambda a^\lambda_{A_{[0]}} -  
{a_{A_{[0]}}}_\delta {a_{A_{[0]}}}_\beta + \sum_{B\not=A}
{\partial \over \partial y^0_A}\Big( \gamma_{\delta\beta} \Big<{\partial 
U_B 
\over\partial {y^0_A}}\Big>_0   - 
4 \Big<{\partial  {V_B}_\beta \over\partial y^\delta_A} \Big>_0\Big)\Big]
+ $$
{}
$$ + 2  \sum^{k}_{l \ge3}  \Big({\partial \over \partial y^0_A}
{L_A}_{\{L\}}(y^0_A)
 + {v_{A_{[0]}}}_\beta \cdot {\partial \over \partial y^0_A}
{Q^\beta_A}_{\{L\}} (y^0_A)\Big)
\cdot  y^{\{L\}}_A + $$
{}
$$ + \Big(\sigma^{\mu\lambda}_A - 2 \gamma^{\mu\lambda} 
\zeta^A_1\Big)\Bigg( {y_A}_\mu \cdot 
\Big< {\partial U_A\over \partial y^\lambda_A}\Big>_0 + 
\int^{y^0_A} \hskip -7pt d t' \Big<{\partial U_A\over \partial
y^\mu_A}\Big>_{0'}\int^{t'} \hskip -7pt d t''
\Big<{\partial U_A\over \partial y^\lambda_A}\Big>_{0''}\Bigg) 
\hskip 1mm + $$
{}
$$  + \sigma^\mu_A\cdot \int^{y^0_A} \hskip -7pt d t'
\Big<{\partial U_A\over \partial y^\mu_A}\Big>_{0'}  -
 2 \zeta^A_1 \cdot  \Big<U_A\Big>_0 + 
{\cal O}(|y^\nu_A|^{k+1}) +{\cal O}(c^{-6}), \eqno(5.42a) $$ 
{}
$$g^{A}_{0\alpha}(y^p_A)  = 4 \gamma_{\alpha\epsilon}
 \Bigg(\sum_B V^\epsilon_B(y^0_A,y^\nu_A) - 
\sum_{B\not=A}\Big[y^\mu_A \Big<{\partial V^{\epsilon}_B\over \partial
y^\mu_A}\Big>_0   + 
\Big<V^{\epsilon}_B\Big>_0 \Big] +
 \sigma^\epsilon_A\Bigg) -$$
{}
$$ -{1\over2} \Big(\gamma_{\alpha\epsilon} \delta^\beta_\lambda + 
\gamma_{\alpha\lambda}\delta^\beta_\epsilon - 
\gamma_{\epsilon\lambda} \delta^\beta_\alpha\Big) \hskip 1mm 
y^\epsilon_A y^\lambda_A\cdot 
\sum_{B\not=A} {\partial \over\partial {y^0_A}}\Big<{\partial  
U_B\over \partial {y^\beta_A}}\Big>_0    + $$
{}
$$+\sum^{k}_{l \ge3}  \Big(\gamma_{\alpha \lambda}
{\partial \over \partial y^0_A}{Q^\lambda_A}_{\{L\}} (y^0_A) + 
\Big[{L_A}_{\{L\}}(y^0_A)  + {v_{A_{[0]}}}_\beta \cdot 
{Q^\beta_A}_{\{L\}} (y^0_A)\Big] {\partial \over \partial y^\alpha_A}\Big)
\cdot  y^{\{L\}}_A  + $$
{}
$$+ \gamma_{\alpha \mu}\Big(\sigma^{\mu\lambda}_A -
 2 \gamma^{\mu\lambda} \zeta^A_1\Big) 
\int^{y^0_A} \hskip -7pt d t' \Big<{\partial U_A\over \partial
 y^\lambda_A}\Big>_{0'}  + {\cal O}(|y^\nu_A|^{k+1})+  {\cal O}(c^{-5}),
 \eqno(5.42b)$$
{}
$$g^{A}_{\alpha\beta}(y^p_A)  =
\gamma_{\alpha\beta} + 
 2 \gamma_{\alpha\beta}\Bigg(\sum_BU_B(y^0_A, y^\nu_A) - 
\sum_{B\not=A}\Big[y^\mu_A \Big< {\partial U_B\over \partial
y^\mu_A}\Big>_0   + 
\Big<U_B\Big>_0 \Big] \Bigg) + {\sigma_A}_{\alpha\beta} + $$ 
{}
$$ +\sum^{k}_{l \ge3}  \Big(\gamma_{\alpha\lambda} {Q^\lambda_A}_
{\{L\}} (y^0_A)  
{\partial \over \partial y^\beta_A} + \gamma_{\beta\lambda}
 {Q^\lambda_A}_{\{L\}} (y^0_A)  
{\partial \over \partial y^\alpha_A} \Big) \cdot 
 y^{\{L\}}_A + {\cal O}(|y^\nu_A|^{k+1})+ {\cal O}(c^{-4}) , \eqno(5.42c)$$

\noindent where the subscript {\small (A)} for the components of the
metric tensor 
specifies that this tensor was obtained by making of use the 
specifically defined transformation functions (5.41). The expressions for 
the functions $W_A$ and $W_B$ were obtained by substituting the
solutions for the 
transforation functions into 
the relations for the $H^{A<4>}_{00}$ and $H^{B<4>}_{00}$ given by the 
eqs.(4.14) and (4.16) correspondingly.  These functions have the following
form:
{}
$$W_A(y^p_A) = U^2_A(y^p_A) +\Psi_A(y^p_A)  
 +  {1\over2}{\partial^2\over\partial {y^0_A}^2} \chi_A(y^p_A) +
  2 a^\lambda_{A_{[0]}}\cdot{\partial \over \partial 
y^\lambda_A} \chi_A (y^p_A) +$$
{}
$$+2\sum_{B\not=A}\Big(U_A(y^p_A) U_B(y^p_A) -\int_A{d^3y'_A
\over{|y^\nu_A - y'^\nu_A|}} 
\rho_A(y^0_A, y'^\nu_A) U_B(y^0_A, y'^\nu_A) \Big)+$$
{}
$$+\sum^k_{l \ge3}  {Q^\lambda_A}_{\{L\}} (y^0_A) 
\int_Ad^3y'_A\rho_A (y_A^0, y_A'^\nu)
{\partial \over\partial y'^\lambda_A}  \Big[{ y^{\{L\}}_A-
y'^{\{L\}}_A\over{|y_A^\nu - y_A'^\nu|}} \Big]  -$$
{}
$$ - f^{\lambda\beta}_A \cdot {\partial^2 \over 
\partial y^\lambda_A \partial y^\beta_A} \chi_A (y^0_A, y^\nu_A)- 
2\zeta_{1A} \cdot U_A(y^0_A, y^\nu_A) + 
O(|y^\nu_A|^{k+1})+ {\cal O}(c^{-6}). \eqno(5.43a)$$ 
{} 
$$W_B(y^p_A) = U_B(y^p_A) \sum_{B'\not= A} U_{B'}(y^p_A) +
\Psi_{B}(y^p_A)   + 
 {1\over2}{\partial^2\over\partial {y^0_A}^2} \chi_B(y^p_A) 
+2 a^\lambda_{A_{[0]}}(y^0_A) \cdot {\partial \over \partial 
y^\lambda_A} \chi_B (y^0_A, y^\nu_A) - $$
{}
$$-2\int_B{d^3y'_A \over{|y^\nu_A - y'^\nu_A|}} \rho_B(y^0_A, y'^\nu_A) 
\sum_{B'}  U_{B'}(y^0_A, y'^\nu_A)+ $$
{}
$$+\sum^k_{l \ge3}  {Q^\lambda_A}_{\{L\}} (y^0_A) \cdot 
\int_Bd^3y'_A \rho_B \Big(y_A^0, y_A'^\nu+y_{BA_0}^\nu(y_A^0)\Big) 
{\partial \over\partial y'^\lambda_A} 
\Big[{ y^{\{L\}}_A-y'^{\{L\}}_A\over{|y_A^\nu - y_A'^\nu|}}\Big] -$$
{}
$$ - f^{\lambda\beta}_A \cdot{\partial^2 \over \partial 
y^\lambda_A \partial y^\beta_A} \chi_B (y^0_A, y^\nu_A) - 
2\zeta_{1A} \cdot U_B(y^0_A, y^\nu_A) + O(|y^\nu_A|^{k+1})+ 
{\cal O}(c^{-6}).\eqno(5.43b)$$ 
 
The expressions  (5.42) are the general solution for 
the field equations of the general theory of relativity, 
which satisfies the generalized Fermi conditions 
eqs.(5.2) in the immediate vicinity of  body {\small (A)}.
This solution reflects the geometrical features of 
the proper {\small RF}$_A$ with respect to the special properties 
of the motion of the $k^{th}$ multipoles of the of  unknown functions
${L_A}_{\{L\}}(y^0_A)$ and  ${Q^\nu_A}_{\{L\}} 
(y^0_A)$ for $l \ge 3$ which will be discussed further.

The transformation  functions eq.(5.41) correspond to 
non-rotating coordinate transformations  between different {\small RF}s   
in the {\small WFSMA}. They were   obtained   by  applying  the 
generalized Fermi conditions  eqs.(5.2).
The set of the resulting formulae eqs.(5.41) together with eqs.(5.12) and
(5.35) represents the generalization of the Poincar\'e group of motion to
the problem  of practical celestial mechanics.
The arbitrary constants $\zeta_A = c^{-2}\zeta^A_1 +
 c^{-4}\zeta^A_2$, $\sigma^\alpha_A$ and $f^{\alpha\beta}_A$
correspond to the maximal number of  Killing vectors ($M=10$) 
in the background pseudo-Euclidean  space-time and the expressions
(5.40)-(5.41) 
represent the ten-parameter  group of motion constructed
 for the dynamic of the celestial bodies in the {\small WFSMA}.
The non-zero parameters describe the 
shift of the origin of the coordinate system, the constant 
spatial rotation of the axes and the relativistic Poincar\'e
rotation. These parameters represent the offset of the origin of
the coordinate system from the center of the field of the body 
under consideration, which  may vary from body to body. 
Moreover, these parameters  lead to the appearance of the proper 
gravitational potential $U_A$ and it's gradients 
$\partial^\alpha U_A$ in the function $L_A$ $(5.41c)$.  
The contribution of this sort could be a useful  tool for some practical 
applications of the atomic time comparison (Brumberg, 1991a). 
This dependence is indicating the fact that the constant part of the proper
gravity
of the body {\small (A)} is also affecting the 
definition of its world-line. This contribution, my be neglected if  one 
will choose these constants in such a way that this influence of the
proper field 
will vanish. In addition, let us mention that  the component of the metric  
tensor $g^A_{00}$ becomes  also dependent
 on these  quantities describing the proper gravitational field, which
  violates the conditions on the metric tensor and the 
coordinate transformations to the proper {\small RF}$_A$ given in the
Section {\small 1}. 
Therefore,  without loosing a generality, in our future 
calculations  we will  eliminate this offset and will set  
all of these parameters to be zero:
{}
$$\zeta_A = c^{-2}\zeta^A_1 +
 c^{-4}\zeta^A_2=\sigma^\alpha_A=f^{\alpha\beta}_A=0. \eqno(5.44)$$
 
In order to find the unknown  functions 
${Q^\alpha_A}_{\{L\}}(y^0_A), {L_A}_{\{L\}}(y^0_A)$ up to the 
$k^{th} \hskip 1mm (k \geq 3)$ order,
one should use the conditions which will contain the spatial 
derivatives from the metric tensor of the $(k-1)$ order. 
Moreover, one   expects to obtain the recurrent formulae which 
would connect the features of transformation of an arbitrary 
$k^{th}$  term with those for the previous  $(k-1)$ terms. 
 Thus, following Synge (1960), one may want to apply some non-local 
geometrical constructions, such as Jacobi equations 
(Manasse \& Misner, 1963) or both Jacobi equations and  Fermi-Walker
 transport (Li \& Ni, 1979a,b). However, 
these  constraints are generally not related to the particular theory
 under consideration, so their  application should  be justified for  
particular theory of gravity under question.
Another method is to use the `external' multipole moments 
as they were defined for the gravitational wave theory by Thorne (1980) 
or Blanchet \& Damour (1986, 1989). Indeed, one could show that the
functions 
${Q^\alpha_A}_{\{L\}}(y^0_A)$ and ${L_A}_{\{L\}}(y^0_A)$ in the 
{\small WFSMA} may be chosen in a such a way 
that the metric tensor in a proper {\small RF}$_A$ eqs.(4.11) 
 corresponding to this choice will accept the desired form. 
The presentation of the transformation functions in the terms of the 
`external' multipole moments  simply corresponds to the specific {\small
RF}
for which {\small KLQ} dynamical parameterization is strictly defined by
this choice.    

\subsection{The Fermi-Normal-Like Coordinates.} 

As we noticed above, in order to determine the metric up to the 
$k^{th}$ multipole contribution, one should apply some 
additional conditions which  enable us to define the 
specific properties of the reference frame with which  we will be dealing.
 For example, we might obtain these functions for the case of
 the motion of the monopole test particle up to the second 
order of a spatial point separation. Assuming  the motion 
of that particle is described by the geodesic equation and the 
deviation of geodesics is governed by the Jacobi equation,
 we might easy obtain the metric tensor in the generalized 
 Fermi normal coordinates (Misner and Manasse, 1963; 
Li \& Ni, 1979;   Dolgov, Khriplovich 1983; Ashby \& Bertotti, 1986;
Marzlin, 1994) up to the second order of the spatial separation 
and presented as follows:
{} 
$$g^{\cal F}_{00}(y^p_{A}) = 1 + H^A_{00}(y^p_A) + \Big<R^B_{0\mu
0\nu}\Big>_0
\cdot y^\mu_A y^\nu_A +  {\cal O}(c^{-6}) +  {\cal O}( |y^\nu_A|^{3}), 
\eqno(5.45a)$$
{}
$$g^{\cal F}_{0\alpha}(y^p_{A}) = H^A_{0\alpha}(y^p_A)+
{2\over3}\Big<R^B_
{0\mu \alpha\nu}\Big>_0  \cdot y^\mu_A y^\nu_A + 
 {\cal O}(c^{-5}) +  {\cal O}( |y^\nu_A|^{3}), \eqno(5.45b)$$
{}
$$g^{\cal F}_{\alpha\beta}(y^p_{A}) = \gamma_{\alpha\beta} +
H^A_{\alpha\beta}(y^p_A)
+ {1\over3}  \Big<R^B_{\alpha\mu\beta\nu}\Big>_0 \cdot y^\mu_A y^\nu_A 
+  {\cal O}(c^{-4}) +  {\cal O}( |y^\nu_A|^{3}), \eqno(5.45c)$$

\noindent where $\Big<R^B_{mnkl}\Big>_0$ are the  components of the 
Riemann tensor eqs.$(G9)$ which is calculated with respect 
to the external gravitational field $H^B_{mn}$ 
and taken on the world-line $\gamma_A$ of the  body {\small (A)} under
consideration. 

Let us mention,  that if the proper gravitational field may be neglected
and the effects due to 
acceleration of the proper {\small RF}$_A$ are also negligible, the
obtained metric tensor
(5.45) will correspond to that of, so-called, Fermi normal coordinates 
constructed in the immediate vicinity of the world-line of an inertial 
observer (Misner {\it et al.}, 1973).
However, for the general case of non-vanishing contributions of the 
proper gravitational field and the accelerated barycentric motion, 
the form of the metric tensor $g^{\cal F}_{mn}$ (5.45) and the
corresponding proper {\small RF},
is   what will be referred to as the Fermi-normal-like coordinates.
 From  these expressions for the metric tensor $g^{\cal F}_{mn}$,
one may see that, in order to obtain this  form of the  metric 
tensor, it is necessary to perform the coordinate transformation
which should contain the terms with the third order of the spatial
 point separation (Li \& Ni, 1979a,b; Zhang, 1985, 1986). 
We will obtain the necessary equations on these functions 
by making of use  the components of the Riemann tensor 
$R_{mnkl}(y^p_A)$ expanded with respect to the spatial separation 
from the world-line of the body {\small (A)} and then
equating  the coefficients proportional to $\sim y^\mu_A y^\nu_A$. 

Thus, the components of the Riemann tensor calculated with respect 
to external gravitational field $H^B_{mn}$ from the relations  eqs.$(G9)$   
might be presented on a world-line of the  body {\small (A)} as follows:  
{}
$$ \Big<R^B_{0\mu0\nu}\Big>_0=   \sum_{B\not=A} \Bigg( -
\Big<{\partial^2  
U_B\over \partial y^\mu_A\partial y^\nu_A}\Big>_0 + \gamma_{\mu\nu} 
{\partial \over\partial {y^0_A}} \Big[\Big<{\partial  U_B\over\partial
{y^0_A}}\Big>_0  - 
2 \Big(\Big<{\partial  {V_B}_{\mu} \over \partial y^{\nu}_A}\Big>_0+
 \Big<{\partial  {V_B}_{\nu} \over \partial y^{\mu}_A}\Big>_0 \Big)\Big] +
$$
{}
$$ +  \Big<{\partial^2  W_B\over \partial y^\mu_A\partial
y^\nu_A}\Big>_0 \Bigg)  
+ \gamma_{\mu\nu} \hskip 1mm {a_{A_{[0]}}}_\lambda
a^\lambda_{A_{[0]}} -
  {a_{A_{[0]}}}_\mu {a_{A_{[0]}}}_\nu  +   {\cal O}(c^{-6}), \eqno(5.46a)$$
{}
$$\Big<R^B_{0\mu\alpha\nu}\Big>_0  =
\gamma_{\mu\alpha}{\dot{a}_{A_{[0]}}}{_{{}_\nu}} -
\gamma_{\mu\nu}{\dot{a}_{A_{[0]}}}{_{{}_\alpha}}+$$
{}
$$+2 \sum_{B\not=A} \Big( \gamma _{\alpha\lambda} 
\Big<{\partial^2 V^{\lambda}_B \over  \partial y^\mu_A \partial
y^\nu_A}\Big>_0
-\gamma _{\nu\lambda}\Big<{\partial^2 V^{\lambda}_B\over  
\partial y^\mu_A\partial y^\alpha_A}\Big>_0  \Big) 
 + {\cal O}(c^{-5}), \eqno(5.46b)$$
{}
$$ \Big< R^B_{\alpha\mu\beta\nu}\Big>_0    = \sum_{B\not=A} \Big(
\gamma _{\alpha\beta} 
\Big<{\partial^2 U_B \over \partial y^\mu_A \partial y^\nu_A}\Big>_0 
  + \gamma _{\mu\nu}\Big<{\partial^2 U_B \over \partial
 y^\alpha_A \partial y^\beta_A}\Big>_0 -$$
{} 
$$ - \gamma _{\beta\mu}\Big<{\partial^2  U_B\over \partial 
y^\alpha_A \partial y^\nu_A}\Big>_0  - 
\gamma _{\alpha\nu}\Big<{\partial^2  U_B\over \partial 
y^\mu_A \partial y^\beta_A}\Big>_0\Big) 
  +  {\cal O}(c^{-4}). \eqno(5.46c)$$

To find the necessary corrections of the third order of the 
transformation functions $Q^\alpha_A$ and $L_A$, let us look  
in the following form:
{}
$$\delta_{\nu_3}Q^\alpha_{A_{[0]}}(y^0_A, y^\nu_A) =$$
{}
$$=\sum_{B \not=A}\Big[c_1 \cdot\gamma^{\alpha\sigma} y^\mu_A 
{y_A}_\mu y^\nu_A\cdot
\Big<{\partial^2  U_B\over \partial y^\nu_A \partial  y^\sigma_A}\Big>_0
 + c_2\cdot y^\alpha_A y^\mu_A y^\lambda_A \cdot
\Big<{\partial^2  U_B\over \partial y^\mu_A \partial  
y^\lambda_A}\Big>_0\Big]+ {\cal O}(|y^\nu_A|^{4}), \eqno(5.47a)$$
{}
$$ \delta_{\nu_3}L_{A_{[0]}}(y^0_A, y^\nu_A) =\sum_{B \not=A}
\Big[q_1\cdot y^\mu_A {y_A}_\mu y^\lambda_A 
\Big<{\partial^2  U^*_B\over \partial y^0_A \partial  y^\lambda_A}\Big>_0
+ 
q_2 \cdot y^\mu_A y^\nu_A y^\lambda_A \Big<{\partial^2  
{V_B}_\lambda\over \partial y^\mu_A \partial  y^\nu_A}\Big>_0\Big] -$$
{}
$$-{v_{A_{[0]}}}_\beta \cdot \delta_{\nu_3}Q^\beta_{A_{[0]}}(y^0_A,
y^\nu_A)
 + {\cal O}(|y^\nu_A|^{4}), \eqno(5.47b)$$ 

\noindent where the constants $c_1, c_2$ and $q_1, q_2$ are   
unknown at the moment.

The expressions for the components of the metric tensor 
$g^{\cal F}_{mn}$ eqs.(5.45) and those for the Riemann tensor 
eqs.(5.46)  will enable us to obtain 
the equations for the determination of the constants 
$c_1, c_2$ and $q_1, q_2$. Thus, from the component 
$g^{\cal F}_{\alpha\beta}$ eq.$(5.45c)$ and relation for the 
$\Big<R^B_{\alpha\mu\beta\nu}\Big>_0$ eq.$(5.46c)$ we will have:
{} 
$$2c_1 + 1 = {1\over3}, \hskip 5mm 2 c_2 ={1\over3},
\qquad   2(c_1 + c_2) = -{1\over3}, $$

\noindent which will give the following values for  these constants: 
{}
$$c_1 = - {1\over3}, \qquad c_2 = {1\over6}. \eqno(5.48)$$ 

Analogously, from the component $g^{\cal F}_{0\alpha}$ eq.$(5.45b)$,
the  relation for the $\Big<R^B_{0\mu\alpha\sigma}\Big>_0$
eq.$(5.46b)$, and the solution for the function $\delta_{\nu_3}
Q^\alpha_{A_{[0]}}$ given by eq.$(5.46a)$ with the obtained $c_1$ and
$c_2$ 
eq.(5.47), we obtains:
{}
$$2q_1 - 1 = -{2\over3}; \hskip 5mm q_1 + {1\over2} = {2\over3}
 \hskip 5mm \Rightarrow \hskip 5mm  q_1 = {1\over6}, $$
{}
$$2q_2 = - {4\over 3}; \hskip 5mm q_2 + 2 = {4\over 3} 
\hskip 5mm \Rightarrow \hskip 5mm q_2 = - {2\over 3}. \eqno(5.49)$$ 

Taking these results into account, the corrections up to the third order 
with respect to the spatial 
point separation to the solutions for the $Q^\alpha_A$ and $L_A$, 
presented  by the equations eqs.(5.47) will take the following form: 
{}
$$\delta_{\nu_3}Q^\alpha_{A_{[0]}}(y^0_A, y^\nu_A) = {1\over6} 
\sum_{B \not=A}\Big[\gamma^{\alpha\sigma} y^\mu_A y_{A_\mu}
y^\nu_A\cdot
\Big<{\partial^2  U_B\over \partial y^\nu_A \partial  y^\sigma_A}\Big>_0
-$$
{}
$$-  2 \cdot y^\alpha_A y^\mu_A y^\lambda_A \cdot
\Big<{\partial^2  U_B\over \partial y^\mu_A \partial  
y^\lambda_A}\Big>_0\Big]+{\cal O}(|y^\nu_A|^4), \eqno(5.50a)$$ 
{}
$$ \delta_{\nu_3}L_{A_{[0]}}(y^0_A, y^\nu_A) = {1\over 6}\sum_
{B \not=A}\Big( y^\mu_A {y_A}_\mu y^\lambda_A 
\Big<{\partial^2  U_B\over \partial y^0_A \partial  y^\lambda_A}\Big>_0 - 
4\cdot y^\mu_A y^\nu_A y^\lambda_A \Big<{\partial^2  
{V_B}_\lambda\over \partial y^\mu_A \partial  y^\nu_A}\Big>_0\Big) -$$ 
{}
$$ -{v_{A_{[0]}}}_\beta \Big(\gamma^{\beta\sigma} y^\mu_A
{y_A}_\mu y^\nu_A\cdot
\Big<{\partial^2  U^*_B\over \partial y^\nu_A \partial 
 y^\sigma_A}\Big>_0 - 2 \cdot y^\beta_A y^\mu_A y^\lambda_A \cdot
\Big<{\partial^2  U^*_B\over \partial y^\mu_A \partial
  y^\lambda_A}\Big>_0\Big)+{\cal O}(|y^\nu_A|^4). \eqno(5.50b)$$ 

\noindent 
By substituting these solutions into the expressions eqs.(5.42) 
 one might get the   metric tensor
in a proper {\small RF}$_A$ of the moving extended body {\small (A)} with  
accuracy up to the second order of the spatial point separation. 
Thus, assuming that all the integration constants   satisfy eq.(5.44),
 one may get the following form of the metric tensor in the 
generalized Fermi normal coordinates: 
{}
$$g^{\cal F}_{00}(y^p_A) = 1  -2U_A(y^p_A)+2W_A(y^p_A)+$$
{}
$$+ \Big(\sum_{B\not=A}
\Big[ - \Big<{\partial^2  U_B \over \partial y^\mu_A\partial
y^\nu_A}\Big>_0 +
\Big<{\partial^2  W_B\over \partial y^\mu_A\partial y^\nu_A}\Big>_0 + 
{\partial  \over\partial {y^0_A} } \Big(\gamma_{\mu\nu} \Big<{\partial 
U_B\over\partial {y^0_A} }\Big>_0  - 
4\Big<{\partial  {V_B}_\mu\over\partial y^\nu_A}\Big>_0 \Big)\Big] +$$
{}
$$ + \gamma_{\mu\nu} \hskip 1mm {a_{A_{[0]}}}_\lambda
a^\lambda_{A_{[0]}} -  
{a_{A_{[0]}}}_\mu {a_{A_{[0]}}}_\nu\Big) \cdot y^\mu_A y^\nu_A +  {\cal
O}(c^{-6}) + 
 {\cal O}(|y^\nu_A|^{3}), \eqno(5.51a) $$
{}
$$g^{\cal F}_{0\alpha}(y^p_A) = 4\gamma_{\alpha\epsilon}
V^\epsilon_A(y^p_A)+
{2\over3}\Big( 2 \sum_{B\not=A} \Big[ \gamma _{\alpha\lambda} 
\Big<{\partial^2  V^{\lambda}_B\over  \partial y^\mu_A \partial
y^\nu_A}\Big>_0 -
\gamma _{\nu\lambda}\Big<{\partial^2  V^{\lambda}_B\over  
\partial y^\mu_A\partial y^\alpha_A}\Big>_0 \Big]+ $$
{}
$$+ \gamma_{\mu\alpha}{\dot{a}_{A_{[0]}}}{_{{}_\nu}}-  
\gamma _{\mu\nu}{\dot{a}_{A_{[0]}}}{_{{}_\alpha}} \Big) \cdot y^\mu_A
y^\nu_A  
 +  {\cal O}(|y^\nu_A|^{3}) +   {\cal O}(c^{-5}), \eqno(5.51b)$$ 
{}
$$g^{\cal F}_{\alpha\beta}(y^p_A) =   
\gamma_{\alpha\beta} \hskip 1mm + 2\gamma_{\alpha\beta}U_A(y^p_A)+
{1\over3}\Big( \sum_{B\not=A} \Big[ \gamma _{\alpha\beta} 
\Big<{\partial^2  U_B\over \partial y^\mu_A \partial y^\sigma_A}\Big>_0  
+ 
\gamma _{\mu\sigma}\Big<{\partial^2  U_B\over \partial y^\alpha_A
 \partial y^\beta_A}\Big>_0 - $$
{}
$$-\gamma _{\beta\mu}\Big<{\partial^2  U_B\over \partial y^\alpha_A 
\partial y^\sigma_A}\Big>_0  - 
\gamma _{\alpha\sigma}\Big<{\partial^2  U_B\over \partial y^\mu_A 
\partial y^\beta_A} \Big>_0\Big]\Big) \cdot y^\mu_A y^\sigma_A 
  +  {\cal O}(|y^\nu_A|^{3}) +   {\cal O}(c^{-4}). \eqno(5.51c)$$ 

Thus we have obtained the form of the metric tensor in  
Fermi-normal-like coordinates and the coordinate transformation, which   
leads to this form as well. These transformations are defined up 
to the third order with respect to the spatial point separation.

A more detailed analysis of the coordinate 
transformation for the extended self-gravitating bodies will be 
performed in the next Section.

\section{General Relativity: 3.  Proper  RF  for the Extended Body.}

In this Section we will generalize the   results obtained   
for the relativistic coordinate transformations (5.40) and will extend
their  
applicability to the problem of motion of a system of {\small N}-
extended-bodies in the  {\small WFSMA}.
The  relations (5.40) were obtained by using the 
generalized Fermi conditions (3.26) and, hence, they are well suited to
describe the motion of the 
system of {\small N} self-gravitating bodies omitting only the  lowest
intrinsic 
multipole moments. To generalize these results in the  case of arbitrarily
shaped 
extended bodies we must use  the more general definition of the   
proper {\small RF} given by the expressions  (3.29). This definition is
based on the 
study of existance of the 
integral conservation  laws for the metric theories of gravity (3.28).
The studies of existance of the 
conservation laws in the general relativity was performed by a number of 
scientists, notably by
Fock (1955) and Chandrasekhar (1965), whose  methods were developed in 
application to the motion of the more general {\small N}-body systems
in the framework of the {\small PPN} formalism 
(Lee {\it et al.}, 1974; Denisov \& Turyshev, 1989; Will, 1993). 
It should be noted that the search for the conservation laws in 
these methods was performed in the barycentric inertial {\small RF}$_0$
and,
 in particular, it was shown that the general relativity in 
the {\small WFSMA} has all ten conservation laws for the closed system
of fields 
corresponding to energy of the system, 
its momentum and angular momentum. The difference of the present 
research from that cited above  is in the fact that we will study the
problem of existence of the 
integral conservation laws in an  accelerated arbitrary {\small KLQ}-
parameterized 
proper {\small RF}$_A$. As a result of our study, we should find  the
conditions, 
necessary to impose on the   transformation functions $K_A, L_A$ and
$Q^\alpha_A$, 
so that  the general relativity in the coordinates of this {\small RF} 
will preserve the existent 
conservation laws for the entire system under consideration. 

\subsection{The Extended-Body Generalization.}
 
It is well-known that  in all metric theories of gravity
the Lagrangian density of matter is the   same functional of 
metric of Riemann space-time $g_{mn}$
and the other fields of matter $\psi_A$. Then the  application of the
method 
of infinitesimal displacements (Bogoljubov \& Shirkov,  1984; Logunov,
1987) 
to the action function of matter in these 
theories together with  the condition that the eq.m. for the fields
$\psi_A$ are 
satisfied, leads to the same covariant equation for the conservation of
density of the
energy-momentum tensor of matter in Riemann space-time:
{}
$$\nabla_k{\hat T}^{mk}=\partial_k{\hat T}^{mk}+ \Gamma^m_{lp}{\hat
T}^{lp}
=0.\eqno(6.1)$$
\noindent Note that this result is independent of the choice of {\small RF}.
 In the case of a system of  bodies formed from   an ideal fluid with the
individual density 
of energy-momentum tensor ${\hat T}^{mn}_B$ of an arbitrary body
{\small (B)} 
in the coordinates  of  its proper {\small RF}$_B$  is given by the
expression (2.1) by
 {}
$${\hat T}^{mn}_B(y^p_B)=\sqrt{-
g}_B\Big(\Big[\rho_{B_0}(1+\Pi)+p\Big]u^mu^n-pg^{mn}_B\Big), 
\eqno(6.2a)$$
\noindent the total density of the energy-momentum tensor of the system
of {\small N} bodies 
in the coordinates $(y^p_A)$ of the proper {\small RF}$_A$ of a particular
body {\small (A)} 
may then   be composed as follows:
{}
$${\hat T}^{mn}(y^p_{A}) = \sum_BJ_B(y^p_A){{\partial
y^m_A}\over{\partial y^k_B}}{{\partial y^n_A}
\over{\partial y^l_B}} \hskip 0.5mm {\hat T}^{kl}_B(y^q_{B}(y^p_A)),
\eqno(6.2b)$$ 
\noindent where $J_B$ is  the Jacobian of the corresponding coordinate
transformation:
{}
$$ J_B(y^p_A)={\rm det} \hskip 1pt ||{{\partial y^m_B}\over{\partial
y^k_A}}||. \eqno(6.2c)$$

In addition, from  the equation (6.1) for an ideal fluid 
model (6.2) we may also  obtain a covariant equation of continuity in
the coordinates $(y^p_A)$  as follows:
{}
$$\nabla^A_k\Big[\sum_B\rho_{B_0}u^k\Big]={1\over \sqrt{-g}_A}
\Big[{\partial {\overline \rho}\over \partial y^0_A}+
{\partial ({\overline \rho} v^\mu)\over \partial
y^\mu_A}\Big]=0,\eqno(6.3)$$
\noindent where $\nabla^A_k$ is the covariant derivative with respect 
to   metric tensor $g_{mn}^A$ of the  proper {\small RF}$_A$. 
The total conserved mass density  of the entire system    in the 
 coordinates $(y^p_A)$ is denoted as
{}   
$${\overline \rho}(y^p_A)=\sum _B  \rho_{B_0}\sqrt{-g}_Bu^0_B
=\sum _B  {\hat \rho}_B(y^p_A)J^{-1}_B{dy^0_B\over dy^0_A},\eqno(6.4)$$ 
\noindent where ${\hat \rho}_B$ is the conserved mass density of the
body {\small (B)}
and all the quantities on the right-hand side of  this
expression are transformed to the coordinates $(y^p_A)$ using the 
standard rules of   relativistic transformations of the 
mechanics of Poincar\'e (Fock, 1955).
Equations (6.1) and (6.3) together with the metric  tensor 
  give all the expressions necessary   for the 
construction of the eq.m.   of the extended bodies composed from
ideal fluid and for   analysis of various general questions.

In order to generalize the results  obtained in the previous Section,
in the case of   arbitrarily composed extended bodies, 
we shall first construct  the components of the density
of the energy-momentum tensor of matter ${\hat T}^{mn}$ to the required
accuracy. 
Thus, from  the definition (6.2)   one may get these 
components in the Newtonian approximation as follows:
{}
$${\hat T}^{00}(y^p_A)= {\overline \rho}\Big(1+{\cal O}(c^{-2})\Big),
\eqno(6.5a)$$
{}
$${\hat T}^{0\alpha}(y^p_A)={\overline \rho}v^\alpha\Big(1+{\cal O}(c^{-
2})\Big),
\eqno(6.5b)$$
{}
$${\hat T}^{\alpha\beta}(y^p_A)={\overline \rho}v^\alpha v^\beta-
\gamma^{\alpha\beta}p +{\hat \rho}{\cal O}(c^{-4}).\eqno(6.5c)$$  

  As a result, the covariant conservation equation (6.1)
for $m=\alpha$ transforms   into the Euler equation  for 
an ideal fluid, while for $m=0$ it transforms into 
the equation for the internal energy $\Pi$ of the {\it local} fields in the
vicinity of the 
body {\small (A)}:
{}
$${\overline \rho}{dv^\alpha \over dy^0_A}=-{\overline \rho} 
\partial^\alpha  {\overline U}+ \partial^\alpha  p  + 
{\overline\rho}\partial^\alpha{\cal O}(c^{-4}),\eqno(6.6a)$$
{}
$$ {\overline \rho}{d\Pi\over dy^0_A}=-p\partial_\mu  v^\mu  
+{\overline\rho}{\cal O}(c^{-5}), \eqno(6.6b)$$
 
\noindent where the total time derivative with respect to the proper time
$y^0_A$ is 
given by the  usual   relation: 
$d/dy^0_A=\partial/\partial y^0_A+v^\mu\partial/\partial y^\mu_A$. 
The total Newtonian potential  of the system in these coordinates 
 was denoted  as   ${\overline U}(y^p_A)$.

In order to apply the conditions (3.29) one must   substitute the 
expression for the total Newtonian potential ${\overline U}$ into (6.6a) 
and integrate this equation over the body {\small (A)}'s compact volume.
However, if we do so  for the potential  from the
solution (5.42), the conditions (3.29) will not be satisfied.
Indeed,  the total Newtonian potential ${\overline U}_{[0]}$ may be 
identified 
 as the terms of order $c^{-2}$ in   the expression   $(5.42a)$ 
for the $g_{00}$ component of the metric tensor   as follows:
{} 
$${\overline U}_{[0]}=\sum_BU_B(y^0_A, y^\nu_A)- \sum_{B\not=A}\Big[
y^\mu_A \Big<{\partial U_B\over \partial y^\mu_A}\Big>_0  
 + \Big<U_B\Big>_0 \Big]. \eqno(6.7) $$
\noindent If one   substitutes this potential  into   equation $(6.6a)$
and  integrates the resultant expression over 
the body  {\small (A)}'s compact volume one obtains:
{}
$${\ddot m}^\alpha_{A_{[0]}} =-\gamma^{\alpha\sigma}\sum_{B\not=A}
\int_Ad^3y'_A{\hat \rho}_A\Big[ {\partial U_B\over \partial y'^\sigma_A}
- 
 \Big<{\partial U_B\over \partial y^\sigma_A}\Big>_0\Big] + 
 {\cal O}(c^{-4})\not=0.\eqno(6.8a)$$
\noindent By expanding the integrand in the expression above in the 
Taylor series with respect to the  spatial  deviation from the supporting
world-line $\gamma_A$ 
(which is given as $\lambda_A\sim{\vec y}_A/|y_{BA_0}|$) one may bring
this result to the following form: 
{}
$${\ddot m}^\alpha_{A_{[0]}}=-
\gamma^{\alpha\sigma}\sum_{B\not=A}\sum_{l\ge1}^k 
{1\over l!}\Big<{\partial^{\{L+1\}} U_B\over \partial
y^\sigma_Ay^{\{L\}}_A}\Big>_0 
\int_Ad^3y'{\hat \rho}_A y'^{\{L\}}_A  + 
 {\cal O}(c^{-4}).\eqno(6.8b)$$

\noindent It is easy to see  that this result   does not satisfy the
requirement for the 
`good' proper {\small RF}  even in the Newtonian order.
The origin of the  {\small RF}, defined this way,  
coincides with the center of inertia  of the {\it local} 
fields in the vicinity of the body under question 
in one particular moment of time only and will
drift  away from it as time progresses. Exactly the same  situation
was encountered  with the solutions in both Brumberg-Kopejkin (Brumberg
\& Kopejkin, 1988a,b) 
and Damour-Soffel-Xu (DSX, 1991-1994; Damour \& Vokrouhlick\'{y},
1995) formalisms.
In both of these methods, the translational motion of   extended bodies in
their proper {\small RF}s  
do not vanish in the Newtonian limit, but rather non-linearly 
depend on the coupling of the intrinsic multipole 
moments with the external gravitational field. 
To solve this problem, the authors of both formalisms  have introduced 
 `external' multipole moments in order to compensate for  the terms  
on the right-hand side of  the expression $(6.8b)$. However, this 
substitution  may not 
be considered as a satisfactory solution to this problem.
The reason for this is that the authors in both approaches were 
trying to describe the motion of   extended bodies using  
methods which were developed to treat the motion of  point-like test
bodies.  
As we already know, to overcome this problem we should develop a
microscopic 
treatment of the matter, the  gravitational field and the field of inertia in
the 
immediate vicinity of the  bodies ({\it i.e.} in their {\it local} region) in
the system.    

In our method, the only step we have to make in order to the
take into account the extent of the bodies, is to change the limiting
procedure
$\big<..\big>_0$ defined by expression (5.7), to an  averaging  over the 
bodies   
volumes\footnote{Note, that this
situation is similar to that from the electrodynamics of continuous media 
where one have to average the field over the body's 
volume (Landau \& Lifshitz, 1987).}. We define 
this new procedure   $\big<..\big>_A$ which, being applied to   any
function $f(y^p_A)$, will
denote  an averaging of this function 
over the body's {\small (A)} three-dimensional compact volume   
in accord to the following formula: 
{}
$$ \Big<f(y^p_A)\Big>_A \equiv {\hat f}(y^0_A)=
{1\over m_A}\int_A d^3y'_A\hskip 2pt{\hat t}^{00}_A(y'^p_A)
f(y^0_A,y'^\nu_A), \eqno(6.9a)$$   
$$m_A =\int_A d^3y'_A{\hat t}^{00}_A(y'^p_A)+{\cal O}(c^{-4}),
\eqno(6.9b)$$
\noindent where ${\hat t}^{00}(y'^p_A)$ is the component of the 
conserved density of the energy-momentum tensor of matter, inertia and  
gravitational field in the {\it local} region of the body {\small (A)} taken
jointly. 
It is easy to see that in the case of a system of {\small N} massive
particles with the total mass density taken to be 
${\overline\rho}(y^p_A)=\sum_Bm_B\delta(y^0_A,{\vec y}_A+{\vec
y}_{BA_0})$,
this new procedure coincides with the  procedure $\big<..\big>_0$ defined 
  by the expression  eq.(5.7). Note  that the new operation
$\big<..\big>_A$ given by eq.(6.9), contrary to that of eq.(5.7),
does not commutatating with the operation of   time defferrentiation.

Because of this change, the   total gravitational  potential 
$\overline{U}(y^p_A)$, which, in the vicinity of the body {\small (A)} 
is  composed from the {\it local} Newtonian potential generated by the 
body {\small (A)} itself and the tidal gravitational 
potential produced by  external  the  sources of gravity, will now  
have the form:
{} 
$$\overline{U}(y^p_A)=\sum_B U_B(y^q_B(y^p_A))-\sum_{B\not=A}
\Big[y^\beta_A\Big<{\partial U_B\over \partial y^\beta_A}\Big>_A +
\big<U_B\big>_A \Big]+ {\cal O}(c^{-4}). \eqno(6.10) $$
 One may make sure that the  expression (6.10) is what we need in order to 
have  the origin  of the proper {\small RF}$_A$ to coincide with the 
{\it local} center of inertia. Indeed, by substituting this result for the
total 
Newtonian potential ${\overline U}$ into the 
equation (6.6a) and integrating the resultant expression over the body 
{\small (A)}'s compact volume, one finds that:
${\ddot m}^\alpha_{A_0}={\cal O}(c^{-4}).$   
Thus the center of inertia of the {\it local} fields, defined as 
the dipole moment of the  fields in the immediate vicinity of the 
body {\small (A)},  moves along a straight line   as
given by the formula: 
$ m^\alpha_{A_0}(y^0_A)={\cal A}^\alpha + {\cal B}^\alpha  y^0_A+ {\cal
O}(c^{-4}) $, where 
${\cal A}^\alpha, {\cal B}^\alpha $ are some constants. One may perform an 
additional infinitesimal post-Galilean transformation (similar to that of
(1.12)) in order to 
make them vanish: ${\cal A}^\alpha={\cal B}^\alpha=0$.  This means that
the 
origin of the proper {\small RF}$_A$ will  coincide with the  center of
inertia 
of the {\it local} fields   and, hence, the 
constructed frame will satisfy the 
definition of a   `good' proper {\small RF} discussed  in the Section {\small
III}. 
 
As a result,  the general form of the coordinate transformations 
between    the coordinates $(x^p)$ of   {\small RF}$_0$   and those  
$(y^p_A)$ of a proper  quasi-inertial   {\small RF}$_A$ of an arbitrary
body {\small (A)}  
for the problem of motion of the {\small N}-extended-body system in the
{\small WFSMA} 
may be presented   as follows:  
{}
$$x^0 = y^0_A+ c^{-2}  K_A(y^0_A, y^\epsilon_A) + 
c^{-4}   L_A(y^0_A, y^\epsilon_A) + {\cal O}(c^{-6}),
\eqno(6.11a)  $$
$$x^\alpha =  y^\alpha_A + y^\alpha_{A_0} (y^0_A) + 
c^{-2} Q^\alpha_A(y^0_A,y^\epsilon_A) + {\cal O}(c^{-4}), 
\eqno(6.11b)$$
\noindent where the barycentric radius-vector $r^\alpha_{A_0}$ 
of the body {\small (A)} in the coordinates of the proper 
{\small RF}$_A$  is decomposed into 
Newtonian and post-Newtonian parts, which are given  
as  follows: 
{}
$$\Big<r^\alpha_{A_0}(y^p_A)\Big>_A = y^\alpha_{A_0} (y^0_A)+
 {1\over m_A c^2}\int_A d^3y'_A{\hat t}^{00} (y'^p_A) 
Q^\alpha_A(y^0_A,y'^\epsilon_A)+{\cal O}(c^{-4}).\eqno(6.12)$$
 
\noindent The transformation  functions $  K_A,   Q^\alpha_A$ and $  L_A$,
in this 
case will take the following form: 
{}
$$K_A(y^0_A,y^\nu_A)  =\int^{y^0_A} \hskip -10pt  dt'  
\Big( \sum_{B\not=A} \Big< U_B\Big>_A- 
{1\over2} v_{A_0}{_{{}_\nu}} v_{A_0}^\nu\Big)  
- v_{A_0}{_{{}_\nu}}\cdot y^\nu_A +  {\cal O}(c^{-4})y^0_A, \eqno(6.13a)$$
{}
$$Q^\alpha_A(y^0_A,y^\nu_A) = - \sum_{B\not=A} \Big( y^\alpha_A 
 y^\beta_A \cdot \Big<\partial_\beta U_B \Big>_A-
{1\over2} {y_A}_\beta  y^\beta_A  
\Big<\partial^\alpha U_B\Big>_A  
 + y^\alpha_A  \Big<U_B\Big>_A \Big)+$$
{}
$$+{y_A}_\beta  \int^{y^0_A} \hskip -10pt dt' 
\Big( {1\over2}a^{[\alpha}_{A_0}  v^{\beta]}_{A_0}  + 
2 \sum_{B\not=A}\Big[
\Big<\partial^{[\alpha} V^{\beta]}_B\Big>_A   
+\Big<\partial^{[\alpha} U_B v^{\beta]}\Big>_A \Big]\Big)- $$
{}
$$- {1\over2} v^\alpha_{A_0} v^\beta_{A_0}{y_A}_\beta  
 + {\underline w}^\alpha_{A_0}(y^0_A) +\sum^{k}_{l = 3}
{Q^\alpha_A}_{\{L\}} (y^0_A) \cdot 
y^{\{L\}}_A+{\cal O}(|y^\nu_A|^{k+1})+ 
{\cal O} (c^{-4}) y^\alpha_A,   \eqno(6.13b)$$ 
{}
$$L_A(y^0_A,y^\nu_A) =  \sum_{B\not=A} \Big( {1\over2} \hskip 2pt
y_{A\beta} y^\beta_A \cdot {\partial \over\partial
y^0_A}\Big<U_B\Big>_A   - 
2\hskip 2pt y^\lambda_A  y^\beta_A \cdot 
\Big[\Big< \partial_\lambda {V_B}_\beta \Big>_A   
+\Big<v_\beta \partial_\lambda U_B \Big>_A \Big] +$$
{}
$$+   \hskip 2pt{v_{A_0}}_\beta \Big[y^\beta_A  y^\lambda_A \cdot 
\Big< \partial_\lambda  U_B \Big>_A - 
{1\over2} y_{A_\lambda} y^\lambda_A \cdot 
\Big< \partial^\beta  U_B \Big>_A \Big] \Big) +$$
{}
$$+ y_{A_\lambda}{v_{A_0}}_\beta \int^{y^0_A} \hskip -10pt dt'
 \Big( {1\over2} a^{[\lambda}_{A_0} v_{A_0}^{\beta]}+ 
2\hskip 2pt\sum_{B\not=A}
\Big[\Big<\partial^{[\lambda} V^{\beta]}_B  \Big>_A   
+\Big<v^{[\beta}\partial^{\lambda]} U_B \Big>_A\Big]\Big)+$$
{}
$$ + y_{A_\beta} \Big[ 2\hskip 2pt v^\beta_{A_0}\sum_{B\not=A} 
\Big<U_B\Big>_A  
 -  4\hskip 2pt\sum_{B\not=A} \Big<V^\beta_B\Big>_A  
 - \dot {\underline w}^\beta_{A_0}(y^0_A)  \Big] - $$
{}
$$ - \int^{y^0_A} \hskip -10pt dt' \Big[ \sum_{B\not=A}
\Big<W_B\Big>_A  + {1\over2}\Big(\sum_{B\not=A} \Big< U_B\Big>_A  - 
 {1\over2}{v_{A_0}}_\beta v^\beta_{A_0}\Big)^2 
  + {v_{A_0}}_\mu\dot{\underline w}^\mu_{A_0}(t')\Big]+$$
{}
$$+ \sum^{k}_{l=3}{L_A}_{\{L\}}(y^0_A) \cdot y^{\{L\}}_A+
  {\cal O}(|y^\nu_A|^{k+1}) +{\cal O}(c^{-6}).\eqno(6.13c)$$ 
\noindent  
One may verify that in the case of the free-falling   massive  test
particle with conserved mass density given as 
${\hat \rho}_A(y^p_A)=m_A\delta({\vec y}_A)$, the  
functions (6.13) will correspond  to the coordinate transformations 
to   the proper {\small RF} defined on the geodesic world line (5.41). 

Note that we have changed the notation $f_{A_{[0]}}$ to  $f_{A_0}$
in the  new expressions eqs.(6.11)-(6.13). This is because   all these 
 quantities are now defined with the procedure eq.(6.9), 
which takes into account the internal structure of the  bodies. 
 As a result,  the   Newtonian acceleration of 
the extended body {\small (A)} with respect to 
the barycentric {\small RF}$_0$ now is given as:
{}
$$a^\alpha_{A_0}(y^0_A)=-\gamma^{\alpha\mu}\sum_{B\not=A}
\Big<{\partial  U_B\over \partial y^\mu_A}\Big>_A+{\cal O}(c^{-
4}).\eqno(6.14)$$   

Furthermore, in order to take into account the extent of the 
bodies and its influence on the post-Newtonian dynamics of the {\small
N}-body system, 
the time-dependent function $w^\alpha_{A_0}$ has been replaced 
by the new function ${\underline w}^\alpha_{A_0}$:
{}
$${\underline w}^\alpha_{A_0}(y^0_A)=w^\alpha_{A_0}+\delta
w^\alpha_{A_0},\eqno(6.15a)$$
\noindent 
where the function $w^\alpha_{A_0}$ is determined as the  solution of 
the following differential equation:
{}
$$\ddot{w}^\alpha_{A_0}(y^0_A)=
\sum_{B\not=A} \Big(\gamma^{\alpha\mu}\Big<{\partial W_B\over
\partial y^\mu_A}\Big>_A+ 
v^\alpha_{A_0}{\partial \over \partial y^0_A} \Big< U_B\Big>_A  - 
 4\hskip 2pt{\partial \over \partial y^0_A}\Big< V^\alpha_B\Big>_A\Big)-
$$
{}
$$- {1\over2}v^\alpha_{A_0}{v_{A_0}}_\beta a^\beta_{A_0} + 
 \hskip 2pt a^\alpha_{A_0}\sum_{B\not=A} \Big<U_B\Big>_A  + $$
{}
$$+{a_{A_0}}_\beta \int^{y^0_A} \hskip -10pt dt' 
\Big( {1\over2}a^{[\alpha}_{A_0}  v^{\beta]}_{A_0}  + 
2 \sum_{B\not=A}\Big[
\Big<\partial^{[\alpha} V^{\beta]}_B\Big>_A   
+\Big<v^{[\beta}\partial^{\alpha]} U_B \Big>_A \Big]\Big)
 + {\cal O}(c^{-6}). \eqno(6.15b)$$
\noindent and the function $\delta w^\alpha_{A_0}$ is  
 unknown at the moment. This function will be determined later, when 
we will apply the conditions  (3.29) in order to make the total 
momentum of the matter, inertia and the gravitational field 
calculated   in the coordinates of the proper
{\small RF}$_A$ vanish in the volume of the body {\small (A)}.

As a result, the `averaged' components of the  metric 
tensor $g^A_{mn}$ in the coordinates $(y^p_A)$
of the proper {\small RF}$_A$ take the following form:
{}
$$g^A_{00}(y^p_A) =1-2{\overline U}+ 2{\overline W} +  
y^\mu_A y^\beta_A \cdot \Big\{ \gamma_{\mu\beta} 
\hskip 1mm {a_{A_0}}_\lambda a^\lambda_{A_0} -  
 \hskip 2pt{a_{A_0}}_\mu {a_{A_0}}_\beta + $$
{}
$$+ \sum_{B\not=A}
{\partial\over\partial y^0_A}\Big(\gamma_{\mu\beta} 
 {\partial  \over\partial y^0_A}\Big<U_B\Big>_A -
2\Big[\Big<\partial_{(\mu} {V_B}_{\beta)}\Big>_A   
+\Big<v_{(\beta}\partial_{\mu)} U_B\Big>_A \Big]\Big)\Big\} +$$
{}
$$ + 2  \sum^{k}_{l\ge3} 
 \Big[ \partial_0 {L_A}_{\{L\}}(y^0_A)
 + {v_{A_0}}_\beta \partial_0 Q^\beta_{A\{L\}} (y^0_A)\Big]\cdot
 y^{\{L\}}_A 
  +  
{\cal O}(|y^\nu_A|^{k+1}) +{\cal O}(c^{-6}), \eqno(6.16a) $$ 
{}
$$g^A_{0\alpha}(y^p_A) = 4\hskip2pt \gamma_{\alpha\epsilon}
{\overline V^\epsilon}-{1\over5} \Big( 
{y_A}_\alpha {y_A}_\beta+{1\over2}
 \gamma_{\alpha\beta} {y_A}_\mu y^\mu_A \Big)  
  \cdot{\dot a}^\beta_{A_0}   + $$
{}
$$+\sum^{k}_{l \ge 3} \Big[ \gamma_{\alpha \lambda}
 \partial_0 {Q^\lambda_A}_{\{L\}} (y^0_A) + 
\Big({L_A}_{\{L\}}(y^0_A)  + {v_{A_0}}_\beta \cdot 
{Q^\beta_A}_{\{L\}} (y^0_A)\Big) 
{\partial \over \partial y^\alpha_A}\Big]
\cdot  y^{\{L\}}_A  + $$
{}
$$+{\cal O}(|y^\nu_A|^{k+1})+  {\cal O}(c^{-5}), \eqno(6.16b)$$
{}
$$g^A_{\alpha\beta}(y^p_A) = 
\gamma_{\alpha\beta}\Big(1 + 2 {\overline U}\Big) +$$
{}
$$+\sum^{k}_{l \ge 3} \Big[ \gamma_{\alpha\lambda} {Q^\lambda_A}_
{\{L\}} (y^0_A)  
{\partial \over \partial y^{\beta}_A}  +
\gamma_{\beta\lambda} {Q^\lambda_A}_
{\{L\}} (y^0_A)  {\partial \over \partial y^\alpha_A}
\Big]\cdot y^{\{L\}}_A + {\cal O}(|y^\nu_A|^{k+1})+ 
{\cal O}(c^{-4}), \eqno(6.16c)$$

\noindent where  the total gravitational  potential ${\overline U}$  
at the vicinity of the body {\small (A)} is composed of the {\it local}
Newtonian 
potential generated by the body {\small (A)} itself and the tidal
gravitational 
potential produced by the  external   sources of gravity is given by
expression the 
(6.10). This potential may now be obtained  from $(6.13a)$ as follows:
{} 
$$\overline{U}(y^p_A)=\sum_B U_B(y^q_B(y^p_A))-{\partial
K_A(y^p_A)\over \partial y^0_A} -
 {1\over2}{v_{A_0}}_\mu v^\mu_{A_0}=$$
{}
$$= \sum_B U_B(y^q_B(y^p_A))-\sum_{B\not=A}
\Big[y^\beta_A\Big<{\partial U_B\over \partial y^\beta_A}\Big>_A +
\big<U_B\big>_A \Big]+ {\cal O}(c^{-4}). \eqno(6.17) $$ 
 
\noindent This potential   is the solution of the 
corresponding Poisson equation in the coordinates $(y^p_A)$: 
{}
$$  \gamma^{\mu\nu}{\partial^2  {\overline U}\over \partial
y^\mu_A\partial y^\nu_A}
= 4\pi{\overline \rho}(y^p_A), \eqno(6.18)$$
\noindent which is 
searched for  together with the   following integral boundary conditions: 
{}
$$\big<{\overline U}\big>_A =
\int_A d^3y'_A\hskip 2pt{\hat\rho}_A(y'^p_A){\overline U}(y'^p_A)=
\int_A d^3y'_A\hskip 2pt{\hat\rho}_A(y'^p_A)U_A(y'^p_A) \eqno(6.19a)$$ 
{}
$$\Big<{\partial  {\overline U}\over \partial y^\mu_A}\Big>_A 
=\int_A d^3y'_A\hskip 2pt{\hat\rho}_A(y'^p_A)
{\partial  {\overline U}(y'^p_A)\over \partial y'^\mu_A}=0. \eqno(6.19b)$$

The quantity  ${\overline V}^\alpha(y^p_A)$, in the expressions $(6.16b)$, 
is the total vector-potential
produced by all the bodies in the system   as seen in the 
coordinates $(y^p_A)$ of the {\small RF}$_A$. 
The averaging procedure (6.9) enables one to define 
this potential as follows:
{}
$${\overline V}^\alpha(y^p_A)=\sum_B V^\alpha_B(y^q_B(y^p_A))-
\sum_{B\not=A}
\Bigg(y^\mu_A \Big[\Big<{\partial V^{\alpha}_B\over \partial
y^\mu_A}\Big>_A+ 
\Big<v^\alpha{\partial U_B\over \partial
y^\mu_A}\Big>_A\Big]+\big<V^\alpha_B\big>_A\Bigg)+ $$
{}
$$+{1\over10}(3y^\alpha_Ay^\lambda_A-
\gamma^{\alpha\lambda}{y_A}_\mu y^\mu_A\Big)
\dot{a}_{A_0}{_{{}_\lambda}}+{\cal O}(c^{-4}). \eqno(6.20) $$

\noindent  It is interesting to note that the vector-potential   now
depends on the
coupling of the  intrinsic motion of matter in the body {\small (A)} 
to the gradient of the external gravitational field. Thus it  can be seen 
from   the   
expression   (6.13) for the function $Q^\alpha_A$, that this  coupling  
 contributes to the corresponding   precession term of the coordinates in 
this {\small RF} relative to the barycentric inertial one.  
This potential  also satisfies the usual Poisson
 equation of the   form:
{}
$$\gamma^{\mu\nu}{\partial^2  {\overline V}^\alpha\over \partial
y^\mu_A\partial y^\nu_A}
  = -4\pi{\overline \rho}(y^p_A)v^\alpha(y^p_A). \eqno(6.21)$$
\noindent Moreover, due to the covariant equation of continuity (6.3), both  
quantities (6.17) and (6.21) are connected  by the following relation:  
{}
$${\partial  {\overline U} \over \partial y^0_A}=
{\partial  {\overline V}^\mu \over \partial y^\mu_A}. \eqno(6.22)$$

Another quantity  we have introduced in the expressions (6.16)  is 
${\overline W(y^p_A)}$. This is the post-Newtonian 
contribution to the component $g_{00}$  of the effective metric tensor in
the 
coordinates $(y^p_A)$  given by $(6.16a)$. 
This  contribution   is given as follows:
{}
$${\overline W(y^p_A)}=\sum_B W_B(y^q_B(y^p_A))-\sum_{B\not=A}
\Big[y^\mu_A \Big<{\partial W_B\over \partial y^\mu_A}\Big>_A +
\big<W_B\big>_A\Big] + {\cal O}(c^{-6}), \eqno(6.23a) $$

\noindent The solution $(6.23a)$ repeats the structure of
the tidal representation of the Newtonian potential (6.17), so it 
could  be considered as the generalized post-Newtonian 
potential in  this {\small RF}.  
 The functions  $W_A$ and $W_B$ in the expression $(6.23a)$ are given by
the relations (5.43) and
they  fully represent the non-linearity of the 
total post-Newtonian gravitational  field in the proper {\small RF}$_A$.
These functions 
contain the contributions of two sorts: (i) the gravitational field produced
by the external 
 bodies in the system {\small (B}$\not=${\small A)}, and (ii) the field of
inertia caused by 
the accelerated   and non-geodesic motion  of 
the proper {\small RF}$_A$. This   happens due to the coupling of the
proper
multipole moments of the body {\small (A)} to  the external gravitational
field as well as  
to  the self-action contributions which are given by the terms with
${Q^\lambda_A}_{\{L\}}$ 
in the  expressions (5.43).
One may obtain the  corresponding Poisson-like equation for this potential 
as well. Thus, directly from the gravitational field equation $(4.4d)$ 
this last equation will take the form:
{}
$$\gamma^{\mu\nu}{\partial^2  {\overline W}\over \partial
y^\mu_A\partial y^\nu_A}=
- 8\pi {\overline \rho}\Big(\Pi-2v_\mu v^\mu+{3p\over{\overline
\rho}}\Big) 
+2\sum_B\Big[\partial^2_{00} U_B+2\partial_\mu U_B
\Big( 2a^\mu_{A_0}+\sum_C\partial^\mu U_C\Big)\Big]- $$
{}
$$-2\sum_B\sum^k_{l\ge3}
{Q^\mu_A}_{\{L\}}(y^0_A)\Big[2\partial^2_{\mu\lambda}U_B
\cdot\partial^\lambda+\partial_\mu U_B\cdot
\partial_\lambda\partial^\lambda\Big]
y^{\{L\}}_A+{\cal O}(|y^\nu_A|^{k+1})+ {\cal O}(c^{-6}). \eqno(6.23b) $$  

We have not yet presented  the last function which is
necessary to complete the  coordinate transformation for the extended
bodies, namely:
the  function $\delta w^\alpha_{A_0}$ from (6.15). To find  this function
one needs to 
apply the procedure for constructing a `good' proper {\small RF}
with the  full post-Newtonian   accuracy. In order to do
this, one must   perform the study of   existance of the conservation laws
 in the proper {\small RF}$_A$ and    define the   conserved quantities
which will
correspond to the energy, momentum and the angular momentum of the {\it
local} fields.
Then, after integrating these quantities over the 
body's compact volume, one must   find the form of the eq.m. for the 
extended bodies in their proper {\small RF}s.  These equations will contain 
the time derivatives of the only unknown function
$\delta w^\alpha_{A_0}$, which should be chosen in such  a way  that the 
conditions (3.29) will be satisfied.

\subsection{Conservation Laws in the Proper RF.}

As we have stated before, our goal is  to construct a formalism  which
will
be useful for calculations in a number of the metric theories of gravity.
This is the reason  why in  our further discussion we will use the method
developed for 
analysis of the conservation laws in the parametrized post-Newtonian
gravity developed by 
Fock and Chandrasekhar (Fock, 1955; Ehlers, 1967; Denisov \& Turyshev,
1989; Will, 1993). 
It is  known that the most important   question for any metric theory of
gravity  
is the presence or absence of laws of conservation 
of energy, momentum and angular momentum for the 
closed system of interacting fields. Strictly speaking, the solution
 of this question requires detailed information regarding the structure of
each metric theory of 
gravitation. It is necessary to know what geometric object has been
 chosen to describe the gravitational field, what geometry is
 natural for it, and what is the form of the equation 
connecting the gravitational field and the metric of the Riemann
space-time. Using the standard methods of theoretical physics, it is then
possible
to give an exhaustive answer to this question. However, such an analysis
can not be 
carried out in a general form for all metric theories of gravity at once.
This  
leaves us with only one option: attempt to obtain some information
regarding the possibility
of the existence of conservation laws in these theories proceeding only
from
the   eq.m. of matter in the {\small WFSMA}. It should be noted  that 
conditions obtained in this way are necessary but not sufficient
to prove the  existence  of integral conservation laws for matter and the 
gravitational field taken jointly  in a particular metric theory of
gravitation.
It is altogether possible that although  the necessary conditions are 
satisfied for some theory of gravitation it  nevertheless may not have
conservation laws 
for closed system of interacting fields. The reason 
for this situation is that quantities that do
not depend  on time, obtained on the basis of post-Newtonian equations of 
motion, may not have the character of integrals of 
the motion for a closed system and hence also   have 
no physical meaning. Therefore, in resolving the question of whether or
not 
conservation laws are present in a particular theory of 
gravitation, the last word can be said only after a complete 
analysis of the theory has been performed.

It is known  that  general relativity in the {\small WFSMA} 
  possesses the integral conservation laws for the energy-momentum
tensor of matter 
and the gravitational field taken jointly. It means that 
the covariant equation of conservation of the 
energy-momentum tensor of matter in Riemann space-time (6.1)
can be identically represented as the covariant conservation law of 
the sum of symmetric energy-momentum 
tensors of the gravitational field $t^{mn}_g$ and matter $t^{mn}_M$ in
space-time
of a  constant curvature:
{}
$$\nabla_k{\hat T}^{mk}=0 \Rightarrow  
{\cal D}_k\Big(t^{mk}_g+t^{mk}_M\Big)=0. \eqno(6.24)$$

It should be especially emphasized that, since in an arbitrary Riemann
space-time the operation of integrating tensors (with the exception of
scalar density)
is meaningless, from a mathematical point of view it follows 
that the presence of some differential conservation equations 
in this case does not guarantee the possibility of 
obtaining corresponding integral conservation laws. The 
possibility of obtaining integral conservation laws 
in a Riemann space-time  is entirely predetermined by its geometry and
closely connected with
the existence of Killing vectors of the given space-time.
Namely, only an equation of the form (6.24) guarantees the 
existence  of all ten integral conservation laws for a closed system
of interacting fields. Indeed, since, in a space-time  of constant curvature
the Killing 
equations ${\cal D}_m\eta_n+{\cal D}_n\eta_m=0$ are completely
integrable 
and their solutions contain the maximal possible
number {\small M}$=10$ of arbitrary parameters (Eisenhart, 1926), 
we have ten independent  Killing vectors in this case. 
Multiplying (6.24) successively by each of these vectors $\eta_k$, we
obtain
{}
$${\cal D}_k\Big[\eta_m\Big(t^{mk}_g+t^{mk}_M\Big)\Big]=
{1\over \sqrt{-\gamma}}\partial_ k\Big[\eta_m 
\Big(t^{mk}_g+t^{mk}_M\Big)\sqrt{-\gamma}\Big]=0. \eqno(6.25)$$
 
Since the left side of this expression is a scalar, we can integrate it over
a three-dimensional
volume (Logunov, 1987) and obtain all ten (the number of independent
Killing vectors) integral conservation laws for a system consisting of
matter, inertia and a gravitational
field taken jointly.

Thus, in general relativity, which    possesses the  integral conservation
laws,
expressions for the integrals of motion of an isolated system can be
determined also from the 
equation of motion of matter eq.(6.1). We shall  find a 
necessary condition which the post-Newtonian 
expansions of this theory in the proper quasi-inertial {\small RF} must
satisfy, and obtain 
post-Newtonian expansions of integrals of 
the motion required for subsequent computation. For this we 
should transform the covariant conservation 
equation (6.1) to the form of eq.(6.24), after which, multiplying this
relation by the 
corresponding Killing vectors of a space of constant 
curvature and integrating over the volume, we may easily
obtain the desired expressions. Since the metric tensor of 
Riemann space-time   in the absence of matter ($\rho_0=p=0$) 
should have  as its limit the pseudo-Euclidean Minkowski metric,
the covariant conservation equation (6.1) should be transformed to the 
conservation law (6.24) just in the pseudo-Euclidean space-time.
Then in the quasi-Cartesian coordinates of the barycentric inertial
{\small RF}$_0$
the expression (6.24) will take  the form:
{}
$$\nabla_k{\hat T}^{mk}=\partial_k\Big(t^{mk}_g+t^{mk}_M\Big)=0.
\eqno(6.26)$$

\noindent We expect that the `good' proper {\small RF} will
resemble the properties of the inertial {\small RF}$_0$, then 
in the  coordinates of this proper {\small RF}
the expression, analogous to that of (6.24), should take  the form 
of the conservation law of the total energy-momentum tensor of 
the fields of inertia, matter  and   gravity  taken jointly:
{}
$$\nabla_k{\hat T}^{mk}(y^p_A)={\partial\over\partial y^k_A}
\Big( t^{mk}_i+t^{mk}_g+t^{mk}_M\Big)=0. \eqno(6.27)$$
 
Knowledge of the metric (6.16) to a post-Newtonian degree
of accuracy makes it possible to determine the components of the energy-
momentum tensor in the 
next approximation. Indeed, using the definition for ${\hat T}^{mn}$ (6.2),
the metric (6.16), 
the expressions for the four-velocity eqs.$(E4)$ and $(E13b)$,  and also
the covariant components of the 
metric tensor $(B5a)$, 
we obtain the following expressions for the components of the density of
the energy-momentum tensor in the 
post-Newtonian approximation in the coordinates of the proper {\small
RF}$_A$:
{}
$${\hat T}^{00}(y^p_A) =  {\overline \rho} \Big[ 1 +\Pi - 
{1\over2}v_\mu  v^\mu    + \overline{U}   +{\cal O}(c^{-4})\Big],
\eqno(6.28a)$$ 
{}
$${\hat T}^{0\alpha}(y^p_A) =  {\overline \rho}  v^\alpha \Big[ 1 +\Pi 
- {1\over2}v_\mu v^\mu   +
{p\over{\overline\rho} } + \overline{U}   +{\cal O}(c^{-4})\Big]
,\eqno(6.28b)$$
{}
$${\hat T}^{\alpha\beta}(y^p_A) = {\overline \rho}  v^\alpha  v^\beta \Big[
1 +\Pi - 
{1\over2}v_\mu  v^\mu  +{p\over{\overline\rho} }  + 
\overline{U} \Big]-p \gamma^{\alpha\beta}+$$
{}
$$+p\sum^k_{l\ge3}
{Q^\mu_A}_{\{L\}}(y^0_A)\Big(\delta^\alpha_\mu\partial^\beta+
\delta^\beta_\mu\partial^\alpha-
\gamma^{\alpha\beta}\partial_\mu\Big)y^{\{L\}}_A
+{\overline \rho}{\cal O}(c^{-4})+{\cal O}(|y^\nu_A|^{k+1} ),\eqno(6.28c)$$ 
\noindent where the total conserved mass density 
of the entire system ${\overline \rho}$ is  given by (6.4).

Furthermore, by using  the solutions for the transformation functions 
(6.11) and (6.13),
from the   expressions $(F2)$ one may obtain the Christoffel
symbols of the Riemann  metric in the proper {\small RF}$_A$ in the form: 
{}
$$\Gamma^{0}_{00}(y^p_{A}) = -{\partial \overline{U} \over \partial
y^0_A}
 +  {\cal O}(c^{-5}), \qquad \Gamma^{0}_{0\alpha}(y^p_{A}) =
 -{\partial \overline{U} \over \partial y^\alpha_A}
 +{\cal O}(c^{-6}), \qquad \Gamma^{0}_{\alpha\beta}(y^p_{A}) =  {\cal
O}(c^{-3}), $$
{}
$$\Gamma^{\alpha}_{00}(y^p_{A}) = \gamma^{\alpha\mu}{\partial\over
\partial y^\mu_A}
\Big[\overline{U}  - \overline{W}  -\overline{U}^2 \Big] 
+ 4{\partial \overline{V}^\alpha \over \partial y^0_A}-
{1\over5}\Big(y^\alpha_Ay^\lambda_A+
{1\over2}\gamma^{\alpha\lambda}y^\mu_A
{y_A}_\mu\Big)\ddot{a}_{A_0}{_{{}_\lambda}}+ \eqno(6.29)$$
{}
$$+ \Big\{ a^\alpha_{A_0} {a_{A_0}}_\lambda - 
\delta^\alpha_\lambda \cdot a^\mu_{A_0} {a_{A_0}}_\mu  +
\sum_{B\not=A}{\partial \over \partial y^0_A}
\Big[2 \Big<{\partial V^{\alpha)}_B\over \partial y^{(\lambda}_A}\Big>_A
+2 \Big<v^{(\alpha}{\partial U_B\over \partial y^{\lambda)}_A}\Big>_A-
\delta^\alpha_\lambda{\partial \over \partial
y^0_A}\Big<U_B\Big>_A\Big]\Big\}y^\lambda_A+$$
{}
$$+ \delta{\ddot w}^\alpha_{A_0} +
\sum^k_{l\ge3}\Big[\partial^2_{00}{Q^\alpha_A}_{\{L\}}(y^0_A)+
{a_{A_0}}_\lambda
{Q^\lambda_A}_{\{L\}}(y^0_A)\partial^\alpha\Big]y^{\{L\}}_A-$$ 
{}
$$-{\partial\overline{U} \over \partial y^\mu_A}
\sum^k_{l\ge3}\Big[{Q^{\mu}_A}_{\{L\}}(y^0_A)\partial^\alpha 
+{Q^\alpha_A}{}_{\{L\}}(y^0_A) \partial^\mu\Big]y^{\{L\}}_A+
{\cal O}(|y^\nu_A|^{k+1})+ {\cal O}(c^{-6}), $$
{}
$$\Gamma^{\alpha}_{0\beta}(y^p_{A}) =\delta^\alpha_\beta 
{\partial {\overline U} \over \partial y^0_A} + 
2\Big( {\partial {\overline V}^\alpha \over\partial y^\beta_A} -  
{\partial {\overline V}_\beta \over\partial {y_A}_\alpha}\Big)  +$$
{}
$$+ \sum^k_{l\ge3}\partial_0{Q^\alpha_A}_{\{L\}}(y^0_A)\partial_\beta
y^{\{L\}}_A+
{\cal O}(|y^\nu_A|^{k+1})+ {\cal O}(c^{-5}), $$
{}
$$\Gamma^{\alpha}_{\beta\omega}(y^p_{A}) = 
 \delta^\alpha_\beta 
{\partial {\overline U} \over \partial y^\omega_A}  + 
\delta^\alpha_\omega {\partial {\overline U} \over \partial
 y^\beta_A}  -\gamma_{\beta\omega}\gamma^{\alpha\sigma}
{\partial {\overline U} \over \partial y^\sigma_A} 
+\sum^k_{l\ge3}{Q^\alpha_A}_{\{L\}}(y^0_A)\partial^2_{\beta\omega}y^{\{L\}}_A +
{\cal O}(|y^\nu_A|^{k+1})+ {\cal O}(c^{-4}). $$

Writing (6.1) for $m=0$ and substituting (6.28) and (6.29) into it, we
obtain
{}
$${\partial\over \partial y^0_A}\Big[{\overline \rho} 
\Big(1+\Pi-{1\over 2}v_\mu v^\mu +{\overline U} \Big)\Big]+
{\partial\over \partial y^\mu_A}\Big[{\overline \rho} v^\mu 
\Big(1+\Pi -{1\over 2}v_\mu v^\mu  +{\overline U} 
 +{p\over {\overline \rho}} \Big)\Big]-$$
{}
$$-{\overline \rho}  {\partial {\overline U} \over \partial y^0_A} -
2{\overline\rho}  v^\mu {\partial{\overline U} \over 
\partial y^\mu_A} ={\overline\rho}{\cal O}(c^{-5}). \eqno(6.30)$$

Thus, to bring this relation to the form of (6.27) it is necessary  to 
transform the last two terms by extracting  from them the partial 
derivatives with respect to time and the three-dimensional divergence. 
Such a transformation cannot be carried out in a unique manner. Therefore,
using the equations $(6.18)$ and (6.21),
we rewrite the given terms in the most general form reflecting this 
ambiguity:
{}
$${\overline\rho}{\partial {\overline U}\over \partial y^0_A}+2{\overline
\rho} v^\mu
{\partial{\overline U} \over \partial y^\mu_A}  =
{\partial \over \partial y^0_A}\Big(a_1{\overline\rho} {\overline U}+
{2a_1-3\over 8\pi}\partial _\mu {\overline U}\partial^\mu {\overline
U}\Big)+$$
{}
$$+{\partial \over \partial y^\mu_A}\Big({1-a_1\over4\pi}\partial^\mu 
{\overline U}{\partial {\overline U}\over \partial y^0_A} +
{a_2\over 4\pi}{\overline U} \partial^\mu 
{\partial {\overline U}\over \partial y^0_A} +(a_1+a_2){\overline
\rho}{\overline  U} v^\mu+$$
{}
$$+{2-a_1-a_2\over 4\pi}\partial_\nu{\overline U} \big[\partial^\nu 
{\overline V}^\mu-\partial^\mu {\overline V}^\nu]\Big), \eqno(6.31)$$
\noindent where $a_1$ and $a_2$ are arbitrary numbers. With
consideration of this relation,
and collecting like terms in (6.30) we get:
{}
$${\partial\over \partial y^0_A}\Big(t^{00}_i+t^{00}_g+t^{00}_M\Big)+
{\partial\over \partial
y^\beta_A}\Big(t^{0\beta}_i+t^{0\beta}_g+t^{0\beta}_M\Big)=
{\overline \rho}{\cal O}(c^{-5}), \eqno(6.32a)$$
\noindent with the following expressions for the 
$(00)$ and $(0\alpha)$ components 
of the density of total energy-momentum tensor: 
{}
$$t^{00}_i+t^{00}_g+t^{00}_M={\overline \rho}\Big(1+\Pi-{1\over 2}v_\mu
v^\mu+
(1-a_1){\overline U}\Big)+{3-2a_1\over 8\pi}
\partial_\mu {\overline U}\partial ^\mu {\overline U}+{\overline
\rho}{\cal O}(c^{-4}),
 \eqno(6.32b)$$
{}
$$t^{0\alpha}_i+t^{0\alpha}_g+t^{0\alpha}_M={\overline \rho}
v^\alpha\Big(1+\Pi-
{1\over 2}v_\mu v^\mu+(1-a_1-a_2){\overline U}+{p\over {\overline
\rho}}\Big)+$$
{}
$$+ {a_1-1\over 4\pi}\partial^\alpha {\overline U}{\partial {\overline
U}\over \partial y^0_A}-
{a_2\over 4\pi}{\overline U}\partial^\alpha {\partial {\overline U}\over
\partial y^0_A}+$$
{}
$$+{a_1+a_2-2\over 4\pi}\partial_\nu{\overline U} \big[\partial^\nu 
{\overline V}^\alpha-\partial^\alpha {\overline V}^\nu]\Big) 
+{\overline \rho}{\cal O}(c^{-5}). \eqno(6.32c)$$

Writing the expression (6.1) for $n=\alpha$ and substituting (6.28) and
(6.29)  into it, we have
{}
$${\partial \over \partial y^0_A}{\hat T}^{\alpha 0} +
{\partial \over \partial y^\mu_A}{\hat T}^{\alpha\mu}  +
{\overline \rho}\partial^\alpha {\overline U}-
 {\overline \rho}\partial^\alpha {\overline W}+
{\overline \rho} \partial^\alpha {\overline U}
\Big(\Pi-{3\over 2} v_\mu v^\mu+{p\over {\overline\rho}}-{\overline
U}\Big)+$$
{}
$$+4{\overline \rho}\Big[{\partial \overline{V}^\alpha\over \partial
y^0_A}-
{1\over20}\ddot{a}_{A_0}{_{{}_\lambda}}\Big(y^\alpha_Ay^\lambda_A+
{1\over2}\gamma^{\alpha\lambda}y^\mu_A {y_A}_\mu\Big)\Big]+$$
{}
$$ + {\overline \rho}\hskip 2pt\Big\{ a^\alpha_{A_0}
a_{A_0}{_{{}_\lambda}} - 
\delta^\alpha_\lambda \cdot a_{A_0}{_{{}_\mu}} a^\mu_{A_0}   +
\sum_{B\not=A}{\partial \over \partial y^0_A}
\Big[2 \Big<{\partial V^{\alpha)}_B\over \partial y^{(\lambda}_A}\Big>_A
+2 \Big<v^{(\alpha}{\partial U_B\over \partial y^{\lambda)}_A}\Big>_A-
\delta^\alpha_\lambda{\partial \over \partial
y^0_A}\Big<U_B\Big>_A\Big]\Big\}y^\lambda_A  + $$ 
{}
$$+4{\overline \rho} v^\mu\Big(\partial_\mu {\overline V}^\alpha -
\partial^\alpha {\overline V}_\mu\Big)+ 
2 {\overline \rho} v^\alpha{\partial {\overline U}\over \partial y^0_A}
+2{\overline \rho} v^\alpha v^\mu\partial _\mu  {\overline U}+ $$
{}
$$+ {\overline \rho}\Big(\delta{\ddot w}^\alpha_{A_0}(y^0_A) +
\sum^k_{l\ge3}\Big[\partial^2_{00}{Q^\alpha_A}_{\{L\}}(y^0_A)+
a_{A_0}{_{{}_\lambda}}
{Q^\lambda_A}_{\{L\}}(y^0_A)\partial^\alpha\Big]y^{\{L\}}_A\Big)-$$ 
{}
$$-{\overline \rho}{\partial\overline{U}\over \partial y^\mu_A}
\sum^k_{l\ge3}\Big[{Q^{\mu}_A}_{\{L\}}(y^0_A)\partial^\alpha 
+{Q^\alpha_A}{}_{\{L\}}(y^0_A) \partial^\mu\Big]y^{\{L\}}_A+ 
2{\overline \rho}v^\mu\sum^k_{l\ge3}\partial_0
{Q^\alpha_A}_{\{L\}}(y^0_A)\partial_\mu y^{\{L\}}_A+$$
{}
$$ +({\overline \rho}v^\mu v^\lambda-\gamma^{\mu\lambda}p)
\sum^k_{l\ge3}{Q^\alpha_A}_{\{L\}}(y^0_A)\partial^2_{\mu\lambda}y^{\{L\}}_A 
={\overline \rho}{\cal O}(|y^\nu_A|^{k+1})+{\overline \rho}{\cal O}(c^{-6}).
\eqno(6.33)$$
\noindent One may note that these expressions are not dependent on the 
function ${L_A}_{\{L\}}$ with $l\ge 3$. This means that, in the post-
Newtonian 
order, the function ${Q^\alpha_A}_{\{L\}}$ with $l\ge3$   only are
responsible for the 
existence of the integrals of motion   in the {\small RF}$_A$ 
 under consideration.  

To reduce this equation to the form (6.27) we use the identities presented
in Appendix H.
Substituting   these into (6.33) and collecting the like terms, we obtain:
{}
$${\partial\over \partial
y^0_A}\Big(t^{\alpha0}_i+t^{\alpha0}_g+t^{\alpha0}_M\Big)+
{\partial\over \partial
y^\beta_A}\Big(t^{\alpha\beta}_i+t^{\alpha\beta}_g+t^{\alpha\beta}_M\Big)=$$
{}
$$=-{\overline \rho}{d^2\over {dy^0_A}^2}\sum^k_{l\ge3}
{Q^\alpha_A}_{\{L\}}(y^0_A)y^{\{L\}}_A 
+{\overline \rho}{\cal O}(|y^\nu_A|^{k+1})+{\overline \rho}{\cal O}(c^{-5}),
\eqno(6.34)$$ 

\noindent with the following expressions for the $(\alpha0)$ and
$(\alpha\beta)$ 
 components of the density of total energy-momentum tensor:

$$t^{\alpha0}_i+t^{\alpha0}_g+t^{\alpha0}_M={\overline \rho}
v^\alpha\Big(1+\Pi-
{1\over 2}v_\mu v^\mu+3 {\overline U}+{p\over \rho}\Big)-$$
{}
$$- {3\over 4\pi}\partial^\alpha {\overline U}{\partial {\overline U}\over
\partial y^0_A}
+{1\over \pi}\partial_\nu{\overline U}\big[\partial^\alpha 
{\overline V}^\nu-\partial^\nu {\overline V}^\alpha]\Big) 
+{\overline \rho}{\cal O}(c^{-5}).\eqno(6.35)$$
{}
$$4\pi\Big(t^{\alpha\beta}_i+t^{\alpha\beta}_g+t^{\alpha\beta}_M\Big)=
\Gamma^{\alpha\beta }\Bigg[ {\overline U}-
{\overline W} +{\overline U}^2+
\Big(y^\mu_A y^\nu_A-\gamma^{\mu\nu} {y_A}_\lambda y^\lambda_A\Big)
a_{A_0}{_{{}_\mu}}
a_{A_0}{_{{}_\nu}}+\Big({y_A}_\mu\delta\ddot{w}^\mu_{A_0}\Big)+$$
{}
$$+ \sum_{B\not=A}{\partial \over \partial y^0_A}
\Big(4y^\lambda_A y^\epsilon_A\Big[\Big<{\partial V_{B\epsilon}\over
\partial y^\lambda_A}\Big>_A
+\Big<v_\epsilon{\partial U_B\over \partial y^\lambda_A}\Big>_A\Big]-
{y_A}_\mu y^\mu_A{\partial \over \partial y^0_A}\Big<U_B\Big>_A\Big)
+2
a_{A_0}{_{{}_\mu}}\sum^k_{l\ge3}{Q^\mu_A}_{\{L\}}(y^0_A)y^{\{L\}}_A\Bigg] + $$
{}
$$+4\Bigg(\partial ^\alpha  {\overline U}{\partial  {\overline V}^\beta
\over \partial y^0_A}+
\partial ^\beta    {\overline U}{\partial   {\overline V}^\alpha\over \partial
y^0_A} -
\gamma^{\alpha\beta }\partial_\lambda{\overline U}
 {\partial {\overline V}^\lambda\over \partial y^0_A}+ 
[\partial^\nu{\overline V}^\alpha-\partial^\alpha{\overline V}^\nu]
[\partial_\nu{\overline V}^\beta -\partial^\beta  {\overline V}_\nu]-$$
{}
$$-\gamma^{\alpha\beta}\Big[\partial^\mu{\overline V}^\nu
\partial_\mu{\overline V}_\nu-\partial^\mu
{\overline V}^\nu\partial_\nu{\overline V}_\mu\Big]\Bigg)+ 
{3\over 2}\gamma^{\alpha\beta }\Big({\partial 
{\overline U}\over \partial y^0_A}\Big)^2+4\pi(1+2{\overline U}){\hat
T}^{\alpha\beta }+$$
{}
$$+{1\over 5}\Big(\gamma^{\alpha\beta}\gamma_{\lambda\mu}-
\delta^\alpha_\lambda\delta^\beta_\mu-
\delta^\alpha_\mu\delta^\beta_\lambda\Big)
\Big(y^\mu_A y^\nu_A+{1\over 2}\gamma^{\mu\nu}{y_A}_\epsilon
y^\epsilon_A\Big)
{\ddot{a}_{A_0}}{_{{}_\nu}}\partial^\lambda{\overline U}+$$
{}
$$+\Big[\delta^\alpha_\lambda\delta^\beta_\mu\gamma_{\sigma\epsilon}-
(\delta^\alpha_\lambda\delta^\beta_\epsilon+
\delta^\alpha_\epsilon\delta^\beta_\lambda)\gamma_{\mu\sigma}+
\gamma^{\alpha\beta}(\gamma_{\lambda\epsilon}\gamma_{\mu\sigma}-
{1\over2}\gamma_{\lambda\mu}\gamma_{\sigma\epsilon})\Big]
\partial^\lambda{\overline U}\partial^\mu{\overline U}\sum^k_{l\ge3}
{Q^\sigma_A}_{\{L\}}(y^0_A)\partial^\epsilon y^{\{L\}}_A+$$
{}
$$+ {1\over2}\partial_\mu{\overline U}\partial^\mu{\overline
U}\sum^k_{l\ge3}
{Q^\beta_A}_{\{L\}}(y^0_A)\partial^\alpha y^{\{L\}}_A
-\partial^\alpha{\overline U}\partial^\mu{\overline U}\sum^k_{l\ge3}
{Q^\beta_A}_{\{L\}}(y^0_A)\partial_\mu y^{\{L\}}_A-$$
{}
$$-p \sum^k_{l\ge3}
{Q^\alpha_A}_{\{L\}}(y^0_A)\partial^\beta y^{\{L\}}_A+
{\overline \rho}{\cal O}(|y^\nu_A|^{k+1})+
{\overline \rho}{\cal O}(c^{-6}). \eqno(6.36)$$

It can be shown that the expression on the right hand side of the 
relation (6.34) cannot 
be represented as four-dimensional divergence of any combination
 of generalized gravitational potentials and characteristics of the ideal
fluid. Then
for arbitrary functions ${Q^\alpha_A}_{\{L\}}$  the expression 
(6.33) cannot be reduced to the form (6.27).
However, since  general relativity  posesses all conservation 
laws, such a reduction is always possible, and it follows that we
 must require that all the  functions ${Q^\alpha_A}_{\{L\}}$ with $l\ge 3$  
vanish:
{}
$${Q^\alpha_A}_{\{L\}}(y^0_A)=0,  \qquad \forall l\ge 3.\eqno(6.37a)$$

In addition to this, as we have noticed earlier, the functions 
${L_A}_{\{L\}}$ with $(l\ge 3)$ do not enter  the eq.m. (6.33) at all
and any choice of these functions will not affect the dynamics of the
system of the extended
bodies in the {\small WFSMA}. 
This suggest that these functions may be considered as the infinitesimal 
gauge functions and, without losing  generality of the description,
 we may set these functions to be zero: 
{}
$${L_A}_{\{L\}}(y^0_A)=0,  \qquad \forall l\ge 3.\eqno(6.37b)$$

Moreover, in correspondence with the definition (6.27)  in 
metric theories of gravitation which
posess all conservation laws, the expression  
 (6.36) must then contain the components
of the complete  energy-momentum tensor of matter and 
gravitational field in pseudo-Euclidean space-time.
Since below we shall mainly be interested only in 
the components $t^{\alpha0}$ of this tensor comparing the expressions for
it given by 
$(6.32c)$ and (6.35), we can see that 
$t^{0\alpha}_i+t^{0\alpha}_g+t^{0\alpha}_M \not=
t^{\alpha0}_i+t^{\alpha0}_g+t^{\alpha0}_M$. 
Therefore, although it is possible to obtain the conservation laws of
energy and momentum,
it is  not yet suffient for 
obtaining the remaining conservation laws for which  it 
is required that the components of the complete energy-momentum tensor 
of the system be symmetric. For our purposes  in order to ensure the
symmetry of the 
complete energy-momentum tensor of the system we should set:
{}
$$ a_1=-2,\qquad a_2=0.\eqno(6.38)$$

\noindent Thus, a necessary (but not sufficient) condition for the
existence of all
conservation laws in any metric theory of gravitation is that relations
(6.37) 
and (6.38)  should hold.  

With consideration of these equalities, the component $t^{00}$ of
$(6.32a)$ of the complete 
energy-momentum tensor will have the form:
{}
$$t^{00}_i+t^{00}_g+t^{00}_M={\overline \rho}\Big(1+\Pi-{1\over 2}v_\mu
v^\mu+
3{\overline U}\Big)+{7\over 8\pi}
\partial_\mu {\overline U}\partial ^\mu {\overline U}+{\overline
\rho}{\cal O}(c^{-4}).
 \eqno(6.39)$$
\noindent This expression can be used to describe the energy distribution
of the 
system is space, while the component $t^{\alpha0}$ of (6.35) can be used
to describe 
the density of momentum. 
Integrating  the expression (6.39) for  the  energy-momentum tensor over 
the body  {\small (A)}'s volume  space and using the trivial relation:
{}
$$\int_A d^3y'_A   \partial_\mu{\overline U} \partial ^\mu {\overline U}
= -4\pi\int_A  d^3y'_A   {\overline \rho}  {\overline U}+
\oint_A  dS^{\mu}_A  \hskip 3pt  {\overline U} \partial_\mu  {\overline U},
\eqno(6.40)$$
\noindent we obtain the following expression for the energy $P^0$ of the
system of matter,
inertia and gravitational field defined in the vicinity of the body {\small
(A)}
 as usual:
{}
$$P^0\equiv m_A= \int_A d^3y'_A  \Big( t^{00}_i+t^{00}_g+t^{00}_M\Big).
\eqno(6.41a)$$
\noindent This corresponds to the following result for the total 
mass of the fields in this {\small RF}$_A$:
{}
$$m_A= \int_A d^3y'_A  {\overline \rho}\Big(1+\Pi-{1\over2} v_\mu
v^\mu-
{1\over 2} {\overline U}\Big)+
{7\over 8\pi}\oint_A  dS^{\mu}_A  \hskip 3pt  
{\overline U} \partial_\mu  {\overline U}=$$
{}
$$=\int_A d^3y'_A    {\hat\rho}_A\Big(1+\Pi-{1\over2}v_\mu v^\mu -
{1\over 2} U_A\Big)+
{7\over 16\pi}\oint_A  dS^{\mu}_A  \hskip 3pt  
\partial_\mu{\overline U}^2+m_A{\cal O}(c^{-3}).\eqno(6.41b)$$

\noindent The obtained result may be presented in terms of the 
unperturbed mass $m_{A(0)}$ of the body {\small (A)} as follows:
{}
$$m_A= m_{A(0)}+
{7\over 16\pi}\oint_A  dS^{\mu}_A  \hskip 3pt  
\partial_\mu{\overline U}^2+m_A{\cal O}(c^{-3}),\eqno(6.42)$$
\noindent where the second term represents the contribution of the 
coupling of the proper gravitational field  of the body under
study to the external gravity.  This term   is zero in the 
case of an isolated body, because one may move the  boundary of
integration to 
infinite distance. Taking into account that  the integrand behaves 
as $r^{-3}$, one makes the conclusion that this integral is zero. 
One loses this useful option in the case of the {\small N}-body system,
and,
due to this reason, we must   take into account such  `surface' effects
in order to  correctly describe the  perturbed motion of the bodies in the
system.

The momentum $P_A^\alpha$ of the system of fields in the coordinates of
this 
{\small RF}$_A$  is determined in an entirely analogous way: 
 by integrating the component $t^{\alpha 0}$ of (6.35) of  the complete 
energy-momentum tensor over the compact volume of the body {\small
(A)}:
{}
$$P_A^\alpha =\int_A d^3y'_A  \Big(
t^{\alpha0}_i+t^{\alpha0}_g+t^{\alpha0}_M\Big). \eqno(6.43a)$$ 
\noindent  Then for the momentum $P_A^\alpha$   we 
obtain the following expression:
{}
$$P^\alpha_A= \int_A d^3y'_A  \Big[{\overline \rho} v^\alpha\Big(1+\Pi-
{1\over 2}v_\mu v^\mu+3 {\overline U}+{p\over \rho}\Big)-$$
{}
$$- {3\over 4\pi}\partial^\alpha {\overline U}{\partial {\overline U}\over
\partial y^0_A}
+{1\over \pi}\partial_\nu{\overline U}\Big(\partial^\alpha 
{\overline V}^\nu-\partial^\nu {\overline V}^\alpha\Big)\Big] 
+{\overline \rho}{\cal O}(c^{-5}).\eqno(6.43b)$$ 

Finally, the  requirement (3.29) may 
be fulfilled by integrating the equation (6.34) over the volume of the body
{\small (A)}
and making such a choice of the function $\delta\ddot w_{A_0}$ such
that corresponding momentum $P^\alpha_A$   in the  
{\small RF}$_A$ will vanish for all times.   
However, as we will see later this requirement is    not  easy to 
satisfy. The problem one is faced with  is that the system of the 
fields and matter overlapping the body {\small (A)} is not a closed system.
This 
system is a part of a bigger ensemble of   celestial bodies which
was initially   taken to be a closed {\small N} body system. 
The definitions for the energy and momentum of the system may not be
given in the 
{\it local} form, instead these quantities are non-zero in all regions of the
system. 
As a result of such a non-locality, one loses the possibility of  
eliminating  the integrals from the three-divergences. 
Thus,  in the analysis of the conservational laws in the gravitational one-
body 
problem, one can integrate such  divergences by using the Stokes theorem
and 
moving the surface of integration at the infinite distance 
(Fock, 1959; Denisov \& Turyshev, 1989; Will, 1993). In the case of  
coordinates originated with the quasi-inertial proper {\small RF}, such an
integration 
is meaningless. Instead, one may integrate the corresponding quantities on
the 
surface of the body under consideration.   
As a result, one may see from the expressions (6.32) that the mass in the
proper {\small RF}
is not a constant anymore. Thus, by  integrating the  expression $(6.32a)$
over the body  
{\small (A)}'s compact volume we obtain:
{}
$${d m_A\over d y^0_A}=  {1\over 4\pi}\oint_A d S^\beta_A    
\Big[3\partial_\beta {\overline U}{\partial {\overline U}\over \partial
y^0_A}
+4\partial_\nu{\overline U}\Big(\partial^\nu
{\overline V}_\beta-\partial_\beta {\overline V}^\nu\Big)\Big] 
+m_A{\cal O}(c^{-5}).\eqno(6.44)$$ 

\noindent The  integral on the right-hand side of the
expression above   vanishes   in the case of an isolated distribution of 
matter, but for the {\small N}-body problem in the quasi-inertial {\small
RF} 
it depends on the 
magnitude of the fields on the surface of the body under study.  
The  analysis of 
the conservation laws is the only way to correctly define the 
important physical quantities such as the mass,   momentum and 
angular momentum  of the field in the {\it local } region of the body.
One expects that, in the immediate vicinity of the origin of the 
coordinate system in the `good' proper {\small RF}, the form 
of these laws should resemble that  which was developed by 
Fock (1955) and Chandrasekhar (1965) for the inertial frames.  
 Therefore we will use the technique which was developed for the 
barycentric approach by modifying it for the case under consideration.

Here we must mention the following   circumstance. 
It follows from eq.(6.35) and eq.(6.43) that, 
in the post-Newtonian approximation,  the density
of the total momentum of the system, in contast to barycentric {\small
RF}$_0$, 
 can be written
in the coordinates of the proper {\small RF} only
in the non-local form (6.35) when the components $t^{\alpha0}$ 
are non-zero, generally speaking, in the 
entire space. Unfortunately, this expression cannot be 
written in the {\it local} form which would be  nonzero
only in the region occupied by the body {\small (A)} because of the 
presence of   external sources of gravity. 
Comparing $(6.32a)$ and (6.40), we can draw an analogous conclusion 
regarding the energy density of the system.  Since the total momentum and
energy of the 
system in the post-Newtonian approximation do not depend on the choice
of the form of 
writing them, the momentum and energy of the gravitational field, which
are non-local 
by their nature,  can be effectively considered in this approximation
by adding {\it local} terms to the energy density of matter.
The latter circumstance is especially convenient in computing the 
motion of complex systems, since it lets us 
distinguish in 
explicit form the total momentum and energy of each of the bodies of the
system.

Therefore, we shall henceforth use the following expression 
for the density of the total momentum   of matter, inertia and the 
gravitational field in the volume occupied by the body:
{}
$${\hat t}^{\alpha0}=  {\overline \rho} v^\alpha\Big(1+\Pi-
{1\over 2}v_\mu v^\mu+3 {\overline U}+{p\over \rho}\Big)-$$
{}
$$- {3\over 4\pi}\partial^\alpha {\overline U}{\partial {\overline U}\over
\partial y^0_A}
+{1\over \pi}\partial_\nu{\overline U}\Big(\partial^\alpha 
{\overline V}^\nu-\partial^\nu {\overline V}^\alpha\Big)\Big] 
+{\overline \rho}{\cal O}(c^{-5}), \eqno(6.45)$$

\noindent and  for the total energy density  we shall use the expression:
{}
$${\hat t}^{00}=  {\overline\rho} \Big(1+\Pi-{1\over2} v_\mu v^\mu 
-{1\over 2}{\overline U}\Big) + 
{7\over 8\pi}\partial_\mu\Big[{\overline U} \partial^\mu  
{\overline U}\Big]+{\overline\rho} {\cal O}(c^{-4}).\eqno(6.46)$$

The relationships obtained will be used in order to define the eq.m. of the 
extended bodies form with respect to the coordinates of the proper
{\small RF}$_A$. 
Note  that by integrating the expressions (6.45) and (6.46) over the 
compact volumes of the bodies in the system, one may obtain the mass and
the momentum of these
bodies measured with respect to the proper {\small RF}$_A$.  Such 
relative quantities 
may be very important in the analysis of the relativistic gravitational
experiments 
in the solar system which we will discuss in the next  Section.
In order to complete the formulation of the coordinate transformations to
the `good' proper 
{\small RF}$_A$ we should present the   function which was not 
yet determined, namely the function $\delta w^\alpha_{A_0}$.
 
\subsection{The Solution for the Function $\delta w^\alpha_{A_0}$.}

To obtain the equation of motion of extended bodies in the gravitational
field   we must first of all bring the covariant conservation equation
(6.33)
to the form
{}
$${\partial\over \partial y^0_A} {\hat t}^{\alpha0}(y^p_A)={\cal
F}^\alpha(y^p_A), \eqno(6.47)$$  
\noindent where ${\hat t}^{\alpha 0}$ is defined by (6.45), 
and ${\cal F}^\alpha$ represents the entire remaining part of (6.33) and
can 
be considered the force density acting on matter. This is exactly the force
we have mentioned 
in   Section {\small 3} while discussing the expressions (3.27). After
performing 
identity transformations   using  eqs.(6.3) and (6.6),  we obtain from
eq.(6.33):
{}
$${\cal F}^\alpha(y^p_A)= -{\overline \rho}\partial^\alpha {\overline U}+
 {\overline \rho}\partial^\alpha {\overline W}-
{\overline \rho} \partial^\alpha {\overline U}
\Big(\Pi-{3\over 2} v_\mu v^\mu+{3p\over {\overline\rho}}+{\overline
U}\Big)+$$
{}
$$+ {\overline \rho} \Big(y^\alpha_Ay^\lambda_A+
{1\over2}\gamma^{\alpha\lambda}{y_A}_\mu y^\mu_A 
\Big){1\over5}\ddot{a}_{A_0}{_{{}_\lambda}}
+{1\over4\pi}{\partial\over\partial y^0_A}
\Big[\partial^\alpha {\overline U}{\partial{\overline U}\over\partial
y^0_A}\Big]-$$
{}
$$- {\overline \rho}y^\lambda_A\hskip 2pt\Bigg( a^\alpha_{A_0}
a_{A_0}{_{{}_\lambda}} - 
\delta^\alpha_\lambda \cdot a^\mu_{A_0} a_{A_0}{_{{}_\mu}}  +
\sum_{B\not=A}{\partial \over \partial y^0_A}
\Big[2 \Big<{\partial V^{\alpha)}_B\over \partial y^{(\lambda}_A}\Big>_A
+2 \Big<v^{(\alpha}{\partial U_B\over \partial y^{\lambda)}_A}\Big>_A-
\delta^\alpha_\lambda{\partial \over \partial
y^0_A}\Big<U_B\Big>_A\Big]\Bigg) - $$ 
{}
$$-{1\over \pi}{\partial\over\partial y^\beta_A}
\Bigg(\partial^\alpha{\overline U}{\partial{\overline
V}^\beta\over\partial y^0_A}+
\partial^\beta{\overline U}{\partial{\overline V}^\alpha\over\partial
y^0_A}-
\gamma^{\alpha\beta}\partial_\nu{\overline U}{\partial{\overline
V}^\nu\over\partial y^0_A}+
 \partial^{[\nu}{\overline V}^{\alpha]}\partial_{[\nu}{\overline V}^{\beta]}
-$$
{}
$$-\gamma^{\alpha\beta} \partial^{[\mu}
{\overline V}^{\nu]}\partial_{[\mu}{\overline V}_{\nu]}+
{1\over2}\gamma^{\alpha\beta}\Big(
{\partial{\overline U}\over\partial y^0_A}\Big)^2+
\pi(1+2{\overline U}){\hat T}^{\alpha\mu}\Bigg)-
{\overline \rho} \delta{\ddot w}^\alpha_{A_0}(y^0_A)  + 
{\overline \rho}{\cal O}(c^{-6}).\eqno(6.48)$$
The eq.m. of each body can be obtained if (6.44) is integrated over the
volume
occupied by this body. In order to find the function $\delta
w^\alpha_{A_0}$
we should  start with finding the eq.m. of the   body {\small (A)} 
relative to its own {\small RF}.
Integrating (6.48) over the $V_A$, we obtain
{}
$${dP^\alpha_A\over d y^0_A}= F^\alpha_A(y^0_A)\eqno(6.49)$$

\noindent where $P^\alpha_A$ given by the expression $(6.43b)$ and 
{}
$$ F^\alpha_A(y^0_A)=\int_A d^3y'_A {\cal F}^\alpha(y^0_A, y'^\nu).
\eqno(6.50)$$

In order to define the function $ \delta w_{A_0}(y^0_A)$ we will require
that
the momentum of the body {\small (A)} in its proper {\small RF} will
vanish. 
This requirement may be fulfilled if  the equation for  $\delta{\ddot
w}_{A_0}(y^0_A)$
 will be chosen in the following form:

$$\delta{\ddot w}_{A_0}(y^0_A)= -\Big<\partial^\alpha {\overline U}
\Big(\Pi-{3\over 2} v_\mu v^\mu+{3p\over {\overline\rho}}+{\overline
U}\Big)\Big>_A+$$
{}
$$+{1\over 5}{\ddot{a}_{A_0}}{_{{}_\mu}}\int_A d^3 y'^\nu_A {\hat
\rho}_A
\Big(y'^\alpha_A y'^\mu_A+{1\over2}\gamma^{\alpha\mu}y'_{A_\lambda}
y_A'^\lambda  \Big)- $$
{}
$$ -{1\over \pi m_A} \oint_A d S^\beta_A 
\Bigg(\partial ^\alpha  {\overline U}{\partial  {\overline V}_\beta \over
\partial y^0_A}+
\partial _\beta    {\overline U}{\partial   {\overline V}^\alpha\over
\partial y^0_A} -
\delta^\alpha_\beta \partial_\lambda{\overline U}
 {\partial {\overline V}^\alpha\over \partial y^0_A}+ 
\partial^{[\nu}{\overline V}^{\alpha]}\partial_{[\nu}{\overline V}_{\beta]}-
$$
{}
$$- {1\over2}\delta^\alpha_\beta \partial^{[\mu}
{\overline V}^{\nu]}\partial_{[\mu}{\overline V}_{\nu]} 
+ {1\over 2} \delta^\alpha_\beta\Big({\partial 
{\overline U}\over \partial y^0_A}\Big)^2\Bigg) 
+ {\cal O}(c^{-6}).\eqno(6.51)$$

As a result, one may  obtain the differential equation for the total  post-
Newtonian   
acceleration $\ddot{{\underline w}}^\alpha_{A_0}$ from (6.13) which is
necessary to apply 
in order to hold the extended  body in the state of equilibrium in its 
proper {\small RF}.  Thus with the help of the expressions
(6.15) and (6.51) we will obtain the following equation:  
{}
$$\ddot{\underline{w}}^\alpha_{A_0}(y^0_A)=
\sum_{B\not=A} \Big(\gamma^{\alpha\mu}\Big<{\partial W_B\over
\partial y^\mu_A}\Big>_A+ 
v^\alpha_{A_0}{\partial \over \partial y^0_A} \Big< U_B\Big>_A  - 
 4\hskip 2pt{\partial \over \partial y^0_A}\Big< V^\alpha_B\Big>_A\Big)-
$$
{}
$$+a_{A_0}{_{{}_\beta}} \int^{y^0_A} \hskip -10pt dt' 
\Big( {1\over2}a^{[\alpha}_{A_0}  v^{\beta]}_{A_0}  + 
2 \sum_{B\not=A}\Big[
\Big<\partial^{[\alpha} V^{\beta]}_B\Big>_A   
+\Big<v^{[\beta}\partial^{\alpha]} U_B \Big>_A \Big]\Big)-$$
{}
$$- {1\over2}v^\alpha_{A_0} v_{A_0}{_{{}_\beta}} a^\beta_{A_0} + 
 \hskip 2pt a^\alpha_{A_0}\sum_{B\not=A} \Big<U_B\Big>_A   -
\Big<\partial^\alpha {\overline U}
\Big(\Pi-{3\over 2} v_\mu v^\mu+{3p\over {\overline\rho}}+{\overline
U}\Big)\Big>_A+$$
{}
$$+{1\over 5}{\ddot{a}_{A_0}}{_{{}_\mu}}\int_A d^3 y'^\nu_A {\hat
\rho}_A
\Big(y'^\alpha_A y'^\mu_A+{1\over2}\gamma^{\alpha\mu}y'_{A_\lambda}
y_A'^\lambda  \Big)- $$
{}
$$ -{1\over \pi m_A} \oint_A d S^\beta_A 
\Bigg(\partial ^\alpha  {\overline U}{\partial  {\overline V}_\beta \over
\partial y^0_A}+
\partial _\beta    {\overline U}{\partial   {\overline V}^\alpha\over
\partial y^0_A} -
\delta^\alpha_\beta \partial_\lambda{\overline U}
 {\partial {\overline V}^\alpha\over \partial y^0_A}+ 
\partial^{[\nu}{\overline V}^{\alpha]}\partial_{[\nu}{\overline V}_{\beta]}-
$$
{}
$$- {1\over2}\delta^\alpha_\beta \partial^{[\mu}
{\overline V}^{\nu]}\partial_{[\mu}{\overline V}_{\nu]} 
+ {1\over 2} \delta^\alpha_\beta\Big({\partial 
{\overline U}\over \partial y^0_A}\Big)^2\Bigg) 
+ {\cal O}(c^{-6}).\eqno(6.52)$$
 
For  practical purposes one may find the value of the surface integrals 
in the expression (6.52) by  performing  the iteration procedure. Thus  
it is easy to show that the lowest multipole moments of the bodies 
will not contribute to  this surface integration. However, the 
general results  will fully depend   on the non-linear 
interaction of the intrinsic multipole moments with the external 
gravity in the {\it local} region at the vicinity of the body under 
consideration. This additional iterative option   will make  all the  results
obtained with the 
proposed formalism  easy to use in the   practical applications.

As one may see, we have reconstructed the  
 post-Newtonian  non-linear group of motion for the {\small WFSMA}. 
Thus, the straight transformation is given by the eqs.(6.11)-(6.13). 
The substitution of the results obtained
 for the transformation functions  in  the relations (3.18)
 will give  the inverse transformation. Finally, the common 
element of this group may be obtained by making of use the 
relations (3.19)-(3.20). These results generalize  and specify those  
obtained by Chandrasekhar \& Contopulos (1967) and given by (1.12). In
this previous  
work  the post-Galilean transformations were  obtained  which preserve
the 
invariancy of the metric tensor. In contrast to these, our
transformations eqs.(6.11)-(6.13), in general, transform the coordinates
in 
different non-inertial {\small RF}s and 
were specifically defined for a system of  self-gravitating extended and
arbitrarily shaped bodies.
Comparison with the Poincar\'e group of motion (1.7) expanded similarly
in 
inverse powers of $c$, shows   that: 
\begin{itemize}
\item[(i).] The spatial part of the 
transformations up to the terms ${\cal O}(c^{-2})$ includes the Lorentzian 
terms and allows, in addition,   infinitesimal   rotations,    
uniform motion, the shift of the origin, and the terms due to the
gravitational 
coupling of the internal multipoles of the extended bodies with the
external gravitation.
\item[(ii).] The temporal part of the transformation includes the
Lorentzian terms up to  
${\cal O}(c^{-4})$ plus additional terms of a purely gravitational nature as
well as the
terms due to  precession of the spatial axes. 
It is the presence of these  gravitational terms in both spatial and
temporal components 
of the transformation that gives the transformation its non-Lorentzian
character.
\end{itemize} 

As one can see   the obtained coordinate transformations are in general
the non-local ones. 
As such, they represent  an important and powerful way to study the 
nature of the multipolar structure of a  system of   
extended bodies and their gravitational  interaction   in the {\small
WFSMA}
of the general theory of relativity. In the next Section we will 
discuss the   generalization of the obtained results  to the case of the 
scalar-tensor theories of gravity.

\section{Parameterized  proper   RF.}
In this Section we will further generalize the results  obtained for the
coordinate transformations and the metric tensor in the proper {\small
RF}
which was obtained in the previous Section.
In order to generalize the results obtained we have applied 
the presented formalism   to the scalar-tensor theories of gravity.
It should be noted  that  considerable interest   has recently been shown in  
the physical processes occurring in the strong gravitational
field regime. However, many modern theoretical models 
which include general relativity as a standard gravity 
theory  are faced with the problem of the unavoidable appearance
of    space-time singularities.  
It is well known that the classical description, provided by general 
relativity, breaks down in a domain where the curvature is large, 
and, hence, a  proper understanding of such regions requires new 
physics (Horowitz \& Myers, 1995).
The tensor-scalar theories of gravity, in which,  the 
usual  for general relativity tensor field  coexists together 
with one or several long-range scalar fields, 
are believed to be the most interesting extension of the 
theoretical foundation of modern gravitational theory. 
The superstring, many-dimensional Kaluza-Klein, and inflationary
 cosmology  theories have
  revived the interest in so-called `dilaton fields', 
{\it i.e.} neutral scalar fields whose background values determine 
the strength of the coupling constants in the effective 
four-dimensional theory. However, although the scalar field naturally 
arises in theory, its existence  leads to a violation of 
the strong equivalence principle and  modification of 
large-scale gravitational phenomena (Damour {\it et al.}, 1990; 
Damour \& Taylor, 1992; Damour \& Esposito-Farese, 1992; 
Damour \& Nordtvedt, 1993; Berkin \& Hellings, 1994; Turyshev, 1996). 
Moreover, the presence of the scalar field 
affects  the equations of motion of the
other matter fields as well (Turyshev, 1996), which makes it interesting
to study 
the opportunities of the advanced dynamical tests of these theories in the 
{\small WFSMA} before they will be applied to the strong-field-regime
research.  
Therefore, the motivation for the present work was to  
perform the similar full-scale analysis of the 
{\small WFSMA} for some tensor-scalar theories of gravity in 
order to generalize the results obtained previously. 

\subsection{Parameterized coordinate transformations.}
In this subsection we will present the results of the  relativistic 
study of the Brans-Dicke theory of gravity (Will, 1993). 
However, concerning the length of the  expressions and also, in order to
avoid
 unreasonable  complication  of the 
discussion in this Section, we will not 
present here the details of these calculations.   Instead, we have
introduced the 
two Eddington parameters $(\gamma, \beta)$ in order to present the
obtained relations in 
more compact form valid for a number of modern metric theories of
gravity.  
This gives us a chance to present the   final results only. 
One may repeat the necessary calculations using the  technique 
 of the general post-Newtonian  power expansions
in the {\small WFSMA} developed in the Appendices.

By taking into account the properties of a scalar-tensor theories of
gravity,
and by applying the rules for constructing 
the proper {\small RF}  presented in the Section {\small 2},
 one may   obtain  the set of differential equations on the transformation  
functions ${\cal K}_A, {\cal Q}^\alpha_A$ 
and ${\cal L}_A$. As a result, the   relativistic coordinate transformation   
between  the coordinates $(x^p)$ of   {\small RF}$_0$   to those  
$(y^p_A)$ of the  proper  quasi-inertial  {\small RF}$_A$ of 
an arbitrary body {\small (A)}  may be given as follows:  
{}
$$x^0 = y^0_A+ c^{-2}{\cal K}_A(y^0_A, y^\epsilon_A) + 
c^{-4} {\cal L}_A(y^0_A, y^\epsilon_A) + {\cal O}(c^{-6}),
\eqno(7.1a)  $$
$$x^\alpha =  y^\alpha_A + y^\alpha_{A_0} (y^0_A) + 
c^{-2}{\cal Q}^\alpha_A(y^0_A,y^\epsilon_A) + {\cal O}(c^{-4}).
\eqno(7.1b)$$

\noindent Analogously to the derivations in the Section {\small 5}, one
may obtain the  
 necessary   corrections of the third order with respect to the spatial
separation  to the functions (7.2).  
Then the parameterized coordinate transformation functions  ${\cal K}_A,
{\cal Q}^\alpha_A$ 
and ${\cal L}_A$ may be given    as   follows:
{}
$${\cal K}_A(y^0_A,y^\nu_A)  =\int^{y^0_A} \hskip -10pt  dt'  
\Big( \sum_{B\not=A} \Big< U_B\Big>_A- 
{1\over2}{v_{A_0}}_\nu v_{A_0}^\nu\Big)  
-{v_{A_0}}_\nu\cdot y^\nu_A +  {\cal O}(c^{-4})y^0_A, \eqno(7.2a)$$
{}
$${\cal Q}^\alpha_A(y^0_A,y^\nu_A) = - \gamma \sum_{B\not=A} \Big(
y^\alpha_A 
 y^\beta_A \cdot \Big<\partial_\beta U_B \Big>_A-
{1\over2} {y_A}_\beta  y^\beta_A  
\Big<\partial^\alpha U_B\Big>_A  
 + y^\alpha_A  \Big<U_B\Big>_A \Big)+$$
{}
$$ - {1\over2} v^\alpha_{A_0} v^\beta_{A_0}{y_A}_\beta 
 + w^\alpha_A(y^0_A) + $$
{}
$$+{y_A}_\beta  \int^{y^0_A} \hskip -10pt dt' 
\Big( {1\over2}a^{[\alpha}_{A_0}  v^{\beta]}_{A_0}  + 
(\gamma+1) \sum_{B\not=A}\Big[
\Big<\partial^{[\alpha} V^{\beta]}_B\Big>_A   
+\Big<\partial^{[\alpha} U_B v^{\beta]}\Big>_A \Big]\Big)-$$
{}
$$+ {1\over6}\gamma 
\sum_{B\not=A}y^\mu_A y^\nu_A
y^\beta_A\cdot\Bigg(\eta^{\alpha\lambda} 
\eta_{\mu\nu}\Big<\partial^2_{\beta\lambda} U_B\Big>_0   - 
2 \delta^\alpha_\beta\Big<\partial^2_{\mu\nu} U_B\Big>_0 \Bigg)
 +{\cal O}(|y^\nu_A|^4)  + {\cal O} (c^{-4}) y^\alpha_A,   \eqno(7.2b)$$ 
{}
$${\cal L}_A(y^0_A,y^\nu_A) =  \sum_{B\not=A} \Big(
{1\over2}\gamma\hskip 2pt
y_{A\beta} y^\beta_A \cdot {\partial \over\partial y^0_A}
\Big<U_B\Big>_A  - 
(\gamma+1)\hskip 2pt y^\lambda_A  y^\beta_A \cdot 
\Big[\Big< \partial_\lambda {V_B}_\beta \Big>_A   
+\Big<v_\beta \partial_\lambda U_B \Big>_A \Big] +$$
{}
$$+ \gamma \hskip 2pt{v_{A_0}}_\beta \Big[y^\beta_A  y^\lambda_A \cdot 
\Big< \partial_\lambda  U_B \Big>_A - 
{1\over2} y_{A_\lambda} y^\lambda_A \cdot 
\Big< \partial^\beta  U_B \Big>_A \Big] \Big) +$$
{}
$$+ y_{A_\lambda}{v_{A_0}}_\beta \int^{y^0_A} \hskip -10pt dt'
 \Big( {1\over2} a^{[\lambda}_{A_0} v_{A_0}^{\beta]}+ 
(\gamma+1) \hskip 2pt\sum_{B\not=A}
\Big[\Big<\partial^{[\lambda} V^{\beta]}_B  \Big>_A   
+\Big<v^{[\beta}\partial^{\lambda]} U_B \Big>_A\Big]\Big)+$$
{}
$$ + y_{A_\beta} \Big[ (\gamma+1)\hskip 2pt
v^\beta_{A_0}\sum_{B\not=A} 
\Big<U_B\Big>_A  
 -  2(\gamma+1)\hskip 2pt\sum_{B\not=A} \Big<V^\beta_B\Big>_A  
 - \dot {w}^\beta_{A_0}(y^0_A)  \Big] - $$
{}
$$ - \int^{y^0_A} \hskip -10pt dt' \Big[ \sum_{B\not=A}
\Big<W_B\Big>_A  + {1\over2}\Big(\sum_{B\not=A} \Big< U_B\Big>_A  - 
 {1\over2}{v_{A_0}}_\beta v^\beta_{A_0}\Big)^2 
  + {v_{A_0}}_\mu\dot{w}^\mu_{A_0}(t')\Big]+$$
{}
$$+{1\over 6}\sum_{B\not=A}
y^\mu_A y^\nu_A y^\beta_A \cdot\Bigg( \gamma \hskip 2pt\eta_{\mu\nu} 
\Big<\partial^2_{0\beta} U_B \Big>_0  
 -2(\gamma+1)\hskip 2pt\Big<\partial^2_{\mu\nu} {V_B}_\beta\Big>_0  -
$$ 
{}
$$ -\gamma \hskip 2pt{v_{A_0}}_\sigma\Big[\eta^{\sigma\lambda} 
\eta_{\mu\nu}\Big<\partial^2_{\beta\lambda} U_B\Big>_0    - 
2 \delta^\sigma_\beta\Big<\partial^2_{\mu\nu} U_B\Big>_0\Big]  
\Bigg)+{\cal O}(|y^\nu_A|^4) +{\cal O}(c^{-6}).\eqno(7.2c)$$  

\noindent Note that, in order to distinguish between the {\small PPN}
parameter $\gamma$ and the Minkowski tensor $\gamma_{mn}$, we  are
using new notation for this
tensor, namely: $\eta_{mn}\equiv\gamma_{mn}=\hbox{\rm diag}(1,-1,-1,-
1)$. 
The time-dependent functions ${Q^\alpha_A}_{\{L\}}$ 
and ${L_A}_{\{L\}}$ in the expressions (7.2) are the 
contributions coming from  the higher multipoles $(l\ge3)$
(both mass and current induced) 
of the external gravitational field generated by the 
 bodies {\small (B}$\not=${\small A)}  in the system.
These functions  enable one to take into account the geometric features of 
the proper {\small RF}$_A$ with respect to three-dimensional  
spatial rotation. The form of these functions may be chosen  arbitrarily.
This freedom
enables one to choose any coordinate dependence for the terms with
$l\ge3$ 
in order to  describe  the motion of the highest monopoles. Moreover, one
may show that, 
even thought the total solution to the metric tensor $g_{mn}(x^p)$ in the  
barycentric inertial  {\small RF}$_0$  
 resembles the form of the one-body solution eqs.(2.5), 
but, if one will express  this solution   
through the proper multipole moments of the bodies, it will contain the 
contributions coming form the functions ${Q^\alpha_A} _{\{L\}}(y^0_A)$ 
and ${L_A} _{\{L\}}(y^0_A)$. 
As a result, the metric tensor in the proper {\small RF}$_A$ 
fully represents the tidal nature of the external gravity in the 
coordinates of this frame.

The  quasi-Newtonian acceleration of the body {\small (A)} with respect
to 
the barycentric {\small RF}$_0$ may be  described as follows
{}
$$a^\alpha_{A_0}(y^0_A)=-\eta^{\alpha\mu}\sum_{B\not=A}
\Big<{\partial  U_B\over \partial y^\mu_A}\Big>_A+{\cal O}(c^{-
6}).\eqno(7.3a)$$
\noindent With  the accuracy  necessary for  future analysis, 
we   present the equation for the  time-dependent function 
${\underline w}_{A_0}^\alpha(y^0_A)$   with
respect to the   time $y^0_A$  as follows: 
{}
$$\ddot{\underline w}^\alpha_{A_0}(y^0_A)=
\sum_{B\not=A} \Big(\eta^{\alpha\mu}\Big<{\partial W_B\over \partial
y^\mu_A}\Big>_0+ 
v^\alpha_{A_0}{\partial \over \partial y^0_A} \Big< U_B\Big>_0  - 
 2(\gamma+1)\hskip 2pt{\partial \over \partial y^0_A}\Big<
V^\alpha_B\Big>_0\Big)-$$
{}
$$- {1\over2}v^\alpha_{A_0}{v_{A_0}}_\beta a^\beta_{A_0} + 
\gamma \hskip 2pt a^\alpha_{A_0}\sum_{B\not=A} \Big<U_B\Big>_0  + $$
{}
$$+{a_{A_0}}_\beta \int^{y^0_A} \hskip -10pt dt' 
\Big( {1\over2}a^{[\alpha}_{A_0}  v^{\beta]}_{A_0}  + 
(\gamma+1) \sum_{B\not=A}\Big[
\Big<\partial^{[\alpha} V^{\beta]}_B\Big>_A   
+\Big<v^{[\beta}\partial^{\alpha]} U_B \Big>_A \Big]\Big) +$$
{}
$$+{1\over 5}{\ddot{a}_{A_0}}{_{{}_\mu}} \int_A  d^3y'_A{\hat \rho}_A
\Big(y'^\alpha_A y'^\mu_A+{1\over2}\gamma^{\alpha\mu}y'_{A_\lambda}
y_A'^\lambda  \Big)  
 + {\cal O}(c^{-6}). \eqno(7.3b)$$
\noindent The  expressions (7.3) are the two parts of the 
force  necessary   to keep the body {\small (A)} in it's orbit (world tube)
in the  {\small N}-body system. 
These expressions are written in the proper time and  if one performs the
coordinate 
transformation from the coordinates $(y^p_A)$   
to those   of   $(x^p)$ for
 all the functions and potentials  entering both equations  (7.3),
and will take into account the lowest intrinsic multipole moments of the
bodies 
only, one will obtain the  simplified equations of motion for the extended 
bodies
(2.14)-(2.20) written in the coordinates $(x^p)$ of the barycentric inertial
{\small RF}$_0$.  

Finally, the   metric tensor, corresponding to  the coordinate  
transformations (7.1)-(7.3),  will take the  form of the   Fermi-normal-
like 
proper {\small RF}$_A$ chosen to study the physical processes in the 
vicinity of the body {\small (A)}: 
{}
$$g^{\cal F}_{00}(y^p_A) = 1-2{\overline U}(y^p_A)+2W_A(y^p_A)+$$
{}
$$+\Bigg( \sum_{B\not=A}\Bigg[\Big<\partial^2_{\mu\nu}  W_B\Big>_0+
{\partial \over \partial y^0_A}\Big(\gamma \hskip 2pt \eta_{\mu\nu} 
\Big<{\partial U_B\over\partial y^0_A}\Big>_0-
(\gamma+1)\Big[ \Big<\partial_{(\nu} V_{B\mu)}\Big>_0+
\Big<\partial_{(\nu} U_B v_{\mu)}\Big>_A \Big]\Big) \Bigg] +$$
{}
$$ + \gamma \hskip 2pt \eta_{\mu\nu} \hskip 1mm {a_{A_0}}_\lambda
a^\lambda_{A_0} -  
(2\gamma-1) \hskip 2pt {a_{A_0}}_\mu 
{a_{A_0}}_\nu\Bigg) \cdot y^\mu_A y^\nu_A  + 
 {\cal O}(|y^\nu_A|^{3}) +  {\cal O}(c^{-6}), \eqno(7.4a) $$
{}
$$g^{\cal F}_{0\alpha}(y^p_A) = 2(\gamma+1)\eta_{\alpha\epsilon}
\Big[V^\epsilon_A(y^p_A)+
\sum_{B\not=A}y^\mu_A\Big<v^\epsilon\partial_\mu U_B\Big>_A\Big]
 +{2\over3}\Bigg(\gamma\hskip 2pt\Big(\eta_{\alpha\mu}
\dot{a}_{A_0}{_{{}_\nu}}-  
\eta_{\mu\nu}\dot{a}_{A_0}{_{{}_\alpha}}\Big) + $$
{}
$$+  (\gamma+1) \sum_{B\not=A} \Big[ \eta_{\alpha\lambda} 
\Big<\partial^2_{\mu\nu}  V^{\lambda}_B\Big>_0  -
\eta_{\nu\lambda}\Big<\partial^2_{\mu\alpha}V^{\lambda}_B\Big>_0
\Big]\Bigg) \cdot y^\mu_A y^\nu_A  
 +  {\cal O}(|y^\nu_A|^{3}) +   {\cal O}(c^{-5}), \eqno(7.4b)$$ 
{}
$$g^{\cal F}_{\alpha\beta}(y^p_A) =   
\eta_{\alpha\beta} \Big(1 + 2 \gamma  U_A(y^p_A)\Big)+
{1\over3} \gamma \sum_{B\not=A}\Big[  \eta_{\alpha\beta} 
\Big<\partial^2_{\mu\nu}U_B\Big>_0    + 
\eta_{\mu\nu} \Big<\partial^2_{\alpha\beta}U_B\Big>_0 - $$
{}
$$-\eta_{\beta\mu}\Big<\partial^2_{\alpha\nu} U_B\Big>_0 - 
\eta_{\alpha\nu} \Big<\partial^2_{\beta\mu} U_B\Big>_0 \Big] 
\cdot y^\mu_A y^\nu_A + {\cal O}(|y^\nu_A|^{3}) +
{\cal O}(c^{-4}), \eqno(7.4c)$$ 

\noindent where total Newtonian potential ${\overline U}$ in the vicinity
of the body {\small (A)} 
is given by expression (6.17). Both  functions $W_A$ and $W_B$  
parameterized by the two  Eddington parameters $\gamma$ and $\beta$:
{}
$$W_A(y^p_A) = \beta \hskip 2pt U^2_A(y^p_A) +\Psi_A(y^p_A) +
2 a^\lambda_{A_0}\cdot{\partial \over 
\partial y^\lambda_A}\chi_A (y^p_A)+{1\over2}{\partial^2\over\partial 
{y^0_A}^2} \chi_A(y^p_A)+$$
{} 
$$+\sum_{B\not=A}\Bigg(2\beta \hskip 2pt U_A(y^p_A) U_B(y^p_A)  
-(3\gamma+1-2\beta)\int_A{d^3y'_A \over{|y^\nu_A - y'^\nu_A|}} 
 \rho_A(y^0_A,y'^\nu_A) U_B (y^0_A,y'^\nu_A)\Bigg)+$$
{}
$$+ {1\over6}\gamma 
\sum_{B \not=A} \Bigg(\eta^{\alpha\lambda} 
\eta_{\mu\nu}\Big<\partial^2_{\beta\lambda} U_B\Big>_0   - 
2 \delta^\alpha_\beta\Big<\partial^2_{\mu\nu} U_B\Big>_0
\Bigg)\times$$
{}
$$\times\int_A d^3y'_A  \rho_A (y_A^0,y_A'^\nu)
  {\partial \over\partial y'^\lambda_A} 
\Big[{y^\mu_A y^\nu_A y^\beta_A-y'^\mu_A y'^\nu_A y'^\beta_A 
\over{|y_A^\nu - y_A'^\nu|}}\Big]
 +{\cal O}(|y^\nu_A|^4)  + {\cal O} (c^{-4}) y^\alpha_A,    \eqno(7.5a)$$  
 {} 
$$W_B(y^p_A) = \beta U_B(y^p_A)\sum_{C\not=A} U_C(y^p_A) +\Psi_B
(y^p_A)  +
2 a^\lambda_{A_0}\cdot 
{\partial \over \partial y^\lambda_A} \chi_B 
(y^p_A)+{1\over2}{\partial^2\over\partial 
{y^0_A}^2} \chi_B (y^p_A) - $$
{}
$$-(3\gamma+1-2\beta)\int_B {d^3y'_A \over{|y^\nu_A-y'^\nu_A|}} 
\rho_B (y^0_A, y'^\nu_A)   \sum_{B'}U_{B'}(y^0_A,y'^\nu_A)+  $$
{}
$$+ {1\over6}\gamma 
\sum_{B \not=A} \Bigg(\eta^{\alpha\lambda} 
\eta_{\mu\nu}\Big<\partial^2_{\beta\lambda} U_B\Big>_0   - 
2 \delta^\alpha_\beta\Big<\partial^2_{\mu\nu} U_B\Big>_0
\Bigg)\times$$
{}
$$\times\int_A d^3y'_A  \rho_B \Big(y_A^0,
y_A'^\nu+y_{BA_0}^\nu(y_A^0)\Big) 
  {\partial \over\partial y'^\lambda_A} 
\Big[{y^\mu_A y^\nu_A y^\beta_A-y'^\mu_A y'^\nu_A y'^\beta_A 
\over{|y_A^\nu - y_A'^\nu|}}\Big]
 +{\cal O}(|y^\nu_A|^4)  + {\cal O} (c^{-4}) y^\alpha_A.   \eqno(7.5b)$$ 

\noindent  The  functions $W_A$ and $W_B $  fully represent the non-
linearity of the 
total post-Newtonian gravitational  field in the Fermi-normal-like
coordinates of the 
proper {\small RF}. As a result we have obtained the  metric tensor in the 
  Fermi-normal-like coordinates and the coordinate transformations 
leading to this form. These transformations are defined up 
to the third order with respect to the spatial coordinates.
Let us note, that as a   partial result of the analysis presented in the
previous Section,
we have shown  that the Fermi-normal-like coordinates do not 
provide one with the conservation laws of the joint density of the 
energy-momentum of matter, inertia and the gravitational field 
in the immediate vicinity of the body under consideration. However, taking 
into 
account the expected accuracy of the radio-tracking data from the future
{\it Mercury Orbiter}
mission, we can neglect the influence of the corresponding effects and,
therefore, 
use the Fermi-normal-like coordinates for out theoretical studies.  
As a result, we will   analyze the motion of the spacecraft 
in orbit around Mercury from the position  of the parameterized 
relativistic gravity.

\subsection{Equations of the Spacecraft Motion.}  
 
We   will now obtain the equations of the spacecraft motion in a Hermean-
centric 
{\small RF}. To do  this, we consider a Riemann space-time whose metric
coincides with the metric of 
 {\small N}  moving extended bodies.  We shall study the motion of a 
point body in the  neighborhood of the  body {\small (A)}. The 
expression for the acceleration of the point body $a^\alpha _{(0)}$ can be
obtained in two ways:
either by using the equations of geodesics of Riemann space-time 
$du^n/ds+\Gamma^n_{mk}u^mu^k=0$ or by computing the acceleration of
the center of 
mass of the extended body and then letting all quantities characterizing
its 
internal structure and proper gravitational field tend to zero. 
In either case  one obtains the same result (Denisov \& Turyshev, 1990).

In order to obtain the Hermean-centric equations of the satellite motion
we will
write out the equations of geodesics to the required degree of accuracy.
For $n=\alpha$ we have:
{}
$${du^\alpha\over ds}+\Gamma^\alpha_{00}u^0u^0+
2\Gamma^\alpha_{0\beta} u^0u^\beta+\Gamma^\alpha_{\mu\beta}u^\mu
u^\beta=
{\cal O}(c^{-6}). \eqno(7.6) $$
\noindent We consider the metric tensor of Riemann space-time  
to be  given by the expressions (7.4) in this case.
It is then possible to find the connection components of Riemann space-
time
needed for subsequent computations:
{}
$$\Gamma^{\alpha}_{00}(y^p_{A}) = \eta^{\alpha\lambda}\Big[ 
{\partial\overline{U}\over \partial y^\lambda_A}
- {\partial W_A\over \partial y^\lambda_A} 
-\gamma   {\partial U^2_A\over \partial y^\lambda_A}\Big] +
2(\gamma+1){\partial V^\alpha_A\over \partial y^0_A}+$$ 
{}
$$ + \Bigg((2\gamma-1) a^\alpha_{A_0} {a_{A_0}}_\mu - 
\gamma\delta^\alpha_\mu \cdot a^\lambda_{A_0} {a_{A_0}}_\lambda -
\eta^{\alpha\lambda}\sum_{B\not=A}\Big[
\Big<\partial^2_{\lambda\mu} W_B\Big>_0+
 2\gamma U_A \Big<\partial^2_{\lambda\mu} U_B\Big>_0\Big]+$$
{}
$$ +\sum_{B\not=A}{\partial \over \partial y^0_A}
\Bigg[(\gamma+1) \Big[ \Big<\partial_{(\mu} V^{\alpha)}_B \Big>_0
+\Big<v^{(\alpha}\partial_{\mu)} U_B\Big>_A\Big] -
\gamma\delta^\alpha_\lambda\Big<{\partial U_B\over \partial y^0_A}
\Big>_0\Bigg]\Bigg) \cdot y^\mu_A +
{\cal O}(|y^\nu_A|^2)+ {\cal O}(c^{-6}), \eqno(7.7a)$$
{}
$$\Gamma^{\alpha}_{0\beta}(y^p_{A}) =\gamma\delta^\alpha_\beta 
{\partial U_A \over \partial y^0_A} + 
(\gamma+1)\delta^\mu_{[\beta}{\partial V^{\alpha]}_A  \over\partial
y^{\mu}_A} + 
\gamma  \eta_{\beta\mu}{\dot a}^{[\mu}_{A_0}y^{\alpha]}_A+$$
{}
$$+(\gamma+1) \sum_{B\not=A}\Big[\Big<\partial^2_{[\beta\mu}
V^{\alpha]}_B\Big>_0
-  \Big<v^{[\alpha}\partial_{\beta]} U_B\Big>_A 
\Big]\cdot y^\mu_A+{\cal O}(|y^\nu_A|^2)+ {\cal O}(c^{-5}), \eqno(7.7b)$$
{}
$$\Gamma^{\alpha}_{\beta\omega}(y^p_{A}) = 
 \gamma\Big(\delta^\alpha_\beta {\partial U_A\over \partial
y^\omega_A}  + 
\delta^\alpha_\omega {\partial U_A\over \partial
 y^\beta_A}  -\eta_{\beta\omega}\eta^{\alpha\lambda}
{\partial U_A\over \partial y^\lambda_A}\Big)+$$
{}
$$ +{1\over3}\gamma\sum_{B\not=A}\Big[\eta_{\beta\mu}
\Big<\partial^\alpha\partial_\omega U_B\Big>_0  +
\eta_{\omega\mu}\Big<\partial^\alpha\partial_\beta U_B\Big>_0+
\delta^\alpha_\omega\Big<\partial^2_{\beta\mu} U_B\Big>_0+
\delta^\alpha_\beta\Big<\partial^2_{\omega\mu} U_B\Big>_0-$$
{}
$$-2\eta_{\beta\omega}\Big<\partial^\alpha\partial_\mu U_B\Big>_0-
2\delta^\alpha_\mu\Big<\partial^2_{\beta\omega} U_B\Big>_0\Big]
\cdot y^\mu_A+{\cal O}(|y^\nu_A|^2)+ {\cal O}(c^{-4}).\eqno(7.7c) $$

To reduce the equation of geodesic motion (7.6) we shall use both
the  expressions above and the definition for the four-vector of velocity in
the form:
{}
$$u^n={dy^n_A\over
dy^0_A}\Big(g_{00}+2g_{0\mu}v^\mu+g_{\mu\lambda}v^\mu
v^\lambda\Big)^{-1/2}.$$
\noindent Then  by taking into account  that $d/ds=u^0d/dy^0_A$ 
(with the components of the three-dimensional velocity
vector of the point body denoted as $v^\alpha_{(0)}=dy^\alpha_A/dy^0_A$)
and  by
  using the Newtonian equation of motion of a point body as:
{}
$$a^\alpha_{(0)}= {d v^\alpha_{(0)}\over dy^0_A}= -\eta^{\alpha\mu}
{\partial {\overline U}\over\partial y^\mu_A}+{\cal O}(c^{-4}),$$
\noindent we may make the following simplification:
{}
$$v^\alpha_{(0)}{d\ln u^0\over dy^0_A}=v^\alpha_{(0)}\Big({\partial
{\overline U}\over\partial y^0_A}+
2v^\mu_{(0)}{\partial {\overline U}\over\partial y^\mu_A}+{\cal O}(c^{-
5})\Big).$$

\noindent Substituting this relation into the equations of motion (7.6), 
we find the acceleration $a^\alpha_{(0)}$ of the point body:
{}
$$a^\alpha_{(0)}=-\eta^{\alpha\mu}{\partial {\overline U}\over\partial
y^\mu_A}
(1-2\gamma U_A)+  \partial^\alpha W_A +
\sum_{B\not=A}\Big<\partial^\alpha\partial_\mu W_B\Big>_0 \cdot
y^\mu_A- $$
{}
$$ -(2\gamma+1)v^\alpha_{(0)}{\partial U_A\over\partial y^0_A}-
2(\gamma+1){\partial   V^\alpha_A \over\partial y^0_A}-2(\gamma+1) 
{v_{(0)}}_\mu\Big[ \partial^{[\mu} V^{\alpha]}_A  +
\sum_{B\not=A}\Big<v^{[\mu}\partial^{\alpha]} U_B\Big>_0\Big]+$$
{}
$$+ \gamma {v_{(0)}}_\lambda v^\lambda_{(0)}\Big[\partial^\alpha U_A+
{2\over3}\sum_{B\not=A}\Big<\partial^\alpha\partial_\mu
U_B\Big>_0\cdot y^\mu_A\Big]- $$
{}
$$-v^\alpha_{(0)}  v^\lambda_{(0)}\Big[2(\gamma+1) \partial_\lambda
U_A+
{2\over3}(\gamma+3)\sum_{B\not=A}\Big<\partial_\lambda\partial_\mu 
U_B\Big>_0\cdot y^\mu_A\Big]+$$
{}
$$+y^\mu_A\Bigg(\gamma\delta^\alpha_\mu{a_{A_0}}_\lambda
a^\lambda_{A_0}-
(2\gamma-1)a^\alpha_{A_0}{a_{A_0}}_\mu +2\gamma v^\lambda_{(0)}
\big[\eta_{\lambda\mu}{\dot a}^\alpha_{A_0}-\delta^\alpha_\mu{{\dot
a}_{A_0}}{_{{}_\lambda}}\big]+$$
{}
$$+\sum_{B\not=A}\Bigg({\partial\over\partial y^0_A}\Big[\gamma
\delta^\alpha_\mu
\Big<{\partial U_B\over \partial y^0_A}\Big>_0-
(\gamma+1) \Big(\Big<\partial_{(\mu}V^{\alpha)}_B\Big>_0+
\Big<v_\mu\partial^\alpha U_B\Big>_0+ 
3\Big<v^\alpha\partial_\mu U_B\Big>_0\Big)\Big]+$$
{}
$$+2(\gamma+1)v^\lambda_{(0)} \Big<\partial^{[\alpha}\partial_\mu
 {V_B}_{\lambda]}\Big>_0+
 {2\over3}\gamma v^\lambda_{(0)}v^\beta_{(0)}
\delta^{[\alpha}_\mu\Big<\partial^2_{\lambda]\beta}U_B\Big>_0
\Bigg)\Bigg)+
{\cal O}(|y^\nu_A|^2)+{\cal O}(c^{-6}).\eqno(7.8)$$

By expanding all the potentials in (7.8) in power series of 
$ 1/y_{BA_0}$ and   retaining    terms with 
$\sim y^\alpha/|y_{BA_0}|$
only  to the required accuracy, we then obtain:
{}
$$a^\alpha_{(0)}=-\eta^{\alpha\mu}{\partial {\overline U}\over\partial
y^\mu_A}+
\delta_A a^\alpha_{(0)}+\delta_{AB} a^\alpha_{(0)}+
\delta_B a^\alpha_{(0)}+\delta_{BC} a^\alpha_{(0)}+{\cal
O}(|y^\nu_A|^2)+{\cal O}(c^{-6}), 
\eqno(7.9)$$

\noindent where the post-Newtonian acceleration $\delta_A
a^\alpha_{(0)}$ 
due to the gravitational field of the 
body {\small (A)} only, may be  given as 
{} 
$$\delta_A a^\alpha_{(0)}=2(\gamma+\beta)U_A \partial^\alpha U_A-
(\gamma+{1\over2})\partial^\alpha\Phi_{1A}+(2\beta-
{3\over2})\partial^\alpha\Phi_{2A}+ $$
{}
$$+(1-\gamma)\partial^\alpha \Phi_{4A}+{1\over2}\partial^\alpha A_A-
(2\gamma+1)v^\alpha_{(0)}
\partial_\mu V^\mu_A +\gamma {v_{(0)}}_\mu v^\mu_{(0)}\partial^\alpha
U_A- $$
{}
$$-2(\gamma+1) v^\alpha_{(0)}  v^\mu_{(0)}\partial_\mu U_A-
2(\gamma+1) 
{v_{(0)}}_\mu\Big[ \partial^\mu V^\alpha_A-\partial^\alpha
V^\mu_A\Big]-$$
{}
$$-2(\gamma+1)\int_A d^3y'_A{\hat\rho}_A v'^\alpha_A
{v'_A}_\mu{(y^\mu_A-y'^\mu_A)
\over|y^\nu_A-y'^\nu_A|}-{1\over2}(4\gamma+3) \int_A d^3y'_A  
{{\hat\rho}_A\partial^\alpha U_A\over|y^\nu_A-y'^\nu_A|}+$$
{}
$$+{1\over2}  \int_A d^3y'_A {\hat\rho}_A\partial_\mu U_A 
{(y^\alpha_A-y'^\alpha_A)(y^\mu_A-y'^\mu_A)\over|y^\nu_A-
y'^\nu_A|}+{\cal O}(c^{-6}).
\eqno(7.10)$$

\noindent  This term  is  known and reasonably well understood (Denissov
\& Turyshev, 1990). 
The term  $\delta_{AB} a^\alpha_{(0)}$ is the 
acceleration due to the interaction of the 
gravitational field of the extended body {\small (A)} with the 
external gravitation in the  {\small N} body system:
{}
$$\delta_{AB} a^\alpha_{(0)}=\sum_{B\not=A}
\Bigg((4\beta-3\gamma-1){m_Am_B\over y_{BA_0}}{n^\alpha\over y^2}
+2(\beta-1){m_Am_B\over y}{N^\mu_{BA_0}\over {y^2_{BA_0}}}
\big(\delta^ \alpha_\mu+n^\alpha n_\mu\big)+$$
{}
$$+{m_Am_B\over y^3_{BA_0}}{\cal P}_{\epsilon\lambda}
\Big((2\beta+{5\over3}\gamma)\eta^{\alpha\epsilon}n^\lambda+
(\beta-{1\over6})n^\alpha n^\epsilon n^\lambda\Big) -
(\gamma+1){m_B\over y^3_{BA_0}}{\cal P}_{\epsilon\lambda}
S^{\alpha\epsilon}_A v^\lambda_{(0)}
+ $$
{}
$$+{m_Am_B\over y^4_{BA_0}}\Big((2\beta+\gamma-
1)\delta^\alpha_\mu+
2(3\beta+ \gamma-1)N^\alpha_{BA_0}{N_{BA_0}}_\mu\Big) y^\mu_A+ $$ 
{}
$$+3\beta {m_Am_B\over y^4_{BA_0}}|y^\nu_A| n^\epsilon n^\lambda
\Big(2\delta^\alpha_\epsilon{N_{BA_0}}_\lambda+ N^\alpha_{BA_0} 
(\eta_{\epsilon\lambda}+5 {N_{BA_0}}_\epsilon
{N_{BA_0}}_\lambda)\Big)\Bigg)+{\cal O}(|y^\nu_A|^2)+ {\cal O}(c^{-6}), 
\eqno(7.11)$$
\noindent where  $S^{\alpha\epsilon}_A$ is the reduced  spin moment of
the 
body {\small (A)} and ${\cal
P}_{\epsilon\lambda}=\eta_{\epsilon\lambda}+
3{N_{BA_0}}_\epsilon{N_{BA_0}}_\lambda$  is the polarizing operator. 
Note  that the combination of the post-Newtonian 
parameters in the first term of the expression $(7.11)$   differs 
from that for the well known Nordtvedt effect (Nordtvedt, 1968b,c; Will,
1993).
This may provide an independent test for the parameters involved.
The reason  that our third term in this expression   differs from 
the analogos  term derived in Ashby \& Bertotti (1986) is that,
in order to obtain this result (7.9)-(7.11), we  used the consistent 
definitions for the 
conserved mass density in the proper {\small RF}$_A$. 
Moreover, in  constructing  the Fermi normal coordinates previous authors 
used incomplete expressions for the spatial coordinate transformations,
which  
  differ  from    eq.(7.2) (specifically in the third order of the spatial
coordinates).
 Note that  if we decide to use our definitions, the 
result cited above will take the form of  (7.11). 
The next term,  $\delta_B a^\alpha_{(0)}$, is the
post-Newtonian acceleration  caused by the other 
bodies in the system on the orbit of the body {\small (A)}
(the effect of the post-Newtonian tidal forces): 
{} 
$$\delta_B a^\alpha_{(0)}=\sum_{B\not=A} y^\mu_A\Bigg(-
{3\over2}{m^2_B\over y^4_{BA_0}}
\Big({1\over3}(14\gamma-4\beta-7)\delta^\alpha_\mu+
3N^\alpha_{BA_0}{N_{BA_0}}_\mu{1\over9}(42\gamma-16\beta-
17)\Big)+$$
{}
$$+{m_B\over y^3_{BA_0}}{\cal P}^\alpha_\mu 
\Big[(\gamma+{1\over2}){v_{BA_0}}_\lambda v^\lambda_{BA_0}+
{3\over2}({v_{BA_0}}_\lambda N^\lambda_{BA_0})^2 +(4\beta-\gamma-
3)E_B\Big]+ $$
{}
$$+\gamma{m_B\over y^3_{BA_0}}{\cal P}_{\epsilon\lambda} 
\Big[\delta^\alpha_\mu v^\epsilon_{BA_0}v^\lambda_{BA_0}-
v^\alpha_{BA_0}v^\epsilon_{BA_0}\delta^\lambda_\mu-
v^\epsilon_{BA_0}{v_{BA_0}}_\mu\eta^{\alpha\lambda}\Big]-$$
{}
$$-{m_B\over
y^3_{BA_0}}\Big(\delta^\alpha_\epsilon\eta_{\lambda\mu}-
3N^\alpha_{BA_0}{N_{BA_0}}_\mu{N_{BA_0}}_\epsilon{N_{BA_0}}_\lambda\Big)
v^\epsilon_{BA_0}v^\lambda_{BA_0}+ $$
{}
$$+{m_B\over y^3_{BA_0}}{\cal P}_{\epsilon\lambda} 
\Big[{2\over3}\gamma {v_{(0)}}_\beta v^\beta_{(0)}
\eta^{\alpha\epsilon}\delta^\lambda_\mu-
{2\over3}(\gamma+3)v^\alpha_{(0)} v^\epsilon_{(0)}\delta^\lambda_\mu+
$$
{}
$$+{2\over3}\gamma v^\beta_{(0)} v^\lambda_{(0)}
[\delta^\alpha_\mu\delta^\epsilon_\beta-
\eta_{\beta\mu}\eta^{\alpha\epsilon}]+
2\gamma v^\beta_{(0)} v^\lambda_{BA_0}
[\delta^\alpha_\mu\delta^\epsilon_\beta-
\eta_{\beta\mu}\eta^{\alpha\epsilon}]+$$
{}
$$+2(\gamma+1)v^\beta_{(0)}\delta^\lambda_\mu[{v_{BA_0}}_\beta\eta^{\alpha\epsilon}-
v^\alpha_{BA_0}\delta^\epsilon_\beta]\Big]\Bigg)+{\cal
O}(|y^\nu_A|^2)+{\cal O}(c^{-6}).
\eqno(7.12)$$

\noindent Finally, the last term in the expression (7.9), $\delta_{BC}
a^\alpha_{(0)}$,
is the contribution to the equation of motion of the non-linear 
gravitational interaction  of the external bodies with each other
given as follows:
{}
$$\delta_{BC} a^\alpha_{(0)}=-\sum_{B\not=A}\sum_{C\not=A,B}
y^\mu_A\Bigg({m_Bm_C\over y^2_{BA_0}y^2_{CA_0}}\Big[3(\gamma-1)
{\cal P}^\alpha_\mu {N_{BA_0}}_\lambda N^\lambda_{CA_0}+$$
{}
$$+(2\beta-1)N^\alpha_{BA_0}{N_{CA_0}}_\mu -
N^\alpha_{CA_0}{N_{BA_0}}_\mu\Big]+
{m_Bm_C\over y^3_{BA_0}y_{CA_0}}\Big[(2\beta-3\gamma+
{1\over2}){\cal P}^\alpha_\mu+$$
{}
$$+{1\over2}\gamma{\cal P}_{\epsilon\lambda} 
\Big(\eta^{\alpha\epsilon}{N_{CA_0}}_\mu
N^\lambda_{CA_0}+\delta^\epsilon_\mu
N^\alpha_{CA_0} N^\lambda_{CA_0}-(\delta^\alpha_\mu+ N^\alpha_{CA_0}
{N_{CA_0}}_\mu)
N^\epsilon_{CA_0} N^\lambda_{CA_0}\Big)\Big]+$$
{}
$$+{m_Bm_C\over y^2_{BA_0}y^2_{CB_0}}\Big({1\over2}
{\cal P}^\alpha_\mu +
\gamma\delta^\alpha_\mu \Big){N_{BA_0}}_\lambda N^\lambda_{CB_0}-$$
{}
$$-(\gamma+{1\over2})(N^\alpha_{BA_0}
{N_{CB_0}}_\mu+{N_{BA_0}}_\mu 
N^\alpha_{CB_0})\Big]+(2\beta-{3\over2}){\cal P}^\alpha_\mu
{m_Bm_C\over y^3_{BA_0}y_{CB_0}}\Bigg)+$$
{}
$$+{\cal O}(|y^\nu_A|^2)+{\cal O}(c^{-6}).\eqno(7.13)$$

Thus, the equations presented in this part are represent the motion of a
test body in the 
Fermi-normal-like coordinates chosen in the proper {\small RF} of a body
{\small (A)}.
Together with the coordinate transformations (7.1)-(7.3)  this is the
general solution of the 
gravitational  {\small N} body problem.

We present here the restricted version of the equations which is
consistent with  
the  expected accuracy for ESA's {\it Mercury Orbiter} mission. This
limited accuracy 
permits us to completely neglect  
contributions  proportional to the spatial coordinates $y^\mu_A$.  
The planeto-centric equations of  satellite motion around  Mercury can be 
represented  by a series of $1/|y_{BA_0}|$ as follows:
{}
$$a^\alpha_{(0)}=-\eta^{\alpha\mu}\Big({\partial U_A\over\partial
y^\mu_A}+
\sum_{B\not=A} \Big[{\partial U_B\over \partial y^\mu_A} -
 \Big<{\partial U_B\over \partial y^\mu_A}\Big>_A\Big] \Big)+\delta_A
a^\alpha_{(0)}+$$
{}
$$+ \sum_{B\not=A}
\Bigg((4\beta-3\gamma-1){m_Am_B\over y_{BA_0}}{n^\alpha\over y^2}
+2(\beta-1){m_Am_B\over y}{N^\mu_{BA_0} \over {y^2_{BA_0}}}
\big(\delta^ \alpha_\mu+n^\alpha n_\mu\big)+$$
{}
$$+{m_Am_B\over y^3_{BA_0}}{\cal P}_{\epsilon\lambda} 
\Big[(2\beta+{5\over3}\gamma)\eta^{\alpha\epsilon}n^\lambda+
(\beta-{1\over6})n^\alpha n^\epsilon n^\lambda\Big]\Bigg)+{\cal
O}(|y^\nu_A|)+{\cal O}(c^{-6}), 
\eqno(7.14)$$
\noindent where index {\small (A)}   denotes the planet Mercury and   the 
post-Newtonian 
acceleration $\delta_A a^\alpha_{(0)}$ is due to the gravitational field of   
Mercury  only. 

Thus, the  formalism  presented in  this Section  could significantly
simplify the general analysis of the 
tracking data for the {\it Mercury Orbiter} mission. We have presented the
Hermean-centric 
equations of the satellite motion, the barycentric equations of the
planet's motion 
in the solar system barycentric {\small RF}, and the coordinate
transformations which   link  these 
equations together. In particular, our analysis had shown that in a  proper
Hermean-centric {\small RF} the 
corresponding equations of the satellite motion  depends on the Mercury's
gravitational field only. 
 This set of equations in well-known 
and  widely in use for studying the dynamics of the test bodies  in the
isolated gravitational 
one body problem. 
  The existence of the external gravitational field  manifests
itself in the form of the tidal forces  only as well as  
determining the dynamic properties of the constructed 
 Hermean-centric proper {\small RF}.  
Note that within the accuracy expected for the future {\it Mercury
Orbiter} mission
one may  completely neglect the post-Newtonian tidal terms. 
However,   while constructing this  {\small RF},
we went further than the expected accuracy of the future experiments.
Indeed, the last term in the equation (7.14) is due to the coordinate 
transformation to the Fermi-normal-like  {\small RF} which may be
chosen in the 
planet's vicinity. One may neglect this term for the solar system motion,
however, 
if one  applies the  presented formalism to the problems of motion with  a
more intensive 
gravitational environment, one will find that this term  may plays a
significant role.
The  application of the results obtained here to the problems of  motion of
the double 
pulsars is currently under study and will be reported elsewhere.

It should be noted at once that the coefficients in front of the 
two terms in the second line of the expression eq.(7.14) prove the
correctness of the 
decomposition of the {\it local} fields in the proper {\small RF}$_A$
which we have performed in the 
end of the Section {\small 4}. Indeed, if even one  of these terms would
have a  non-zero value,
this  would mean that the metric tensor of the {\it local} problem would
depend on the
external gravity not through the relativistic 
tidal-like potential, which is of   second order with respect to the spatial 
coordinate   $\sim  y^{\{2\}}_A$, 
but instead, this dependence will be at least of order $\sim y^{\{1\}}_A$. 
As a result, this new dependence  may lead  to  
a violation of the Strong Equivalence Principle {\small (SEP)}. 
There are certainly no worries for the  general theory of relativity  for 
 which the {\small PPN} parameters have the values: $\gamma=\beta=1$.
However, the theories 
having a different means of $\gamma$ and $\beta$  may predict a new
effects in motion of the 
satellites due to the corresponding {\small SEP} violation.    
At this point we have all the necessary equations in order to discuss this
and other  
gravitational experiments with the future {\it Mercury Orbiter } mission.

\subsection{Gravitational Experiment for Post 2000 Missions}
 
Mercury is the closest to the Sun of all the planets of the 
terrestrial group  and because of its
 unique location and orbital parameters, it is 
well suited to relativistic 
gravitational experiments. The short period of its solar orbit
 allows experiments over several orbital revolutions and its high 
eccentricity and inclination allow various effect to be well separated.
 In this Section  we will   discuss the possible gravitational
experiments  for the {\it Mercury Orbiter} mission.
Analysis  performed in this Section  is directed towards the future
mission, so we 
will show which relativistic effects may be measured  and how 
accurately.
 
It is generally considered that   processing the data  
from orbiters is more complicated than that of landers.
This is because of the need to convert from the measured earth-
spacecraft distance to the 
desired earth-planet distance. This involves determining the orbit of the
spacecraft about the 
planetary center of mass, which requires solving from the tracking data  
for a number of spatial harmonics of the gravitational field, 
solving for radiation pressure, and other such effects.
The other non-gravitational perturbations, such as firing   attitude control
jets which have unbalanced forces,  are also frequently present 
which further complicates the analysis. The 
orbit determination of the {\it Mariner 9}, for example,  was 
substantially 
affected not only   by these factors, but also by  the fact that the 
spacecraft was placed on the 12 hr period orbit with low periapsis.
Thus, in order to precisely describe the motion of the {\it Mercury Orbiter} 
relative to Earth,
 one should solve two problems, namely: (i) the problem
of the satellite motion about the Mercury's center of inertia in the
Hermean-centric frame, and (ii) the 
relative motion of the 
both planets - Earth and Mercury - in the solar system barycentric {\small
RF}$_0$. 
Our analysis is intended to  provide a complete solution of these two
problems.

In order to study the relativistic effects in the motion of the 
{\it Mercury Orbiter} satellite, we  separate these effects into
the three following groups: 
\begin{itemize}
\item[(i).] The effects due to   Mercury's motion with 
respect to the solar system barycentric {\small RF}$_0$.

\item[(ii).] Effects  in the satellite's motion with respect to the 
Hermean-centric {\small RF}.

\item[(iii).] Effects due to the dragging of the inertial frames.  
\end{itemize}

The effects of the first group 
are standard and, with the  accuracy anticipated for the future {\it
Mercury Orbiter} 
mission, most of them may be obtained directly from the 
Lagrangian function (2.9) or from the equations of motion (2.14)-(2.20). 
The effects of the second group can be discussed 
based on the equations (7.14). And finally, the effects of the last group 
can be discussed based on the coordinate transformation rules 
given by the eqs.(7.2). In the last case, however, we employ 
a simplified version of these transformations, due to the limited expected 
accuracy ($\sim 1$ m) of the Mercury ranging data.  Thus, in the future
discussion 
we will use the following expression for the temporal components:
{} 
$$x^0(y^0_A, y^\mu_A)=y^0_A+c^{-
2}\Bigg(\int^{y^0_A}\Big[\sum_{B\not=A} {m_B\over y_{BA_0}}
\Big(1+(I^{\lambda\mu}_A+I^{\lambda\mu}_B){{N_{BA_0}}_\lambda{N_{BA_0}}_\mu\over y^2_{BA_0}}\Big)-$$
{}
$$-{1\over2}{v_{A_0}}_\mu v^\mu_{A_0}\Big]dt'-{v_{A_0}}_\mu
y^\mu_A\Bigg)+{\cal O}(c^{-4}),
\eqno(7.15)$$
\noindent where  $I^{\mu\nu}_C$ represents the {\small STF} intrinsic
quadrupole moments 
of the bodies. Note that the terms contained in the function ${\cal
L}_A\sim   {\cal O}(c^{-4})$ 
 will contribute to the 
post-Newtonian  redshift. However, it will not be possible to perform the 
redshift
experiment with the   accuracy anticipated for the mission, and therefore
this term was omitted.
The corresponding expression for the spatial components of the coordinate
transformation 
is given by: 
{}
$$x^\alpha(y^0_A, y^\mu_A)=y^\alpha_{A_0}(y^0_A)+y^\alpha_A+
c^{-2}\Bigg( {y_A}_\mu\Big[ \int^{y^0_A}\Omega_A^{\alpha\mu}(t') dt'-
{1\over 2} v^\alpha_{A_0} v^\mu_{A_0}-
\gamma \eta^{\alpha\mu}\sum_{B\not=A}{m_B\over y_{BA_0}}\Big]-$$
{}
$$-\sum_{B\not=A}{m_B\over y^2_{BA_0}}\Big[y^\alpha_A {y_A}_\mu
N^\mu_{BA_0}  -
{1\over2}{y_A}_\mu y^\mu_A
N^\alpha_{BA_0}\Big]+w^\alpha_A(y^0_A)\Bigg)+
{\cal O}(|y_A|^3)+{\cal O}(c^{-4}), \eqno(7.16a)$$
\noindent with the precession angular velocity tensor 
$\Omega_A^{\alpha\beta}$ given as follows:
{}
$$\Omega_A^{\alpha\beta}(y^0_A)=\sum_{B\not=A}
\Big[(\gamma+{1\over2}){m_B\over
y^2_{BA_0}}N^{[\alpha}_{BA_0}v^{\beta]}_{A_0}-
(\gamma+1){m_B\over y^2_{BA_0}}N^{[\alpha}_{BA_0}v^{\beta]}_{b_0}+ $$
{}
$$+(\gamma+1){m_B\over 2y^3_{BA_0}}{\cal P}^{[\alpha}_{\lambda}
(S^{\beta]\lambda}_A+S^{\beta]\lambda}_B)\Big],
\eqno(7.16b) $$
where  $S^{\mu\nu}_C$ is the {\small STF} intrinsic spin moment    of the
bodies.

We mention that by means of a topographic Legendre expansion complete
through the
second degree and order, the systematic error in Mercury radar
ranging has been reduced significantly (Anderson  {\it et al.}, 1995).
However, a {\it Mercury Orbiter} is required before significant
improvements in relativity tests become possible.  Currently, the
precession rate of Mercury's perihelion, in excess of the 530
arcsec per century ($''$/cy) from planetary perturbations, is 
43.13 $''$/cy with a realistic standard error of 0.14  $''$/cy
 (Anderson  {\it et al.}, 1991).  After taking into account a small
excess precession from  solar oblateness, Anderson  {\it et al.} find
that this result is consistent with general relativity.  Pitjeva
(1993) has obtained a similar result but with a smaller estimated
error of 0.052  $''$/cy.  Similarly, attempts to detect a
time variation in the gravitational constant $G$ using Mercury's
orbital motion have been unsuccessful, again consistent with
general relativity.  The current result (Pitjeva, 1993) is
$\dot{G}/G = (4.7 \pm 4.7) \times 10^{-12} \hbox{ yr}^{-1}$.  

\subsubsection{Mercury's Perihelion Advance.}

Based on   Mercury's barycentric equations of motion  
one may study the  phenomenon of Mercury's perihelion advance.
The secular trend in Mercury's perihelion\footnote{ 
It should be noted that  the {\it Mercury Orbiter} itself, 
being placed in orbit around Mercury,   will experience the 
phenomenon of periapse advance as well. 
However, we expect that uncertainties in Mercury's gravity field will
mask the 
relativistic precession, at least at the  level of interest for 
ruling out alternative gravitational theories.}
 depends on the linear combination of the  {\small PPN} 
parameters $\gamma$ and $\beta$ and the solar quadrupole 
coefficient $J_{2\odot}$ (Nobili \& Will, 1986; 
Heimberger {\it et al.}, 1990; Will, 1993):
{}
$$\dot{\pi} =  (2+2\gamma-\beta){  \mu_\odot n_M \over  a_M(1-e^2_M)} + 
 {3\over 4}\Big({R_\odot \over a_M}\Big)^2 
{J_{2\odot}n_M\over (1-e^2_M)^2 } (3\cos^2 i_M-1), 
\hskip 10pt ''/\hbox{cy} \eqno(7.17a)$$

\noindent where $a_M, n_M, i_M$ and $e_M$ are the 
mean distance, mean motion, inclination and eccentricity of Mercury's
orbit.
The parameters $\mu_\odot$ and $R_\odot$ are the solar gravitational
constant  and  
radius respectively. 
For   Mercury's orbital parameters  one obtains:
{}
$$\dot{\pi}  = 42 {''\hskip -3pt .}98 \hskip 2pt
\Big[ \hskip 1pt{1\over 3}(2+2\gamma-\beta) +0.296 
\cdot {J_{2\odot}}\times 10^4\Big], \hskip 10pt ''/\hbox{cy} 
\eqno(7.17b)$$
 
\noindent Thus, the accuracy of the relativity tests on the {\it Mercury
Orbiter}
 mission will depend on   our knowledge of the solar gravity field.
The major source of  uncertainty in these measurements is 
the solar quadrupole  moment $ {J_2}_\odot$. 
  As evidenced by the oblateness of the photosphere 
(Brown  {\it et al.}, 1989) and perturbations in frequencies of solar
oscillations, the internal structure of the Sun is slightly
aspherical.  The amount of this asphericity is uncertain. 
It has been suggested that it could be significantly larger than
calculated for a simply rotating star, and that the internal 
rotation rate varies with the
solar cycle (Goode \& Dziembowski, 1991).  Solar oscillation data
suggest that most of the Sun rotates slightly slower than the
surface except possibly for a more rapidly rotating core (Duvall \&
Harvey, 1984).  An independent measurement  of $ {J_2}_\odot$
performed  with the {\it Mercury Orbiter} 
would provide a valuable direct confirmation of the indirect
helioseismology value $(2\pm0.2)\times 10^{-7}$.  
Furthermore, there are suggestions of a
rapidly rotating core, but helioseismology determinations are 
limited by uncertainties at depths below 0.4 solar radii 
(Libbrecht \& Woodard, 1991).  

The {\it Mercury Orbiter} will help us understand  this
asphericity and  independently will enable us to gain some     
important data on the properties of the solar interior and the 
features of it's rotational motion.  Preliminary 
analysis of a {\it Mercury Orbiter} mission suggests 
that $J_{2\odot}$ would  be measurable to at best
 $\sim 10^{-9}$  (Ashby  {\it et al.}, 1995) or about 1\% of the expected  
$J_{2\odot}$ value. This should be compared with the  present $10\%$   
solar oscillation determination (Brown  {\it et al.}, 1989).

\subsubsection{The Redshift Experiment.}

Another important experiment that could be performed on a {\it Mercury
Orbiter} mission is a test of the solar gravitational redshift.
This would require  a stable frequency standard to be  flown on the
spacecraft.
The experiment would provide a fundamental test of the theory of
general relativity and the Equivalence Principle upon which it
and other metric theories of gravity are based (Will, 1993).
Because in general relativity the gravitational redshift of an
oscillator or clock depends upon its proximity to a massive body
(or more precisely the size of the Newtonian potential at its
location), a frequency standard at the location of Mercury would
experience a large, measurable redshift due to the Sun.
With  the result  for the   function
${\cal K}_A$ given eqs. $(7.2a)$ and $(7.15)$ in hand,  one can obtain  the 
corresponding Newtonian proper frequency variation  between the
barycentric 
standard of time and that of the satellite  (the  terms with  the magnitude
up to $10^{-12}$), 
given as:
{} 
$${dx^0\over dy^0_{(0)}}=1+ {\mu_\odot\over R_M} +{\mu_M\over
y_{(0)}}+{1\over2c^2}({\vec v}_M  +
{\vec v}_{(0)} )^2-{\mu_\odot\over R^3_M}({\vec R}_M {\vec y}_{(0)}) 
+{\cal O}(c^{-4}),
 \eqno(7.18a)$$
\noindent where $(y^0_{(0)},{\vec y}_{(0)})$ are the 
four-coordinates of the spacecraft in the Hermean-centric {\small RF}
and ${\vec v}_{(0)}$ is the spacecraft orbital   velocity. 
One can see that  the eccentricity of Mercury's orbit would be highly
effective in
varying the solar potential at the clock, thereby producing a
distinguishing signature in the redshift. The anticipated frequency 
variation between perihelion and aphelion is to first-order in
 eccentricity:   
{}
$$\Big({\delta f\over f_0}\Big)_{e_M}=   
{2\mu_\odot e_M \over  a_M}.\eqno(7.18b)$$

\noindent This contribution  
is quite considerable and is calculated to be 
$(\delta f/f_0)_{e_M}=1.1\times 10^{-8}$. 
Its   magnitude, for instance, 
at a radio-wave length $\lambda_0=3$ cm ($f_0=10$ GHz) 
is $ (\delta f)_{e_M}=  110$ Hz.  We would also benefit
from the short orbital period of Mercury, which would permit the
redshift signature of the Sun to be measured several times over
the duration of the mission. If the spacecraft tracking and modelling 
are of sufficient precision to determine the 
spacecraft position relative to the sun to $100 \hskip 1pt \rm m$ 
(a conservative estimate) then 
a frequency standard with $10^{-15}$ fractional frequency stability 
$\delta f/f=10^{-15}$ would be able to measure the redshift to 1 part in
$10^7$ or better.
This stability is within the capability of proposed spaceborne trapped-ion
(Prestage {\it et al.}, 1992)
or H-maser  clocks (Vessot {\it et al.}, 1980; Walsworth {\it et al.},
1994).

\subsubsection{The SEP Violation Effect.}

Besides the   Nordtvedt   effect 
(for more details see Anderson, Turyshev {\it et al.} (1996)), there  exists
an 
interesting possibility for testing the {\small SEP} violation effect  
by studying spacecraft   motion in orbit around Mercury. The corresponding 
equation of motion is given by eq.(7.14).   
As one can see, the  two terms in the second line of this equation   vanish
 for   general relativity, but for scalar-tensor theories,
they   become responsible for  
small deviations of the spacecraft  motion from the support  geodesic.
Both of these effects, if 
they exist,  are due to non-linear coupling of the gravitational field of
Mercury to 
external gravity. They come from the    expression for  $W_A$ given by
eq.$(7.5a)$,
which is the {\it local} post-Newtonian contribution to the $g_{00}$
component  of 
the metric tensor in the proper {\small RF}.

The   first of these terms may be interpreted as a
dependence of the locally measured   gravitational constant 
on the  external gravitational environment and may be expressed in the 
vicinity of body {\small (A)} as follows:
{}  
$$G_A=G_0\Big[1-(4\beta-3\gamma-1)\sum_{B\not=A}{m_B\over
y_{BA_0}} \Big]. \eqno(7.19)$$

\noindent  In the case of a satellite around Mercury, the main contribution
to this effect  
 comes from the Sun\footnote{Note  that this  combination of  {\small
PPN}  parameters  differs 
from the one presented for a similar effect in (Will, 1993). The reason for
this  is that, 
in this case the   transformations  in the form of eqs.(7.2) 
 let us  define transformation rules of the   metric tensor between
the barycentric  and a proper  planeto-centric  {\small RF}s and, hence, to 
obtain the correct and complete equations of geodesic motion in the
quasi-inertial 
Hermean-centric {\small RF}$_M$.}. Because of the high  eccentricity of
Mercury's orbit,
the periodic changing of the sun's {\it local} gravitational potential may
produce an observable effect,
which can be modeled by a  
periodic time variation in the effective {\it local} gravitational constant:
{}
$$ \Big[{\dot{G}\over  G}\Big]_{\rm period}=(4\beta-3\gamma-1)
\Big[{\mu_\odot\over a_M(1-e_M^2)}\Big]^{3\over2}
{c e_M\sin\phi(t)\over a_M(1-
e_M^2)}\big(1+e_M\cos\phi(t)\big)^2,\eqno(7.20)$$
\noindent which gives the following estimate for this effect  on 
Mercury's orbit:
{}
$$\Big[{\dot{G}\over  G}\Big]_{\rm period}\approx(4\beta-3\gamma-1) 
\times 1.52\times 10^{-7}\sin\phi(t) 
\qquad   {\rm yr^{-1}}.\eqno(7.21)$$

Note that this  effect eq.(7.21)    is fundamentally
different from that introduced by Dirac's  hypothesis of possible  
time dependence  of the gravitational constant (Pitjeva, 1993).
As one can see from expression (7.21), the characteristic time   
in this case is  Mercury's siderial period.
This short period may be considered as an advantage from the
experimental point of view.
In addition,   the results of the redshift experiment could help in 
confident studies of this effect. Recently a different combination 
of the post-Newtonian parameters 
entering in Nordtvedt effect, $\eta=4\beta-\gamma-3$, was  measured at 
$\eta\le 10^{-4}$
(Dickey  {\it et al.}, 1994). This means  that, in order to 
obtain the  comparable accuracy for the combination of parameters
eq.(7.21), 
one should perform the   Mercury  gravimetric measurements  
 on the level no less precise than $\big[{\dot{G}/ G}\big]_{\rm period} 
\approx 10^{-11} \hskip 3pt  {\rm yr}^{-1}$. 
Recently a   group at the University of Colorado has analyzed 
a  number of gravitational experiments possible  with future 
  Mercury   missions (Ashby {\it et al.}, 1995).
Using a modified worst case error analysis, this group suggestes that
after one year  
of ranging between Earth and Mercury (and assuming a 6 cm rms error), the 
fractional accuracy of determination of the sun's gravitational constant
$m_\odot G$ is
expected to be of order $\sim 2.1 \times 10^{-11}$. 
Moreover, even higher accuracy could be achieved  with a Mercuty  lander
as proposed by  
Ashby {\it et al.} (1995). This suggests that the experiment 
for determination of the effect eq.(7.21) may be feasible with 
the {\it Mercury Orbiter} mission. 
      
Another interesting effect on the satellite's orbit may be derived 
from the eq.(7.14) in the form of the following acceleration term: 
{}
$$\delta {\vec a}_{(0)SEP}= 2(\beta-1){m_Mm_\odot\over yR_M^2} 
\big({\vec N_M}- {\vec n} ({\vec n}\cdot{\vec N_M})\big), \eqno(7.22)$$
\noindent where  $R_M$ is  Mercury's heliocentric radius-vector
 and $\vec{N}_M$ is the unit vector along this direction. 
This effect is very small for the  orbit proposed for ESA's {\it Mercury
Orbiter} mission.
However, one can show that there exist two resonant orbits for a satellite
around Mercury, 
either with the orbital frequency $\omega_{(0)}$ equal to    
Mercury's siderial frequency $\omega_M$: 
$\omega_{(0)}\approx\omega_M$ or at one third of
this frequency $\omega_{(0)}\approx \omega_M/3$.
For these resonant orbits, the corresponding experiment 
could provide  an independent direct test of the  parameter $\beta$.

\subsubsection{The Precession Phenomena.}

In addition to the perihelion advance, while constructing the 
Hermean proper {\small   RF}$_M$, one should take into account  several 
precession phenomena included in the transformation function ${\cal
Q}^\alpha_A$
and associated with the angular momentum of the bodies. 
As one may see directly from  eqs.$(7.2b)$ and $(7.16a)$,  besides the
obvious  
special relativistic contributions, the    post-Newtonian 
transformation of the spatial coordinates 
contains   terms due to the non-perturbative  influence of the
gravitational 
field. This non-Lorentzian behavior of 
the post-Newtonian transformations was discussed first by
Chandrasekhar \& Contopulos (1967)
for the case of post-Galilean transformations. Our derivations differ from
the latter by 
taking into account the acceleration of the proper {\small RF} and by
including 
the infinitesimal precession of the coordinate axes with the angular
velocity  
tensor $\Omega^{\alpha\beta}_M$ given as follows: 

$$\Omega_M^{\alpha\beta}=\sum_{B\not=M}
\Big[(\gamma+{1\over2}){\mu_B\over
y^2_{BM_0}}N^{[\alpha}_{BM_0}v^{\beta]}_{M_0}-
(\gamma+1){\mu_B\over y^2_{BM_0}}N^{[\alpha}_{BM_0}v^{\beta]}_{B_0}+
$$
{}
$$+(\gamma+1){\mu_B\over 2y^3_{BM_0}}{\cal P}^{[\alpha}_{\lambda} 
(S^{\beta]\lambda}_M+S^{\beta]\lambda}_B)\Big],\eqno(7.23) $$

\noindent where, as before, the  subscript {\small (M)} 
stays for the Mercury and summation is performed 
over the bodies of the solar system.
This expression re-derives and generalizes the result for the precession
of  
the spin of a gyroscope  $\vec{s_0}$ attached to a test body orbiting 
a gravitating primary. Previously this result was obtained from the 
theory of  
Fermi-Walker transport (Will, 1993).  Indeed, in accord  with  
eq.$(7.16a)$, 
this spin (or coordinate axes of a proper Hermean {\small RF}$_M$)
will precess  with respect to a distant standard of 
rest such as quasars or distant galaxies. The motion of the 
spin vector of a gyroscope can be described by the relation: 
{}
$${d  \vec{s_0}\over dt}=[ \vec{\Omega}_M \times 
\vec{s_0}].\eqno(7.24)$$

\noindent By keeping   the leading contributions only and neglecting the
influence of the 
Mercury's intrinsic spin moment we obtain from the expression (7.23)  
 the angular velocity $\vec{\Omega}_M$ in the  following form:
{}
$${\vec \Omega}_M= (\gamma+{1\over2}){\mu_\odot\over R_M^3}[{\vec
R}_M\times {\vec v}_{M_0}]-
(\gamma+1){\mu_\odot\over R^3_M}[{\vec R}_M\times {\vec v}_\odot]+ $$
{}
$$+(\gamma+1){\mu_\odot\over 2R^3_M}\Big(\vec{S_\odot}-
{3(\vec{S_\odot}\vec{N}_M)\vec{N}_M}\Big), \eqno(7.25)$$

\noindent where $\vec{v}_{M_0}$  and $\vec{v}_\odot$  are  Mercury's and
the Sun's 
barycentric orbital velocities and 
${\vec S}_\odot$ is the solar intrinsic spin moment. 

The first term in eq.(7.25)   is known as  geodetic precession (De-Sitter,
1916). 
This term arises in any non-homogeneous gravitational field  because of
the 
parallel transport of a direction defined by $\vec{s_0}$ in (7.24). It can be 
viewed as spin precession caused by a  coupling 
between the particle velocity $\vec{v}_{M_0}$ and the static  
part of the space-time geometry.
For  Mercury orbiting the Sun this precession has the form:
{}  
$$\vec{\Omega}_G= (\gamma +{1\over 2})
{\mu_\odot\over R^3_M}(  \vec{R}_M\times \vec{ v}_{M_0}). \eqno(7.26)$$
\noindent 
This effect could be studied for the {\it Mercury Orbiter}, which, being 
placed in orbit around  Mercury is in effect a gyroscope orbiting the Sun.
Thus, if we    introduce the angular momentum per unit mass, 
$ \vec{ L}= \vec{R}_M\times  \vec{v_{M_0}}$, 
of   Mercury in solar orbit, the equation (7.26) shows
that $ \vec{ \Omega}_G$ is directed along the pole of the ecliptic, 
in the direction of $ \vec{ L}$. The vector $ \vec{\Omega}_G$ has a
constant part
{}
$$ \vec{\Omega}_0={1\over 2}(1+2\gamma) 
 {\mu_\odot\omega_M\over a_M}=  {1+2\gamma\over 3}\cdot 0.205  
 \hskip 10pt  { ''/\hbox{yr}}, \eqno(7.27a) $$

\noindent with a significant correction   due to the eccentricity $e_M$
of the Mercury's orbit,
{}
$$ \vec{\Omega}_1 \cos \omega_M t=
 {3\over 2}(1+2\gamma) {\mu_\odot\omega_M\over a_M} e_M\cos
\omega_M t_0= 
{1+2\gamma\over 3}\cdot0.126 \cos \omega_M t_0
 \hskip 10pt  {''/\hbox{yr}}, \eqno(7.27b) $$

\noindent where $\omega_M$ is   Mercury's siderial frequency,   
$t_0$ is reckoned from a perihelion passage; 
$a_M$ is the semimajor axis of  Mercury's orbit. 

Geodetic precession has been studied for the motion of lunar perigee 
and its existence   was first confirmed with
an accuracy of 10\% (Bertotti {\it et al.}, 1987).
Two other groups have analyzed the lunar laser-ranging data more
completely to estimate the deviation of the lunar orbit from the  
predictions of general relativity  
 (Shapiro {\it et al.}, 1988; Dickey {\it et al.}, 1989). 
Geodetic precession has been confirmed  within a standard deviation of
2\%.  
The precession of the orbital plane proposed for ESA's {\it Mercury 
orbiter} (periherm at 400 km altitude, apherm at 16,800 km, period 13.45
hr and latitude
of  periherm at +30 deg) would  include a contribution of 
order 0.205 ${''/\hbox{yr}}$  from the geodetic precession.
We recommend this
precession   be included in future studies of the 
{\it Mercury Orbiter} mission.

The third term in  expression (43) is known as 
Lense-Thirring precession $\vec{\Omega}_{LT}$. 
This term gives the 
relativistic precession of the gyroscope's spin 
$\vec{s}_0$ caused by the intrinsic angular momentum 
$\vec{S}$ of the central body.
This effect is responsible for a small perturbation in the  orbits 
of artificial satellites around the Earth (Tapley  {\it et al.}, 
1972;  Ries  {\it et al.}, 1991).
However,  our preliminary studies indicate that 
this effect is so small for the satellite's orbit around Mercury 
that will be masked by uncertainties in the orbit's inclination.

\section{Discussion: Relativistic Astronomical RFs.}
\vskip 11pt\noindent
In this Section   we will discuss some questions of the 
practical application of the  results presented in the report. Let us
mention that presently 
the radio sources seems to be able to provide one with 
a more stable and precise reference measurements needed for a reliable 
navigation in the outer space.  
This makes it   reasonable to  construct the future astronomical  {\small
RF}s   based 
upon the radio source catalogues which are expected to be an essential
part
 of the future relativistic navigation  in the solar system and beyond
(Standish, 1995).
Moreover, as we know the accuracy of the {\small VLBI} timing
measurements has  improved rapidly over
the last few years and is presently a few tens of a picosecond (ps). It is
important in 
precise measurements such as these that inadequate modelling not
contribute to the 
inaccuracy of the results. 
We believe that the results obtained in this report 
are ready to be   used directly in application to this and many other
problems 
of the relativistic observations in the solar system.

The {\small KLQ} parameterized theory of the astronomical {\small RF}s 
discussed in this report  enable one to perform  the necessary 
calculations in the 
most arbitrary form valid for many theories of gravity. 
The different physical aspects of choosing a well defined 
{\it local} {\small RF} in a curved space-time   
has been discussed in many publications. In  summary, in  modern 
astronomical practice there are two 
physically different types of relativistic  
{\small RF}s  which  are extensively in use, namely:

\begin{itemize}
\item[{\small I}.] The set of inertial {\small RF}s, which includes:
\begin{enumerate}
\item[(i).] The asymptotic inertial    {\small RF}.
\item[(ii).] The barycentric inertial   {\small RF}.
\end{enumerate}
\item[{\small II}.] The set of observer's quasi-inertial proper {\small
RF}s, which consist of:
\begin{enumerate}
\item[(i).]  The bodycentric  {\small RF}, constructed for a particular 
extended body in the system.  
\item[(ii).] The satellite {\small RF}, 
defined on the geodetic world-line of a test
particle orbiting the   body  under consideration. 
\item[(iii).] The topocentric   {\small RF}, 
which is defined on the surface of the body  under study. \end{enumerate}
\end{itemize}

\noindent  The main difference between these  two classes of the  {\small
RF}s   is that, unlike   the 
frames of the first type, which are inertial, the observer's frame  is, in
general,   non-inertial. 
Such  an hierarchy of frames in the {\small WFSMA}, if needed, may be
extended 
to a larger scale of motion.  
The barycentric {\small RF}$_0$ could also be used (with some
cosmological assumptions) 
as an analog of the rest frame of the universe for the description of the
galactic and  extra-galactic 
motion. One may find a more detailed discussion of this hierarchy of the   
{\small RF}s  in applications to  
  problems of  the modern astronomical practice in (Brumberg, 1991;
Voinov, 1994; Folkner {\it et al.}, 1994).
Theoretically, the {\small RF} and the set of coordinates selected may be
arbitrary. 
The relativistic terms in the equations of motion, the light time
equations, and 
the transformation from coordinate  time to physically measurable time 
will
vary with the  {\small RF} and coordinates selected. In general, the
numerical 
values of various constants, obtained  by fitting the theory to
observations, will also vary.
However, the numerical values of the computed observable are independent
of the {\small RF}
and the {\small CS} selected (Moyer, 1971).

While the properties of an inertial {\small RF}s from the first set of the
frames listed above are 
well understood and widely accepted in many areas of modern
astronomical practice, below we shall
concentrate our attention on the properties of the   relativistic {\small
RF}s from the  second 
set, namely we will be interested in construction of the geocentric, the
satellite 
and the topocentric frames. The logic of construction of these 
frames is quite simple: Due to the fact that the geocentric frame was 
previously well
justified physically and  explicitly constructed from the mathematical
stand point,
the construction of the two remaining frames will be made based on 
these established properties of the geocentric {\small RF}.
Indeed, the proper  {\small RF}$_A$ of an extended body {\small (A)}  
 contains all the information about the proper gravitational 
field of the body {\small (A)} as well as the explicit information about the 
external gravity.
Then, we will give the definition of the satellite and the topocentric
{\small RF}s 
considering that the properties of the geocentric {\small RF} are already
known.
Moreover, we will present the results   generalized  on the case of the 
scalar-tensor theories of gravity
and  will include in the analysis the two Eddington parameters $(\gamma,
\beta)$.

\subsection{The Geocentric Proper RF.}

The properties of construction the geocentric {\small RF} 
were discussed in Section {\small 6} of the present report 
and below we will present the final results only. 
Thus, the  form of the coordinate transformations 
between    the coordinates $(x^p)$ of the barycentric inertial  {\small
RF}$_0$   and those  
$(y^p_A)$ of a proper  quasi-inertial   {\small RF}$_A$ of an arbitrary
body {\small (A)}  
for the problem of motion of the {\small N}-extended-body system in the
{\small WFSMA} 
was obtained   as follows:  
{}
$$x^0 = y^0_A+ c^{-2}  K_A(y^0_A, y^\epsilon_A) + 
c^{-4}   L_A(y^0_A, y^\epsilon_A) + {\cal O}(c^{-6}),
\eqno(8.1a)  $$
$$x^\alpha =  y^\alpha_A + y^\alpha_{A_0} (y^0_A) + 
c^{-2} Q^\alpha_A(y^0_A,y^\epsilon_A) + {\cal O}(c^{-4}), 
\eqno(8.1b)$$
\noindent We will present the results corresponding to 
the coordinate transformations to the {\small RF}, which   has 
all ten   parameters $\zeta_A, \sigma^\alpha_A$ and $f^{\alpha\beta}_A$ 
of the constructed group of motion 
vanish, which is  given by the eq.(5.44) as:
{}
$$\zeta_A = c^{-2}\zeta^A_1 +
 c^{-4}\zeta^A_2=\sigma^\alpha_A=f^{\alpha\beta}_A=0.$$ 
\noindent Moreover, we shall be interested in  such a {\small RF}s
which preserves  all ten  existing conservation 
laws of the {\it local} gravity, inertia and matter, so   
that we  require that the conditions eqs.(6.37) hold, namely:
{}
$${Q^\alpha_A}_{\{L\}}(y^0_A)=
{L_A}_{\{L\}}(y^0_A)=0,  \qquad \forall l\ge 3.$$
\noindent With these conditions, the transformation  
functions $K_A, Q^\alpha_A$ and $L_A$, take the following form: 
{}
$$K_A(y^0_A,y^\nu_A)  =\int^{y^0_A} \hskip -10pt  dt'  
\Big[ \sum_{B\not=A} \Big< U_B\Big>_A- 
{1\over2} v_{A_0}{_{{}_\nu}} v_{A_0}^\nu\Big] 
- v_{A_0}{_{{}_\nu}}\cdot y^\nu_A +  {\cal O}(c^{-4})y^0_A, \eqno(8.2a)$$
{}
$$Q^\alpha_A(y^0_A,y^\nu_A) = \gamma\sum_{B\not=A} \Big[ 
 \Big({1\over2} \eta^{\alpha\lambda} y_{A_\mu}  y^\mu_A  
-y^\alpha_A y^\lambda_A\Big)\Big<\partial_\lambda U_B\Big>_A  
- y^\alpha_A  \Big<U_B\Big>_A \Big]- 
{1\over2}v^\alpha_{A_0} v^\mu_{A_0} y_{A_\mu}+$$
{}
$$+{y_A}_\beta  \int^{y^0_A} \hskip -10pt dt' 
\Big( {1\over2}a^{[\alpha}_{A_0}  v^{\beta]}_{A_0}  + 
(\gamma+1) \sum_{B\not=A}\Big[
\Big<\partial^{[\alpha} V^{\beta]}_B\Big>_A   
+\Big<\partial^{[\alpha} U_B v^{\beta]}\Big>_A \Big]\Big) +$$
{}
$$+ w^\alpha_{A_0}(y^0_A) + 
{\cal O} (c^{-4}) y^\alpha_A,   \eqno(8.2b)$$ 
{}
$$L_A(y^0_A,y^\nu_A) =  \sum_{B\not=A} \Bigg( \gamma{1\over2} \hskip
2pt
y_{A\beta} y^\beta_A \cdot {\partial \over\partial
y^0_A}\Big<U_B\Big>_A   - 
(\gamma+1)\hskip 2pt y^\lambda_A  y^\beta_A \cdot 
\Big[\Big< \partial_\lambda {V_B}_\beta \Big>_A   
+\Big<v_\beta \partial_\lambda U_B \Big>_A \Big] +$$
{}
$$+  \gamma{v_{A_0}}_\beta \Big(y^\beta_A  y^\lambda_A 
 - {1\over2} \eta^{\beta\lambda} y_{A_\mu} y^\mu_A\Big) \cdot 
\Big< \partial_\lambda  U_B \Big>_A  \Bigg) +$$
{}
$$+ y_{A_\lambda}{v_{A_0}}_\beta \int^{y^0_A} \hskip -10pt dt'
 \Big( {1\over2} a^{[\lambda}_{A_0} v_{A_0}^{\beta]}+ 
(\gamma+1)\hskip 2pt\sum_{B\not=A}
\Big[\Big<\partial^{[\lambda} V^{\beta]}_B  \Big>_A   
+\Big<v^{[\beta}\partial^{\lambda]} U_B \Big>_A\Big]\Big)+$$
{}
$$ + y_{A_\beta} \Big[ (\gamma+1)\hskip 2pt
v^\beta_{A_0}\sum_{B\not=A} 
\Big<U_B\Big>_A  
 -  2(\gamma+1)\hskip 2pt\sum_{B\not=A} \Big<V^\beta_B\Big>_A  
 - \dot w^\beta_{A_0}(y^0_A)  \Big] - $$
{}
$$ - \int^{y^0_A} \hskip -10pt dt' \Big[ \sum_{B\not=A}
\Big<W_B\Big>_A  + {1\over2}\Big(\sum_{B\not=A} \Big< U_B\Big>_A  - 
 {1\over2}{v_{A_0}}_\beta v^\beta_{A_0}\Big)^2 
  + {v_{A_0}}_\mu\dot w^\mu_{A_0}(t')\Big]+
{\cal O}(c^{-6})y^0_A.\eqno(8.2c)$$ 
\noindent The  equations for the functions $a^\alpha_{A_0}$ 
and $\ddot{w}_{A_0}$ was 
given previously by the equations (7.3) and  (6.52) respectively.

The  transformations eqs.(8.2) produce the metric tensor $g^A_{mn}$ of
 the geocentric {\small RF} with the following  components:
{}
$$g^A_{00}(y^p_A) =1-2{\cal U}+ 2{\cal W} +{\cal O}(c^{-6}), \eqno(8.3a) $$ 
{}
$$g^A_{0\alpha}(y^p_A) = 4\hskip2pt \eta_{\alpha\epsilon}{\cal
V}^\epsilon
+  {\cal O}(c^{-5}), \eqno(8.3b)$$
{}
$$g^A_{\alpha\beta}(y^p_A) = 
\eta_{\alpha\beta}\Big(1 + 2 \gamma {\cal U}\Big)  + 
{\cal O}(c^{-4}), \eqno(8.3c)$$

\noindent where for the brevity of the future discussion we introduced 
the following notations
for the generalized  gravitational potentials in this {\it local} frame:
{}
$${\cal U}(y^p_A)\equiv\overline{U}(y^p)= \sum_B U_B(y^q_B(y^p_A))-
\sum_{B\not=A}
\Big[y^\beta_A\Big<{\partial U_B\over \partial y^\beta_A}\Big>_A +
\big<U_B\big>_A \Big], \eqno(8.4a) $$ 
{}
$${\cal W}(y^p_A)=\sum_B W_B(y^q_B(y^p_A))-\sum_{B\not=A}
\Big[y^\mu_A \Big<{\partial W_B\over \partial y^\mu_A}\Big>_A +
\big<W_B\big>_A\Big] +$$
{}
$$+ {1\over2}\hskip 2pt  y^\mu_A y^\beta_A \hskip 1pt \Bigg(\gamma
\eta_{\mu\beta} 
\hskip 1mm {a_{A_0}}_\lambda a^\lambda_{A_0} - 
(2\gamma-1)\hskip 2pt{a_{A_0}}_\mu {a_{A_0}}_\beta +  $$
{}
$$+ \sum_{B\not=A}
{\partial\over\partial y^0_A}\Big(\gamma\eta_{\mu\beta} 
 {\partial  \over\partial y^0_A}\Big<U_B\Big>_A -
 (\gamma+1)\Big[\Big<\partial_{(\mu} {V_B}_{\beta)}\Big>_A   
+\Big<v_{(\beta}\partial_{\mu)} U_B\Big>_A
\Big]\Big)\Bigg),\eqno(8.4b)$$
{}
$${\cal V}^\alpha(y^p_A)={1\over 2}(\gamma+1)
\Bigg(\sum_B V^\alpha_B(y^q_B(y^p_A))-$$
{}
$$- \sum_{B\not=A}
\Big[y^\mu_A \Big(\Big<{\partial V^{\alpha}_B\over \partial
y^\mu_A}\Big>_A+ 
\Big<v^\alpha{\partial U_B\over \partial y^\mu_A}\Big>_A\Big)+
\big<V^\alpha_B\big>_A\Big]\Bigg)+\gamma{1\over4}(y^\alpha_Ay^\lambda_A-
 {1\over2}\eta^{\alpha\lambda}{y_A}_\mu y^\mu_A\Big)
\dot{a}_{A_0}{_{{}_\lambda}}, \eqno(8.4c)$$
\noindent where the both   functions $W_A$ and $W_B$ are   given by the
expressions eqs.(7.5).
 
The presented  expressions for the geocentric proper {\small RF} are
taking into account 
the proper gravitational field of the body {\small (A)}, the   {\it external}
gravity, and the dynamical properties of the inertial sector of the {\it
local} space-time. 
This presentation of the {\it local} metric eqs.(8.3), will enable  us to
simplify the 
discussion of the results obtained for the two 
other important quasi-inertial frames which are  widely in use for many
practical 
applications of the modern astronomy - the satellite and the topocentric
ones. 

\subsection{The Satellite Proper RF.}

The motion of an artificial satellite may be presented as the motion of a
test particle,
which is moving along the geodetic world-line in the effective space-time
with the  
metric tensor given by the eqs.(8.3). 
This means that,  in order to define the coordinate transformations 
and the metric tensor of a satellite {\small RF}$_{(0)}$, we can apply the
conditions 
eqs.(3.26) or those of eqs.(5.2).
By performing the same calculations as in the Section {\small 5} for the
test 
particle, we can obtain the   post-Newtonian  
dynamically non-rotating coordinate transformations linking together   
the coordinates $(y^p_A)$ of the geocentric 
quasi-inertial   {\small RF}$_A$  and those  $(z^p)$
of the proper quasi-inertial {\small RF}$_{(0)}$. 
These  transformations may be obtained in the familiar form:
{}
$$y^0_A = z^0 + c^{-2}K_{(0)}(z^0, z^\nu) + 
c^{-4}  L_{(0)}(z^0, z^\epsilon) + {\cal O}(c^{-6})z^0, \eqno(8.5a)  $$
$$y^\alpha_A =  z^\alpha  + z^\alpha_{(0)}(z^0) + 
c^{-2}  Q^\alpha_{(0)}(z^0, z^\nu) + {\cal O}(c^{-4})z^\alpha.
\eqno(8.5b)$$
The solutions for the   transformation functions $K_{(0)}, Q^\alpha_{(0)}$
and $L_{(0)}$ 
was chosen with the same conditions as those for the functions eqs.(8.2),
namely:
the corresponding group parameters $\zeta_{(0)}, \sigma^\alpha_{(0)}$
and $f^{\alpha\beta}_{(0)}$ 
are taken to be zero and the requirement of 
preserving  all the conservation laws in the satellite's 
{\it local} vicinity is fulfilled. 
The resultant functions was obtained  as follows:
{}
$$K_{(0)}(z^0,z^\nu)  =\int^{z^0} \hskip -10pt  dt'  
\Big[   \Big<{\cal U}\Big>_{(0)'}-  
{1\over2}{v_{(0)}}_\nu v^\nu_{(0)}\Big]  
-{v_{(0)}}_\nu\cdot z^\nu +  {\cal O}(c^{-4})z^0, \eqno(8.6a)$$
{}
$$  Q^\alpha_{(0)}(z^0,z^\nu) = 
\gamma  \Big({1\over2} \eta^{\alpha\beta}z_\mu  z^\mu - z^\alpha 
 z^\beta \Big)   \Big<{\partial {\cal U}\over \partial z^\beta}\Big>_{(0)}- 
\gamma  z^\alpha  \Big<{\cal U}\Big>_{(0)} - 
{1\over2} v^\alpha_{(0)} v^\beta_{(0)}   z_\beta  +$$
{}
$$+z_\beta \int^{z^0} \hskip -10pt dt' \Big[ {1\over2}   a^{[\alpha}_{(0)}
 v_{(0)}^{\beta]}   + 
2   \Big< \partial^{[\alpha} {\cal V}^{\beta]}  \Big>_{(0)'}\Big]  
 +w^\alpha_{(0)}(z^0)  + {\cal O} (c^{-4}) z^\alpha,   \eqno(8.6b)$$ 
{}
$$  L_{(0)}(z^0,z^\nu) =   \gamma {1\over2} z_\beta 
z^\beta{\partial  \over \partial z^0} \Big<{\cal U}\Big>_{(0)} - 
2 z^\lambda   z^\beta    \Big<{\partial_\lambda} 
{\cal V}_\beta\Big>_{(0)} 
+\gamma {v_{(0)}}_\beta \Big(z^\beta   z^\lambda    
- {1\over2}\eta^{\beta\lambda} z_\mu z^\mu\Big) 
 \Big<{\partial {\cal U}\over \partial z^\lambda} \Big>_{(0)}   +$$
{}
$$+{v_{(0)}}_\beta z_\lambda  \int^{z^0} \hskip -10pt dt'
 \Big[ {1\over2} a^{[\lambda}_{(0)}  v_{(0)}^{\beta]} + 
2 \Big<\partial^{[\lambda}  {\cal V}^{\beta]}\Big>_{(0)'} \Big]   
 + z_\beta  \Big( (\gamma+1)  v^\beta_{(0)}  \Big<{\cal U}\Big>_{(0)}
- 4\Big<{\cal V}^{\beta}\Big>_{(0)} - \dot {w}^\beta_{(0)}\Big) - $$
{}
$$-\int^{z^0} \hskip -10pt dt' \Big[\Big<{\cal R}\Big>_{(0)'} +
{1\over2}\Big(  \Big<{\cal U}\Big>_{(0)'} - {1\over2}{v_{(0)}}_\beta 
v^\beta_{(0)}\Big)^2 + 
v_{(0)}{_{{}_\mu}}\dot{w}^\mu_{(0)}\Big]   +{\cal O}(c^{-
6})z^0,\eqno(8.6c)$$ 

\noindent where the quantities $v_{(0)}^\alpha$ and $a_{(0)}^\alpha$ are
the 
geocentric velocity and   acceleration of the spacecraft respectively. 
The notation $\Big< f\Big>_{(0)}$, analogously to that of eq.(5.7), 
denotes the limiting procedure of taking the value of the function $f(z^p)$
on the geodetic 
world-line of an artificial satellite, where   $z^\alpha\rightarrow 0$. 
The equations for both time-dependent functions 
$z^\alpha_{(0)}$ and $w^\alpha_{(0)}$ may be determined  
similarly to those presented in the Section {\small 5}.
Thus, the equation eq.(5.4) provides us with the usual relation for the 
Newtonian acceleration   $a^\alpha_{(0)}$ 
of the center of inertia of a test body:
{}
$$a^\alpha_{(0)}(z^0)=-\eta^{\alpha\mu} 
\Big<{\partial  {\cal U}\over \partial z^\mu}\Big>_{(0)}+
{\cal O}(c^{-4}).\eqno(8.7)$$   
\noindent Analogously, the function $w^\alpha_{(0)}$ is determined 
as the  solution of the   equation below:
{}
$$\ddot{w}^\alpha_{(0)}(z^0) =  
 \eta^{\alpha\mu}\Big< {\partial {\cal R}\over \partial z^\mu}\Big>_{(0)}
+ v^\alpha_{(0)}{\partial\over \partial z^0} \Big<{\cal U}\Big>_{(0)}+ 
a^\alpha_{(0)}  \Big<{\cal U} \Big>_{(0)}- 
4   {\partial  \over \partial z^0} \Big<{\cal V}^{ \alpha}\Big>_{(0)}  - $$
{}
$$-{1\over2}v^\alpha_{(0)}{v_{(0)}}_\beta  a^\beta_{(0)}   + 
{a_{(0)}}_\lambda  \int^{z^0} \hskip -10pt dt' 
\Big[ {1\over2} a^{[\alpha}_{(0)}  v^{\lambda]}_{(0)} 
  +2   \Big<\partial^{[\alpha}{\cal V}^{\lambda]}\Big>_{{(0)}'} \Big]
 + {\cal O}(c^{-6}). \eqno(8.8a)$$
 
\noindent where function ${\cal R}\sim {\cal O}(c^{-4})$ is defined 
in the same way as the functions
$W_A$ and $W_B$ eqs.(5.43) from the result of the fields decomposition 
in the 
{\it local} quasi-inertial {\small RF}$_{(0)}$ of a satellite. By repeating
this decomposition 
as it was presented in the  Section {\small 4}, one may obtain  this 
function as follows:
{}
$${\cal R}(z^p) = {\cal W}(z^p)
+2 a^\lambda_{(0)}   \sum_B {\partial \over \partial 
z^\lambda} \chi_B (z^0, z^\nu) + 
{\cal O}(c^{-6}),\eqno(8.8b)$$ 

At this point we may present the   form 
of the metric tensor in the proper {\small RF}$_{(0)}$ of an artificial
satellite 
defined on the geodetic world-line with 
the generalized Fermi conditions (3.26). Thus, by substituting 
the solutions obtained for the functions  $K_{(0)}, Q^\alpha_{(0)}$ and
$L_{(0)}$ 
 into general form of the metric tensor
$g_{mn}(z^p)$  in the expressions for the metric in a proper {\small
RF}$_{(0)}$  
given by the relations eqs.(4.11), 
we will obtain this tensor in the following form: 
{}
$$g^{(0)}_{00}(z^p) = \hskip 5pt 1  
 -2 {\overline{\cal U}}_{(0)} + 2{\overline{\cal R}}_{(0)} 
 +{\cal O}(c^{-6}), \eqno(8.9a) $$ 
{}
$$g^{(0)}_{0\alpha}(z^p)  = 4 \eta_{\alpha\epsilon}
{\overline{\cal V}}^\epsilon_{(0)}  +  {\cal O}(c^{-5}),
 \eqno(8.9b)$$
{}
$$g^{(0)}_{\alpha\beta}(z^p)  =
\eta_{\alpha\beta}\Big(1 + 2 \gamma  {\overline{\cal U}}_{(0)}\Big)+
 {\cal O}(c^{-4}). \eqno(8.9c)$$
 
\noindent The expressions  (8.9) are the general solution for 
the field equations of the general theory of relativity, 
which satisfies the generalized Fermi conditions 
 in the immediate vicinity of a dimensionless test body.
By the  definition, the proper gravity of the test body is negligibly 
small, then the effective Newtonian potential in the vicinity of the
satellite 
may be presented as follows:
{}
$${\overline{\cal U}}_{(0)}(z^p)= {\cal U}(z^0, z^\nu)-  \Big[
z^\mu \Big<{\partial {\cal U}\over \partial z^\mu}\Big>_{(0)}   + 
\Big<{\cal U}\Big>_{(0)} \Big].\eqno(8.10)$$
\noindent In addition, the  functions ${\overline{\cal R}}_{(0)}$ and 
${\overline{\cal V}}^\alpha_{(0)}$ 
were obtained in the following form:
{} 
$${\overline{\cal R}}_{(0)}(z^p)= {\cal R}(z^0, z^\nu) -   
\Big[z^\mu \Big<{\partial {\cal R}\over \partial z^\mu}\Big>_{(0)}   
+ \Big<{\cal R}\Big>_{(0)} \Big]  + {1\over2}\hskip 2pt z^\mu z^\beta 
 \hskip 1pt  \Big[ \gamma\eta_{\mu\beta} 
\hskip 1mm {a_{(0)}}_\lambda a^\lambda_{(0)}  - $$
{}
$$- (2\gamma-1){a_{(0)}}_\mu {a_{(0)}}_\beta +  
{\partial \over \partial z^0}\Big( \gamma\eta_{\mu\beta}{\partial  
\over\partial {z^0}}\Big<{\cal U}\Big>_{(0)}   - 
2 \Big<\partial_{(\mu}  {\cal V}_{\beta)}\Big>_{(0)}\Big)\Big],  
\eqno(8.11)$$
\noindent and 
{}
$${\overline{\cal V}}^\alpha_{(0)}(z^p)=  {\cal V}^\alpha(z^0, z^\nu) - 
 \Big[z^\mu \Big<{\partial {\cal V}^{\alpha}\over \partial
z^\mu}\Big>_{(0)}   + 
\Big<{\cal V}^{\alpha}\Big>_{(0)} \Big] +\gamma{1\over4} \Big(z^\alpha
z_\beta   - 
{1\over2} \delta^\alpha_\beta  z_\mu z^\mu\Big){\dot a}^\beta_{(0)}.
\eqno(8.12)$$

\subsection{The Topocentric Proper RF.}

The construction of the topocentric {\small RF} required a little bit 
more sophisticated analysis. Thus, we have to specify where this 
frame is located on the surface of the extended body, and 
what point will be considered as the  origin of the coordinates. 
In order to find the dynamical conditions 
necessary for construction the transformation functions (analogous to
those 
given by the  eqs.(3.26)-(3.29)), one   should  made  an explicit 
relativistic analysis of the constrained motion of the 
tracking station placed on the earth's surface. 
This analysis should provide one with the detailed description of the 
problem of static equilibrium of a test particle on the surface
of an extended body which interior characterizes by the energy-momentum
tensor 
$T^{mn}$ and the corresponding equation of state $p(\rho)$. 
Likely, the present accuracy of the 
topocentric radio-metric measurements  do not requires this level of 
generality. This permits us to neglect the  geometry of the 
tracking station, its weight,  and instead, to consider the law of the 
relativistic motion of an atomic time standard only. 
Then the answer to the second part of the above  
question is simple: the origin  of the coordinates of the 
topocentric {\small RF} coincides with the atomic time standard 
 which is used as the physically measurable time $\tau$. The world-line 
of the clocks may be considered as the geodetic line of the   massless test
particle.
This suggests that, in order to find the form of 
the corresponding coordinate transformation functions, one can apply the
same 
generalized Fermi conditions (3.26). 

As a result, the general form of the coordinate transformations 
between    the coordinates $(y^p_A)$ of  geocentric {\small RF}$_A$    and
those  
$(\zeta^p)\equiv(\tau, \zeta^\nu)$ of a   topocentric   one in the {\small
WFSMA} 
may be presented   as follows:  
{}
$$ y^0_A = \tau+ c^{-2}  K_{S_0}(\tau, \zeta^\epsilon) + 
c^{-4}   L_{S_0}(\tau, \zeta^\epsilon) + {\cal O}(c^{-6}),
\eqno(8.13a)  $$
$$ y^\alpha_A =  \zeta^\alpha  + \zeta^\alpha_{S_0} (\tau) + 
c^{-2} Q^\alpha_{S_0}(\tau,\zeta^\epsilon) + {\cal O}(c^{-4}). 
\eqno(8.13b)$$
 \noindent Where we, as before, have neglected  the associated group
parameters  
 $\zeta^{S_0}, \sigma^\alpha_{S_0}$ and $f^{\alpha\beta}_{S_0}$ 
 and require that the constructed frame should 
preserve have all the conservation laws in its  immediate vicinity. 
The transformation  functions $K_{S_0}, Q^\alpha_{S_0}$ and 
$L_{S_0}$, in this case will take the following form: 
{}
$$K_{S_0}(\tau,\zeta^\nu)  =\int^{\tau} \hskip -10pt  dt'  
\Big[   \Big< {\cal U}\Big>_{S'_0}- 
{1\over2} v_{S_0}{_{{}_\nu}} v_{S_0}^\nu\Big]  
- v_{S_0}{_{{}_\nu}}\cdot \zeta^\nu +  {\cal O}(c^{-4})\tau, \eqno(8.14a)$$
{}
$$Q^\alpha_{S_0}(\tau,\zeta^\nu) =
\gamma\Big({1\over2}\eta^{\alpha\beta}
{\zeta}_\mu  \zeta^\mu-  \zeta^\alpha
\zeta^\beta\Big) \Big<\partial_\beta  {\cal U} \Big>_{S_0}-
\gamma \zeta^\alpha  \Big< {\cal U}\Big>_{S_0} - 
{1\over2} v^\alpha_{S_0} v^\beta_{S_0}{\zeta}_\beta+$$
{}
$$+{\zeta}_\beta  \int^{\tau} \hskip -10pt dt' 
\Big[ {1\over2}a^{[\alpha}_{S_0}  v^{\beta]}_{S_0}  + 
2  \Big<\partial^{[\alpha} {\cal  V}^{\beta]}\Big>_{S'_0}\Big] 
 + w^\alpha_{S_0}(\tau) + 
{\cal O} (c^{-4}) \zeta^\alpha,   \eqno(8.14b)$$ 
{}
$$L_{S_0}(\tau,\zeta^\nu) =    \gamma {1\over2} \hskip 2pt
\zeta_{\mu} \zeta^\mu  {\partial \over\partial \tau}
\Big< {\cal U}\Big>_{S_0}   - 
2\hskip 2pt \zeta^\lambda   \zeta^\beta  
 \Big< \partial_\lambda  {\cal V}_\beta \Big>_{S_0} +   
\gamma  {v_{S_0}}_\beta \Big(\zeta^\beta  \zeta^\lambda  - 
{1\over2} \eta^{\beta\lambda}\zeta_\mu  \zeta^\mu \Big) 
\Big< \partial_\lambda   {\cal U}\Big>_{S_0}  + $$
{}
$$+\zeta_\lambda{v_{S_0}}_\beta \int^{\tau} \hskip -10pt dt'
 \Big[ {1\over2} a^{[\lambda}_{S_0} v_{S_0}^{\beta]}+ 
2 \Big<\partial^{[\lambda} {\cal  V}^{\beta]}  \Big>_{S'_0}\Big]+ 
 \zeta_\beta  \Big(  (\gamma+1) v^\beta_{S_0}  
\Big< {\cal U}\Big>_{S_0} - 4\hskip 2pt \Big<{\cal V}^\beta\Big>_{S_0}  
 - \dot { w}^\beta_{S_0} \Big) - $$
{}
$$ - \int^{\tau} \hskip -10pt dt' \Big[  
\Big< {\cal Z}\Big>_{S'_0}  + {1\over2}\Big( \Big< {\cal U}\Big>_{S'_0}  - 
 {1\over2}{v_{S_0}}_\beta v^\beta_{S_0}\Big)^2 
  + {v_{S_0}}_\mu\dot{ w}^\mu_{S_0}\Big]  +{\cal O}(c^{-
6})\tau,\eqno(8.14c)$$ 
\noindent where the new quantities $v_{S_0}^\alpha$ and
$a_{S_0}^\alpha$ are the 
geocentric velocity and acceleration of a particular  point $S_0$ 
on the surface  of an extended body   under question. 
The notation $\Big<f\Big>_{S_0}$ reflects that this quantity was defined 
in a  particular point $S_0$ on the surface $S_A$ of an 
extended body {\small (A)}. 
The   Newtonian acceleration of the clock with respect to 
the geocentric {\small RF}$_A$  is given as:
{}
$$a^\alpha_{S_0}(\tau)=-\eta^{\alpha\mu} 
\Big<{\partial  {\cal U}\over \partial \zeta^\mu}\Big>_{S_0}+{\cal O}(c^{-
4}).\eqno(8.15)$$   
\noindent
Furthermore, the function $w^\alpha_{S_0}$ is determined as the  solution
of 
the following differential equation:
{}
$$\ddot{w}^\alpha_{S_0}(\tau)=
  \eta^{\alpha\mu}\Big<{\partial  {\cal Z}\over \partial
\zeta^\mu}\Big>_{S_0}+ 
v^\alpha_{S_0}{\partial \over \partial \tau} \Big<  {\cal U}\Big>_{S_0}  +
  a^\alpha_{S_0} \Big< {\cal U}\Big>_{S_0}- 
 4\hskip 2pt{\partial \over \partial \tau}\Big< {\cal 
V}^\alpha\Big>_{S_0} -$$
{}
$$- {1\over2}v^\alpha_{S_0}{v_{S_0}}_\beta a^\beta_{S_0}  +
{a_{S_0}}_\beta \int^{\tau} \hskip -10pt dt' 
\Big[ {1\over2}a^{[\alpha}_{S_0}  v^{\beta]}_{S_0}  + 
2  \Big<\partial^{[\alpha}  {\cal V}^{\beta]}\Big>_{S'_0}\Big]
 + {\cal O}(c^{-6}), \eqno(8.16)$$
\noindent where the function ${\cal Z}\sim{\cal O}(c^{-4})$ 
was defined similarly to   the function 
${\cal R}$ from eqs.$(8.8b)$:
{}
$${\cal Z}(\zeta^p) = {\cal W}(\zeta^p)
+2 a^\lambda_{S_0}   \sum_B {\partial \over \partial 
\zeta^\lambda} \chi_B (\tau,\zeta^\nu) + 
{\cal O}(c^{-6}),\eqno(8.17)$$ 

\noindent As a result, the  components of the  metric 
tensor $g^{S_0}_{mn}$ in the coordinates
$(\zeta^p)\equiv(\tau,\zeta^\nu)$
of the topocentric {\small RF} take the following form:
{}
$$g^{S_0}_{00}(\zeta^p) =1-2{\overline {\cal U}}_{S_0}+ 
2{\overline {\cal Z}}_{S_0} + {\cal O}(c^{-6}), \eqno(8.18a) $$ 
{}
$$g^{S_0}_{0\alpha}(\zeta^p) = 4\hskip2pt \eta_{\alpha\epsilon}
{\overline {\cal V}}^\epsilon_{S_0}  +  {\cal O}(c^{-5}), \eqno(8.18b)$$
{}
$$g^{S_0}_{\alpha\beta}(\zeta^p) = 
\eta_{\alpha\beta}\Big(1 + 2 \gamma {\overline {\cal U}}_{S_0}\Big) + 
{\cal O}(c^{-4}), \eqno(8.18c)$$
 
 The obtained expressions  (8.18) represent the general solution for 
the field equations of the general relativity on the surface of an extended
body 
in the {\small WFSMA}. 
The effective Newtonian potential in the vicinity of the antenna 
may be presented as follows:
{}
$${\overline{\cal U}}_{S_0}(\zeta^p)= {\cal U}(\tau,\zeta^\nu)-
{\partial K_S(\tau,\zeta^\nu)\over \partial \tau} -
 {1\over2}{v_{S_0}}_\mu v^\mu_{S_0}=$$
{}
$$= {\cal U}(\tau, \zeta^\nu)-  \Big[
\zeta^\mu \Big<{\partial {\cal U}\over \partial \zeta^\mu}\Big>_{S_0}   + 
\Big<{\cal U}\Big>_{S_0} \Big].\eqno(8.19)$$
\noindent The  functions ${\overline{\cal Z}}_{S_0}$ and 
${\overline{\cal V}}^\alpha_{S_0}$ were obtained in the following form:
{} 
$${\overline{\cal Z}}_{S_0}(\zeta^p)= {\cal R}(\tau, \zeta^\nu) -   
\Big[\zeta^\mu \Big<{\partial {\cal R}\over \partial
\zeta^\mu}\Big>_{S_0}   
+ \Big<{\cal R}\Big>_{S_0} \Big]  + 
{1\over2}\hskip 2pt \zeta^\mu \zeta^\beta  \hskip 1pt  
\Big[ \gamma \eta_{\mu\beta} 
\hskip 1mm {a_{S_0}}_\lambda a^\lambda_{S_0}  - $$
{}
$$-  (2\gamma-1){a_{S_0}}_\mu {a_{S_0}}_\beta +  
{\partial \over \partial \tau}\Big( \gamma \eta_{\mu\beta}{\partial  
\over\partial \tau}\Big<{\cal U}\Big>_{S_0}   - 
2 \Big<\partial_{(\mu}  {\cal V}_{\beta)}\Big>_{S_0}\Big)\Big],  
\eqno(8.20)$$
{}
$${\overline{\cal V}}^\alpha_{S_0}(\zeta^p)=  {\cal
V}^\alpha(\tau,\zeta^\nu) - 
 \Big[\zeta^\mu \Big<{\partial {\cal V}^{\alpha}\over \partial
\zeta^\mu}\Big>_{S_0}  + 
\Big<{\cal V}^{\alpha}\Big>_{S_0} \Big] +\gamma {1\over4}
\Big(\zeta^\alpha \zeta_\beta   - 
{1\over2} \delta^\alpha_\beta  \zeta_\mu \zeta^\mu\Big)
 {\dot a}^\beta_{S_0}. \eqno(8.21)$$

It should be stressed that the more detailed analysis is necessary for the 
final solution of the problem of the relativistic astronomical
measurements 
performed from the topocentric {\small RF}.
However,  we believe that the presented general approach, incorporated in
the 
new formalism, enables one to construct the topocentric proper reference  
frame   with the  well defined physical properties.
Moreover, the  accuracy of  the theoretical expressions  obtained here, is 
far beyond that achieved in the real astronomical practice. 
This suggests that  the presented formulae could be used for  quite a long 
time before the practical needs will
require  theoreticians to reconsider the presented results in order to 
achieve the higher accuracy of the physical modelling of the 
relativistic measurements.  

\subsection{Discussion.}
 
It is generally understood that any {\small RF} is not a physical substance 
but rather a conventional artifact. 
The main reason we need  a {\small RF}s is that they are   convenient in 
exchanging the observational data and one's
discoveries and opinions, which are the starting points in doing a
scientific research. 
In this sense, the most important character of the {\small RF} is that it is
widely accepted and
is related clearly with the other existing  references. In addition, it is
desirable to
represent the actual phenomenon precisely. If the first point is respected,
what we should do in the
days of advanced electronic/computational environment is to thrust a
movement toward the
standardization which never means the exclusion of other points of view.
Rather it
should be understood as only a scale which enables us to express 
observation/theoretical quantities in a concise manner. 

An application of atomic frequency standards is the establishment of
atomic time scales.  International Atomic Time is the official basis by
which events are dated.  However, the need to distinguish between
theoretical times and their realizations, the need for a relativistic
treatment and the survival of previous astronomical times generate a
complex situation.  Specific problems raised by time scales, the
relationships they have with one another and with the successive
definitions of the second should be  examined in more details.
Thus, currently employed definitions of ephemeris astronomy and the
system of astronomical constants
are based on Newtonian mechanics with its absolute time and absolute
space. To avoid any
relativistic ambiguities in applying new IAU (1991) resolutions on {\small
RF}s and time
scales one should specify the astronomical constructions and definitions
of constants to make them
consistent with general relativity.
 However, up to this time, the VSOP theories of the motion of 
the planets were constructed on the base of the
integration of the Lagrange's differential equations (Brumberg {\it et al.}, 
1993).
 The development of the perturbative function
included the mutual perturbations of the bodies and was performed up to
the third order of the
perturbative masses using the Newtonian perturbative function. 
The relativistic contributions to the equations of motion 
 were limited to the Schwarzschild problem. The accuracy reached by
such solutions is only a few mas for the inner planets and less for the
outer ones. 
Due to the fact that the present astrometric accuracies had reached the
mas level, 
the mutual relativistic perturbations of the planets must be included 
in the ephemeris constructions. 
 
In this report we addressed these and other problems of the modern
astronomy and 
have presented the theoretical foundation, necessary for   conducting the 
relativistic measurements in the curved space-time in the {\small
WFSMA}. 
Our approach in naturally incorporates the general properties of the 
dynamical {\small RF} into the hierarchy of the relativistic {\small RF}s
and the 
time scales. Moreover, we obtained  the new relation between the 
time scales, which was obtained  to the 
fourth order in 1/c, c being the velocity of light in the vacuum. The
accuracy of these expressions is at the ps level which is the  future
requirement in many different applications. Thus, this formulation leads
to
improved relations between barycentric and geocentric quantities.  
These expressions will be useful in 
converting the numerical values of some astronomical constants
determined 
in the old IAU time scale TDB. The obtained results naturally contain an
exhaustive
information about the multipolar structure of the gravitational field 
in the {\small N}-body system and enable one to model the experimental
situation with a very high accuracy.  Because of this,
we anticipate the that the results presented in this report
may be immediately applied in the following 
important areas of the modern astronomy and astrophysics:

\begin{itemize}
\item[(i).] The precise {\small VLBI} timing measurements.
\item[(ii).] The precise radiometric navigation of the future space
missions and the 
corresponding data  analysis.
\item[(iii).] The more precise analysis of the binary system dynamics
including the 
modelling of the coalescing experiments and the studies of the 
gravitational wave 
physics.
\end{itemize}    

Let us mention that there are some problems which remains to be 
unsolved.  Thus, it is known that the rotational motion of extended bodies
in 
general relativity is a complicated problem which has
no complete solution up to now. This is also true, because the modern
observational 
accuracy of the geodynamical observations makes it necessary to have a
rigorous 
relativistic model of Earth's rotation. 
Currently employed solution for the Earth's rotation problem is valid for 
restricted intervals of time. 
Moreover, there is an urgent  necessity to elaborate a theory of 
nutation-precession matching the accuracy of very modern
techniques as {\small VLBI} and {\small LLR}. To do this, one would have
to  model  the
transfer function leading to theoretical determination of the nutation 
coefficients when including predominant geophysical characteristics 
(elastic mantle, coupling at core-mantle boundary, free core
nutation, free inner core nutation etc...). 
Furthermore, reductions of measurements included relativistic
corrections, effects of propagation of
electromagnetic signals in the Earth troposphere and in the solar corona
with simultaneous
evaluation of parameters of corona model from general fitting. 
The presented formalism provides one with the necessary basis in
studying this 
problem from the very general positions and could serve as the foundation
for the 
future theoretical analysis.

As a result, an astronomical reference system may be defined as a set of
the 
transformation functions  and constants including  the physically 
well defined set of the {\small RF}s and their mutual relationships,
time arguments, ephemerides, and the standard constants and algorithms.
The extragalactic, or radio,  {\small RF} will be the basic frame for the 
development of the future ephemeris (Standish, 1995).
Achieving milli to microarcsecond accuracies at optical wavelengths will
reduce the disparity between optical, radar, and radio {\small RF}
determinations. Thus, the relationships and identifications of common
sources should be much more accurate.  Another significant change should
be
the ability to determine distances, and thus space motions on a
three-dimensional basis, rather than the current two-dimensional basis of
proper motions.
Improvements in ephemerides provide the opportunity to investigate the
difference between atomic and dynamical time, the relationship between
the
dynamical and extragalactic {\small RF} and the values of precession
and nutation.
Also, the relationships between the bright and faint optical catalogs, 
the infrared, and extragalactic {\small RF}s should be better determined.  
The theory of the relativistic astronomical {\small RF}s
presented in this report was developed in order to serve exactly the
mentioned
above needs and it will be used in the future analysis of these problems of  
fundamental  importance.

In order to accomplish these goals, our future efforts will be directed
onto 
the  finalizing the transcription of the results obtained on  the 
language of the practical applications. We will   establish  the necessary
relativistic
 measurements models and will implement these results into existing 
computer software codes,
as well as will  perform the detailed analysis of the real data from the 
space gravitational experiments.
The analysis of the  above mentioned problems  from the new positions 
of the presented  theory of the relativistic astronomical {\small RF}s, 
will   be the subject  for the specific studies and future publications.

\vskip 15pt
\noindent {\Large \bf Appendix A: Generalized Gravitational Potentials.}
\vskip 11pt \noindent 
 The generalized gravitational potentials for the non-radiative 
problems in the  {\small WFSMA} are given in   Will (1993) as:
{}
$$U(z^p)  = \int { d^3z' \rho_0 (z'^p)\over |z^\nu - z'^\nu|}, 
\qquad V^\alpha(z^p)  = - \int d^3z' {\rho_0(z'^p)v^\alpha (z'^p)\over 
|z^\nu-z'^\nu|},$$
{}
$$W^\alpha(z^p)  = \int d^3z' \rho_0(z'^p)v_\mu(z'^p)
{(z^\alpha-z'^\alpha)(z^\mu-z'^\mu)\over|z^\nu-z'^\nu|},$$
{}
$$A(z^p)  = \int d^3z' \rho_0(z'^p){ [v^\mu(z'^p)(z^\mu-z'^\mu)]^2
\over|z^\nu-z'^\nu|^3},
\qquad \chi(z^p)  = -\int  d^3z'  \rho_0 (z'^p)  |z^\nu - z'^\nu|, $$ 
{}
$$U^{\alpha\beta}(z^p)  = \int d^3z' \rho_0(z'^p){(z^\alpha-z'^\alpha)
(z^\beta-z'^\beta)\over|z^\nu-z'^\nu|^3},$$
{}
$$\Psi(z^p)= -(\gamma+1)\Phi_1 -  (3\gamma+1-2\beta)\Phi_2-
 \Phi_3 -3\gamma\hskip 1pt\Phi_4, $$

\noindent where the other potentials are given as  follows:
{}
$$\Phi_1(z^p)  = - \int d^3z'  {\rho_0(z'^p) v_\lambda 
(z'^p)v^\lambda(z'^p)\over 
|z^\nu  - z'^\nu|},\qquad \Phi_{2} (z^p) 
= \int d^3z' {\rho_0(z'^p)U(z'^p)\over |z^\nu - z'^\nu |}, $$
{}
$$\Phi_{3} (z^p) =  \int  d^3z' { \rho_0(z'^p)\Pi(z'^p)\over 
|z^\nu  -z'^\nu|}, \qquad \Phi_{4} (z^p) = 
\int d^3z' {\rho_0(z'^p)  p(\rho(z'^p))\over |z^\nu- z'^\nu|},$$
{}
$$\Phi_w(z^p)  = \int\hskip -3pt\int d^3z'  d^3z''\rho_0(z'^p)\rho_0(z''^p)
{(z^\beta-z'^\beta)\over|z^\nu-z'^\nu|^3} 
\Big[{(z_\beta-z''^\beta)\over|z'^\nu-z''^\nu|}
-{(z'^\beta-z''^\beta)\over|z^\nu-z''^\nu|}
\Big]. $$

\noindent In order to indicate the functional dependence in the potentials 
introduced above, we have  used the following notation: 
$(z^p)\equiv (z^0, z^\nu)$. Then for any function $f$
one will have: $f(z^p)=f(z^0,z^\nu)$ and $f(z'^p)=f(z^0,z'^\nu)$.

\vskip 11pt \noindent 
{\Large \bf Appendix B:  Power Expansions of the General Geometric
\hfill 

\hskip 29mm  Quantities.}
\vskip 11pt \noindent 
In this Appendix we will present the expansion of some physical 
quantities with respect to powers of the small parameter $c^{-1}$. We
will use 
 these   expansions  for  linearizing of the gravitational field 
equations of the metric theories of gravity in the {\small WFSMA}. 

\vskip 5pt
\noindent {\bf  B.1.    Expansion for the Metric Tensor $g_{mn}$.}
\vskip 5pt \noindent 
The post-Newtonian expansion for the 
metric tensor $g_{mn}$ with respect to the powers of the   small  
parameter $c^{-1}$ in the coordinates $(z^p)$ of  an arbitrary 
{\small RF}  (either barycentric inertial 
{\small RF}$_0$ or proper {\small RF}$_A$ non-inertial one)
 may be presented as follows:
{}
$$g_{00} = 1 + c^{-2} g^{<2>}_{00} + c^{-4}g^{<4>}_{00} + 
{\cal O}(c^{-6}), \eqno(B1a) $$
{}
$$g_{0\alpha} = c^{-3}g^{<3>}_{0\alpha} + c^{-5}g^{<5>}_{0\alpha}
+{\cal O}(c^{-7}), \eqno(B1b)$$
{}
$$g_{\alpha\beta} = \gamma_{\alpha\beta} + 
c^{-2}g^{<2>}_{\alpha\beta} + c^{-4}g^{<4>}_{\alpha\beta} + 
{\cal O}(c^{-6}),\eqno(B1c)$$

\noindent where  $\gamma_{\alpha\beta}$ is the spatial part of the
 background  metric $\gamma_{mn}$.  The notations 
$g^{<k>}_{mn}, \hskip 2mm (k = 1,2,3...)$ at the 
right-hand side  of the expressions  $(B1)$ are the terms of the
expansion of $g_{mn}$ with the order of magnitude  
$\epsilon^k \sim c^{-k}$ respectively. In some calculations, 
we will omit the multipliers $c^{-k}$ in order to achieve  
brevity in the expressions.
It should be noted that   reversing the sign of
 the time $z^0 \rightarrow - z^0$, corresponds to 
the change of the sign of the small 
parameter $\epsilon$. Because of this reason in
the  expressions for $g_{00}$ $(B1a)$ and $g_{\alpha\beta}$ $(B1c)$ 
only the terms with  even powers of the 
small parameter $c^{-1}$ have been taken into account, 
and in the expressions for $g_{0\alpha}$ $(B1b)$ only the odd ones
are used. The fact that in 
expression  $(B1b)$ the term $g^{<1>}_{0\alpha}$ is 
absent, is quite natural. Indeed, even the main  expansion
 for $g_{0\alpha}$  (Newtonian) should not be less than the second order 
with respect to the small parameter $c^{-1}$ (Will, 1993). 
In our further calculations we will not be investigating the processes
of generating of the gravitational waves by the system of the
astronomical 
bodies, so our expressions for the component $g_{00}$ 
in the expressions $(B1)$, do not  contain the term of order 
${\cal O}(c^{-5})$.
However, one may easily reconstruct all the 
calculations to account for this term as well. 

\pagebreak \noindent
{\bf B.2.   Expansion for the $ det \hskip 2pt \big[g_{mn}\big]$ 
and $g^{mn}$.}
\vskip 5pt \noindent In some calculations we will need the  relations for
the 
determinant of the metric  tensor $g = det\hskip 2pt \big[g_{mn}\big]$ and 
the inverse metric $g^{mn}$. 
>From the expressions eqs.$(B1)$ one may obtain the following 
relations which are valid for any {\small RF}:
{}
$$ g =  - 1 - g^{<2>}_{00} + g^{<2>}_{11} + g^{<2>}_{22} + 
g^{<2>}_{33} - g^{<4>}_{00} + g^{<4>}_{11} + 
g^{<4>}_{22} + g^{<4>}_{33} + $$
{}
$$ + g^{<2>}_{00}\Big(g^{<2>}_{11} + g^{<2>}_{22} + 
g^{<2>}_{33}\Big) -  g^{<2>}_{11}g^{<2>}_{22} - 
g^{<2>}_{11}g^{<2>}_{33} - $$
{}
$$ -g^{<2>}_{22}g^{<2>}_{33} + g^{<2>}_{12}{}^2 + 
g^{<2>}_{13}{}^2 + g^{<2>}_{23}{}^2 + {\cal O}(c^{-6}), \eqno(B2)$$
and
{}
$$g^{00} = 1 + g^{<2> 00} + g^{<4> 00} + {\cal O}(c^{-6}), $$
{}
$$g^{0\alpha} = g^{<3> 0\alpha} + g^{<5> 0\alpha} +
{\cal  O}(c^{-7}), $$
{}
$$g^{\alpha\beta} =  \gamma^{\alpha\beta} + 
g^{<2>  \alpha\beta} + g^{<4> \alpha\beta} + 
{\cal O}(c^{-6}),\eqno(B3)$$
\noindent 
where the components of inverse metric $g^{<k>mn}$ 
are given as follows:
{}
$$g^{<2> 00} = \hskip 1mm - \hskip 1mm g^{<2>}_{00}, 
\qquad 
g^{<2> \alpha\beta} = \hskip 1mm - \hskip 1mm 
\gamma^{\alpha\mu}\gamma^{\beta\nu}g^{<2>}_{\mu\nu},$$
{}
$$g^{<3> 0\alpha} = \hskip 1mm - \hskip 1mm 
\gamma^{\alpha\nu}g^{<3>}_{0\nu}, \qquad 
g^{<4> 00} = \hskip 1mm (g^{<2>}_{00})^2 - g^{<4>}_{00},$$
{}
$$g^{<4> \alpha\beta} = - \hskip 1mm \gamma^{\alpha\mu}
\gamma^{\beta\nu}g^{<4>}_{\mu\nu} \hskip 1mm  + \hskip 1mm 
\gamma^{\alpha\sigma}\gamma^{\beta\lambda}\gamma^{\mu\nu}
g^{<2>}_{\sigma\mu}g^{<2>}_{\lambda\nu},$$
{}
$$g^{<5> 0\alpha} = - \hskip 1mm \gamma^{\alpha\mu}
g^{<5>}_{0\mu} + \gamma^{\alpha\mu}g^{<2>}_{00}
g^{<3>}_{0\mu} + 
\gamma^{\alpha\lambda}\gamma^{\mu\nu}g^{<2>}_{\lambda\mu}
g^{<3>}_{0\nu}. \eqno(B4)$$

 \vskip 5pt \noindent
{\bf B.3. Expansion for the $\hat{g}^{mn}=\sqrt{-g} g^{mn}$.}
\vskip 5pt \noindent 
For some practical applications we will need the   
expansions for the components of density of the metric 
tensor $\hat{g}^{mn} = \sqrt{-g} 
g^{mn}$ as well. One may easy obtain those from the expressions 
eqs.$(B2)$-$(B4)$ in the following form:
{}
$$\hat{g}^{00} = 1 + \hat{g}^{<2> 00} + 
\hat{g}^{<4> 00} + {\cal O}(c^{-6}),  $$
{}
$$\hat{g}^{0\alpha} = \hat{g}^{<3> 0\alpha} + 
\hat{g}^{<5> 0\alpha} + {\cal O}(c^{-7}),  $$
{}
$$\hat{g}^{\alpha\beta} =  \hat{\gamma}^{\alpha\beta} + 
\hat{g}^{<2> \alpha\beta} + \hat{g}^{<4> \alpha\beta} + 
{\cal O}(c^{-6}),\eqno(B5a)$$
\noindent  
with the components of $\hat{g}^{mn}$ are given as:
{}
$$\hat{g}^{<2> 00} =  g^{<2> 00} + 
{1\over2} A^{<2>},$$
{}
$$\hat{g}^{<4> 00} =  g^{<4> 00} + 
{1\over2}g^{<2> 00}  A^{<2>} + {1\over2}\Big( A^{<4>}
 - {1\over4} (A^{<2>})^2\Big),$$
{}
$$\hat{g}^{<3> 0\alpha} =  g^{<3> 0\alpha},
 \qquad 
\hat{g}^{<5> 0\alpha} =  g^{<5>  0\alpha} + 
{1\over2} g^{<3> 0\alpha}A^{<2>},$$
{}
$$\hat{g}^{<2>  \alpha\beta} =  g^{<2> \alpha\beta} +
 {1\over2} \gamma^{\alpha\beta} A^{<2>},$$
{}
$$\hat{g}^{<4> \alpha\beta} =  g^{<4> \alpha\beta} +
 {1\over2} g^{<2> \alpha\beta}A^{<2>} + 
{1\over2}\gamma^{\alpha\beta}\Big( A^{<4>} - 
{1\over4} (A^{<2>})^2\Big). \eqno(B5b)$$
\noindent  
In the expressions  $(B5b)$ we have introduced the 
following notations:
{}
$$A^{<2>} = g^{<2>}_{00} - g^{<2>}_{11} - g^{<2>}_{22}
 - g^{<2>}_{33}, $$
{}
$$A^{<4>} = g^{<4>}_{00} - g^{<4>}_{11} - g^{<4>}_{22} 
- g^{<4>}_{33} - $$
{}
$$- g^{<2>}_{00} \Big(g^{<2>}_{11} + g^{<2>}_{22}+
g^{<2>}_{33}\Big) + $$
{}
$$ + g^{<2>}_{11}g^{<2>}_{22}  + g^{<2>}_{11}g^{<2>}_{33}
  + g^{<2>}_{22} g^{<2>}_{33} -$$ 
{}
$$- (g^{<2>}_{12})^2 -  (g^{<2>}_{13})^2 - (g^{<2>}_{23})^2.
 \eqno(B6)$$
 \vskip 5pt \noindent
{\bf B.4. Expansion for the Gauge Conditions.} 
 \vskip 5pt \noindent
 The covariant de Donder  gauge conditions are given by eq.(3.6) as follows:
{}
$${\cal D}_{m}\Big({\sqrt{-g}}g^{mn}(z^p)\Big) = 0, \eqno(B7a)$$ 
\noindent 
or equivalently 
{}
$${\partial \over \partial z^m} \Big({\sqrt{-g}}g^{mn}(z^p)\Big)
 + \gamma^{n}_{kl}(z^p){\sqrt{-g}}g^{kl}(z^p) = 0,
\eqno(B7b)$$
where $\gamma^{n}_{kl}(z^p)$ is the Christoffel symbols
 with respect to the background metric $\gamma_{kl}(z^p)$ in 
 coordinates $(z^p)$ of an arbitrary {\small RF}. 
The relations eqs.$(B5)$-$(B6)$ enables one to find the expressions
for  linearized gauge conditions $(B7b)$.  Thus, for $n = 0$, 
we will have:
{}
$${1\over2}{\partial \over \partial z^0} \Big(\gamma^
{\epsilon\nu} g^{<2>}_{\epsilon\nu} - g^{<2>}_{00}\Big) - 
 \gamma^{\epsilon\nu}{\partial \over \partial z^\epsilon} 
g^{<3>}_{0\nu}  + \gamma^0_{00}{}^{<3>}(z^p) + $$
{}
$$+ \gamma^{\mu\nu}\gamma_{\mu\nu}^{0 <3>}(z^p) = {\cal O}(c^{-5}).
 \eqno(B7c)$$
\noindent  
For $n = \alpha$ we will obtain: 
{}
$${1\over2}\gamma^{\alpha\lambda}{\partial \over \partial
 z^\lambda} \Big(g^{<2>}_{00} + 
\gamma^{ \epsilon\nu} g^{<2>}_{\epsilon\nu}\Big) - 
 \gamma^{\epsilon\nu}\gamma^{\mu\alpha}{\partial \over 
\partial z^\epsilon} g^{<3>}_{\nu\mu} + 
\gamma^{\alpha <3>}_{00}(z^p) + $$
{}
$$+\gamma^{\mu\nu}\gamma^{\alpha <3>}_{\mu\nu}(z^p) = {\cal O}(c^{-4}),
 \eqno(B7d)$$
 
\noindent where the $\gamma^m_{kl}(z^p)$ is the components of the 
Christoffel symbols with respect to   Riemann-flat 
non-inertial background metric  $\gamma_{mn}(z^p)$ in coordinates
$(z^p)$.
 One may easily see that for any non-inertial {\small RF} these 
components may produce a non-vanishing contribution to the 
gauge conditions $(B7)$. This property will be in use in order to  
write the field equations in an arbitrary {\small RF}.  

 \vskip 5pt \noindent
 {\bf B.5. Expansion of the Christoffel Symbols}.
\vskip 5pt \noindent

\noindent  One could easily find the expansion of the connection 
components $\Gamma^k_{mn}$ with respect to the small
 parameter $c^{-1}$ and present those
in terms of expansions $g^{<k>}_{mn}$. Thus, defining 
$\Gamma^k_{mn}$ as usual: 
{}
$$\Gamma^{k}_{nm}(z^p) = {1\over2} g^{kp}(z^p)\Big(
{\partial_n} g_{mp}(z^p) + {\partial_m} g_{pn}(z^p)
 - {\partial_p}g_{mn}(z^p)\Big),$$
\noindent where ${\partial_n}={\partial /\partial z^n}$, 
in  coordinates $(z^p)$  of 
an arbitrary  {\small RF} from the relations  $(B1),(B3)$-$(B4)$ we will
have
the following expressions for the components of the Christoffel
symbols with respect to the powers of the small parameter $c^{-1}$:
{}
$$\Gamma^{0}_{00}(z^p) = {1\over 2} \partial_0 g^{<2>}_{00} + 
{1\over 2} \partial_0 g^{<4>}_{00} + $$
{}
$$+{1\over2}\Big(g^{<2> 00} \partial_0 g^{<2>}_{00}- 
g^{<3> 0\mu}\partial_\mu g^{<2>}_{00}\Big) + {\cal O}(c^{-7}),
 \eqno(B8a)$$
{}
$$\Gamma^{0}_{0\alpha}(z^p) = {1\over 2} \partial_\alpha g^{<2>}_{00}
 + {1\over 2} \Big(\partial_\alpha g^{<4>}_{00} +
g^{<2> 00} \partial_\alpha g^{<2>}_{00}\Big) + {\cal O}(c^{-6}),
 \eqno(B8b)$$
{}
$$\Gamma^{0}_{\alpha\beta}(z^p) = {1\over 2} \Big(\partial_\alpha 
g^{<3>}_{0\beta} +\partial_\beta g^{<3>}_{0\alpha}- 
\partial_0 g^{<2>}_{\alpha\beta}\Big) + {\cal O}(c^{-5}), \eqno(B8c)$$
{}
$$\Gamma^{\alpha}_{00}(z^p) = - {1\over 2} \gamma^{\alpha\mu} 
\partial_\mu g^{<2>}_{00} -
{1\over 2} \gamma^{\alpha\mu}\partial_\mu g^{<4>}_{00} + 
\gamma^{\alpha\mu}\partial_0 g^{<3>}_{0\mu} -$$
{}
$$-{1\over 2} g^{<2> \alpha\mu}\partial_\mu g^{<2>}_{00}+ 
{\cal O}(c^{-6}), \eqno(B8d)$$
{}
$$\Gamma^{\alpha}_{0\beta}(z^p) = {1\over 2} \gamma^{\alpha\mu}
 \partial_\beta g^{<3>}_{0\mu} +
{1\over 2} \gamma^{\alpha\mu}\partial_0 g^{<2>}_{\beta\mu} - 
{1\over2}\gamma^{\alpha\mu}\partial_\mu g^{<3>}_{0\beta} +
 {\cal O}(c^{-5}), \eqno(B8e)$$
{}
$$\Gamma^{\alpha}_{\beta\nu}(z^p) = 
\gamma^{\alpha(0)}_{\beta\nu}+{1\over 2} \gamma^{\alpha\mu}
  \Big(\partial_\beta g^{<2>}_{\mu\nu} +
\partial_\nu g^{<2>}_{\mu\beta} - \partial_\mu g^{<2>}
_{\beta\nu}\Big) + {\cal O}(c^{-4}), \eqno(B8f)$$
 
\noindent where $\gamma^{\alpha(0)}_{\beta\nu}$ are the Christoffel 
symbols  in coordinates of Galileian inertial {\small RF}. 
One may make them vanish by choosing quasi-Cartesian coordinates.

 \vskip 5pt \noindent
{\bf B.6. Expansion  for  the Ricci Tensor $R_{mn}$.}
\vskip 5pt \noindent

\noindent By making of use the expressions eqs.$(B8)$ one may 
also find the relations for  the expanded components 
of Ricci tensor $R_{mn}(z^p)$ in coordinates 
$(z^p)$ of an arbitrary {\small RF}. This tensor is defined as follows:
{}
$$R_{mn}(z^p) =g^{kp} R_{kmnp}(z^p)= {\partial_p}\Gamma^p_{mn} 
- {\partial_n}\Gamma^p_{mp} + \Gamma^l_{mn}\Gamma^p_{lp} -
 \Gamma^l_{mp} \Gamma^p_{ln}. $$

Then, in quasi-Cartesian coordinates of 
an arbitrary {\small RF} one may obtain the expanded 
components of the Ricci tensor as follows:
{}
$$R_{00}(z^p) = - {1\over2} \gamma^{\mu\lambda} 
\partial^2_{\mu \lambda} g^{<2>}_{00}  
- {1\over2} \gamma^{\mu\lambda}\partial^2_{\mu\lambda} g^{<4>}_{00} 
+\gamma^{\mu\lambda}\partial^2_{0\mu} g^{<3>}_{0\lambda} -$$ 
{}
$$ - {1\over2} \partial_\mu \Big( g^{<2> \mu\lambda} 
 \partial_\lambda g^{<2>}_{00}\Big) - 
{1\over2} \gamma^{\mu\lambda}\partial^2_{00} 
 g^{<2>}_{\mu\lambda}-$$  
{}
$$- {1\over4} \gamma^{\mu\lambda}  \gamma^{\sigma\nu} 
 \partial_\mu g^{<2>}_{00}\partial_\lambda g^{<2>}_{\sigma\nu} 
+ {1\over4} \gamma^{\mu\lambda} \partial_\mu g^{<2>}_{00} 
 \partial_\lambda g^{<2>}_{00} + {\cal O}(c^{-6}), \eqno(B9a)$$
{} 
$$R_{0\alpha}(z^p) = {1\over2} \gamma^{\mu\lambda} 
\partial^2_{0\mu} g^{<2>}_{\alpha\lambda}  
-{1\over2}\gamma^{\mu\lambda}\partial^2_{0\alpha} 
 g^{<2>}_{\mu\lambda}+$$ 
{}
$$+{1\over2}\gamma^{\mu\lambda}\partial^2_{\alpha\mu}
g^{<3>}_{\lambda0} 
 - {1\over2} \gamma^{\mu\lambda}
\partial^2_{\mu\lambda}g^{<3>}_{0\alpha} 
+ {\cal O}(c^{-5}), \eqno(B9b)$$
{}
$$R_{\alpha\beta}(z^p) = - {1\over2} \gamma^{\mu\lambda} 
\partial^2_{\mu \lambda} g^{<2>}_{\alpha\beta}  
+ {1\over2} \gamma^{\mu\lambda} \partial^2_{\mu \alpha} 
g^{<2>}_{\lambda\beta} 
+  {1\over2}\gamma^{\mu\lambda} \partial^2_{\mu \beta} 
g^{<2>}_{\lambda\alpha} -  $$
{}
$$ - {1\over2}\partial^2_{\alpha\beta} g^{<2>}_{00}-
{1\over2}\gamma^{\mu\lambda}\partial^2_{\alpha\beta}  
g^{<2>}_{\mu\lambda} + {\cal O}(c^{-4}). \eqno(B9c)$$

 \vskip 5pt \noindent
{\bf B.7. Expansion of an Arbitrary Energy-Momentum Tensor $T^B_{mn}$.}
\vskip 5pt \noindent

\noindent At this point, the precise
definition for the  energy-momentum  tensor of the matter distribution
$T^B_{mn}$ is not
 important. For the future analysis we will accept the most general
assumptions concerning this
quantity. Namely, we will work with such   energy-momentum tensors 
$T^B_{mn}$, 
the temporal, the temporal-spatial and the spatial components of which  
may be presented in terms of the order of magnitude
as follows: $T^B_{mn}(y^p_B) = \Big({\cal O}(1), {\cal O}(c^{-1}), 
 {\cal O}(c^{-2})\Big).$

 The construction of the iterative
 scheme  is required to perform the power expansion of the 
energy-momentum tensor of matter $T^{mn}$ as well. Suppose 
that $T^{mn}$ may be expanded with respect to the small
 parameter $c^{-1}$ as follows:
{}
$$T^{00} = T^{<0> 00} + T^{<2> 00} +  {\cal O}(c^{-4}), \eqno(B10a)$$
{}
$$T^{0\alpha} = T^{<1> 0\alpha} + T^{<3> 0\alpha} +  {\cal O}(c^{-5}),
  \eqno(B10b)$$
{}
$$T^{\alpha\beta} = T^{<2> \alpha\beta} + T^{<4> \alpha\beta}
 +  {\cal O}(c^{-6}).  \eqno(B10c)$$ 

Then, by taking into account the expressions $(B1)$ and with the help of
relations
$(B10)$ we will  get the  inverse tensor  $T_{mn}$ as follows: 
{}
$$T_{00} = T^{<0>}_{00} + T^{<2>}_{00} +  
{\cal O}(c^{-4}); \eqno(B11a)$$
{}
$$T_{0\alpha} = T^{<1>}_{0\alpha} + T^{<3>}_{0\alpha} +  
{\cal O}(c^{-5}); \eqno(B11b)$$
{}
$$T_{\alpha\beta} = T^{<2>}_{\alpha\beta} + 
T^{<4>}_{\alpha\beta} +  {\cal O}(c^{-6}), \eqno(B11c)$$
\noindent  
where 
{}
$$T^{<0>}_{00} = T^{<0> 00}, \qquad T^{<2>}_{00} = T^{<2> 00} + 
2 \hskip 1mm g^{<2>}_{00}\hskip 1mm  T^{<0> 00},$$ 
{}
$$T^{<3>}_{0\alpha} = g^{<3>}_{0\alpha} T^{<0> 00} + \Big(
g^{<2>}_{\alpha\mu} +
 \gamma_{\alpha\mu} g^{<2>}_{00} \Big) \hskip 1mm T^{<1> 0\beta},$$
{}
$$ T^{<1>}_{0\alpha} = \gamma_{\alpha\beta}\hskip 1mm T^{<1> 0\beta},
\qquad 
 T^{<2>}_{\alpha\mu} = \gamma_{\alpha\mu}\gamma_{\beta\lambda}
\hskip 1mm T^{<2> \mu\lambda}. \eqno(B11d)$$

Concluding this part, we will present   the 
expression for the right-hand side of the 
Hilbert-Einstein field equations eqs.(4.1), which is given as follows:
{}
$$S_{mn} = T_{mn} - {1\over2} g_{mn} \hskip 1mm T. \eqno(B12)$$

By substituting the expressions eqs.$(B1),(B11)$ into definition  $(B12)$ 
we will obtain the  expansions
for the quantity $S_{mn}$ in the {\small WFSMA}:  
{}
$$S_{00} = {1\over2} T^{<0> 00} +{1\over2} \Big(T^{<2> 00} +
 2 \hskip 1mm g^{<2>}_{00}\hskip 1mm T^{<0> 00}- 
\gamma_{\mu\lambda}\hskip 1mm T^{<2> \mu\lambda}\Big) + {\cal O}(c^{-
4}),
 \eqno(B13a)$$
{}
$$S_{0\alpha} = \gamma_{\alpha\lambda}\hskip 1mm T^{<3> 0\lambda} + 
 {\cal O}(c^{-4}), \eqno(B13b)$$
{}
$$S_{\alpha\beta} = - {1\over2}\gamma_{\alpha\beta} \hskip 1mm 
T^{<0> 00} + 
\Big(\gamma_{\alpha\mu}\gamma_{\beta\lambda} - 
{1\over2}\gamma_{\alpha\beta}\gamma_{\mu\lambda}\Big)\hskip 1mm 
T^{<2> \mu\lambda}-$$
{}
$$ - {1\over2}\Big(\gamma_{\alpha\beta}\hskip 1mm T^{<2> 00} + 
2\hskip 1mm  \gamma_{\alpha\beta}\hskip 1mm g^{<2>}_{00}
\hskip 1mm T^{<0> 00}+ 
g^{<2>}_{\alpha\beta}\hskip 1mm T^{<0> 00}\Big) 
 + {\cal O}(c^{-4}). \eqno(B13c)$$

\vskip 11pt\noindent 
{\Large\bf Appendix C: Transformation Laws of the Coordinate Base\hfill

\hskip 29mm Vectors.}
\vskip 11pt\noindent  
 In this Appendix  we will present the transformation rules
for the coordinate base vectors 
 under the general post-Newtonian coordinate transformations, which
 were discussed  in   Section {\small 3}.

\vskip 5pt \noindent 
{\bf C.1. Direct Transformation  of the Coordinate Base Vectors.}
\vskip 5pt\noindent  
According to the transformation rules of the solutions 
of the field equations $h^{mn}_{(0)}$ and an arbitrary  energy-momentum 
tensor $T^{mn}$ given by eqs.(3.1)-(3.4), 
in order to develop the consistent perturbation theory 
for   N-body problem in the {\small WFSMA}, one   needs to have the 
post-Newtonian expansions for the following derivatives: 
{}
$${\partial x^k\over\partial y^m_A}, \hskip 6mm 
{\partial y^l_B\over\partial x^n},\hskip 6mm  
{\partial y^k_B\over\partial y^m_A}.$$
\noindent  
These derivatives   form the transformation matrix
$\lambda^{\overline{k}}_l$ of the coordinate bases  while the transition 
between the different coordinate systems is performed. 
Thus, for the transition from the barycentric {\small RF}$_0$ coordinate
 base $e_m = \partial/\partial x^n$ to the body-centric one 
 $e^A_m = \partial/\partial y^m_A$, the transformation matrix is
defined as usual: $e_m = e^A_{\overline{p}}{\partial x^{\hat p}/\partial
y^m_A}
 =   e^A_{\overline{p}} \lambda^{\overline{p}}_{A m}$. 
Then, making of use the transformations eqs.$(3.5)$ it is easy to get: 
$$
\lambda^{\overline{0}}_{A 0} (y^p_A)  = 
{\partial x^0\over\partial y^0_A} 
= 1 + 
{\partial\over\partial y^0_{A}} K_{A} (y^0_A,y^\nu_A) + 
{\partial\over\partial y^0_{A}} L_{A} (y^0_A,y^\nu_A) + {\cal O}(c^{-6}),
 \eqno(C1a)$$
{}
$$
\lambda^{\overline{0}}_{A \alpha}(y^p_A)  = 
{{\partial x^0}\over{\partial y^\alpha_A}} = 
 {\partial\over\partial y^\alpha_{A}} K_{A} (y^0_A,y^\nu_A) + 
{\partial\over\partial y^\alpha_{A}} L_{A} (y^0_A,y^\nu_A) + 
{\cal O}(c^{-5}), \eqno(C1b) $$
{}
$$
\lambda^{\overline{\alpha}}_{A 0} (y^p_A) = 
{\partial x^\alpha\over\partial y^0_A} = v^\alpha_{A_0}(y^0_A) +
 {\partial\over\partial y^0_{A}} Q^\alpha_{A} 
(y^0_A,y^\nu_A) + {\cal O}(c^{-5}), \eqno(C1c) $$
{}
$$
\lambda^{\overline{\alpha}}_{A \mu} (y^p_A) = 
{ \partial x^\alpha\over \partial y^\mu_A}  = 
\delta^\alpha_\mu  +
 {\partial\over\partial y^\mu_{A}} Q^\alpha_{A} 
(y^0_A,y^\nu_A) + {\cal O}(c^{-4}). \eqno(C1d)$$

By using the expressions  $(C1)$,  one could obtain the determinant of 
this transformation matrix  as follows:
{}
$$det\Big[\lambda^{\overline{p}}_{A m} (y^p_A)\Big] = 1+  
 {\partial\over\partial y^0_{A}} K_{A} (y^0_A,y^\nu_A) - $$
{}
$$- v^\lambda_{A_0} 
{\partial\over\partial y^\lambda_{A}} K_{A} (y^0_A,y^\nu_A) + 
{\partial\over\partial y^\mu_{A}} Q^\mu_{A} (y^0_A,y^\nu_A)  
+ {\cal O}(c^{-4}). \eqno(C2)$$

\noindent The condition $ det \Big[ \lambda_{A m}^{\overline{p}} 
(y^p_A) \Big] = 0$ gives  the boundary of 
validity  of these transformations in application of those to  
 constructing a proper {\small RF}$_A$.

 \vskip 5pt \noindent
{\bf C.2.  Transformation of the 
Background Metric $\gamma_{mn}$.}
\vskip 5pt 
\noindent The relations $(C1)$ are the useful tool for the calculations 
of   metric tensor $\gamma^{A}_{kl}(y^p_A)$ of the background
space-time in the non-inertial proper {\small RF}$_A$ from the 
eqn.(3.4). The  transformation rule for these 
components is given by the  usual expression:
{}  
$$\gamma^{A}_{mn}(y^p_A) = {{\partial x^k}\over{\partial y^m_{A}}}
{{\partial x^l}\over{\partial y^n_{A}}}\hskip1mm 
\gamma_{kl}(x^s(y^p_A)). \eqno(C3)$$ 

\noindent Then, with the help of the  relations $(C1)$ 
the temporal-spatial components of the background metric could be 
presented as:
{}
$$\gamma^A_{0\alpha}(y^p_A) = \gamma^{A<1>}_{0\alpha}(y^p_A)+
\gamma^{A<3>}_{0\alpha}(y^p_A) + {\cal O}(c^{-5})=$$
{}
$$= v^\alpha_{A_0}(y^0_A)  + 
{\partial\over\partial y^\alpha_A} K_A (y^0_A,y^\nu_A)+$$
{}
$$ + {\partial\over\partial y^\alpha_A} L_A (y^0_A,y^\nu_A) + 
{\partial\over\partial y^\alpha_A} K_A(y^0_A,y^\nu_A)
 {\partial\over\partial y^0_A} K_A (y^0_A,y^\nu_A) + $$
{}
$$+
{v_{A_0}}_\nu  {\partial\over\partial y^\alpha_{A}}Q^\nu_{A} 
(y^0_A,y^\nu_A) + 
\gamma_{\alpha\nu}{\partial\over\partial y^0_{A}} Q^\nu_{A} 
(y^0_A,y^\nu_A) + {\cal O}(c^{-5}).\eqno(C4)$$

\noindent The expression $(C4)$ contains  the terms of two orders of 
magnitude: $c^{-1}$ and $c^{-3}$.
However,   as we discussed in Appendix B, in the post-Newtonian
 approximation  for any arbitrary {\small RF} one  expects 
that these components of the background metric tensor to be of
order:  $g_{0\alpha}(y^p_A)\sim {\cal O}(c^{-3})$. 
This gives the following condition for  the function $K_{A}$: 
{}
$$v^\alpha_{A_0}(y^0_{A}) + {\partial\over\partial y^\alpha_{A}} 
K_A(y^0_A,y^\nu_A) = {\cal O}(c^{-3}) \eqno(C5a)$$

\noindent Then, by formally integrating this last equation, we
may find the following expression for the  function $K_A$:
{}
$$K_{A} (y^0_A,y^\nu_A) = P_{A} (y^0_{A}) - 
 {v_{A_0}}_\mu   \cdot y^\mu_{A}  + 
{\cal O}(c^{-4}) y^0_{A} \eqno(C5b)$$

The result $(C5)$ considerably simplifies the calculation  of the 
transformation rules   between the different {\small RF}. Thus, taking 
into 
account the relation $(C3)$, one may obtain the following expression 
for the tensor $\gamma^A_{kl}(y^p_A)$: 
{}
$$\gamma^{A}_{00}(y^0_A, y^\nu_{A}) = 1 + 2 {\partial\over\partial 
y^0_{A}} K_{A} (y^0_A,y^\nu_A) + 
{v_{A_0}}_\beta v_{A_0}^\beta + $$   
{}
$$ + 2 {\partial\over\partial y^0_{A}} L_{A} (y^0_A,y^\nu_A) +
 \Big({{\partial\over\partial y^0_{A}}} K_{A} (y^0_A,y^\nu_A)\Big)^2+$$   
{}
$$+ 2 {v_{A_0}}_\beta {\partial\over\partial y^0_{A}} Q^\beta_{A}
 (y^0_A,y^\nu_A) + {\cal O}(c^{-6}), \eqno(C6a)$$
{}
$$\gamma^{A}_{0\alpha}(y^0_A, y^\nu_{A}) = {\partial\over
\partial y^\alpha_{A}} L_{A} (y^0_A,y^\nu_A) -
 {v_{A_0}}_\alpha {\partial\over\partial y^0_{A}} K_{A}
(y^0_A,y^\nu_A) + $$ 
{}
$$ + {v_{A_0}}_\nu (y^0_{A}){\partial\over\partial y^\alpha_{A}}
Q^\nu_{A} (y^0_A,y^\nu_A) + 
\gamma_{\alpha\nu}{\partial\over\partial y^0_{A}} Q^\nu_{A}
 (y^0_A,y^\nu_A) + {\cal O}(c^{-5}), \eqno(C6b)$$
{}
$$\gamma^{A}_{\alpha\beta}(y^0_A, y^\nu_{A}) = \gamma_{\alpha\beta} + 
{v_{A_0}}_\alpha  {v_{A_0}}_\beta  +$$
{}
$$+ \gamma_{\alpha\nu}{\partial\over\partial y^\beta_{A}} Q^\nu_{A} 
(y^0_A,y^\nu_A) +
 \gamma_{\beta\nu}{\partial\over\partial y^\alpha_{A}} Q^\nu_{A}
(y^0_A,y^\nu_A) + {\cal O}(c^{-4}). \eqno(C6c) $$

The relations $(C6)$ are the {\small KLQ} parametrization of the
 metric  $\gamma^{A}_{mn}$, which 
form the background  Riemann-flat space-time in the proper 
{\small RF}$_A$: 
{}
$$R_{klmn}(\gamma^{A}_{st})=0.$$ 

\noindent The functions $K_A, L_A \hbox{ and } Q^\alpha_A$ 
will be chosen in order to separate the forces 
of inertia  from the gravitational forces  which are measured by   
  the observer in this {\small RF}.
 
The  relations $(C5)$ are a useful tool for simplifying  
 the result  $(C2)$ as well. Thus, for determinant of the transformation
matrix
 we will get following expression: 
{}
$$det\Big[\lambda_{A m}^{\overline{p}}(y^p_A) \Big] = 
1 +  {\partial\over\partial y^0_{A}} K_{A} (y^0_A,y^\nu_A) + $$
{}
$$+   {v_{A_0}}_\lambda  v^\lambda_{A_0}   + 
{\partial\over\partial y^\mu_{A}} Q^\mu_{A} (y^0_A,y^\nu_A)  
+{\cal  O}(c^{-4}). \eqno(C7)$$

\pagebreak \noindent 
{\bf C.3. Inverse Transformation of the Coordinate Base Vectors.} 
\vskip 5pt
\noindent Using the transformation rule for the base vectors
$e^p_A=\partial/\partial y^p_A$ 
of the proper {\small RF}$_A$ to those $e^p=\partial/\partial
x^{\overline{p}}$ of the inertial
barycentric {\small RF}$_0$  
given by the expressions  (3.18), one easily obtains the inverse
transformation 
matrix $\lambda^n_{A \overline{m}}(x^p) ={{\partial y^n_{A}}/{\partial
x^m}}$
for this transition as well:
{}
$${{\partial y^0_{A}}
\over{\partial x^0}} = 1 - 
{\partial\over\partial x^0} K_{A} \Big(x^0, x^\nu -
 y^\nu_{A_0}(x^0)\Big)  - $$
{}
$$-{\partial\over\partial x^0} L_{A} \Big(x^0, x^\nu - 
y^\nu_{A_0}(x^0)\Big) + {\partial\over\partial x^0}\Big[{1\over2}
{\partial\over\partial x^0} K^2_{A} \Big(x^0, x^\nu -
 y^\nu_{A_0}(x^0)\Big)  - $$
{}
$$-{v_{A_0}}_\beta (x^0) \cdot Q^\beta_{A} \Big(x^0, x^\nu - 
y^\nu_{A_0}(x^0)\Big)\Big] + {\cal O}(c^{-6}), \eqno(C8a)$$
{}
$$ 
{\partial y^0_A\over\partial x^\alpha} = {v_{A_0}}_\alpha(x^0)- 
{\partial\over\partial x^\alpha}L_{A} 
\Big(x^0, x^\nu-y^\nu_{A_0}(x^0)\Big)+$$ 
{}
$$+ {\partial\over\partial x^\alpha}\Big[{1\over2}
{\partial\over\partial x^0} K^2_{A} \Big(x^0, x^\nu -
y^\nu_{A_0}(x^0)\Big) -$$
{}
$$ - {v_{A_0}}_\beta (x^0) \cdot Q^\beta_{A} \Big(x^0, x^\nu - 
y^\nu_{A_0}(x^0)\Big)\Big] + {\cal O}(c^{-5}), \eqno(C8b)$$
{}
$${{\partial y^\alpha_A}\over{\partial x^0}} = - v^\alpha_{A_0}(x^0) 
+ {\partial\over\partial x^0}\Big[ -Q^\alpha_{A}\Big(x^0, 
x^\nu - y^\nu_{A_0}(x^0)\Big) + $$
{}
$$+
{v^\alpha_{A_0}}(x^0) \cdot K_{A} \Big(x^0, x^\nu - 
y^\nu_{A_0}(x^0)\Big)\Big] + {\cal O}(c^{-5}), \eqno(C8c)$$
{}
$$
{{\partial y^\alpha_A}\over{\partial x^\beta}} =
 \delta^\alpha_\beta  + {\partial\over\partial x^\beta}\Big[- Q^\alpha_{A}
 \Big(x^0, x^\nu - y^\nu_{A_0}(x^0)\Big) +$$
{}
$$+ 
{v^\alpha_{A_0}}(x^0) \cdot K_{A} \Big(x^0, x^\nu - 
y^\nu_{A_0}(x^0)\Big)\Big] + {\cal O}(c^{-4}), \eqno(C8d)$$

\noindent where we partially have taken the result ($C5b$) into account
in a form of the relation: 
{}
$${\partial\over\partial x^\alpha} K_{A} \Big(x^0, x^\nu -
 y^\nu_{A_0}(x^0)\Big) = 
{\partial\over\partial y^\alpha_{A}} K_{A} (y^0_{A}, y^\nu_{A}) + 
O(c^{-3}) = - {v^\alpha_{A_0}}(x^0) + {\cal O}(c^{-3}).$$

\vskip 5pt\noindent 
{\bf C.4.  Mutual Transformation Between the Two Quasi-Inertial RFs.}
\vskip 5pt
\noindent  The expressions for the transformation of the base vectors 
between two quasi-inertial {\small RF} $(e^p_B)$ and $(e^p_A)$   may be 
obtained from the relations (3.19). These transformations are given as:
{}
$${{\partial y^0_{B}}\over{\partial y^0_{A}}} = 1 + 
{\partial\over\partial y^0_{A}} K_{BA} (y^0_{A}, y^\nu_A) +  
 {\partial\over\partial y^0_{A}} L_{BA} (y^0_{A}, y^\nu_{A}) + 
{\cal O}(c^{-6}), \eqno(C9a)$$
{}
$$ {\partial y^0_B\over\partial y^\alpha_A} = - 
{v_{BA_0}}_\alpha(y^0_A)+  
{\partial\over\partial y^\alpha_{A}} L_{BA} (y^0_{A}, y^\nu_{A})  + 
{\cal O}(c^{-5}), \eqno(C9b)$$
{}
$${ \partial y^\alpha_B \over \partial y^0_A} =  
v^\alpha_{BA_0}(y^0_{A}) +
 {\partial\over\partial y^0_{A}} Q^\alpha_{BA} (y^0_{A}, 
y^\nu_{A}) + {\cal O}(c^{-5}), \eqno(C9c)$$
{}
$${\partial y^\alpha_B\over\partial y^\beta_A} = 
\delta^\alpha_\beta  + 
{\partial\over\partial y^\beta_A} Q^\alpha_{BA} (y^0_A, 
y^\nu_A)+ {\cal O}(c^{-4}), \eqno(C9d)$$

\noindent where the functions $K_{BA},  L_{BA}$ and $ Q^\alpha_{BA}$
are defined by the expressions (3.20). 
>From these  expressions  $(C9)$,  one may obtain the determinant of 
the transformation matrix $\lambda_{BA m}^{\overline{p}}  
(y^p_A)$ for the  transformations between 
two different proper {\small RF}s as follows:
{}
$$det\Big[ \lambda_{BA m}^{\overline{p}} (y^p_A) \Big] = 1+  
 {\partial\over\partial y^0_{A}} K_{BA} (y^0_A,y^\nu_A) + $$
{}
$$+ v^\lambda_{{BA}_0}(y^0_A)  {v_{{BA}_0}}_\lambda(y^0_A)  + 
{\partial\over\partial y^\mu_A} Q^\mu_{BA} (y^0_A,y^\nu_A)  
+ {\cal O}(c^{-4}) \eqno(C10)$$
\noindent 
The condition $ det \Big[\lambda_{BA m}^{\overline{p}} 
(y^p_A)\Big] = 0$ gives  the boundary of 
validity  of these  transformations.

\vskip 11pt\noindent 
{\Large\bf Appendix D: Transformations of Some Physical 
Quantities \hfill 

\hskip 3cm and Solutions.} 
\vskip 11pt\noindent 
In this Appendix we will present the transformation laws 
 for the gauge conditions, the components of  
Ricci tensor, the components of an arbitrary  energy-momentum  
tensor of matter of the matter distribution 
 $T^{mn}$, for the unperturbed solutions of the field equations 
$h^{(0)}_{mn}$ and for the interaction term  $h^{int}_{mn}$ 
eqs.(3.1)-(3.4) under the general   coordinate transformations 
discussed in Section {\small 3} of this report.
 
\vskip 5pt \noindent
{\bf D.1. Transformation of the Gauge Conditions}.
\vskip 5pt 
\noindent  With the help of eqs.$(F1)$ and the  
 expansion of the metric tensor $g_{mn}$ given by eqs.$(B7)$, 
we may obtain the relations for the gauge conditions 
expanded in a power series of the small parameter $c^{-1}$.

\begin{itemize}
\item[(i).] In Cartesian coordinates of the inertial
  {\small RF}$_0$  the background space-time  may be taken in a simple
form
of Minkowski  metric: 
$\gamma^{(0)}_{mn} = ( 1, -1, -1, -1)$. Then the power expansion
 of the gauge condition eqs.$(B7)$ may be presented 
for $n = 0$ as follows:
{}
$${1\over2}{\partial \over \partial y^0_A} \Big(\gamma^
{\epsilon\nu} g^{<2>}_{\epsilon\nu}(x^p)- g^{<2>}_{00}(x^p)\Big) - 
 \gamma^{\epsilon\nu}{\partial \over \partial y^\epsilon_A} 
g^{<3>}_{0\nu}(x^p)  = {\cal O}(c^{-5}), \eqno(D1a) $$
{}
And for $n = \alpha$: 
{}
$${1\over2}\gamma^{\alpha\lambda}{\partial \over \partial 
y^\lambda_A} \Big(g^{<2>}_{00}(x^p) + 
\gamma^{ \epsilon\nu} g^{<2>}_{\epsilon\nu}(x^p)\Big) - $$
{}
$$- \gamma^{\epsilon\nu}\gamma^{\mu\alpha}{\partial \over 
\partial y^\epsilon_A} g^{<3>}_{\nu\mu}(x^p)
  =  {\cal O}(c^{-4}) \eqno(D1b)$$

\item[(ii).] Analogously one may obtain the 
expressions for the gauge conditions in a coordinates $(y^p_A)$ 
 of the proper {\small RF}$_A$ of body {\small (A)}. For $n = 0$:
{}
$${1\over2}{\partial \over \partial y^0_A} \Big(
 \gamma^{\epsilon\nu} g^{<2>}_{\epsilon\nu}(y^p_A) - 
g^{<2>}_{00}(y^p_A)\Big) - 
 \gamma^{\epsilon\nu}{\partial \over \partial y^\epsilon_A}
g^{<3>}_{0\nu}(y^p_A)+$$
{}
$$ + {\partial^2 K_{A}\over \partial {y^0_A}^2} + 
\gamma^{\nu\lambda}{\partial^2 L_A\over \partial y^\nu_A
\partial^\lambda_A}+ v_{A_0}{_{{}_\mu}}\Big({ a^{\mu}_{A_0} +
\gamma^{\nu\lambda}
{\partial^2 Q^\mu_{A}\over \partial y^\nu_A\partial^\lambda_A}}\Big)  = 
 {\cal O}(c^{-5}) \eqno(D2a) $$
\noindent  
For $n = \alpha$: 
{}
$${1\over2}\gamma^{\alpha\lambda}{\partial \over \partial
 y^\lambda_A} \Big(g^{<2>}_{00}(y^p_A) + 
 \gamma^{ \epsilon\nu} g^{<2>}_{\epsilon\nu}(y^p_A)\Big) - 
 \gamma^{\epsilon\nu}\gamma^{\mu\alpha}{\partial \over 
\partial y^\epsilon_A} g^{<3>}_{\nu\mu}(y^p_A)+$$ 
{}
$$ + \Big(a^{\alpha}_{A_0} + 
\gamma^{\nu\lambda}{\partial^2 Q^\alpha_A\over \partial y^\nu_A
\partial^\lambda_A}\Big)  = {\cal O}(c^{-4}). \eqno(D2b)$$ 
\end{itemize}

\vskip 5pt \noindent
{\bf D.2. Transformation  of the   Ricci Tensor $R_{mn}$.}
\vskip 5pt 

\noindent  With the help of the expansion for the components of 
Ricci tensor eqs.$(B9)$, one may 
obtain  those in coordinates of the different {\small RF}s.
\begin{itemize}
\item[ (i).] Thus, by  making of use the relations for 
the covariant de Donder gauge conditions in coordinates $(x^p)$ of 
inertial {\small RF}$_0$ eqs.$(D1)$  one may present the  components for  
Ricci tensor  in the following form:
{}
$$R_{00}(x^p) = - {1\over2}\gamma^{\nu\lambda}{\partial^2 
\over \partial x^\nu\partial x^\lambda} g^{<2>}_{00}(x^p) - 
{1\over2} \gamma^{\nu\lambda}{\partial^2 \over \partial
 x^\nu\partial x^\lambda} g^{<4>}_{00}(x^p) - $$
{}
$$- {1\over2} {\partial^2 \over \partial{x^0}^2} g^{<2>}_{00}(x^p) 
+ {1\over2} \gamma^{\lambda\mu}\gamma^{\nu\delta} 
g^{<2>}_{\lambda\nu}(x^p)
{\partial^2 \over \partial x^\mu\partial x^\delta} 
g^{<2>}_{00}(x^p) + $$
{}
$$+{1\over2} \gamma^{\lambda\nu}{\partial \over \partial 
x^\lambda} g^{<2>}_{00}(x^p)
{\partial \over \partial x^\nu} g^{<2>}_{00}(x^p) +
{\cal O}(c^{-6}),\eqno(D3a)$$
{}
$$R_{0\alpha}(x^p) = - {1\over2} \gamma^{\nu\lambda}
{\partial^2 \over \partial 
x^\nu\partial x^\lambda}g^{<3>}_{0\alpha}(x^p)  + 
 {\cal O}(c^{-5}),\eqno(D3b)$$
{}
$$R_{\alpha\beta}(x^p)= - {1\over2} \gamma^{\nu\lambda}
{\partial^2 \over \partial x^\nu\partial x^\lambda}
g^{<2>}_{\alpha\beta}(x^p) +   {\cal O}(c^{-4}).\eqno(D3c)$$

\item[ (ii).]  From the relations $(D2)$ and
 with the help of the expressions for		 
 Ricci tensor given by eqs.$(B9)$  one may get those in  
coordinates $(y^p_A)$ of the proper {\small RF}$_A$ as well: 
{}
$$R_{00}(y^p_A) = - {1\over2}\gamma^{\nu\lambda}{\partial^2
 \over \partial y^\nu_A \partial y^\lambda_A} g^{<2>}_{00}(y^p_A) - 
{1\over2} \gamma^{\nu\lambda}{\partial^2 \over \partial 
y^\nu_A\partial y^\lambda_A} g^{<4>}_{00}(y^p_A) - $$
{}
$$- {1\over2} {\partial^2 \over \partial {y^0_A}^2} 
g^{<2>}_{00}(y^p_A) + {1\over2} \gamma^{\lambda\mu}\gamma^{\nu\delta}
g^{<2>}_{\lambda\nu}(y^p_A)
{\partial^2 \over \partial y^\mu_A\partial y^\delta_A} 
g^{<2>}_{00}(y^p_A)+ $$
{}
$$+{1\over2} \gamma^{\lambda\nu}{\partial \over \partial
 y^\lambda_A} g^{<2>}_{00}(y^p_A)
{\partial \over \partial y^\nu_A} g^{<2>}_{00}(y^p_A) + $$
{}
$$ + {\partial \over \partial y^0_A}\Bigg({\partial^2 K_A
\over \partial {y^0_A}^2} + 
\gamma^{\nu\lambda}{\partial^2 L_{A} \over \partial 
y^\nu_A\partial y^\lambda_A}
+ v_{A_0}{_{{}_\mu}}\Big({ a^{\mu}_{A_0} + \gamma^{\nu\lambda}
{\partial^2 Q^\mu_{A}\over \partial y^\nu_A\partial y^\lambda_A}}
\Big)\Bigg) +  $$
{}
$$ + {1\over2} \Big({ a^{\mu}_{A_0} + \gamma^{\nu\lambda}
{\partial^2 Q^\mu_{A}\over \partial y^\nu_A\partial y^\lambda_A}}
\Big) \hskip 1mm
{\partial \over \partial y^\mu_A} g^{<2>}_{00}(y^p_A) + 
 {\cal O}(c^{-6}),\eqno(D4a)$$
{}
$$R_{0\alpha}(y^p_A) = - {1\over2} \gamma^{\nu\lambda}
{\partial^2 \over \partial 
y^\nu_A\partial y^\lambda_A}g^{<3>}_{0\alpha}(y^p_A) +  
{1\over2} \gamma_{\alpha\mu}{\partial \over \partial y^0_A}
\Big({ a^{\mu}_{A_0} + \gamma^{\nu\lambda}
{\partial^2 Q^\mu_{A}\over\partial y^\nu_A\partial y^\lambda_A}
}\Big) + $$
{}
$$ + {1\over2}{\partial \over \partial y^\alpha_A}
\Bigg({\partial^2 K_A\over \partial {y^0_A}^2} + 
\gamma^{\nu\lambda}{\partial^2 L_A\over \partial y^\nu_A
\partial y^\lambda_A}
+{v_{A_0}}_\mu\Big(a^\mu_{A_0}+\gamma^{\nu\lambda}
{\partial^2 Q^\mu_A\over \partial y^\nu_A\partial y^\lambda_A}
\Big)\Bigg)  +  {\cal O}(c^{-5}),\eqno(D4b)$$
{}
$$R_{\alpha\beta}(y^p_A) = - {1\over2} \gamma^{\nu\lambda}
{\partial^2 \over \partial y^\nu_A\partial y^\lambda_A}
g^{<2>}_{\alpha\beta}(y^p_A) + $$
{}
$$+ {1\over2}\Big(\gamma_{\mu\beta}{\partial \over \partial 
y^\alpha_A} + \gamma_{\mu\alpha}{\partial \over \partial
y^\beta_A}\Big)
\Big({ a^{\mu}_{A_0} + \gamma^{\nu\lambda}{\partial^2 Q^\mu_{A}\over
 \partial y^\nu_A\partial y^\lambda_A}}\Big) +
 {\cal O}(c^{-4}).\eqno(D4c)$$
\end{itemize}

\vskip 5pt \noindent 
{\bf D.3. Transformation Law for an Arbitrary 
Energy-Momentum Tensor  $T^{mn}$.}
\vskip 5pt 

\noindent In this part we will present the power expansion for the 
components of the $S_{mn} = T_{mn} - {1\over2} g_{mn} T$ 
defined by the equation  $(B12)$. 

\begin{itemize}
\item[(i).] By assuming that each body {\small (B)} in the system 
may be described by the reduced energy-momentum tensor $S^B_{mn}$, one
may
easily obtain the total energy-momentum tensor $S_{mn}$ for the entire
system. 
Thus, in the coordinates of the inertial {\small RF}$_0$ this tensor 
may be presented as follows:
{}
$$S_{mn}(x^p) = \sum_{B} {{\partial y^k_B}\over
{\partial x^m}}{{\partial y^l_B}\over{\partial x^n}}
\hskip 0.5mm S^B_{kl}(y^q_{B} (x^p)).$$

\noindent Then, with this relation above and from the eqs.$(B10)$-
$(B11),(B13)$ 
and eqs.$(C6)$ for the coordinate transformations to the barycentric
inertial {\small RF}$_0$, we will obtain the following result:
{}
$${S}_{00}(x^p) = \hskip 1mm {1\over2} \hskip 1mm \sum_{B}
\hskip 1mm \Bigg( T_{B}^{<0> 00}(y^q_{B}(x^p)) + 
T_B^{<2> 00}(y^q_B(x^p))+$$ 
{}
$$+ 2 \hskip 1mm g_{00}^{<2>}
(y^p_B(x^q))\cdot T_B^{<0> 00}(y^p_B(x^q))-  
\gamma_{\epsilon\nu}T_{B}^{<2> \epsilon\nu}(y^p_B(x^q))- $$
{}
$$-\Big[2{\partial\over\partial x^0}K_{B} \Big(x^0, x^\nu -
 y^\nu_{B_0}(x^0)\Big) + 
{v_{B_0}}_\nu (x^0) v^\nu_{B_0}(x^0)\Big] \cdot T_B^{<0> 00}
(y^q_B(x^p))-$$ 
{}
$$- 4 \hskip 1mm \gamma_{\epsilon\mu} v^\epsilon_{B_0}(x^0)
\hskip 1mm T_{B}^{<1> 0\mu}(y^q_B(x^p))\Bigg) 
 +  {\cal O}(c^{-4}), \eqno(D5a)$$
{}
$${S}_{0\alpha}(x^p) =   \sum_{B}\gamma_{\alpha\mu}\Bigg(
 T_{B}^{<1> 0\mu}(y^q_B(x^p)) 
+ v^\mu_{B_0}(x^0)\hskip 1mm  T_{B}^{<0> 00}(y^q_{B}(x^p))
\Bigg) +  {\cal O}(c^{-3}), \eqno(D5b) $$
{}
$${S}_{\alpha\beta}(x^p) = - {1\over2}\gamma_{\alpha\beta}
\sum_B \hskip 1mm T^{B  <0> 00}(y^q_{B}(x^p)) +  {\cal O}(c^{-2}).
 \eqno(D5c) $$

\item[(ii).] One may obtain the relation for the energy-momentum 
tensor of entire system in the coordinates of proper {\small RF}$_A$ as
follows:
{}
$$S_{mn}(y^p_{A}) = \sum_{B} {{\partial y^k_B}\over{\partial
y^m_A}}{{\partial y^l_B}
\over{\partial y^n_A}} \hskip 0.5mm S^B_{kl}(y^q_{B}(y^p_A)) 
=S^A_{mn}(y^p_A)+ \sum_{B\not=A} {{\partial y^k_B}\over{\partial
y^m_A}}{{\partial y^l_B}
\over{\partial y^n_A}} \hskip 0.5mm S^B_{kl}(y^q_{B}(y^p_A))$$

Then, making of use the formula above and 
from the eqs.$(B10)$-$(B11)$ and $(B13)$, but in this case eqs.$(C7)$, 
the expression for the quantity  $S_{mn}$ in coordinates 
$(y^p_{A})$ of the proper {\small RF}$_A$ may be presented as follows:
{}
$${S}_{00}(y^p_{A}) = {1\over2}\Bigg( T_A^{<0> 00}
(y^p_{A}) + T_A^{<2> 00}(y^p_{A}) +$$
{}
$$+ 2 \hskip 1mm g_{00}^{<2>}(y^p_{A}) \hskip 1mm T_{A}^{<0> 00}
(y^p_{A}) - 
\gamma_{\epsilon\nu}T_{A}^{<2> \epsilon\nu}(y^p_{A})\Bigg) + $$
{}
$$ + {1\over2}\sum_{B\not=A}\Bigg( T_B^{<0> 00}(y^q_{B}
(y^p_{A})) + T_B^{<2> 00}(y^q_B(y^p_{A})) + $$
{}
$$+2 g^{<2>}_{00}(y^q_B(y^p_{A}))\hskip 1mm T_{B}^{<0> 00}
(y^q_{B}(y^p_{A}))- 
\gamma_{\epsilon\nu}T_B^{<2> \epsilon\nu}(y^q_B(y^p_{A})) + $$
{}
$$ + \Big[ 2 {\partial\over\partial y^0_A} K_{BA} (y^0_{A}, y^\nu_A)  - 
 v^\nu_{BA_0}(y^0_A){v_{BA_0}}_\nu (y^0_A)\Big] 
\cdot T_B^{<0> 00}(y^q_B(y^p_A)) +$$
{}
$$ + 4 \gamma_{\epsilon\mu} v^\epsilon_{BA_0}(y^0_{A}) \hskip 1mm 
T_B^{<1> 0\mu}(y^q_{B}(y^p_{A})) \Bigg) +  {\cal O}(c^{-4}),
 \eqno(D6a)$$
{}
$$S_{0\alpha}(y^p_{A}) =  \gamma_{\alpha\mu}T_A^
{<1> 0\mu}(y^p_{A}) +$$ 
{}
$$+\sum_{B\not=A}\gamma_{\alpha\mu}\Bigg( T_B^{<1> 0\mu}
(y^q_{B}(y^p_{A}))
{} - v^\mu_{BA_0}(y^0_A) T_B^{<0> 00}(y^q_{B}(y^p_A))
\Bigg) +  {\cal O}(c^{-3}), \eqno(D6b)$$
{}
$$S_{\alpha\beta}(y^p_A) = - {1\over2}\gamma_
{\alpha\beta}\Bigg( T_A^{<0> 00}(y^p_A) + 
\sum_{B\not=A} T_B^{<0> 00}(y^q_{B}(y^p_A))
 \Bigg) + {\cal O}(c^{-2}). \eqno(D6c)$$
\end{itemize}

\vskip 5pt \noindent
{\bf D.4. Transformation of the Unperturbed Solutions $h^{(0)}_{mn}$.} 
\vskip 5pt 
\noindent In this chapter we will obtain the transformation rules for  
unperturbed solutions.
\begin{itemize}
\item[(i).]	Using the following notation for the second  term in the
expression (3.1):
{}
$$H^{(0)}_{mn}(x^p) = \sum_B{{\partial y^k_B}\over
{\partial x^m}}{{\partial y^l_B}\over{\partial x^n}}
\hskip 0.5mm h^{(0) B}_{kl} (y^q_B (x^p)),$$

\noindent from the equations eqs.$(C6)$, we will obtain the 
relations for the components $H^{(0)}_{mn}(x^p)$ in coordinates of the
inertial {\small RF}$_0$ as follows:
{}
$$H^{(0)}_{00}(x^p) = \sum_{B} \Bigg(h^{(0) <2>}_{B 00} 
(y^q_B(x^p)) + h^{(0) <4>}_{B 00} (y^q_{B} (x^p)) -$$
{}
$$- 2{\partial\over\partial x^0}K_B\Big(x^0, x^\nu - 
y^\nu_{B_0}(x^0)\Big) \cdot  h^{(0) <2>}_{B 00} (y^q_{B} (x^p)) -$$
{}
$$ - 2 v^\epsilon_{B_0}(x^0)  \cdot h^{(0) <3>}_{B0\epsilon} (y^q_{B} (x^p))
+ 
v^\epsilon_{B_0}(x^0)v^\nu_{B_0}(x^0) \cdot  h^{(0) <2>}_
{B\epsilon\nu} (y^q_{B} (x^p))\Bigg)  + {\cal O}(c^{-6}), \eqno(D7a)$$
{}
$$H^{(0)}_{0\alpha}(x^p) = \sum_{B} \Bigg(h^{(0) <3>}_
{B 0\alpha}(y^q_B(x^p)) + $$
{}
$$ + {v_{B_0}}_\alpha(x^0) \cdot h^{(0) <2>}_{B 00}(y^q_{B}
 (x^p)) - 
v^\epsilon_{B_0}(x^0) \cdot  h^{(0) <2>}_{B\epsilon\alpha} 
 (y^q_{B} (x^p))\Bigg)  + {\cal O}(c^{-5}), \eqno(D7b)$$ 
{}
$$H^{(0)}_{\alpha\beta}(x^p) = \sum_{B} h^{(0) <2>}_
{B\alpha\beta} (y^q_{B}(x^p))  +  {\cal O}(c^{-4}). \eqno(D7c)$$

\item[(ii).] The transformed components of $H^{(0)}_{mn}(y^p_A)$
 in coordinates $(y^p_A)$ of the proper {\small RF}$_A$ are defined as
in the eqn.(3.4):
{}
$$H^{(0)}_{mn}(y^p_{A}) = {{\partial x^k}\over{\partial y^m_A}}
{{\partial x^l}\over{\partial y^n_A}}H^{(0)}_{kl}(x^q(y^p_A)) =
\sum_{B} {{\partial y^k_B}\over{\partial y^m_A}}{{\partial y^l_B}
\over{\partial y^n_A}} \hskip 0.5mm h^{(0) B}_{kl} (y^q_{B}
 (y^p_A)) = $$
{}
$$ = h^{(0) B}_{mn} (y^p_{A}) + \sum_{B\not=A} {{\partial y^k_B}
\over{\partial y^m_A}}{{\partial y^l_B}\over{\partial y^n_A}}
\hskip 0.5mm h^{(0) B}_{kl} (y^q_{B} (y^p_A)).$$

\noindent 
Then, for these components,   from the relations 
 $(C7)$ one may obtain the following result:
{}
$$H^{(0)}_{00}(y^p_A) = h^{(0) <2>}_{A 00} (y^p_A) + 
H^{(0) <4>}_{A 00}(y^p_A)+$$ 
{}
$$ + \sum_{B\not=A} \Bigg(h^{(0) <2>}_{B 00} (y^q_{B}(y^p_A)) + 
h^{(0) <4>}_{B 00} (y^q_{B} (y^p_A)) + $$
{}
$$+ 2 {\partial\over\partial y^0_A} K_{BA}(y^0_A, 
y^\nu_A)\cdot h^{(0) <2>}_{B 00} (y^q_{B} (x^p)) + 
2  v^\epsilon_{BA_0}(y^0_A) \cdot h^{(0) <3>}_{B 0\epsilon}
(y^q_{B} (y^p_A)) + $$
{}
$$+v^\epsilon_{BA_0}(y^0_A)v^\nu_{BA_0}(y^0_A)\cdot h^{(0) <2>}_
{B \epsilon\nu} (y^q_B (y^p_A))\Bigg) +  {\cal O}(c^{-6}), \eqno(D8a)$$ 
{}
$$H^{(0)}_{0\alpha}(y^p_A) = h^{(0) <3>}_{A 0\alpha} (y^q_{A}) + 
\sum_{B\not=A} \Bigg(h^{(0) <3>}_{B 0\alpha}(y^q_{B}(y^p_A)) - $$
{}
$$- {v_{BA_0}}_\alpha(y^0_A) \cdot h^{(0) <2>}_{B 00} (y^q_{B} (y^p_A)) + 
v^\epsilon_{BA_0}(y^0_A) \cdot h^{(0) <2>}_{B \epsilon\alpha} (y^q_{B} 
(y^p_A)) \Bigg) + {\cal O}(c^{-5}), \eqno(D8b)$$ 
{}
$$H^{(0)}_{\alpha\beta}(y^p_A) = h^{(0) <2>}_{A \alpha\beta}(y^p_A) + 
\sum_{B\not=A} h^{(0) <2>}_{B \alpha\beta} (y^q_B(y^p_A)) +  {\cal O}(c^{-
4}).
  \eqno(D8c)$$ 

\end{itemize}
 \vskip 5pt \noindent
{\bf D.5. Transformation Rules for the Interaction Term $h^{int}_{mn}$.}
\vskip 5pt 

\noindent  The components of the interaction 
term $h^{int}_{mn}(x^s(y^p_{A}))$ in coordinates $(y^p_A)$ 
of the   proper {\small RF}$_A$ are given as follows:
{}
$$ h^{int}_{mn}(y^p_A) = {{\partial x^k}\over{\partial y^m_{A}}}
{{\partial x^l}\over{\partial y^n_{A}}}
\hskip1mm h^{int}_{mn}(x^s(y^p_{A})). \eqno(D9a)$$

\noindent 
By making of use the expressions $(C6)$, the components of
 $h^{int}_{mn}$  will take the  following form:
{}
$$ h^{int}_{00}(y^p_A) = h^{int <4>}_{00}(x^s(y^p_A)) + 
{\cal O}(c^{-6}),\eqno(D9b)$$
{}
$$ h^{int}_{0\alpha}(y^p_A) = O(c^{-5}),
  \hskip 8mm h^{int}_{\alpha\beta}(y^p_A) =  {\cal O}(c^{-4}).  \eqno(D9c)$$

\vskip 5pt\noindent  {\bf D.6. Transformation for 
the  Energy-Momentum Tensor of a Perfect Fluid}.
\vskip 5pt 
 
\noindent Let us define the model of the matter distribution of a body
{\small (B)} 
in its proper {\small RF}$_B$ by the tensor density ${\hat T}^{mn}_B$
given by
{}
$${\hat T}^{mn}_B(y^p_B)=\sqrt{-g}\Big(\Big[\rho_{B_0}(1+\Pi)+p\Big]
u^mu^n-pg^{mn}\Big),\eqno(D10)$$
 
\noindent where all the quantities entering the formula 
above are calculated in the coordinates $(y^p_B)$
of the non-inertial proper {\small RF}$_B$.
 Then, one may obtain the following post-Newtonian 
expansion of the tensor $T^{mn}_B$ in the  
coordinates  $(y^p_B)$ of the  proper {\small RF}$_B$: 
{}
$$T^{00}(y^p_B) =  \rho_{B_0}\Big[1 - v_\mu v^\mu + \Pi
 + 2\Big(\sum_C U_C-{\partial K_B\over \partial y^0_B} -
 {1\over2}{v_{B_0}}_\mu v^\mu_{B_0}\Big)+{\cal O}(c^{-4})\Big],
\eqno(D11a)$$ 
{}
$$T^{0\alpha}(y^p_B) =  \rho_{B_0}v^\alpha\Big[ 1 - v_\mu v^\mu + \Pi +  
2\Big(\sum_C U_C-{\partial K_B\over \partial y^0_B} -
 {1\over2}{v_{B_0}}_\mu v^\mu_{B_0}\Big)+ {p\over\rho_{B_0}} + 
 {\cal O}(c^{-4})\Big],\eqno(D11b)$$
{}
$$T^{\alpha\beta}(y^p_B)  =  \rho_{B_0} v^\alpha v^\beta - 
p \gamma^{\alpha\beta} + \rho  {\cal O}(c^{-4}).\eqno(D11c)$$ 
 
Then, by  using  these relations one may easily obtain the 
expressions for the right-hand side of the Hilbert-Einstein 
gravitational field equations 
in a form of the  quantity $S_{mn}$ defined by eqs.(4.1),$(B13)$ and
 eqs.$(D5)$-$(D6)$. 

\begin{itemize}
\item[(i).] From the eqs.$(D5)$ and with the 
help of the expressions eqs.$(D10)$ we 
will obtain components of the quantity $S_{mn}$  in coordinates $(x^p)$ 
of  barycentric inertial {\small RF}$_0$ as follows: 
{}
$$S_{00}(x^0, x^\nu) = {1\over2} \sum_B \rho_{B_0}
\big(y^q_B(x^p)\big)\times$$
{}
$$\times\Big[ \hskip 1mm 1 \hskip 1mm + 
\hskip 1mm  \Pi\hskip 1mm  - \hskip 1mm 2\sum_{B'}U_{B'} 
 - 2 v_\mu(x^p)v^\mu(x^p) + {3 p\over \rho} +  {\cal O}(c^{-4})\Big],
\eqno(D12a)$$
{}
$$S_{0\alpha}(x^0, x^\nu) = 
\gamma_{\alpha\epsilon} \sum_B \rho_B\big(x^0, x^\nu - 
y^\nu_{B_0}(x^0)\big)\Big[v^\epsilon(x^p) + {\cal O}(c^{-
3})\Big],\eqno(D12b)$$
{}
$$S_{\alpha\beta}(x^0, x^\nu) = - {1\over2}\gamma_{\alpha\beta}
 \sum_B \rho_B\big(x^0, x^\nu - y^\nu_{B_0}(x^0)\big)\Big[ 1 + 
  {\cal O}(c^{-2})\Big],\eqno(D12c)$$
\noindent where the total mass density of the system was denoted as:
{}
$$\rho=\sum_B\rho_{B_0}.$$ 

\item[(ii).] Analogously, but with the help of the
 expressions eqs.$(D6)$, we may get the  
relations for tensor $S_{mn}$ in the coordinates $(y^p_A)$ 
of proper {\small RF}$_A$:
{}
$$S_{00}(y^0_A, y^\nu_A) = {1\over2} \sum_B \rho_B
\big(y^q_B(y^p_A)\big) \Big[\hskip 1mm 1 \hskip 1mm + 
\hskip 1mm  \Pi \hskip 1mm -\hskip 1mm 2\sum_{B'}U_{B'} +$$
{}
$$+2 {\partial  K_A\over \partial y^0_A} - 
2 v_\mu(y^p_A) v^\mu(y^p_A) + {v_{A_0}}_\mu v^\mu_{A_0} + 
{3 p\over \rho} +  {\cal O}(c^{-4})\Big], \eqno(D13a)$$
{}
$$S_{0\alpha}(y^0_A, y^\nu_A) = 
\gamma_{\alpha\epsilon} \sum_B \rho_B\big(y^0_A, y^\nu_A +
 y^\nu_{BA_0}(y^0_A)\big)\Big[v^\epsilon(y^p_A) +
 {\cal O}(c^{-3})\Big],\eqno(D13b)$$
{}
$$S_{\alpha\beta}(y^0_A, y^\nu_A) = - {1\over2}\gamma_{\alpha\beta}
 \sum_B \rho_B\big(y^0_A, y^\nu_A + 
y^\nu_{BA_0}(y^0_A)\big)\Big[1+{\cal O}(c^{-2})\Big]. \eqno(D13c)$$
\end{itemize} 

It should be noted  that the functional dependance of the 
densities in the expressions $(D12)$-$(D13)$ reflects the positions of all
the bodies with  respect to  different  {\small RF}s  in the sense of 
the Dirac's delta function.

\vskip 14pt
\noindent{\Large\bf Appendix E: Transformations of the Gravitational
Potentials.} 
\vskip 11pt \noindent
  To establish the transformation properties 
of the unperturbed solutions for $h^{(0)A}_{mn}$ (given in  Appendix A) 
for  transitions from the coordinates $(y^p_A)$ of the 
proper {\small RF}$_A$ to those of 
the barycentric {\small RF}$_0$ (and backwards), one should take into
account  that
 these solutions   contain  the integrals over the 
3-volumes of the bodies. 
Because of this reason, we should  first derive the 
transformation laws for generalized gravitational potentials. 
The powerful technique for obtaining these  rules was elaborated 
for some special cases of transformations earlier by 
 Chandrasekhar \& Contopulos (1967) (see also,  
Brumberg \& Kopejkin, 1988a; Will, 1993). It was noted 
that the transformation of the integrands should 
include the point transformation combined 
with the Lie transfer from one hypersurface to another.
 This transfer should be produced along the integral curves of 
the vector field of 
the body's matter four-velocity. The most sophisticated 
transformation at the post-Newtonian level is required for 
the Newtonian potential $U_B$.   We will extend 
this technique to the general case of the coordinate 
transformations which was discussed in Section {\small 3} 
and in   Appendix C. 

For the transformation from the proper {\small RF}$_A$ 
to the barycentric one {\small RF}$_0$ with the help of expressions (3.18)
one may 
establish the relationship between the observer's spatial coordinates and
those 
of the integrating point  as follows:
{} 
$${1\over |y^\nu_B - y'^\nu_B|} = 
{1\over |x^\nu - x'^\nu|} \cdot \Bigg( 1  - 
 v_\beta(x^0, x'^\nu) {v_{B_0}}_\lambda(x^0)
{(x^\beta - x'^\beta) (x^\lambda - x'^\lambda) 
\over{|x^\nu - x'^\nu|^2}} + $$  
{}
$$ + \Big[ Q^\beta_B \Big(x^0, x^\nu-
y^\nu_{B_0}(x^0)\Big) - Q^\beta_{B} \Big(x^0, 
x'^\nu-y^\nu_{B_0}(x^0)\Big)\Big]\cdot
{(x_\beta - x'_\beta)\over{|x^\nu - x'^\nu|^2}}+
{\cal O} (c^{-4})\Bigg). \eqno(E1a)$$

\noindent By using the same procedure as above, from
eqs.(3.19)-(3.20) we may obtain the expression for 
the observer's spatial 
coordinates and those of the integrating point while the 
transformation  between two proper {\small RF}s (corresponding two 
the bodies {\small (A)} and {\small (B)}) is being performed: 
{}
$${1\over |y^\nu_B - y'^\nu_B|} = 
{1\over |y^\nu_A - y'^\nu_A|} \cdot \Bigg[ 1  + $$
{}
$$+ \left({v_{A_0}}_\beta(y^0_A) + v_\beta(y^0_A, 
y'^\nu_A) \right){v_{BA_0}}_\lambda(y^0_A) \cdot
 {(y^\beta_A - y'^\beta_A) (y^\lambda_A - y'^\lambda_A) 
\over  |y^\nu_A - y'^\nu_A|^2 } + $$  
{}
$$ + \Bigg(\Big[Q^\beta_{A} (y^0_A, y^\nu_A) - Q^\beta_{A}
(y^0_A, y'^\nu_A)\Big]-\Big[Q^\beta_B\Big(y^0_A+y^\nu_{BA_0} 
(y^0_A)\Big) - $$
{}
$$-Q^\beta_B\Big(y^0_A,y'^\nu_A+ 
 y^\nu_{BA_0}(y^0_A)\Big)\Big]\Bigg) \cdot
 {(y^\beta_A - y'^\beta_A) \over{|y^\nu_A - 
y'^\nu_A|^2}}+    {\cal O}(c^{-4})\Bigg]. \eqno(E1b)$$

For the transformation of the integrand we should 
take into account the property of the invariant 
elementary volume (Kopejkin, 1988; Will, 1993):
{}
$$d^3y'_B \cdot \sqrt{-g(y^p_B)} u^0(y^p_B)= 
d^3x'  \cdot {\sqrt{-g(x^p)}u^0(x^p)}, \eqno(E2) $$
\noindent where $\sqrt{-g}$  is the determinant of the metric 
tensor and $u^0$   is the temporal component of 
the invariant four-velocity. 

>From the expressions $(B2)$ and the components of the metric tensor of
order 
$\sim c^{-2}$ in the different {\small RF}s (given by (4.8) and (4.11)) we 
will get: 
{}
$$\sqrt{-g(x^p)} = 1 + 2\sum_B U_B (y^q_B(x^p))+
   {\cal O}(c^{-4}), \eqno(E3a)$$
\noindent and 
{}
$$\sqrt{-g(y^p_A)} = 1 + {\partial\over \partial y^0_A}
 K_{A} (y^p_A) + 
{v_{A_0}}_\beta(y^0_A)  v_{A_0}^\beta(y^0_A)+$$
$$+{\partial\over\partial y^\beta_A}Q^\beta_A(y^p_A) +
2\sum_B U_B(y^q_B(y^p_A))+{\cal O}(c^{-4}). \eqno(E3b)$$   
\noindent
The components of the  invariant four-velocity are 
defined as follows: 
{}
$$u^k(z^p) = v^k(z^p)\Big[ g_{00}(z^p) + 2 g_{0\epsilon}
(z^p) v^\epsilon(z^p) +
 g_{\nu\epsilon}(z^p)v^\epsilon (z^p)v^\nu(z^p) \Big]^{-1/2},
 \eqno(E4)$$

\noindent where $v^k(z^p)=dz^k_B/dz^0=
(1, \dot{z}^\epsilon_B).$ From this last expression 
and eqs.(4.8) and (4.11) 
one may obtain the relations  for the 
component $u^0$ in the coordinates of the barycentric and 
the observer's proper  {\small RF}  as follows:
{}
$$u^0(x^p) = 1 + \sum_B U_B (y^p_B(x^p))- 
{1\over2}{v_\beta(x^p)} v^\beta(x^p)+   {\cal O}(c^{-4}), 
\eqno(E5a)$$
\noindent and
{}
$$u^0(y^p_A) = 1 + \sum_B U_B\big(y^q_B(y^p_A)\big)- 
 {1\over2}{v_\beta(y^p_A)}{v^\beta(y^p_A)} - $$
{}
$$-{1\over2}{v_{A_0}}_\beta(y^0_{A})  v_{A_0}^\beta(y^0_{A})
-{\partial\over\partial y^0_A} K_A(y^p_A)+  
 {\cal O}(c^{-4}). \eqno(E5b)$$
\noindent 
Then making of use the expression eqs.$(E3),(E5)$
we will have 
{}
$$\sqrt{-g(x^p)}u^0(x^p) = 1 + 3\sum_B U_B  
(y^p_B(x^p))-{1\over2}{v_\beta(x^p)} v^\beta(x^p)
 +    {\cal O}(c^{-4}) \eqno(E6a)$$ 
\noindent and
{}
$$\sqrt{-g(y^p_A)} u^0(y^p_A) = 1 + 3\sum_B U_B
(y^k_B(y^p_A))-{1\over 2}{v_\beta(y^p_A)}{v^\beta(y^p_A)} + $$
{}
$$+{\partial\over\partial 
y^\beta_A}Q^\beta_A (y^p_A) + 
{1\over2}{v_{A_0}}_\beta(y^0_A)  v_{A_0}^\beta(y^0_A)
  + {\cal O}(c^{-4}).
\eqno(E6b)$$

>From the relation $(E2)$ the following transformation 
laws for the elementary volume may be established:
{}
$$d^3y'_B = d^3x'  
{\sqrt{-g(x^p)}u^0(x^p)\over \sqrt{-g(y^k_B(x^p))}
 u^0(y^k_B(x^p))} = $$ 
{}
$$ = d^3x'  \Bigg(1 - {v_{B_0}}_\beta(x^0) 
 v^\beta(x^0, x'^\nu) - 
{\partial\over \partial x'^\beta}Q^\beta_{B} 
\Big(x^0, x'^\nu-y^\nu_{B_0}(x^0)\Big) + 
   {\cal O}(c^{-4})\Bigg), \eqno(E7 )$$ 
\noindent and 
{}
$$d^3y'_B = d^3y'_A  {\sqrt{-g(y^p_A)}
u^0(y^p_A)\over \sqrt{-g(y^k_B(y^p_A))} u^0(y^k_B(y^p_A))} = $$ 
{}
$$ = d^3y'_A  \Bigg(1 + {v_{BA_0}}_\beta(y^0_A) 
 \Big( v^\beta_{A_0}(y^p_{A_0}) + v^\beta
(y^0_A, y'^\nu_A)\Big) + $$
{}
$$+ 
{\partial\over \partial y'^\beta_A}\Big(Q^\beta_{A} 
(y^0_A, y'^\nu_A)- Q^\beta_{B} \Big(y^0_A, y'^\nu_A + 
y^\nu_{BA_0}(y^0_A)\Big)\Big)
 +   {\cal O}(C^{-4})\Bigg). \eqno(E7b)$$ 

Since the quantities $ \rho_B(z^0, z^\nu), \Pi(z^0, z^\nu)$ 
and $p(z^0, z^\nu)$ from the potentials defined 
in Appendix A are all measured 
in the co-moving {\it local} quasi-inertial frames, 
they are transformed as  scalars and for any given 
element of fluid  the following relations are hold:
{}
 $$  \rho_B(x^0, x^\nu) = 
\rho_B(y^k_B(x^p)),\qquad  \Pi(x^0, x^\nu) =  
\Pi(y^k_B(x^p)), \qquad
  p(x^0, x^\nu) = p(y^k_B(x^p)). \eqno(E8)$$

Finally, the  expressions  $(E1),(E7)$-$(E8)$ will enable  
one to present the transformation law for the 
Newtonian potential as follows:
{}
$$U_B(y^0_B,y^\nu_B) = U_B(x^0, x^\nu) + 
v_{B_0}^\beta (x^0) 
\cdot {\partial^2\over\partial x^0 \partial
 x^\beta}\chi_B(x^0, x^\nu)+$$
{}
$$+\int_B d^3x' \rho_B\Big(x^0, x'^\nu -
 y^\nu_{B_0}(x^0)\Big) {\partial \over\partial x'^\lambda}
\Big[{ Q^\lambda_{B} \Big(x^0, x^\nu- 
y^\nu_{B_0}(x^0)\Big) - Q^\lambda_B
\Big(x^0, x'^\nu-y^\nu_{B_0}(x^0)\Big)
\over{|x^\nu - x'^\nu|}}\Big] +$$
{}
$$+ {\cal O}(c^{-6}), \eqno(E9a)$$
\noindent and 
{}
$$U_B(y^0_B,y^\nu_B) = U_B(y^0_A, y^\nu_A)-$$
{}
$$- v_{BA_0}^\beta (y^0_A) \Bigg({\partial\over\partial y^0_A} -
 v_{A_0}^\mu (y^0_A) {\partial\over\partial y^\mu_A}\Bigg)
{\partial\over \partial y^\beta_A}\chi_B\Big(y^0_A, 
y^\nu_A \Big)-$$  
{}
$$-\int_Bd^3y'_A
\rho_B\Big(y^0_A, y'^\nu_A +
 y^\nu_{BA_0}(y^0_A)\Big) {\partial \over\partial y'^\lambda_A} 
\Bigg( \Big[{ Q^\lambda_{A} (y^0_A, y^\nu_A) - 
Q^\lambda_{A} (y^0_A, y'^\nu_A) \over{|y^\nu_A - y'^\nu_A|}}\Big] -$$
{}
$$-  \Big[{ Q^\lambda_B \Big(y^0_A, y^\nu_A+ 
y^\nu_{BA_0}(y^0_A)\Big) - 
Q^\lambda_B\Big(y^0_A, y'^\nu_A+y^\nu_{BA_0}(y^0_A)\Big)
\over{|y^\nu_A - y'^\nu_A|}}  \Big]\Bigg) +  
  {\cal O}(c^{-6}). \eqno(E9b)$$

\noindent The Newtonian potential and the super-potential in the formulae
above 
are given as follows:
{}
$$U_B(z^0, z^\nu) = \int_B {d^3z'  \over |z^\nu -z'^\nu|}  
\hskip 1mm \rho_B(y'^q_B(z'^p)) +   {\cal O}(c^{-6})\eqno(E9c)$$
{}
$$\chi_B (z^0, z^\nu) = - \int_B d^3z'  \rho_B(y'^q_B(z'^p))
  \cdot |z^\nu -z'^\nu| +   {\cal O}(c^{-4})L_B^2.\eqno(E9d)$$
\noindent where $L_B$ is the proper dimensions of the body {\small (B)}.
 
In order to establish the transformation properties for the potentials: 
{}
$$V^\alpha_B(z^0, z^\nu), \qquad \Phi_{1B}
(z^0, z^\nu) \quad \hbox{ and } \quad {\partial^2\over\partial {z^0}^2}
 \chi_B(z^0, z^\nu),$$ 

\noindent one should find the transformation rules for the spatial
 components of the four-velocity $u^k(z^p)$ while 
transiting between {\small RF}s. 
Let $u^k(x^p) $ and $u^k(y^p_A)$ are four-velocities of 
matter measured in two different {\small RF}s under consideration.
Since they are related by the usual tensorial law:
{}
$$u^m(y^p_B)=u^k(x^p){\partial y^m_B\over    
\partial x^k} \hskip 10mm
\Rightarrow  \hskip 10mm{dy^m_B \over ds}  = 
{\partial y^m_B \over \partial x^k} {dx^k\over ds},
 \eqno(E10)$$ 
{}
\noindent the following expression for the transformation 
of the invariant four-velocity may be obtained:
{}
$$u^0(y^p_B) = {dy^0_B \over ds}  = {\partial y^0_B
 \over \partial x^k} {dx^k\over ds}, \eqno(E11a)$$ 
{}
$$ u^\epsilon(y^p_B)= {dy^\epsilon_B\over ds}= 
u^0(y^p_B)v^\epsilon (y^p_B)=   
{\partial y^\epsilon_B \over \partial x^k} 
{dx^k\over ds}. \eqno(E11b)$$
 
\noindent The last two equations are providing one with the 
result for transformation of the 3-velocity as follows:
{}
$$v^\epsilon(y^p_B) = {u^0(x^q)\over u^0(y^p_B(x^q))}
\Big({\partial y^\epsilon_B \over \partial x^0} + 
v^\nu(x^p) {\partial y^\epsilon_B \over \partial x^\nu}  
\Big). \eqno(E12)$$ 
 
\noindent By collecting together the expressions $(E5)$ and $(C8)$
and substituting those into eq.$(E12)$, one may get the
relation between the components of the velocity while
the  transformation from the proper {\small RF}$_A$ to the barycentric
 one {\small RF}$_0$ is performed.
With the required accuracy this result may be 
written as follows:
{}
$$v^\alpha(y^q_B(x^p)) = v^\alpha(x^p) - 
v^\alpha_{B_0}(x^0) +    {\cal O}(c^{-3}). \eqno(E13a)$$ 
\noindent
Analogously, but with the help of the equation $(C9)$,  one obtains the
relations for velocities
 in two different proper {\small RF}s:
{}
$$v^\alpha(y^q_B(y^p_A)) = v^\alpha(y^p_A) + 
v^\epsilon_{BA_0}(y^0_A)+ \Big({\partial \over \partial y^0_A}
+v^\lambda(y^p_A){\partial \over \partial
y^\lambda_A}\Big)\Big[Q^\alpha_A(y^p_A)-
Q^\alpha_B(y^p_A)\Big]-$$
{}
$$-\Big(v^\alpha(y^p_A) + v^\epsilon_{BA_0}(y^0_A)\Big)
\Bigg({\partial \over \partial y^0_A}\Big[K_A(y^p_A)-
K_B(y^p_A)\Big]-v^\lambda(y^p_A) {v_{BA_0}}_\lambda(y^0_A)\Bigg)-$$
{}
$$-a^\alpha_{B_0}(y^0_A)\Big[K_A(y^p_A)-
K_B(y^p_A)\Big]+{\cal O}(c^{-5}). \eqno(E13b)$$
\noindent
Then, based on  the expressions $(E1)$ and $(E13)$,
we may get the   expression for the 
transformation law for the vector-potential $V^\epsilon_B$:
{}
$$V^\epsilon_B(y^0_B, y^\nu_B) = V^\epsilon_B(x^0, x^\nu) 
 + v^\epsilon_{B_0}(x^0) \cdot U_B(x^0, x^\nu) + 
   {\cal O}(c^{-5}),\eqno(E14a)$$
\noindent and
{}
$$V^\epsilon_B(y^0_B, y^\nu_B) = V^\epsilon_B(y^0_A,y^\nu_A)
 - v^\epsilon_{BA_0}(y^0_A) \cdot U_B(y^0_A, y^\nu_A)
 +    {\cal O}(c^{-5}).\eqno(E14b)$$
\noindent
>From the expressions  $(E1)$ and $(E13)$ we will obtain
 the relation for the potential $\Phi_{1B}$:
{}
$$\Phi_{1B}(y^0_B, y^\nu_B) = \Phi_{1B}(x^0, x^\nu) +
4{v_{B_0}}_\epsilon (x^0)\cdot V^\epsilon_B (x^0,x^\nu)+$$ 
{}
$$+2 {v_{B_0}}_\epsilon (x^0) v^\epsilon_{B_0}(x^0) \cdot
 U_B(x^0, x^\nu)+{\cal O}(c^{-6}),\eqno(E15a)$$
\noindent and
{}
$$\Phi_{1B}(y^0_B, y^\nu_B) = \Phi_{1B}(y^0_A, y^\nu_A) -
 4 {v_{BA_0}}_\epsilon (y^0_A) \cdot V^\epsilon_B(y^0_A,y^\nu_A) +$$ 
{}
$$+2 {v_{BA_0}}_\epsilon (y^0_A) v^\epsilon_{BA_0}(y^0_A) 
\cdot U_B(y^0_A, y^\nu_A) +   {\cal O}(c^{-6}).\eqno(E15b)$$
{}
Finally, for the transformation of the superpotential $\chi_B$
from $(E1),(E12)$ and $(C8)$-$(C9)$ one  obtains: 
{}
$${\partial^2\over\partial {y^0_B}^2} \chi_B(y^0_B, 
y^\nu_B) =  {\partial^2\over\partial {x^0}^2} 
\chi_B(x^0,x^\nu) + 
a^\epsilon_{B_0}(x^0){\partial \over\partial x^\epsilon}
 \chi_B(x^0, x^\nu) + $$
{}
$$ + 2 v^\epsilon_{B_0}(x^0){\partial^2\over
{\partial x^\epsilon \partial x^0}} \chi_B(x^0,x^\nu) + $$
{}
$$+
v^\epsilon_{B_0}(x^0)v^\lambda_{B_0}(x^0){\partial^2 
\over{\partial x^\epsilon \partial x^\lambda}} \chi_B(x^0, x^\nu) +
 {\cal O}(c^{-6}),\eqno(E16a)$$
\noindent and 
{}
$${\partial^2\over\partial {y^0_B}^2} 
\chi_B(y^0_B, y^\nu_B) =  {\partial^2\over\partial 
{y^0_A}^2} \chi_B(y^0_A,y^\nu_A) - 
a^\epsilon_{BA_0}(y^0_A){\partial \over\partial 
y^\epsilon_A} \chi_B(y^0_A, y^\nu_A) - $$
{}
$$ -2v^\epsilon_{BA_0}(y^0_A){\partial^2\over
{\partial y^\epsilon_A \partial y^0_A}}
\chi_B(y^0_A,y^\nu_A) +$$ 
{}
$$+ v^\epsilon_{BA_0}(y^0_A) v^\lambda_{BA_0}(y^0_A)
{\partial^2 \over{\partial y^\epsilon_A \partial y^\lambda_A}}
 \chi_B(y^0_A, y^\nu_A) +  {\cal O}(c^{-6}). \eqno(E16b)$$

\pagebreak \noindent
{\Large \bf Appendix F:   Christoffel Symbols in the  Proper     RF$_A$.}
\vskip 11pt \noindent
  In this Appendix we will 
present some   expressions, which are  
in use in various parts of the present paper.

\vskip 5pt \noindent {\bf   F.1. Christoffel Symbols with Respect
to the Background Metric $\gamma^A_{mn}$.}
\vskip 5pt 

\noindent The connection components (or, so-called, Christoffel symbols)
 for the background metric $\gamma_{lm}(y^p_{A})$  in the coordinates
$(y^p_A)$ of proper {\small RF}$_A$ are defined as usual:
{}
$$\gamma^{kA}_{nm}(y^p_{A}) = {1\over2} \gamma^{kp}_A
(y^p_{A})\Big({\partial^A_n}\gamma^A_{mp}(y^p_{A})
 + {\partial^A_m}\gamma_{pn}(y^p_{A}) - {\partial^A_p} 
\gamma_{mn}(y^p_{A})\Big),$$

\noindent where    
$ \partial^A_p = \partial/\partial y^p_A$. Then from 
the eqs.$(C6)$ one may obtain expressions for the Christoffel symbols
for the  metric tensor $\gamma_{lm}(y^p_{A})$  as follows:
{}
$$\gamma^{0A}_{00} (y^p_{A}) = {\partial^2  K_{A}\over \partial
 {y^0_A}^2} + {a_{A_0}}_\epsilon  {v^\epsilon_{A_0}} + 
{\partial^2 L_A\over \partial {y^0_A}^2} - a^\epsilon_{A_0}
{\partial L_A\over \partial y^\epsilon_A}  - $$ 
{}
$$ - \Big({\partial^2 K_A\over \partial {y^0_A}^2}+ 
{a_{A_0}}_\epsilon {v^\epsilon_{A_0}}\Big)
\Big({\partial K_A\over \partial y^0_A}  + {v_{A_0}}_\epsilon 
{v^\epsilon_{A_0}}\Big) +  
{v_{A_0}}_\epsilon \Big({\partial^2 Q^\epsilon_A \over \partial {y^0_A}^2}
 -{a_{A_0}}_\epsilon {\partial Q^\epsilon_A\over \partial y^0_A}
\Big) +  {\cal O}(c^{-7}), \eqno(F1a)$$
{}
$$\gamma^{0A}_{0\alpha}(y^p_{A}) = - {a_{A_0}}_\alpha  + 
{\partial^2 L_{A}\over \partial y^0_A \partial y^\alpha_A} +
{ v_{A_0}}_\epsilon {\partial^2 Q^\epsilon_{A}\over \partial y^0_A 
\partial y^\alpha_A}  +{a_{A_0}}_\alpha \Big({\partial K_A\over \partial
y^0_A} +
 {v_{A_0}}_\epsilon {v^\epsilon_{A_0}}\Big) + 
  {\cal O}(c^{-6}),\eqno(F1b)$$
{}
$$\gamma^{0 A}_{\alpha\beta}(y^p_{A}) = {\partial^2 
\over \partial y^\alpha_A \partial y^\beta_A}\Big(L_{A} +
 {v_{A_0}}_\lambda Q^\lambda_{A}\Big) +  O(c^{-6}),\eqno(F1c)$$
{}
$$\gamma^{\alpha A}_{00}(y^p_{A}) = a^\alpha_{A_0} + 
\Big({\partial^2 Q^\alpha_{A}\over \partial {y^0_A}^2} 
-a^\lambda_{A_0}{\partial Q^\alpha_{A}\over\partial y^\lambda_A}  
\Big) -  v^\alpha_{A_0}\Big({\partial^2 K_{A}\over \partial {y^0_A}^2}
 + {a_{A_0}}_\lambda{v^\lambda_{A_0}}\Big) 
 +  {\cal O}(c^{-6}),\eqno(F1d)$$
{}
$$\gamma^{\alpha A}_{0\beta}(y^p_{A}) = v^\alpha_{A_0}
{a_{A_0}}_\beta  + 
{\partial^2 Q^\alpha_{A}\over \partial y^0_A \partial y^\beta_A}
 +  {\cal O}(c^{-5}),\eqno(F1e)$$
{}
$$\gamma^{\alpha A}_{\beta\omega}(y^p_{A}) = 
\gamma^{\alpha<0>}_{\beta\omega} + 
{\partial^2 Q^\alpha_{A} \over \partial y^\beta_A \partial 
y^\omega_A}+  {\cal O}(c^{-4}). \eqno(F1f)$$

\noindent where $\gamma^{\alpha<0>}_{\beta\omega}$ is the
 Christoffel symbols in coordinates of Galilean inertial {\small RF} 
(by choosing the quasi-Cartesian coordinates we 
may make these symbols vanish: $\gamma^{\alpha<0>}_{\beta\omega} = 0
$).

\vskip 5pt \noindent {\bf F.2. Christoffel Symbols with Respect to
the  Riemann Metric $g^A_{mn}$.} 
\vskip 5pt

\noindent From the expressions eqs.$(B8)$ one may also obtain 
the connection components $\Gamma^{n}_{kl}(y^p_{A})$ 
with respect to total Riemann metric $g_{lm}(y^p_{A})$ in
 coordinates $(y^p_A)$ of the {\small RF}$_A$:
{}
$$\Gamma^{0}_{00}(y^p_{A}) = {\partial \over \partial
 y^0_A}\Bigg({\partial K_{A}\over \partial y^0_A} + 
{1\over2}{v_{A_0}}_\epsilon {v^\epsilon_{A_0}} - 
\sum_B U_B(y^0_A, y^\nu_A)\Bigg) + $$
{}
$$ +{\partial \over \partial y^0_A} \Bigg( 
{\partial  L_{A}\over \partial y^0_A} + 
{1\over2}\Big({\partial K_{A}\over \partial y^0_A}\Big)^2 
+ {v_{A_0}}_\epsilon {\partial Q^\epsilon_{A}\over \partial
 y^0_A} +{1\over2} H^{<4>}_{00}(y^0_A, y^\nu_A)\Bigg) - $$ 
{}
$$ - {\partial \over \partial y^0_A}\Bigg({\partial K_{A}
 \over \partial y^0_A} + 
{1\over2}{v_{A_0}}_\epsilon  {v^\epsilon_{A_0}} - 
\sum_B U_B(y^0_A, y^\nu_A)\Bigg)^2 - $$
{}
$$ - \Bigg[{\partial L_{A}\over \partial y^\epsilon_A}
-{v_{A_0}}_\epsilon{\partial K_{A} \over \partial y^0_A} 
+ {\partial Q_{A \epsilon}\over \partial y^0_A} + 
{v_{A_0}}_\lambda {\partial Q^\lambda_{A}\over \partial
 y^\epsilon_A}
+  4 \hskip 1mm \gamma_{\epsilon\lambda}\sum_{B}
V^\lambda_B(y^0_A, y^\nu_A) \Bigg] \times $$
{}
$$\times \Big( a^\epsilon_{A_0} + \gamma^
{\epsilon\nu}\sum_B {\partial \over \partial 
y^\nu_A}U_B(y^0_A, y^\nu_A)\Big) + 
 {\cal O}(c^{-7}), \eqno(F2a)$$
{}
$$\Gamma^{0}_{0\alpha}(y^p_{A}) = - {a_{A_0}}_\alpha - 
\sum_B{\partial \over \partial y^\alpha_A}
U_B(y^0_A, y^\nu_A) +$$
{}
$$ +{\partial \over \partial y^\alpha_A} \Bigg(
 {\partial  L_{A} \over \partial y^0_A}+ 
{1\over2}\Big({\partial K_{A}\over \partial y^0_A}
\Big)^2 + {v_{A_0}}_\epsilon {\partial Q^\epsilon_{A}\over 
\partial y^0_A} +
{1\over2} H^{<4>}_{00}(y^0_A, y^\nu_A)\Bigg) + $$ 
{}
$$ + 2 \Bigg({\partial K_{A}\over \partial y^0_A} + 
{1\over2}{v_{A_0}}_\epsilon  {v^\epsilon_{A_0}} - 
\sum_B U_B(y^0_A, y^\nu_A)\Bigg)\Bigg( {a_{A_0}}_\alpha +  \sum_B
{\partial \over
 \partial y^\alpha_A}U_B(y^0_A, y^\nu_A)\Bigg) +
 {\cal O}(c^{-6}), \eqno(F2b)$$
{} 
$$\Gamma^{0}_{\alpha\beta}(y^p_{A}) = {\partial^2
 \over \partial y^\alpha_A \partial y^\beta_A}\Big(L_{A} +
{v_{A_0}}_\lambda Q^\lambda_{A}\Big) + $$
{}
$$ +   \sum_B \Bigg(  2\gamma _{\beta\lambda}
 {\partial \over \partial y^\alpha_A} V^\lambda_B
 (y^0_A, y^\nu_A) + 
2 \gamma _{\alpha\lambda}{\partial \over 
\partial y^\beta_A} V^\lambda_B (y^0_A, y^\nu_A)  
-  \gamma _{\alpha\beta} {\partial \over \partial
 y^0_A} U_B (y^0_A, y^\nu_A)\Bigg)
  +  {\cal O}(c^{-5}), \eqno(F2c)$$
{}
$$\Gamma^{\alpha}_{00}(y^p_{A}) =  a^\alpha_{A_0} + 
\gamma^{\alpha\lambda}\sum_B {\partial \over \partial
 y^\lambda_A}U_B(y^0_A, y^\nu_A) -$$
{}
$$ - \gamma^{\alpha\lambda}{\partial \over \partial
 y^\lambda_A} \Bigg( {\partial  L_{A}\over \partial y^0_A} + 
{1\over2}\Big({\partial K_{A}\over \partial y^0_A}\Big)^2
 + {v_{A_0}}_\epsilon {\partial Q^\epsilon_{A}\over 
\partial y^0_A} +{1\over2} H^{<4>}_{00}(y^0_A, y^\nu_A)\Bigg) + $$ 
{}
$$ + {\partial \over \partial y^0_A}
\Bigg(\gamma^{\alpha\lambda}{\partial \over 
\partial y^\lambda_A}\Big(
L_A+{v_{A_0}}_\nu Q^\nu_A\Big)- 
v^\alpha_{A_0}{\partial K_{A}\over \partial y^0_A}  +
 {\partial Q^\alpha_{A}\over \partial y^0_A} + 
 4 \sum_{B}V^\alpha_B(y^0_A, y^\nu_A) \Bigg) - $$
{}
$$ - \Bigg(v^\alpha_{A_0}v^\mu_{A_0} + 
\gamma^{\mu\lambda}{\partial Q^\alpha_{A}\over \partial y^\lambda_A} +
 \gamma^{\alpha\lambda}{\partial Q^\mu_{A}\over \partial y^\lambda_A}
+ 
2\gamma^{\alpha\mu}\sum_{B}U_B(y^0_A, y^\nu_A) \Bigg)\times$$
{}
$$\times \Big({a_{A_0}}_\mu+\sum_B {\partial
 \over \partial y^\mu_A}U_B(y^0_A, y^\nu_A)\Big) + 
 {\cal O}(c^{-6}), \eqno(F2d)$$
{}
$$\Gamma^{\alpha}_{0\beta}(y^p_{A}) =
 v^\alpha_{A_0}{a_{A_0}}_\beta+
{\partial^2 Q^\alpha_{A}\over \partial y^0_A \partial
 y^\beta_A}  + $$
{}
$$ + \sum_B \Bigg(  2{\partial \over
\partial y^\beta_A} V^\alpha_B(y^0_A,y^\nu_A)-  
2\gamma^{\alpha\mu}\gamma_{\beta\nu}{\partial 
\over \partial y^\mu_A} V^\nu_B (y^0_A, y^\nu_A)  
+  \delta ^\alpha_\beta {\partial \over \partial
 y^0_A} U_B (y^0_A, y^\nu_A)\Bigg) + 
 {\cal O}(c^{-5}), \eqno(F2e)$$
{}
$$\Gamma^{\alpha}_{\beta\omega}(y^p_{A}) = 
\gamma^{\alpha<0>}_{\beta\omega} + 
{\partial^2 Q^\alpha_{A} \over \partial y^\beta_A \partial
 y^\omega_A} + $$ 
{}
$$ + \sum_{B} \Bigg(\delta ^\alpha_\beta 
{\partial \over \partial y^\omega_A} U_B 
(y^0_A, y^\nu_A) + 
\delta^\alpha_\omega {\partial \over \partial
 y^\beta_A} U_B (y^0_A, y^\nu_A)  -\gamma_{\beta\omega}
\gamma^{\alpha\lambda}
{\partial \over \partial y^\lambda_A} 
U_B (y^0_A, y^\nu_A) \Bigg) +
 {\cal O}(c^{-4}).\eqno(F2f) $$

\noindent where the quantity  $H^{<4>}_{00}(y^0_A, y^\nu_A)$
 comes from the relations for the metric tensor
 in coordinates of proper  {\small RF}$_A$ eqn.(4.11) and is given by the  
relation (4.12). 

\vskip 11pt
\noindent{\Large \bf Appendix G:  The Component $g^A_{00}$ and the
Riemann Tensor.} 
\vskip 11pt\noindent 
  In this Appendix we will present the expressions
 for the flat metric $\gamma^A_{00}(y^p_A)$, 
"inertial friction" term  and the interaction term 
$h^{int <4>}_{00}(x^p)$.

\vskip 5pt \noindent 
{\bf G.1.  The Form of the Component $\gamma^A_{00}$.} 
\vskip 5pt
 
\noindent By substituting in the relations  eqs.$(C6)$
 the solutions for the transformation functions
$K_A, L_A$ and $Q^\alpha_A$ which are given by the
 expressions eqs.(5.11),(5.12),(5.23),(5.34) and (5.35),
one  obtains the following relations for the 
components of the metric $\gamma^A_{00}(y^p_A)$:
{}
$$\gamma^{A <2>}_{00}(y^0_A, y^\nu_A) = 
{\partial\over\partial y^0_{A}} K_{A} (y^0_A,y^\nu_A) + 
{v_{A_0}{_{{}_\beta}}(y^0_{A})} v_{A_0}^\beta(y^0_{A}) =  $$   
{}
$$ = \sum_{B\not=A}\Big[ \Big<U_B\Big>_0 + 
y^\lambda_A \Big<{\partial U_B\over \partial y^\lambda_A}\Big>_0\Big] 
+ \zeta^A_1 +  {\cal O}(c^{-4}), \eqno(G1a)$$
{}
$$ \gamma^{A <4>}_{00}(y^0_A, y^\nu_A) = 
2{\partial\over\partial y^0_{A}} L_{A} (y^0,y^\nu) + 
\Big({\partial\over\partial y^0_{A}} K_{A} 
(y^0,y^\nu)\Big)^2  + 2 {v_{A_0}}_\beta(y^0_{A}){\partial\over\partial
 y^0_{A}} Q^\beta_{A} (y^0,y^\nu) = $$
{}
$$=y^\mu_A y^\beta_A \cdot \Bigg(\gamma_{\mu\beta} 
\hskip 1mm {a_{A_0}}_\lambda a^\lambda_{A_0} -  
{a_{A_0}}_\mu {a_{A_0}}_\beta +  
\sum_{B\not=A}{\partial\over\partial {y^0_A}}\Big[ \gamma_{\mu\beta}
\Big<{\partial U_B\over\partial {y^0_A}}\Big>_0 - 
4 \Big<{\partial {V_B}_\beta\over\partial y^\mu_A}\Big>_0\Big]\Bigg) +
$$
{}
$$ - y^\mu_A \Big<{\partial W^{B}_{00}\over \partial y^\mu_A}\Big>_0 
  - \Big<W^{B}_{00}\Big>_0 + 2 \zeta^A_2 + $$
{}
$$ + 2  \sum^k_{l \ge3}  
 \Bigg({\partial \over \partial y^0_A}{L_A}_{ \{L\}}(y^0_A)
 + {v_{A_0}}_\mu \cdot{\partial \over \partial y^0_A} {Q^\mu_A}_{\{L\}}
(y^0_A)\Bigg)
\cdot y^{\{L\}}_A  + {\cal O}(c^{-6}) +
 {\cal O}|y^\nu_A|^{k+1}),\eqno(G1b)$$
{}
$$\gamma^{A <3>}_{0\alpha} (y^0_A, y^\nu_{A})  = 
{\partial\over\partial y^\alpha_{A}} L_{A} (y^0,y^\nu) -
{v_{A_0}}_\alpha(y^0_A){\partial\over\partial y^0_{A}}
 K_A(y^0,y^\nu)  + $$
{}
$$+ {v_{A_0}}_\nu (y^0_{A}){\partial\over\partial 
y^\alpha_{A}}Q^\nu_{A} (y^0,y^\nu) + 
\gamma_{\alpha\nu}{\partial\over\partial y^0_{A}}
 Q^\nu_{A} (y^0,y^\nu) = $$ 
{}
$$ = - 4 \gamma_{\alpha\mu} \sum_{B\not=A}
\Big[y^\lambda_A  
\Big<{\partial V^{\mu}_B\over \partial y^\lambda_A}\Big>_0 + 
\Big<V^\mu_B\Big>_0\Big]  + \sigma^A_\alpha - $$
{}
$$-{1\over2} \big(\gamma_{\alpha\epsilon} 
\delta^\beta_\lambda + \gamma_{\alpha\lambda}\delta^\beta_\epsilon - 
\gamma_{\epsilon\lambda} \delta^\beta_\alpha\big) 
\hskip 1mm y^\epsilon_A y^\lambda_A \cdot 
\sum_{B\not=A} {\partial  \over\partial {y^0_A}}
\Big<{\partial U_B \over\partial {y^\beta_A}}\Big>_0+ $$
{}
$$ +\sum^{k}_{l \ge3}  \Bigg({\partial \over \partial y^0_A}
{Q^\alpha_A}_{\{L\}} (y^0_A) 
\cdot y^{\{L\}}_A   + \Big({L_A}_{\{L\}}(y^0_A)  + 
{v_{A_0}}_\nu \cdot {Q^\nu_A}_{\{L\}} (y^0_A) \Big)\cdot 
{\partial \over \partial y^\alpha_A} \Big( 
y^{\{L\}}_A\Big)\Bigg) + $$
{}
$$ +  {\cal O}(c^{-5}) +  {\cal O}(|y^\nu_A|^{k+1}), 
\eqno(G1c)$$
{}
$$\gamma^{A <2>}_{\alpha\beta}(y^0_A, y^\nu_{A}) = 
{v_{A_0}}_\alpha(y^0_A)  {v_{A_0}}_\beta(y^0_{A})  
+ \gamma_{\alpha\nu}{\partial\over\partial y^\beta_{A}}
 Q^\nu_{A} (y^0_A,y^\nu_A) +
 \gamma_{\beta\nu}{\partial\over\partial y^\alpha_{A}}
 Q^\nu_{A} (y^0_A,y^\nu_A) = $$
{}
$$ = -2 \gamma_{\alpha\beta} \sum_{B\not=A}\Big[  
y^\lambda_A \Big<{\partial U_B\over \partial y^\lambda_A}\Big>_0 + 
\Big<U_B\Big>_0\Big]  + \sigma^A_{\alpha\beta} + $$ 
{}
$$ +\sum^{k}_{l \ge3}  \Big(\gamma_{\alpha\nu} 
{Q^\nu_A}_{\{L\}} (y^0_A)  
{\partial \over \partial y^\beta_A} + \gamma_{\beta\nu} 
 {Q^\nu_A}_{\{L\}} (y^0_A)  
{\partial \over \partial y^\alpha_A} \Big) \cdot 
y^{\{L\}}_A +  {\cal O}(c^{-4}) +  {\cal O}(|y^\nu_A|^{k+1}).
\eqno(G1d)$$

\vskip 5pt \noindent {\bf G.2. Lemma.}
The following relation hold for 
any values $k$: 
{}
$$a^{\{K\}} -  b^{\{K\}}= \sum^k_{s=1} (-)^{k-s}
 {P^{k-s+1}_k \over{(s-1)!(k-s+1)!}}a^{\{S-1\}}
\Big(a^\nu - b^\nu\Big)^{\{K-S+1\}}, \eqno(G2)$$

\noindent where $a^{\{K\}} = a^{\nu_1} a^{\nu_2}...a^{\nu_k}$
and $P^p_n$ - is the operation  of all the possible 
arrangements of $p$ different objects from $n$ ones. 

The formula $(G2)$ may be proved by the direct verification for 
several arbitrary values of $s$. Thus, for $s = 1$ and $s =2$ 
this formula  is trivial. For $s = 3$ and $s = 4$ from the 
right-hand side of the equation $(G2)$ one may check   
 that these relations are hold as well.

Indeed, by the straightforward calculation we will have the 
following result for $s=3$:
{}
$$ (a^{\nu_1}-b^{\nu_1}) (a^{\nu_2}  -  
b^{\nu_2}) (a^{\nu_3}-b^{\nu_3})
- a^{\nu_1}(a^{\nu_2}-b^{\nu_2}) (a^{\nu_3}  - 
b^{\nu_3})-$$
{}
$$- a^{\nu_2}(a^{\nu_3}-b^{\nu_3}) 
(a^{\nu_1}-b^{\nu_1}) - 
a^{\nu_3}(a^{\nu_1}-b^{\nu_1}) (a^{\nu_2}  - 
b^{\nu_2}) +$$ 
{}
$$+  a^{\nu_1}a^{\nu_2}(a^{\nu_3}-b^{\nu_3})  + 
a^{\nu_2}a^{\nu_3}(a^{\nu_1} -b^{\nu_1})  + 
a^{\nu_3}a^{\nu_1}(a^{\nu_2} -b^{\nu_2})  =$$
{}
$$=a^{\nu_1}a^{\nu_2}a^{\nu_3}-b^{\nu_1}b^{\nu_2}b^{\nu_3}=
 a^{\{3\}}-b^{\{3\}}.$$ 
\noindent 
And for $s = 4$
{}
$$ - (a^{\nu_1}-b^{\nu_1}) (a^{\nu_2}  -  
b^{\nu_2}) (a^{\nu_3}-b^{\nu_3})(a^{\nu_4}  - 
b^{\nu_4}) 
+ a^{\nu_1}(a^{\nu_2}-b^{\nu_2}) (a^{\nu_3}  - 
b^{\nu_3})(a^{\nu_4}-b^{\nu_4})  + $$
{}
$$ + a^{\nu_2} (a^{\nu_3}-b^{\nu_3})(a^{\nu_4} 
-b^{\nu_4}) (a^{\nu_1}-b^{\nu_1})  + 
a^{\nu_3}(a^{\nu_4}-b^{\nu_4}) (a^{\nu_1}  - 
b^{\nu_1}) (a^{\nu_2}-b^{\nu_2})(a^{\nu_3}-
b^{\nu_3})  +$$
{}
$$+a^{\nu_4}(a^{\nu_1}-b^{\nu_1}) (a^{\nu_2}  - 
b^{\nu_2}) (a^{\nu_3}-b^{\nu_3})- 
a^{\nu_1}a^{\nu_2}(a^{\nu_3}-b^{\nu_3})(a^{\nu_4} 
-b^{\nu_4})  - 
a^{\nu_1}a^{\nu_3}(a^{\nu_2}-b^{\nu_2})(a^{\nu_4} 
-b^{\nu_4}) - $$
{}
$$ - a^{\nu_1}a^{\nu_4}(a^{\nu_2}  -  
b^{\nu_2}) (a^{\nu_3}-b^{\nu_3})-
a^{\nu_2}a^{\nu_3}(a^{\nu_1}-b^{\nu_1}) (a^{\nu_4}
-b^{\nu_4}) - 
a^{\nu_2}a^{\nu_4}(a^{\nu_1}-b^{\nu_1}) (a^{\nu_3}
-b^{\nu_3}) - $$
{}
$$ - a^{\nu_3}a^{\nu_4}(a^{\nu_1}  -  
b^{\nu_1}) (a^{\nu_2}-b^{\nu_2})+
a^{\nu_1}a^{\nu_2}a^{\nu_3}(a^{\nu_4}-b^{\nu_4})+
a^{\nu_2}a^{\nu_3}a^{\nu_4}(a^{\nu_1}-b^{\nu_1})+ 
a^{\nu_3}a^{\nu_4}a^{\nu_1}(a^{\nu_2}-b^{\nu_2})  +$$
{}
$$+ a^{\nu_4}a^{\nu_1}a^{\nu_2}(a^{\nu_3}-b^{\nu_3})  = 
a^{\nu_1}a^{\nu_2}a^{\nu_3}a^{\nu_4}-
b^{\nu_1}b^{\nu_2}b^{\nu_3}b^{\nu_4}=a^{\{4\}}-b^{\{4\}}.$$

\noindent Then by the induction,  one may extrapolate the 
validity of the expression $(G2)$ for any $s > 4$.

 Making of use the relation $(G2)$ we will simplify
the form of some   expressions for the metric tensor in the 
proper {\small RF}$_A$ and the interaction term in the 
coordinates $(x^p)$ of the barycentric inertial {\small RF}$_0$. 
Let us present two expressions which will be 
 necessary for the future analysis. The following integral is easy to 
calculate in the form:  
{}
$$\int_B d^3y'_A  
\rho_B\Big(y^0_A, y'^\nu_A + y^\nu_{BA_0}(y^0_A)\Big) 
{\partial \over\partial y'^\lambda_A}\Big[{
 y^{\{K\}}_A  -  y'^{\{K\}}_A  
\over{|y^\nu_A - y'^\nu_A|}}\Big] = $$
{}
$$ = \sum^k_{s=1} {(-)^{k-s+1}P^{k-s+1}_k 
\over{(s-1)!(k-s+1)!}}\cdot  y^{\{S-1\}}_A  \times$$
{}
$$\times\int_B d^3y'_A 
\cdot  \rho_B\Big(y^0_A, y'^\nu_A + y^\nu_{BA_0}(y^0_A) \Big) 
{\partial \over\partial y'^\lambda_A} \Big[ {\Big(y^\nu_A -y'^\nu_A
\Big)^{\{K-S+1\}}
\over{|y^\nu_A-y'^\nu_A|}} \Big]=$$
{}
$$ =\sum^k_{s=1}  {(-)^{k-s+1}P^{k-s+1}_k \over
{(s-1)!(k-s+1)!}}\cdot y^{\{S-1\}}_A \hskip -5pt
\cdot{\partial \over\partial y^\lambda_A} 
Z(y^p_A)_B^{(K-S+1)}. \eqno(G3)$$ 
 
The same quantity will have the following form in the  coordinates 
$(x^p)$ of the inertial {\small RF}$_0$:
 
$$\int_Bd^3x' \rho_B\Big(x^0, x'^\nu - 
y^\nu_{B_0}(x^0)\Big) \times$$
{}
$$\times{\partial \over\partial x'^\lambda}\Big[ {\Big(x^\nu-
y^\nu_{B_0}(x^0)\Big)^{\{K\}}- 
\Big(x'^\nu-y^\nu_{B_0}(x^0)\Big)^{\{K\}}\over{|x^\nu - x'^\nu|}}
\Big] = $$
{}
$$ =\sum^k_{s=1}{(-)^{k-s+1}P^{k-s+1}_k \over
{(s-1)!(k-s+1)!}}\Big(x^\nu-y^\nu_{B_0}(x^0)
\Big)^{\{S-1\}} \times$$
{}
$$\times\int_Bd^3x' 
\cdot \rho_B\Big(x^0, x'^\nu - y^\nu_{B_0}
(x^0)\Big) \cdot {\partial \over\partial x'^\lambda}
\Big[{\big(x^\nu-x'^\nu\big)^{\{K-S+1\}}\over{|x^\nu - x'^\nu|}}\Big]=$$
{}
$$= \sum^k_{s=1} {(-)^{k-s+1}P^{k-s+1}_k \over
{(s-1)!(k-s+1)!}}\Big(x^\nu-y^\nu_{B_0}(x^0)
\Big)^{\{S-1\}} \hskip -5pt
\cdot{\partial \over\partial x^\lambda} Z(x^p)_B^{(K-S+1)},
 \eqno(G4)$$
\noindent where potential $Z(x^p)_B^{(S)}$ was defined as 
{}
$$Z(z^p)_B^{(S)}=\int_B{ d^3z' \over{|z^\nu - z'^\nu|}}  \cdot 
\rho_B\Big(z^0, z'^\nu - z^\nu_{B_0}(z^0)\Big)
\cdot \big(z^\nu-z'^\nu\big)^{\{S\}}. \eqno(G5)$$

\vskip  5pt \noindent  
{\bf G.3. The Form of the `Inertial Friction' Term.} 
\vskip 5pt

\noindent The following  term in the temporal component of the 
metric tensor $g^A_{00}(y^0_A,y^\nu_A)$  eq.$(4.16b)$
 has the meaning of the gravitational inertial friction:
{}
$$\int_Bd^3y'_A  \rho_B
\Big(y^0_A, y'^\nu_A + y^\nu_{BA_0}(y^0_A)\Big) 
 {\partial \over\partial y'^\lambda_A} \Big[{ Q^\lambda_{A}
(y^0_A,y^\nu_A) -
 Q^\lambda_{A}(y^0_A, y'^\nu_A)\over{|y^\nu_A - y'^\nu_A|}}\Big].$$
{}
\noindent The substitution in this relation the obtained
  function $Q^\alpha_A$, will enable us
 to present  the `inertial friction' term as follows:
{}
$$\int_Bd^3y'_A  \rho_B\Big(y^0_A, y'^\nu_A + y^\nu_{BA_0}(y^0_A)\Big) 
 {\partial \over\partial y'^\lambda_A} \Big[ {Q^\lambda_{A} (y^0_A,
y^\nu_A) -
 Q^\lambda_{A}(y^0_A, y'^\nu_A)\over{|y^\nu_A - y'^\nu_A|}}\Big]=$$
{}
$$ = 2 U_B(y^0_A,y^\nu_A)\sum_{B'\not=A}\Big[   
\Big<U_{B'}\Big>_0 +  y^\mu_A \Big<{\partial U_{B'}\over \partial
y^\mu_A}\Big>_0\Big]+$$
{}
$$+{1\over 2} v_{A_0}^\lambda(y^0_A)v^\beta_{A_0}
(y^0_A) \cdot {\partial^2 \over \partial
 y^\lambda_A \partial y^\beta_A} 
\chi_B (y^0_A,y^\nu_A) + $$
{}
$$  +  {5\over2}a^\lambda_{A_0}(y^0_A) \cdot 
{\partial \over \partial y^\lambda_A} \chi_B 
(y^0_A,y^\nu_A)  
- f^{\lambda\beta}_A \cdot {\partial^2 \over 
\partial y^\lambda_A \partial y^\beta_A} \chi_B
 (y^0_A,y^\nu_A) + $$
{}
$$  + \sum^{k}_{l\ge3}  {Q^\lambda_A}_{\{L\}} (y^0_A)
\cdot\sum^l_{s=1}{{(-)}^{l-s+1} P^{l-s+1}_l  
\over (s-1)!(l-s+1)! }\cdot y^{\{S-1\}}_A \hskip -5pt
\cdot{\partial \over\partial y^\lambda_A}
 Z(y^q_A)_B^{(L-S+1)} +$$
{}
$$+{\cal O}(c^{-4}) +{\cal O}(|y^q_A|^{k+1}).\eqno(G6)$$

\vskip  5pt \noindent 
{\bf G.4. The Form of the Interaction Term.} 
\vskip 5pt

\noindent Making of use the solutions for the 
functions $K_A, L_A$ and $Q^\alpha_A$ of the coordinate 
transformation,  one may also obtain the form 
 of the interaction term $h^{{\it int}<4>}_{00}$ 
in any coordinate system. 
Thus, for example, in coordinates of inertial {\small RF}$_0$
from eq.(4.6) we will have the following expression:
{}
$$h^{{\it int}<4>}_{00}(x^0,x^\nu) = 
4\int_B{d^3x' \over{|x^\nu - x'^\nu|}} 
\sum_B\rho_B\Big(x^0, x'^\nu - y^\nu_{B_0}(x^0)\Big) 
\sum_C U_C(x'^p)  + $$
{}
$$ +  2\Big(\sum_B U_B(x^p)\Big)^2 + 
2\sum_B\Bigg[2 a^\lambda_{B_0}(x^0) \cdot{\partial \over \partial
 x^\lambda} \chi_B (x^0,x^\nu) -$$
{}
$$- 2 a^\lambda_{B_0}(x^0) \cdot \Big(x_\lambda - 
{y_{B_0}}_\lambda(x^0)\Big)\cdot U_B(x^0,x^\nu) - 
\hskip 1mm $$
{}
$$ + \sum^{k}_{l\ge3}  {Q^\lambda_B}_{\{L\}} 
(x^0)\sum^l_{s=1} {(-)^{l-s+1}P^{l-s+1}_l \over{(s-1)!(l-s+1)!}} 
\Big(x^\nu-y^\nu_{B_0}(x^0)\Big)^{\{S-1\}} \hskip -5pt
\cdot{\partial \over\partial x^\lambda} Z(x^q)_B^{(P-S+1)} -$$ 
{}
$$ - 2\zeta^B_1 \cdot U_B(x^0,x^\nu) - 
f^{\lambda\beta}\cdot{\partial^2 \over\partial
 x^\lambda \partial x^\beta} \chi_B (x^0, x^\nu)\Bigg]  
 +{\cal O}(c^{-4}) + {\cal O}\big( \big|x'^\nu-y^\nu_{B_0}(x^0)
\big|^{k+1}\big). \eqno(G7)$$

\vskip  5pt \noindent {\bf G.5.  The Form of the  Riemann Tensor 
in  the Proper {\small RF}$_A$.} 
\vskip 5pt

\noindent We are using the following notation for the components of the
Riemann 
tensor:
{}
$$R^k_{mnp} = {\partial_p}\Gamma^k_{mn} - {\partial_n}
\Gamma^k_{mp} + \Gamma^l_{mn}\Gamma^k_{lp} -
 \Gamma^l_{mp} \Gamma^k_{ln}, \eqno(G8) $$

\begin{itemize}
\item[(i).] By making of use the expressions for the metric 
tensor $g^A_{mn}(y^p_A)$ given by the eqs.(6.7) one will obtain  the
following  post-Newtonian expansions
of the components of this tensor   in an arbitrary 
{\small RF} $(z^p)$:
{}
$$  R_{0\alpha0\beta}(z^p) = 
{\partial \over \partial z^\beta}\Gamma_{\alpha0}^{0<2>}+
g_{00}^{<2>}{\partial \over \partial z^\beta}\Gamma_{\alpha0}^{0<2>}+
{\partial \over \partial z^\beta}\Gamma_{\alpha0}^{0<4>}-
{\partial \over \partial z^0}\Gamma_{\alpha\beta}^{0<3>}+$$
{}
$$ +\Gamma_{\alpha0}^{0<2>}\Gamma_{\beta0}^{0<2>}-
\Gamma_{\alpha\beta}^{\nu<2>}\Gamma_{\nu0}^{0<2>}+{\cal O}(c^{-
6}),\eqno(G9a)$$
{}
$$  R_{0\nu\alpha\beta}(z^p) = 
{\partial \over \partial z^\beta}\Gamma_{\nu\alpha}^{0<3>}-
{\partial \over \partial z^\alpha}\Gamma_{\nu\beta}^{0<3>}+
{\cal O}(c^{-5}), \eqno(G9b)$$
{}
$$  R_{\alpha\mu\beta\sigma}(z^p) = \gamma_{\alpha\lambda}\Bigg(
{\partial \over \partial z^\sigma}\Gamma_{\mu\beta}^{\lambda<2>}-
{\partial \over \partial z^\beta}\Gamma_{\mu\sigma}^{\lambda<2>}
\Bigg)+{\cal O}(c^{-4}).\eqno(G9c)$$

\item[(ii).] By making of use the  expressions $(G9)$ one may
obtain the components of the Riemann tensor 
$R_{mnpk}$  in the  coordinates $(x^p)$ of the barycentric 
inertial {\small RF}$_0$ as follows:
{}
$$  R_{0\alpha0\beta}(x^0, x^\nu) = - \sum_B 
{\partial^2 U_B\over \partial x^\alpha \partial x^\beta}   +
{1\over2}{\partial^2 H^{<4>}_{00}\over \partial x^\alpha 
\partial x^\beta} +  \sum_B \Bigg( {\partial U_B\over \partial 
x^\alpha} {\partial U_B\over \partial x^\beta}  
- \gamma_{\alpha\beta} \gamma^{\mu\nu}{\partial U_B
\over \partial x^\mu} 
{\partial U_B\over \partial x^\nu} \Bigg) - $$
{}
$$- \sum_B \Bigg(2 \gamma _{\beta\lambda} 
{\partial^2  V^\lambda_B\over \partial x^0\partial x^\alpha}
 + 2\gamma _{\alpha\lambda}{\partial^2 V^\lambda_B \over \partial
x^0\partial
 x^\beta} - \gamma _{\alpha\beta} {\partial^2 U_B\over \partial {x^0}^2} 
\Bigg)  + {\cal O}(c^{-6}), \eqno(G10a)$$
{}
$$R_{0\mu\alpha\beta}(x^0, x^\nu) =   2 \sum_B \Bigg( \gamma
_{\alpha\lambda} 
{\partial^2  V^\lambda_B\over \partial x^\mu\partial x^\beta}
  -\gamma _{\beta\lambda} 
{\partial^2  V^\lambda_B\over \partial x^\mu\partial y^\alpha_A} 
\Bigg)+$$
{}
$$ + \sum_B  \Bigg( \gamma _{\mu\beta} 
{\partial^2 U_B \over  \partial x^0\partial x^\alpha}  -
\gamma _{\mu\alpha}{\partial^2 U_B \over  \partial x^0\partial
x^\beta}\Bigg)
  + {\cal O}(c^{-5}), \eqno(G10b)$$
{}
$$  R_{\alpha\mu\beta\sigma} (x^0, x^\nu)  =\sum_B \Bigg( \gamma
_{\alpha\beta} 
{\partial^2 U_B\over \partial x^\mu  \partial x^\sigma}
 + \gamma _{\mu\sigma}{\partial^2 U_B\over \partial x^\alpha 
\partial x^\beta}  - $$
{}
$$ - \gamma _{\beta\mu}{\partial^2 U_B\over \partial 
x^\alpha  \partial x^\sigma}  - 
\gamma _{\alpha\sigma}{\partial^2 U_B\over \partial x^\mu  
\partial x^\beta}\Bigg)
  + {\cal O}(c^{-4}). \eqno(G10c)$$
 
\item[(iii).] By making of use the expressions for the
 connection components $\Gamma^k_{mp}$ presented 
by the relations eqs.$(F2)$, one may obtain the components 
of the Riemann tensor $R_{mnpk}$  in the  coordinates 
$(y^p_A)$ of the proper {\small RF}$_A$ as follows:
{}
$$ \Big< R_{0\alpha0\beta}\Big>_0
= - \sum_B \Big<{\partial^2 U_B\over \partial y^\alpha_A
\partial y^\beta_A}\Big>_0  +
{1\over2}\Big<{\partial^2H^{<4>}_{00}\over \partial y^\alpha_A
\partial y^\beta_A}\Big>_0 + $$
{}
$$ +   \Big<{\partial U_A \over \partial y^\alpha_A} 
{\partial U_A\over \partial y^\beta_A}\Big>_0  - 
\gamma_{\alpha\beta} \gamma^{\mu\nu}\Big<{\partial U_A
\over \partial y^\mu_A} 
{\partial U_A\over \partial y^\nu_A}\Big>_0  
 +\gamma_{\alpha\beta}a^\lambda_{A_0}{a_{A_0}}_\lambda -
{a_{A_0}}_\alpha{a_{A_0}}_\beta-$$
{}
$$ - \sum_B {\partial \over \partial y^0_A}\Bigg( 2 \gamma
_{\beta\lambda} 
\Big<{\partial  V^{\lambda}_B\over  \partial y^\alpha_A}\Big>_0 
+ 2\gamma _{\alpha\lambda}\Big<{\partial  V^{\lambda}_B\over  
\partial y^\beta_A}\Big>_0   -  \gamma _{\alpha\beta} \Big<{\partial 
U_B\over 
\partial {y^0_A}}\Big>_0  \Bigg)
  + {\cal O}(c^{-6}), \eqno(G11a)$$
{}
$$\Big<R_{0\mu\alpha\beta}\Big>_0  =  \sum_B \Bigg( 2\gamma
_{\alpha\lambda} 
\Big<{\partial^2 V^{\lambda}_B \over  \partial y^\mu_A\partial  
y^\beta_A}\Big>_0 
 -2\gamma _{\beta\lambda} \Big<{\partial^2 V^{\lambda}_B
\over  \partial y^\mu_A\partial y^\alpha_A}\Big>_0+$$
{}
$$ +  \gamma _{\mu\beta} 
\Big<{\partial^2 U_B\over  \partial y^0_A\partial y^\alpha_A}\Big>_0 
 -\gamma _{\mu\alpha}\Big<{\partial^2 U_B\over  \partial y^0_A\partial
y^\beta_A}\Big>_0
\Bigg)+ {\cal O}(c^{-5}), \eqno(G11b)$$
{}
$$\Big<  R_{\alpha\mu\beta\sigma}\Big>_0 =\sum_B \Bigg( \gamma
_{\alpha\beta} 
\Big<{\partial^2 U_B\over \partial y^\mu_A \partial y^\sigma_A}\Big>_0 + 
\gamma _{\mu\sigma}\Big<{\partial^2 U_B\over \partial y^\alpha_A 
\partial y^\beta_A}\Big>_0 - $$
{}
$$ - \gamma _{\beta\mu}\Big<{\partial^2 U_B\over \partial 
y^\alpha_A \partial y^\sigma_A}\Big>_0 - 
\gamma _{\alpha\sigma}\Big<{\partial^2 U_B\over \partial y^\mu_A 
\partial y^\beta_A}\Big>_0\Bigg)
  + {\cal O}(c^{-4}). \eqno(G11c)$$
\end{itemize}
 
It is interesting to note  that in the case when the {\it local} gravity
produced by 
the body {\small (A)} under consideration may be neglected, 
the Riemann curvature tensor $(G10)$ is formed only by 
  the gravitational field of the other bodies in the system.  
This suggests that  one may extend the generalized Fermi conditions in the
{\it local} 
region   of body {\small (A)} (or at the immediate vicinity of it's 
world-line $\gamma_A$, given by  the relations (5.2)), as follows:
{} 
$$g_{mn}(y^p_A)   =  
\hskip 2mm g^{(loc)}_{mn}(y^p_A)   + 
\delta g^{(ext)}_{mn}(\sim |y^\alpha_A|^2) +{\cal O}(|y^\alpha_A|^3),
 \eqno(G12a)$$ 
{}
$$\Gamma^{k}_{mn}(y^p_A)    = 
\hskip 2mm \Gamma^{k(loc)}_{mn}(y^p_A)  
+\delta\Gamma^{k(ext)}_{mn}(\sim |y^\alpha_A|)  +
{\cal O}(|y^\alpha_A|^2),   \eqno(G12b)$$
{}
$$ R_{mnkl}(y^p_A) = \hskip 2mm  R_{mnkl}(y^0_A)\Big|_{\gamma_A} + 
{\cal O}(|y^\alpha_A|), \eqno(G12c)$$ 
{}
\noindent where superscript {\it ext} denotes the   external 
sources of gravity.
The relations $(G12)$ are summarizing our expectations  based on the 
Equivalence Principle about the {\it local} gravitational environment 
of the self-gravitating and arbitrarily shaped extended bodies. 

\vskip 11pt \noindent
{\Large\bf Appendix H:  Some Important Identities.} 
\vskip 11pt \noindent
In this Appendix we will present some identities, necessary to reduce the 
expressions in Section {\small 6}. We will use the definition for the 
total mass density of the system ${\overline\rho}$ in the coordinates
$(y^p_A)$ as given by (6.4); 
for the total Newtonian potential ${\overline U}$ as given by (6.17); for
the total vector-potential 
of the system ${\overline V}^\alpha$ as given by the expression (6.20).
Then, one  may obtain the required identities simply  by using  the 
eq.m. (6.6), the Poisson equations for the potentials ${\overline U}$ and 
${\overline V}^\alpha$ (6.18), (6.21) respectievly, and with the help of  
expression (6.22).
{}
$${\overline\rho}{\partial {\overline V}^\alpha\over \partial y^0_A}+
{\overline\rho}v^\beta\Big[\partial_\beta{\overline V}^\alpha-
\partial^\alpha{\overline V}_\beta\Big]=$$
{}
$$={1\over 4\pi}{\partial\over\partial y^0_A}\Bigg(-
\partial^\alpha{\overline U}
{\partial{\overline U}\over\partial y^0_A}+
\partial_\nu{\overline U}\Big[\partial^\alpha {\overline V}^\nu-
\partial^\nu{\overline V}^\alpha\Big]\Bigg) +$$
{}
$$+{1\over 4\pi}{\partial\over\partial y^\beta_A}
\Bigg(\partial^\alpha{\overline U}{\partial{\overline
V}^\beta\over\partial y^0_A}+
\partial^\beta{\overline U}{\partial{\overline V}^\alpha\over\partial
y^0_A}-
\gamma^{\alpha\beta}\partial_\nu{\overline U}{\partial{\overline
V}^\nu\over\partial y^0_A}+
\Big[\partial^\nu{\overline V}^\alpha-\partial^\alpha{\overline
V}^\nu\Big]
\Big[\partial_\nu{\overline V}^\beta-\partial^\beta{\overline V}_\nu\Big]
-$$
{}
$$-\gamma^{\alpha\beta}\Big[\partial^\mu{\overline
V}^\nu\partial_\mu{\overline V}^\nu-
\partial^\mu{\overline V}^\nu\partial_\nu{\overline V}_\mu\Big]+
{1\over2}\gamma^{\alpha\beta}\left(
{\partial{\overline U}\over\partial y^0_A}\right)^2\Bigg),\eqno(H1)$$
{}
$${\overline\rho}v^\alpha\Big({\partial {\overline U}\over \partial
y^0_A}+ 
  v^\mu \partial_\mu{\overline U} \Big)  =
{\partial \over \partial y^0_A}\Big( {\overline\rho}v^\alpha {\overline
U}\Big)+
 {\partial\over\partial y^\beta_A}\Big[{\overline\rho}v^\alpha v^\beta 
{\overline U}-p{\overline U}\gamma^{\alpha\beta}\Big]+
p\partial^\alpha  {\overline U}+{\overline\rho}
{\overline U}\partial^\alpha {\overline U}, \eqno(H2)$$
{}
$${\overline\rho}\partial^\alpha {\overline V}^\beta=
{1\over 4\pi}\partial_\nu\Big[
 \partial^\nu{\overline U} \partial^\alpha{\overline V}^\beta +
\partial^\alpha{\overline U} \partial^\nu{\overline V}^\beta
-\gamma^{\alpha\nu}\partial^\mu{\overline U} \partial^\mu{\overline
V}^\beta
\Big]+{\overline\rho}v^\beta\partial^\alpha {\overline U}, \eqno(H3)$$
{}
$${\overline\rho}_0\partial^\alpha f={1\over
2\pi}\partial_\beta\Gamma^{\alpha\beta}(f)-
{1\over 4\pi}\partial^\alpha {\overline U}\partial_\mu\partial^\mu f,
\eqno(H4)$$ 
\noindent where $\Gamma^{\alpha\beta}$ defined as follows:
{}
$$\Gamma^{\alpha\beta}(f)={1\over2}\Big[\partial^\alpha f
\partial^\beta{\overline U}+
\partial^\beta f \partial^\alpha{\overline U}-\gamma^{\alpha\beta}
\partial^\nu f \partial^\nu{\overline U}\Big]. \eqno(H5)$$
The following  identities are also easy to verify:
{}
$${\overline\rho}\delta\ddot{w}^\alpha_{A_0}= 
{\overline\rho}\partial^\alpha\Big({y_A}_\mu\delta\ddot{w}^\mu_{A_0}\Big)=
{1\over 2\pi}\partial_\beta\Gamma^{\alpha\beta}
\Big({y_A}_\mu\delta\ddot{w}^\mu_{A_0}\Big),  \eqno(H6)$$
{}
$${\overline \rho}\hskip 2pt\Big( a^\alpha_{A_0} {a_{A_0}}_\lambda - 
\delta^\alpha_\lambda \cdot a^\mu_{A_0} {a_{A_0}}_\mu
\Big)y^\lambda_A= 
{1\over 2\pi}\partial^\alpha {\overline U} \cdot{a_{A_0}}_\lambda 
a^\lambda_{A_0}+ $$
{}
$$+{1\over 4\pi}\partial_\beta\Gamma^{\alpha\beta}
\Big[\Big(y^\lambda_A y^\mu_A-\gamma^{\lambda\mu} {y_A}_\nu
y^\nu_A\Big) 
{a_{A_0}}_\lambda {a_{A_0}}_\mu \Big], \eqno(H7)$$
{}
$$ + {\overline \rho}\hskip 2pt\Bigg( 
\sum_{B\not=A}{\partial \over \partial y^0_A}
\Big[2 \Big<{\partial V^{\alpha)}_B\over \partial y^{(\lambda}_A}\Big>_A
+2 \Big<v^{(\alpha}{\partial U_B\over \partial y^{\lambda)}_A}\Big>_A-
\delta^\alpha_\lambda{\partial \over \partial
y^0_A}\Big<U_B\Big>_A\Big]\Bigg)y^\lambda_A= $$ 
{}
$$={1\over4\pi}\partial_\beta\Gamma^{\alpha\beta}\Bigg[
\sum_{B\not=A}{\partial \over \partial y^0_A}
\Big(4y^\lambda_A y^\epsilon_A\Big[\Big<{\partial V_{B\epsilon}\over
\partial y^\lambda_A}\Big>_A
+\Big<v_\epsilon{\partial U_B\over \partial y^\lambda_A}\Big>_A\Big]-$$
{}
$$-{y_A}_\mu y^\mu_A{\partial \over \partial
y^0_A}\Big<U_B\Big>_A\Big)\Bigg] 
-{1\over4\pi}\partial^\alpha{\overline U}\cdot
{\partial^2\over\partial {y^0_A}^2}\Big<U_B\Big>_A,\eqno(H8)$$
{}
$${\overline \rho}
\ddot{a}_{A_0}{_{{}_\lambda}}\Big(y^\alpha_Ay^\lambda_A+
{1\over2}\gamma^{\alpha\lambda}y^\mu_A {y_A}_\mu\Big)
=-{5\over4\pi}\partial^\alpha{\overline
U}\cdot\ddot{a}_{A_0}{_{{}_\lambda}}y^\lambda_A+$$
{}
$$+{1\over4\pi}{\partial\over \partial y^\beta_A}\Big[
\Big(\gamma^{\alpha\beta}\gamma_{\lambda\mu}-
\delta^\alpha_\lambda\delta^\beta_\mu-
\delta^\alpha_\mu\delta^\beta_\lambda\Big)
\Big(y^\mu_A y^\nu_A+{1\over 2}\gamma^{\mu\nu}{y_A}_\epsilon
y^\epsilon_A\Big)
{\ddot{a}_{A_0}}{_{{}_\nu}}\partial^\lambda{\overline U}\Big].\eqno(H9)$$

>From the equation for the potential ${\overline W}$  
  $(6.23a)$ and with the help of $(H4)$ one obtains:
{}
$${\overline\rho}\partial^\alpha {\overline W}=
{1\over 2\pi}\partial_\beta\Gamma^{\alpha\beta}({\overline W})+
{1\over 2\pi}\partial^\alpha {\overline U}\Big[
4\pi\sum_B {\overline\rho}_B\Big(\Pi-2v_\mu v^\mu+{3p\over
{\overline\rho}}\Big)-   $$
{}
$$-\sum_B\partial^2_{00}U_B-
2\sum_B \partial_\lambda U_B \Big(2a_{A_0}^\lambda+\sum_{B'} 
\partial^\lambda U_{B'}\Big)+$$
{}
$$+\sum_{l\ge3}{Q^\alpha_A}_{\{L\}}(y^0_A)\Big(2\sum_B
\partial^2_{\epsilon\lambda}U_B\partial^\lambda+
\sum_B\partial_\epsilon
U_B\partial_\mu\partial^\mu\Big)y^{\{L\}}_A\Big]+
{\overline \rho}{\cal O}(|y^\nu_A|^{k+1})+
{\overline \rho}{\cal O}(c^{-6}). \eqno(H10)$$

The following identity may be written in a two different ways. 
In order to reflect this  ambiguity we present it  as follows:

$${\overline\rho}{\partial {\overline U}\over \partial y^0_A}+2{\overline
\rho} v^\mu
{\partial{\overline U} \over \partial y^\mu_A}  =
{\partial \over \partial y^0_A}\Big(a_1{\overline\rho} {\overline U}+
{2a_1-3\over 8\pi}\partial _\mu {\overline U}\partial^\mu {\overline
U}\Big)+$$
{}
$$+{\partial \over \partial y^\mu_A}\Big({1-a_1\over4\pi}\partial^\mu 
{\overline U}{\partial {\overline U}\over \partial y^0_A} +
{a_2\over 4\pi}{\overline U} \partial^\mu 
{\partial {\overline U}\over \partial y^0_A} +
(a_1+a_2){\overline \rho}{\overline  U} v^\mu+$$
{}
$$+{2-a_1-a_2\over 4\pi}\partial_\nu{\overline U} \big[\partial^\nu 
{\overline V}^\mu-\partial^\mu {\overline V}^\nu]\Big), \eqno(H11)$$
\noindent where $a_1$ and $a_2$ are arbitrary numbers. 

One can verify the correctness of following identities necessary to 
reduce  the terms in the equation (6.32) which  
contain the functions ${Q^\alpha_A}_{\{L\}}$ with $l\ge3$:
{}
$${\overline\rho}{a_{A_0}}_\nu \sum_{l\ge3}^k\Big[ 
{Q^\alpha_A}_{\{L\}}(y^0_A)\partial^\alpha y^{\{L\}}_A\Big] = $$ 
{}
$$={1\over2\pi}\partial_\beta\Gamma^{\alpha\beta}
\Big({a_{A_0}}_\nu\sum_{l\ge3}^k\Big[ 
{Q^\nu_A}_{\{L\}}(y^0_A)y^{\{L\}}_A\Big]\Big)-
{1\over4\pi}\partial^\alpha {\overline U}
\Big[{a_{A_0}}_\nu\sum_{l\ge3}^k  
{Q^\nu_A}_{\{L\}}(y^0_A)\cdot\partial_
\lambda\partial^\lambda y^{\{L\}}_A\Big], \eqno(H12)$$
{}
$${\overline\rho}  \sum_{l\ge3}^k\partial^2_{00} 
{Q^\alpha_A}_{\{L\}}(y^0_A)\cdot y^{\{L\}}_A + 
2{\overline\rho} v_\mu \sum_{l\ge3}^k\partial_0 
{Q^\alpha_A}_{\{L\}}(y^0_A)\cdot \partial^\mu y^{\{L\}}_A+$$
{}
$$+\partial_0({\overline\rho}v^\lambda)  \sum_{l\ge3}^k 
{Q^\alpha_A}_{\{L\}}(y^0_A)\cdot\partial_\lambda y^{\{L\}}_A=
{\partial\over \partial y^\lambda_A}\Big[{\overline\rho} v^\lambda  
\sum_{l\ge3}^k\partial_0 {Q^\alpha_A}_{\{L\}}(y^0_A)
\cdot y^{\{L\}}_A\Big]+$$
{}
$$+{\partial\over \partial y^0_A}\Big[{\overline\rho}  
\sum_{l\ge3}^k\partial_0 {Q^\alpha_A}_{\{L\}}(y^0_A)\cdot y^{\{L\}}_A+
{\overline\rho} v^\lambda \sum_{l\ge3}^k 
{Q^\alpha_A}_{\{L\}}(y^0_A)\cdot \partial_\lambda
y^{\{L\}}_A\Big],\eqno(H13) $$
{}
$${\overline\rho}  \partial^\alpha{\overline U}
\sum_{l\ge3}^k{Q^\mu_A}_{\{L\}}(y^0_A)\cdot \partial_\mu y^{\{L\}}_A- 
{\overline\rho} \partial^\mu {\overline U} \sum_{l\ge3}^k 
{Q^\mu_A}_{\{L\}}(y^0_A)\cdot \partial^\alpha y^{\{L\}}_A-$$
{}
$$-{1\over4\pi}\partial^\alpha {\overline U}  \sum_{l\ge3}^k   
{Q^\nu_A}_{\{L\}}(y^0_A)\Big[2\sum_B\partial^2_{\nu\lambda}U_B\cdot\partial^\lambda y^{\{L\}}_A+
\partial_\nu U_B\cdot\partial_\mu\partial^\mu  y^{\{L\}}_A\Big]=$$
{}
$$ ={1\over 4\pi}{\partial\over \partial y^\beta_A}
\Bigg(\sum_{l\ge3}^k{Q^\lambda_A}_{\{L\}}(y^0_A)
\Big[\partial^\beta {\overline U}\partial^\alpha {\overline
U}\cdot\partial^\lambda y^{\{L\}}_A
- \Big(\partial^\beta {\overline U}\cdot\partial^\alpha y^{\{L\}}_A
+\partial^\alpha {\overline U} \cdot\partial^\beta y^{\{L\}}_A\Big)
\partial_\lambda  {\overline U}\Big]+$$
{}
$$+\gamma^{\alpha\beta}\partial_\lambda  
{\overline U}\sum_{l\ge3}^k\partial_\mu y^{\{L\}}_A\Big(
\partial^\mu {\overline U}{Q^\lambda_A}_{\{L\}}(y^0_A)-
{1\over2}\partial^\lambda{\overline U}{Q^\mu_A}_{\{L\}}(y^0_A)\Big)+$$
{}
$$+\partial_\lambda {\overline
U}\sum_{l\ge3}^k{Q^\beta_A}_{\{L\}}(y^0_A)\Big(
{1\over2}\partial^\lambda{\overline U} \partial^\alpha y^{\{L\}}_A- 
\partial^\alpha {\overline U}\partial^\lambda
y^{\{L\}}_A\Big)\Bigg),\eqno(H14)$$ 
{}
$${\partial\over \partial y^0_A}\Big[{\overline\rho}  
\sum_{l\ge3}^k\partial_0 {Q^\alpha_A}_{\{L\}}(y^0_A)\cdot y^{\{L\}}_A+
{\overline\rho} v^\lambda \sum_{l\ge3}^k 
{Q^\alpha_A}_{\{L\}}(y^0_A)\cdot \partial_\lambda y^{\{L\}}_A\Big]+$$
{}
$$+{\partial\over \partial y^\lambda_A}\Big[{\overline\rho} v^\lambda 
\sum_{l\ge3}^k\partial_0 {Q^\alpha_A}_{\{L\}}(y^0_A)\cdot y^{\{L\}}_A+
({\overline\rho} v^\lambda v^\mu-\gamma^{\lambda\mu}p)
 \sum_{l\ge3}^k {Q^\alpha_A}_{\{L\}}(y^0_A)\cdot 
\partial_\lambda y^{\{L\}}_A\Big]=$$
{}
$$={\overline \rho}{d^2\over {dy^0_A}^2}\Big[\sum^k_{l\ge3}
{Q^\alpha_A}_{\{L\}}(y^0_A)y^{\{L\}}_A\Big]-
{\partial\over \partial y^\beta_A}\Big[p \sum_{l\ge3}^k 
{Q^\alpha_A}_{\{L\}}(y^0_A)\cdot \partial^\beta y^{\{L\}}_A\Big].
  \eqno(H15)$$ 

\vskip 14pt \noindent
{\Large\bf Appendix I:  Astrophysical Parameters Used in the Report.}
\vskip 11pt \noindent
In this Appendix we  present the astrophysical parameters used in the 
calculations of the gravitational effects for the {\it Mercury Orbiter}
mission 
in  Section {\small 7} of this report:
{}
$$\hbox{Solar radius}:  R_\odot  = 695\hskip 2pt  980 \hbox{ km},$$  
$$\hbox{Solar gravitational constant}:  \mu_\odot  = {G M_\odot\over
c^2}=1.4766 \hbox{ km},$$  
$$\hbox{Solar quadrupole coefficient (Brown {\it et al.}, 1989)}: 
J_{2\odot} = (1.7\pm0.17)\times 10^{-7},$$ 
$$\hbox{Solar rotation period}:  \tau_\odot =  25.36 \hbox{ days},$$ 
$$\hbox{Mercury's mean distance}:  a_M  = 0.3870984  \hbox{ AU}=
57.91\times 10^6 \hbox{ km},$$ 
$$\hbox{Mercury's radius}: R_M  = 2 \hskip 2pt 439  \hbox{ km},$$ 
$$\hbox{Mercury's gravitational constant}:  \mu_M = {G M_M\over c^2}=
1.695\times 10^{-7}\mu_\odot,$$
$$\hbox{Mercury's sidereal period}: T_M  = 0.241  \hbox{ yr} =87.96 \hbox{
days},$$ 
$$\hbox{Mercury's rotational period}: \tau_M  =  59.7 \hbox{ days },$$ 
$$\hbox{Eccentricity of Mercury's orbit}: e_M  =  0.20561421,$$ 
$$\hbox{Jupiter's gravitational constant}:  \mu_J = 9.547\times 10^{-
4}\mu_\odot,$$ 
$$\hbox{Jupiter's sidereal period}: T_J  = 11.865  \hbox{ yr},$$ 
$$\hbox{Astronomical Unit}: AU = 1.49597892(1)\times 10^{13}  \hbox{
cm}.$$   

\pagebreak\noindent
{\Large \bf Acknowledgments}
\vskip 11pt \noindent
The author wishes to thank   John D. Anderson, Ephraim L. Akim, Bruno
Bertotti,
Luc Blanchet, Thibault Damour, Victor I. Denissov, Francis C. W. Everitt,  
William M. Folkner, Mark Gross, Ronald W. Hellings, 
Juri F. Koljuka, Sergei M. Kopejkin, Theodore D. Moyer, Kenneth L.
Nordtvedt, Wei-Tou Ni, 
Ojars J. Sovers, Peter K. Silaev, E. Myles Standish, Jr., Kip S. Thorne, 
Robert F. C. Vessot, 
and Jim G. Williams  for  valuable discussions and  comments on this
paper. 
The author is especially   indebted to John D. Anderson for his  warm
hospitality at the JPL.
The author  also very grateful to Catherine J. Anderson  for  editing  
 the manuscript and  to Sergei V. Zinoukov for his  encouragement during
the 
preparation of this manuscript. 
This research was supported by the  National Research Council and was 
carried out at the Jet Propulsion Laboratory,  California Institute of
Technology, under a 
contract with National Aeronautic and Space Administration. 

\vskip 13pt
\noindent{\Large \bf References}
\def\ref{\vskip 5pt \par \hangindent 15pt\noindent} 
\vskip 2pt 
\ref Akim, Eh. L., Brumberg, V. A., Kislik, M. D., Koljuka, Ju. F., Krasinsky, G.
A., Pitjeva, E. V., 
     Shishov, V. A., Stepanianz, V. A., Sveshnikov, M. L., and Tikhonov, V. F.:
1986, 
     in: {\it  Relativity in Celestial Mechanics and Astrometry} (IAU
Symposium 114), 
     eds. J. Kovalevsky and V. A. Brumberg, Reidel: Dortrecht,  p.63.

\ref Allen, C. W.: 1985, {\it Astrophysical Quantities}. Third ed., London:
Athlone press.

\ref Anderson, J. D., Levy, G. S., and Renzetti, N. A.: 1986, in: {\it Relativity
in Celestial Mechanics and 
     Astrometry} (IAU Symposium 114), eds. J. Kovalevsky and V. A.
Brumberg,  Reidel: Dorthecht, p.329.
\ref Anderson, J. D.: 1989,  {\it Gravitational Experiments on a Solar Probe
     Mission:  Scientific Objectives and Technology Considerations.}
     In: {\it Relativistic Gravitational Experiments in Space.}, edited by
     R. W. Hellings, NASA.
\ref Anderson, J. D., Slade,  M. A.,  Jurgens, R. F., Lau, E. L.,  
     Newhall, X X and Standish, E. M., Jr.: 
     1991, {\it Proc. Astron. Soc. Australia,} {\bf 9}, p.324. 
\ref Anderson, J. D., Lau, E. L., Krisher, T. P., Discus, D. A., Rosenbaum, D. C.,
     Teplitz, V. L.:  1995, {\it ApJ.}, {\bf 448}, p.885.
\ref Anderson, J. D.,   Gross, M.,  Nordtvedt  K., 
     and Turyshev,  S. G.: 1996,  {\it ApJ}, {\bf 459}, p.365.
\ref Anderson, J. D.,   Turyshev, S. G.,   Asmar, S. W.,   Konopliv, A. S.,
     Krisher, T. P.,   Lau, E. L.,   Maleki, L.,   Prestage, J. D.,   Sjogren, W. L.
     and   Bird, M. K.: 1996,  To appear in {\it Planetary and Space Sciences}.

\ref Anderson, J. L. and DeCanio, T. C.: 1975, {\it GRG}, {\bf 6}, p.197.
\ref Anderson, J. L., Kates, R. E., Kegeles, L. S.,  Madonna, R. G.: 1982, 
    {\it Phys. Rev. D}{\bf25}, p.2038.

\ref Ashby, N and Bertotti, B.: 1984, {\it Phys. Rev. Lett.}, {\bf 52}, p.485.
\ref Ashby, N and Bertotti, B.: 1986, {\it Phys. Rev. D}{\bf34}, p.2246.
\ref Ashby, N., Shahid-Saless, B.: 1990, {\it Phys. Rev. D}{\bf42}, p.1118.
\ref Ashby, N., Bender, P. L. and Wahr, J. M.: 1995, private communication.

\ref Baierlein, R.: 1967, {\it Phys. Rev.}, {\bf 162}, p.1275.

\ref Bardas, B. {\it et al.}: 1989, in: {\it Proc. 5th Marcel Grossmann
Meeting on General Relativity}.
     Eds. D. G. Blair and M. J. Buckingham (Singapore: World Scientific), part
B, p.1633.

\ref Barker, B. M., and  O'Connel R. F.: 1975, {\it Phys. Rev. D}{\bf12}, p.329.

\ref Berkin,  A. L., and  Hellings, R. W.: 1994 {\it Phys. Rev. D}{\bf 49},
p.6442.  

\ref Bertotti, B.: 1954, {\it  Nuovo Cimento}, {\bf 12}, p.226.
\ref Bertotti, B.: 1986, in :{\it  Relativity in Celestial Mechanics and 
     Astrometry} (IAU Symposium 114), 
     eds. J. Kovalevsky and V. A. Brumberg,  Reidel: Dorthecht, p.233.
\ref Bertotti, B., Ciufolini, I.,  and Bender, P.: 1987, {\it Phys. Rev.
D}{\bf58}, p.1062. 
\ref Bertotti, B. and Grishchuk L. P.: 1990, {\it Class. Quant. Grav.,} {\bf 7},
p.1733. 

\ref Bogoljubov, N. N., Shirkov, D. V.: 1984, {\it Introduction to Quantum
Field Theory},
     2nd edition.  Nauka: Moscow (in Russian).

\ref Border, J. S.  {\it et al.}: 1982, in: {\it Proc. of the 1982 AIAA/AAS
Astrodynamics Conf.},
     San Diego, CA, paper 82-1471. 

\ref Bini, D., Carini, P., Jantzen, R. T., Wilkins, D.: 1994,  
        {\it Phys. Rev. D.}{\bf49}, p.2820.

\ref Blanchet, L. and Damour, T.:  1986, {\it Phil. Trans. Roy. Soc. London}
{\bf A320}, p.379.
\ref Blanchet, L. and Damour, T.: 1989, {\it Ann. Inst. Henri Poincar\'e},
{\bf 50}, p.77.
\ref Blanchet, L., Damour, T., Iyer, B. R., Will, C. M. and Wisemann, A. G.:
1995, 
    {\it Phys. Rev. Lett.}, {\bf 74},  p.3515.

\ref Braginsky, V. B. and Panov, V. I.:  1972, {\it Zh. Eksp. Teor. Fiz.} {\bf
61}, p.873. 
    (1972, {\it Sov. Phys.   JETP }{\bf 34}, p.463.)
\ref Braginsky, V. B.: 1994, {\it Class. Quant. Grav.}, {\bf 11}, A1. 

\ref Brown, T.M., Christensen-Dalsgaard, J.,  Dziembowski, W.A.,
     Goode,  P., Gough,  D.O. and  Morrow, C.A.: 1989, {\it ApJ}, {\bf343},
p.526.

\ref Brumberg, V. A.: 1958, {\it Bull. Inst. Theor. Astron. Akad. Sci. USSR.},
{\bf 6}(10), p.773. 
\ref Brumberg, V. A.: 1972, {\it Relativistic Celestial Mechanics.} Nauka:
Moscow. (in Russian).
\ref Brumberg, V. A.: 1991a, {\it Essential Relativistic Celestial
Mechanics.} Hilger, Bristol.
\ref Brumberg, V. A.: 1991b, in: {\it Reference Systems}, the proceedings
of the 127th Colloquium of the IAU,
     Virginia Beach, 1990, eds. J. A. Hughes, C. A. Smith, G. H. Kaplan, p.36.
\ref Brumberg, V. A.: 1992,  {\it A{\rm\&}A}, {\bf257}, p.777.
\ref Brumberg, V. A., and  Kopejkin, S. M.: 1988a, {\it Nuovo Cimento B}{\bf
103}, p.63.
\ref Brumberg, V. A., and  Kopejkin, S. M.: 1988b, in: {\it Reference
Systems}, eds. J. Kovalevsky, 
     I. I. Muller and  B. Kolachek,  Reidel: Dortrecht, p.115.
\ref Brumberg, V. A., Bretagnon, P., and Francou, G.:  1993,  {\it
A{\rm\&}A},  {\bf 275}, p.651.

\ref Campbell, J. K. and Synnott, S. P.: 1985,  {\it A{\rm\&}A}, {\bf 90},
p.364.

\ref Ciufolini, I.: 1986, {\it Phys. Rev. Lett.}, {\bf56}, p.278.

\ref Chandler, J. F.,  Reasenberg, R. D. and  Shapiro I. I.: 1994, {\it
BAAS},{\bf26}, p.1019. 
\ref Chandler, J. F.,  Reasenberg, R. D: 1990, in {\it Inertial Coordinate
System on the Sky}, 
     eds. J. H. Lieske and V. K. Abalakin. Kluwer: Dortrecht, p. 217. 

\ref Chandrasekhar, S.: 1965, {\it ApJ.}, {\bf 142}, p.1488.
\ref Chandrasekhar, S., and  Contopulos, G.: 1967, {\it Proc. Roy. Soc.
London} {\bf A298}, p.123.

\ref Damour, T.: 1983, in: {\it Gravitational Radiation}, eds. N. Deruelle and
T. Piran (Les Houches 1982), p.59.
\ref Damour, T.: 1986, in: {\it Gravitation in Astrophysics}, eds. B. Carter
and J. B. Hartle 
     (Carg\`ese 1986),   Plenum Press, New York,  p.3. 
\ref Damour, T.: 1987, in: {\it Three Hundred Years of Gravitation}, eds. S.
W. Hawking and W. Israel,  
     Cambridge University Press, p.128.
\ref Damour, T., Gibbons,  G. W.,  and  Gundlach, G.: 1990,  {\it Phys. Rev.
Lett.}, {\bf 64}, p.123. 
\ref Damour, T. and Sch$\ddot{\rm a}$fer G.: 1991,  {\it Phys. Rev. Lett.},
{\bf 66},  p.2549. 
\ref Damour, T., Soffel, M., and Xu, C.: 1991,  {\it Phys. Rev. D}{\bf43}, 
p.3273.
\ref Damour, T.: 1992, {\it Class. Quant. Grav.}, {\bf 9},  S55. 
\ref Damour, T. and Esposito-Farese, G.: 1992, {\it Class. Quant. Grav.},
{\bf 9}, p.2093. 
\ref Damour, T., Soffel, M., and Xu, C.: 1992,  {\it Phys. Rev. D}{\bf45}, 
p.1017. 
\ref Damour, T. and  Taylor, J. H.: 1992,  {\it Phys. Rev. D}{\bf45}, p.1840.
\ref Damour, T., Soffel, M., and Xu, C.: 1993, {\it Phys. Rev. D}{\bf 47},
p.3436.
\ref Damour, T. and  Nordtvedt, K.: 1993, {\it Phys. Rev. D}{\bf48}, p.3436.  
\ref Damour, T., and  Polyakov, A. M.: 1994, {\it GRG}, {\bf26},  p.1171.
\ref Damour, T., Soffel, M., and Xu, C.: 1994,  {\it Phys. Rev. D}{\bf 49},
p.618.   
\ref Damour, T. and Vokrouhlick\'{y}, D.: 1995, {\it Phys. Rev. D}{\bf 52},
p.4455.
 
\ref D'Eath, P. D.: 1975a, {\it Phys. Rev. D}{\bf11}, p.1387.
\ref D'Eath, P. D.: 1975b, {\it Phys. Rev. D}{\bf12}, p.2183.
\ref D'Eath, P. D., Payne, P. N.: 1992, {\it Phys. Rev. D}{\bf46}, p.658.

\ref De Donder,  Th.: 1921, {\it La Grvifique Einshtenienne}. Gauthier-
Villars: Paris.
\ref De Donder,  Th.: 1926, {\it Theorie des Champs Gravifique}. Gauthier-
Villars: Paris.

\ref Denisov, V. I., and  Turyshev, V. G.: 1989, {\it Theor. Mat. Fiz.}, {\bf
81}, p.1222 (in Russian).
\ref Denisov, V. I., and  Turyshev, V. G.: 1990, {\it Theor. Mat. Fiz.}, {\bf
83}, p.429 (in Russian).
\ref Denisov, V. I., Zaitcev, A. L., Kopejkin, S. M., Laptev, N. V., Logunov, A.
A., and  Turyshev, V. G.:
     1989, {\it Gravitational Experiments in Space}, Centr. Mash. Constr.
Inst. TR 2889 (in Russian).

\ref De Sitter, W.: 1916, {\it Mon. Not. Roy. Astron. Soc.} 
     {\bf 76},  p.699.

\ref Dickey, J. O., Newhall, X X and Williams, J. G.: 1989, 
     {\it Adv. in Space Res.,} {\bf 9}, p.75.
\ref Dickey, J. O. {\it et  al.}: 1994, {\it Science}, {\bf 265}, p.482.

\ref Dirac, P. A. M.: 1937,  {\it Nature}, {\bf 139}, p.323.

\ref Dolgov, A. D. and  Khriplovich, I. B.: 1983, {\it GRG}, {\bf15}, p.1033.

\ref Duan, Y.,  Zhang, S., and  Jiang, L.: 1992, {\it GRG}, {\bf 24}, p.1033.

\ref Duvall, T.L., Jr. and  Harvey, J.W.: 1984,  {\it  Nature}, {\bf 310},  p.19.

\ref Einstein, A.: 1915a, {\it Preuss. Akad. Wiss. Berlin, Sitzber.,} p.778.
\ref Einstein, A.: 1915b,  {\it Preuss. Akad. Wiss. Berlin, Sitzber.,} p.799.
\ref Einstein, A., Infeld, L., and Hoffmann, B.: 1938, {\it Ann. Math.},
{\bf41}, p.65.

\ref Eisenhart, L. P.: 1926, {\it Riemannian geometry}. Princeton Univ.
Press, Princeton, N.J. 

\ref Ehlers, J., ed.: 1967, {\it Relativity Theory and Astrophysics. 3.
Stellar Structure.}
     Proceedings of the Summer Seminar, Ithaca, New York, 1965. 
     American Mathematical Society, Providence, Rhode Island.

\ref Fermi, E.: 1922a, {\it Atti. Accad. Naz. Lincei Rend. Cl. Sci. Fis. Mat.
Nat.} {\bf 31}, p.21.
\ref Fermi, E.: 1922b, {\it Atti. Accad. Naz. Lincei Rend. Cl. Sci. Fis. Mat.
Nat.} {\bf 31},  p.51.

\ref Fock,  V. A.: 1939, {\it Journ. of Phys.}, {\bf 1}, p.81. 
\ref Fock,  V. A.: 1955, {\it The Theory of Space, Time and Gravitation}.
Fizmatgiz. Moscow (in Russian).
     English translation: 1959,  Pergamon, Oxford.
\ref Fock,  V. A.: 1957, {\it Rev. Mod. Phys.}, {\bf 29}, p.325. 

\ref Folkner, W. M. {\it et al.}:  1994, {\it A{\rm\&}A}, {\bf 287}, p.279.  

\ref Fujimoto, M., and Grafarend, E.: 1986, in: {\it Relativity in Celestial
Mechanics
     and Astrometry.} ({\small IAU} Symposium 114), eds. J. Kovalevsky and
V. A. Brumberg, 
     Reidel: Dorthecht, p.269.

\ref Fukushima, T., Fujimoto, M. K., Kinoshita, H., and Aoki, S.: 1986, {\it
Celest. Mech.}, {\bf 36}, p.215.
\ref Fukushima, T.:  1988, {\it Celest. Mech.}, {\bf 44}, p.61.
\ref Fukushima, T.: 1991a,  {\it A{\rm\&}A}, {\bf 244}, L11.
\ref Fukushima, T.: 1991b, in: {\it Reference Systems}, the proceedings of
the 127th Colloquium of the IAU,
     Virginia Beach, 1990, eds. J. A. Hughes, C. A. Smith, G. H. Kaplan, p.27.
\ref Fukushima, T.: 1995a,  {\it A{\rm\&}A}, {\bf 294}, 895.
\ref Fukushima, T.: 1995b, {\it  A numerical scheme to integrate the
rotational motion of the rigid body}.
     Submitted to  {\it A{\rm\&}A}.  

\ref Guinot, B., and Seidelmann, P. K.: 1988, {\it A{\rm\&}A}, {\bf 194},
p.304.

\ref Goode, P.R. and Dziembowski, W.A.:  1991, {\it Nature}, {\bf349}, 
p.223. 

\ref Gladman, B. and Duncan, J.: 1990, {\it ApJ.}, {\bf 100}(5), p.1680.

\ref Habib, S., Holz, D. E., Kheyfets, A., Matzner, R. A., Miller, W. A., 
     Tolman, B. W.: 1994, {\it Phys. Rev. D} {\bf 50}, p.6068.

\ref Hawking, S. W., and Israel, W., eds.: 1987, {\it Three Hundred  Years of
Gravitation}.  
     Cambridge  Univ. Press, Cambridge. 

\ref Hellings, R. W.: 1986, {\it AJ}, {\bf 91}, p.650.

\ref Heimberger, J., Soffel, M. H., and Ruder, H.: 1990, 
     {\it Celest. Mech. and Dynam. Astron.},   {\bf 47}, p.205. 

\ref Herring, T. A.: 1995,  {\it Rev.Geophys.}, {\bf 33}, p.345.
\ref Herring, T. A.: 1996, {\it Scient. Amer.}, {\bf 274}, p.44.

\ref Holman, M. J. and Wisdom, M.: 1993, {\it ApJ.}, {\bf 105}(5), p.1987.

\ref Horowitz, G. T.,   Myers, R.: 1995, {\it GRG} {\bf 27}, p.915.

\ref Hilbert, D.: 1915, {\it Konigl. Gesell. d. Wiss. G$\ddot{\rm o}$ttingen,
Nachr. Math.-Phys. Kl.,} p.395.

\ref Huang, C.,  Ries, J. C., Tapley,  B. D. and  Watkins, M. M.: 1990, {\it
Celest. Mech. and Dynam. 
     Astron.}, {\bf 48}: p.167. 

\ref Ivanitskaja, O. S.: 1979, {\it Lorentz Basis and Gravitational 
     Effects in Einsteinian Theory of Gravitation.} Nauka and Tekhnika,
Minsk (in Russian).

\ref Jacobs, C. S. {\it et al.}: 1993, {\it Adv. in Space Res.}, {\bf 13}, p.161.

\ref Kates, R. E.:  1980a, {\it Phys. Rev. D}{\bf22}, p.1853.
\ref Kates, R. E.:  1980b, {\it Phys. Rev. D}{\bf22}, p.1871.
\ref Kates, R. E., Madonna, R. G.: 1982, {\it Phys. Rev. D}{\bf25}, p.2499.

\ref Kopejkin, S. M.: 1985, {\it Astron. Zh.}, {\bf 62}, p.889 (in Russian). 
\ref Kopejkin, S. M.: 1987, {\it Trans. Sternberg State Astron. Inst.},
{\bf59}, p.53 (in Russian). 
\ref Kopejkin, S. M.: 1988, {\it Celest. Mech.}, {\bf 42}, p.87. 

\ref Konopliv A. S., and   Sjogren, W. L.: 1994,  {\it Icarus}, {\bf 112},  p.42.
\ref Konopliv A. S., Lawson, C. L.,  Sjogren, W. L.: 1995, to be published in: 
     {\it Proc. of  XXI General Assembly of IUGG}, Boulder, CO, July 2-14,
1995.

\ref Krisher, T. P.: 1993, {\it Phys. Rev. D} {\bf48}, p.4639.
\ref Krisher, T. P., Morabito,  D. D., and  Anderson, J. D.: 1993,
     {\it Phys. Rev. Lett.}, {\bf 70}, p.2213.
 
\ref Klioner, S. A. and Kopejkin, S. M.: 1992, {\it AJ}, {\bf104 },  p.897.
\ref Klioner, S. A.: 1993,  {\it A{\rm\&}A}, {\bf 279}, p.273.
\ref Klioner, S. A. and Voinov, A. V.: 1993, {\it Phys. Rev. D}{\bf48}, 
p.1451.

\ref Landau, L. D. and Lifshitz, E. M.: 1988, {\it The Classical Theory of
Fields}. 
     7-th edition. Nauka: Moscow (in Russian).

\ref Lee, D. L., Lightmann, A. P., and Ni, W.-T.: 1974, {\it Phys. Rev. D}{\bf
10}, p.1685.

\ref Lebach, D. E.,  {\it et al.}: 1995, {\it Phys. Rev. Lett.}, {\bf75}(8),
p.1439. 

\ref Lestrade, J. F. and Chapront-Touze', M.: 1982,  {\it A{\rm\&}A}, {\bf
116}, p.75. 

\ref Libbrecht, K. G. and  Woodard, M. F.: 1991, {\it Science}, {\bf253},
p.152.

\ref Li, W.-Q., and Ni W.-T.: 1978, {\it Chinese Journ. of Phys.}, {\bf 16},
p.214.
\ref Li, W.-Q., and Ni W.-T.: 1979a, {\it J. Math. Phys.}, {\bf 20}, p.1473.
\ref Li, W.-Q., and Ni W.-T.: 1979b, {\it J. Math. Phys.}, {\bf 20}, p.1925.

\ref Logunov, A. A.: 1987, {\it The Lectures on the theory of relativity and
gravitation:
     The modern analysis of the problem.}, Moscow, Nauka (in Russian).

\ref LoPresto, J. C., Schrader,  C., and Pierce, A. K.: 1991,
     {\it ApJ}, {\bf 376}, p.757.
 
\ref Lorentz, H. A.,   Einstein, A.,   Minkowski, H., and  Weyl, H.: 1923, 
     {\it The Principle of Relativity: A Collection of Original Memoirs},
     Methuen, London. Paperback reprint, Dover, New York.

\ref Manasse, F. K., and Misner, C. W.: 1963, {\it J. Math. Phys.}, {\bf 4},
p.735.
\ref Manasse, F. K.: 1963, {\it J. Math. Phys.}, {\bf 4}, p.746.

\ref Martin, C. F., Torrence, M. H., and Misner, C. W.: 1985, {\it J. of
Geophys. Res.}, {\bf 90}, p.9403.

\ref Marzlin, K. P.: 1994, {\it Phys. Rev. D}{\bf 50}, p.888. 

\ref Minkowski, H.: 1908, Address Delivered at the 80-th Assembly of
German Natural Scientists and Physicians,
     Cologne. English translation, {\it Space and time}, in {\it The Principle
of Relativity},
     Dover, New York, 1923.

\ref Misner, C. W., Thorne,  K. S., and Wheeler, J. A.: 1973, {\it Gravitation}. 
     San Francisco: W. H. Freeman \& Co.

\ref Moyer, T. D.: 1971, in {\it Mathematical Formulation of the Double
Precision Orbit Determination Program
     (DPODP)}, Jet Propulsion Laboratory  Technical Report 32-1527,
Pasadena, CA.
\ref Moyer, T. D.: 1981, {\it Celest. Mech.}, {\bf 23}, pp.33-57.

\ref Nerem, R. S., Jeleki, C., Kaula, W. M.: 1995, {\it J. of Geophys. Res.},
{\bf 100}, p.15053.

\ref Newhall, X X., Standish, E. M. and Williams, J. G.: 1983,  {\it
A{\rm\&}A}, {\bf 125},  p.150.

\ref Ni, W.-T.: 1977, {\it Chinese Journ. of Phys.}, {\bf 15}, p.51.
\ref Ni, W.-T.,  and Zimmermann, M.: 1978, {\it  Phys. Rev. D}{\bf17},
p.1473.

\ref Nobili, A., and Will,  C. M.: 1986,  {\it Nature}, {\bf 320}, p.39.   

\ref Nordtvedt, K., Jr.: 1968a, {\it Phys. Rev. D}{\bf169}, p.1014.
\ref Nordtvedt, K., Jr.: 1968b, {\it Phys. Rev. D}{\bf 169}, p.1017.
\ref Nordtvedt, K., Jr.: 1968c, {\it Phys. Rev. D}{\bf 170}, p.1186.
\ref Nordtvedt, K., Jr.: 1994,  {\it ApJ.}, {\bf 437}, p.529.
\ref Nordtvedt, K., Jr.: 1995,  {\it Icarus}, {\bf 114}, p.51.

\ref Papapetrou, A.: 1948, {\it Proc. Roy. Irish. Acad.} {\bf A52}, p.11.
\ref Papapetrou, A.: 1951, {\it Proc. Roy. Soc. London}  {\bf A209}, p.248.

\ref Pars, L. A.: 1965, {\it Analytical Dynamics}. John Wiley \& Sons, Inc.
New York.

\ref Pitjeva, E. V.: 1993,  {\it Celest. Mech. and Dynam. Astron.},
          {\bf 55}, p.313.        

\ref Poincar\'e,  H.: 1904, {\it Bull. des Sciences Math.}, {\bf28}, p.302.

\ref Prestage, J. D., Tjoelker, R. L., Dick, G. J., and Maleki, L.:
     1992, {\it J. of Mod. Optics}, {\bf 39}, p.221.

\ref Rapp, R. H., Wang, Y. M., and Pavlis, N. K.: 1991, {\it The Ohio State
geopotential and 
     sea surface topography harmonic coefficient models}, Rep.410, 
     Dep. of Geod. Sci. and Surv., Ohio State Univ., Columbus.

\ref Ries, J. C., Huang, C., Watkins, M. M., and Tapley, B. D.: 1991, 
    {\it J. of the Astronaut. Sci.},  {\bf 39}, p.173.

\ref Robertson, H. P. and Noonan, T. W.: 1968, 
    {\it Relativity and Cosmology}. Saunders, Philadelphia.

\ref Roll, P. G., Krotkov, R., and Dicke, R. H.: 1964, 
    {\it Ann. Phys. (N.Y.)}, {\bf 26}, p.442. 

\ref Sard, R. D.: 1970, {\it Relativistic Mechanics}. W. A. Benjamin, Inc.
New York.

\ref Soffel, M., Ruder, H. and Schneider, M.: 1986,  {\it A{\rm\&}A}, {\bf
157}, p.357.  
\ref Soffel, M. H.: 1989, {\it Relativity in Astrometry, Celestial Mechanics
and Geodesy}. Berlin: Springer-Verlag.
\ref Soffel, M., M$\ddot{\rm u}$ller, J., Wu, X., and Xu, C.:  1991,  in: {\it
Reference Systems}, 
     the proceedings of the 127th Colloquium of the IAU,
     Virginia Beach, 1990, eds. J. A. Hughes, C. A. Smith, G. H. Kaplan, p.351.  
\ref Soffel, M. H. and Brumberg, V. A.: 1991, {\it Celestial Mech. and Dynam.
Astron.}, {\bf 52}, p.355. 

\ref Sovers, O. J., Jacobs, C. S.: 1994, in {\it Observational Model and
Parameter 
     Partials for the JPL VLBI Parameter Estimation Software "MODEST" -
1994}, 
     Jet Propulsion Laboratory  Technical Report 83-39, Rev. 5, Pasadena,
CA.

\ref Seidelmann, P. K., and Fukushima, T.: 1992,  {\it A{\rm\&}A}, {\bf
265}, p.833.

\ref Shapiro, I. I. {\it et al.}: 1976, {\it Phys. Rev. Lett.}, {\bf 36}, p.555.
\ref Shapiro, I. I., Reasenberg, R. D., Chandler, J. F., Babcock, R. W.: 1988,
     {\it Phys. Rev. Lett.}, {\bf 61}, p.2643.

\ref Shapiro, S. L., and Teukolsky, S. A.: 1986a, {\it ApJ.}, {\bf 307}, p.575.
\ref Shapiro, S. L., and Teukolsky, S. A.: 1986b, in: 
     {\it Dynamical Spacetimes and Numerical Relativity}, ed. J. M.
Centrella. 
     Cambridge University Press: Cambridge. 

\ref Shahid-Salees, B., Hellings R. W., and Ashby, N.: 1991, {\it Geophys.
Res. Lett.}, 
    {\bf 18}, p.1139.
\ref Shahid-Salees, B.: 1992, {\it Phys. Rev. D}{\bf 46}, p.5404.

\ref Schubert, G., Limonadi, D., Anderson,  J. D., Campbell, J. K., 
     Giampieri, G.: 1994, {\it Icarus}, {\bf 111},  p.433.

\ref Standish, E. M. Jr., Keesey, M. S. W., and Newhall, X X: 1976, JPL
Development Ephemeris {\bf 96}, 
     Jet Propulsion Laboratory, TR 32-1603, p.30. 
\ref Standish, E. M. Jr. {\it et al.}: 1992, in {\it Exp. Suppl. 
     to the Astron. Almanac}, p.279,   Univ. Sci. Books, Mill Valley. 
\ref Standish, E. M. Jr.: 1994, {\it Icarus}, {\bf 108}, p.180.
\ref Standish, E. M. Jr.,  Newhall, X X, Williams, J. G., and Folkner, W. M.:
1995,
     {\it JPL Planetary and Lunar Ephemeris, DE403/LE403}, 
     Jet Propulsion Laboratory  IOM \# 314.10-127.
\ref Standish, E. M. Jr.: 1995,  {\it Astronomical and Astrophysical
Objectives of Sub-Milliarcsecond 
     Optical Astrometry. IAU-SYMP}, {\bf166}, eds. E. H\"{o}g and P. K.
Seidelmann. p.109.

\ref Suen, W. M.: 1986, {\it Phys. Rev. D}{\bf 34}, p.3617.

\ref Synge, J. L.: 1960, {\it Relativity: the General Theory.} 
     (Amsterdam: North-Holland).

\ref Tapley, B. D.: 1972. {\it  Statistical Orbit Determination Theory},  
     in: {\it NATO  Advanced Study Institute in Dynamical Astronomy,} 
     ed. B. D. Tapley and V. Szebehely.   D. Reidel, p.396.

\ref Thomas, L. H.: 1927,  {\it Philos. Mag.}, {\bf 3},  p.1.

\ref Thomas, J. B.: 1975, {\it AJ}, {\bf 80}, p.405.

\ref Thorne, K. S.: 1980, {\it Rev. Mod. Phys.}, {\bf 52}, p.299.  
\ref Thorne, K. S. and  Hartle, J. B.: 1985, {\it Phys. Rev. D}{\bf31}, p.1815. 
\ref Thorne, K. S., Price, R. H., and Macdonald, D. A., eds.: 1988, 
     {\it Black Holes: The Membrane Paradigm.} Yale University Press.
\ref Thorne, K. S.: 1989, {\it Gravitational Radiation: A New Window Onto
the Universe}.
     Cambridge University Press: Cambridge.

\ref Turyshev, S. G.: 1990, {\it The Motion of  Extended Self-Gravitating
Bodies}, Ph.D. Thesis. 
     Moscow State University, Moscow, Russia  (unpublished).
\ref Turyshev, S. G.: 1994,  
     in: {\it Proc. of the {\small VII}-th Marcel Grossmann Meeting on
General Relativity}, 
     Stanford (to be published).
\ref Turyshev, S. G.:  1996,  {\it Canad. Journ. of Phys.}, {\bf 74}, p.17.
\ref Turyshev, S. G. {\it et al.}:  1996,  {\it To be published}.

\ref Voinov, A. V.: 1990, {\it Manuscripta Geodaetica}, {\bf15}, p.65.

\ref Vincent, M. A.: 1986,  {\it Celest. Mech.}, {\bf 39}, p.15.

\ref Vessot, R. F. C.  {\it et al.}: 1980, {\it Phys. Rev. Lett.},
     {\bf 45}, p.2081.
\ref Walsworth, R. L., Mattison, E. M.,  Vessot, R. F. C., 
     Silvera, I. F.: 1994, {\it Physica B}, {\bf 194}. p.915.

\ref Will, C. M.: 1971, {\it ApJ.}, {\bf 169}, p.125.
\ref Will, C. M. and Nordtvedt, K., Jr.: 1972, {\it ApJ.}, {\bf 177}, p.757.
\ref Will, C. M.: 1993, {\it Theory and Experiment in Gravitational
Physics}, 
     (Rev. Ed.), Cambridge  Univ. Press, Cambridge, England.

\ref Williams, J. G. {\it et al.}: 1976, {\it  Phys. Rev. Lett.}, {\bf 36}, p.551.
\ref Williams, J. G., Dickey, J. O., Newhall X X, and Standish, E. M: 1991, 
     in: {\it Reference Systems}, the proceedings of the 127th Colloquium of
the IAU,
     Virginia Beach, 1990, eds. J. A. Hughes, C. A. Smith, G. H. Kaplan, p.146. 
\ref Williams, J. G., Newhall X X, and Dickey, J. O.: 1996, 
     To be published in the {\it Phys. Rev.  D}.

\ref Zel'manov, A. L.: 1956, {\it Dokl. Acad. Sci. USSR}, {\bf 107}, p.815. 

\ref Zhang, X. H.: 1985, {\it Phys. Rev. D}{\bf 31}, p.3130. 
\ref Zhang, X. H.: 1986, {\it Phys. Rev. D}{\bf 34}, p.991.  
 
\end {document}